%% file: main.tex
\documentclass[twocolumn]{aastex701}

\usepackage{amsmath}
\usepackage{xspace}
\usepackage{array}
\usepackage{ulem}  
\usepackage{tabulary}
\usepackage{multirow}
\usepackage{hyperref}

\usepackage[utf8]{inputenc}
\usepackage{tabularx}
\usepackage{booktabs}

\newcommand{\obja}{J1025+1402}
\newcommand{\objb}{J1047+0739}
\newcommand{\objc}{J1022+0841}

\newcommand{\ha}{H$\alpha$}
\newcommand{\hb}{H$\beta$}
\newcommand{\hei}{\ion{He}{1} $\lambda$10833}
\newcommand{\caT}{Ca T}
\usepackage[encapsulated]{CJK}

\newcommand{\xj}[1]{{#1}}
\newcommand{\xjnew}[1]{{#1}}

\newcommand{\egg}{{\it The Egg}}

\received{XXX}
\revised{YYY}
\accepted{ZZZ}

\submitjournal{ApJ}

\shorttitle{Local LRDs}
\shortauthors{Lin et al.}

\graphicspath{{./}{figures/}}

\begin{document}

\title{The Discovery of Little Red Dots in the Local Universe: Signatures of Cool Gas Envelopes}

\input{00_Authorship}

\begin{abstract}
JWST observations have revealed a population of high-redshift ``little red dots" (LRDs) that challenge conventional AGN models.  
We report the discovery of three local LRDs at $z = 0.1$--$0.2$, initially selected from the SDSS database, with follow-up optical/near-IR spectroscopy and photometry. They exhibit properties fully consistent with those of high-redshift LRDs, including broad hydrogen and helium emission lines, compact morphologies, V-shaped UV-optical SED, declining near-IR continua, and no significant variability. Two sources were targeted but not detected in X-rays with statistical significance. All three sources show blue-shifted  \ion{He}{1} absorption, while two exhibit H$\alpha$ and Na D absorption lines. 
We detect full Balmer and Paschen line series in all three objects, along with abundant narrow [\ion{Fe}{2}] emission in two. The emission line analyses suggest narrow lines originate from AGN-powered, metal-poor regions with minimal dust; broad lines come from inner regions with exceptionally high density or atypical dust properties; and [\ion{Fe}{2}] emission arises from dense gas between broad and narrow-line regions.  
One of our objects, J1025+1402 (nicknamed \egg), shows extremely high equivalent width Na D, \ion{K}{1}, and \ion{Ca}{2} triplet absorption lines, along with other potential low-ionization absorption features,   suggesting the presence of a cool ($\sim$5000 K), metal-enriched gas envelope. The optical/near-IR continua of these LRDs are also consistent with theoretical models featuring an atmosphere around black holes. The WISE-detected IR emission is consistent with weak dust emission of $T \sim 10^2-10^3$ K.  We propose a conceptual model consisting of a largely thermalized cool-gas envelope surrounding the central black hole and an extended emission line region with high-density outflowing gas to explain the observed properties of these local LRDs.
\end{abstract}

\keywords{high-redshift --- active galactic nuclei --- black holes}

\input{01_Introduction}

\input{02_Selection}

\input{03_Observation}

\input{04_OverviewProperties}

\input{05_J1025}

\input{06_Summary}

\newpage
\begin{acknowledgments}
\input{99_Acknowledgement}

\end{acknowledgments}

\appendix
\input{99_Appendix}
\bigskip

\bibliography{main}{}
\bibliographystyle{aasjournalv7}

\end{document}

%% file: 00_Authorship.tex

\author[0000-0001-6052-4234]{Xiaojing Lin}
\affiliation{Department of Astronomy, Tsinghua University, Beijing 100084, China}
\affiliation{Steward Observatory, University of Arizona, 933 N Cherry Ave, Tucson, AZ 85721, USA}
\email[show]{xiaojinglin.astro@gmail.com}

\author[0000-0003-3310-0131]{Xiaohui Fan}
\affiliation{Steward Observatory, University of Arizona, 933 N Cherry Ave, Tucson, AZ 85721, USA}
\email[show]{xfan@arizona.edu}

\author[0000-0001-8467-6478]{Zheng Cai}
\affiliation{Department of Astronomy, Tsinghua University, Beijing 100084, China}
\email[]{zcai@tsinghua.edu.cn}


\author[0000-0002-1620-0897]{Fuyan Bian}
\affiliation{European Southern Observatory, Alonso de C\'ordova 3107, Casilla 19001, Vitacura, Santiago 19, Chile}
\affiliation{Chinese Academy of Sciences South America Center for Astronomy, National Astronomical Observatories, CAS, Beijing 100101, China}
\email[]{}

\author[0000-0003-2488-4667]{Hanpu Liu}
\affiliation{Department of Astrophysical Sciences, Princeton University, 4 Ivy Lane, Princeton, NJ 08544, USA}
\email[]{}

\author[0000-0002-4622-6617]{Fengwu Sun}
\affiliation{Center for Astrophysics $|$ Harvard \& Smithsonian, 60 Garden St., Cambridge, MA 02138, USA}
\email[]{}

\author[0000-0002-0463-9528]{Yilun Ma (\begin{CJK*}{UTF8}{gbsn}马逸伦\ignorespacesafterend\end{CJK*})}
\affiliation{Department of Astrophysical Sciences, Princeton University, 4 Ivy Lane, Princeton, NJ 08544, USA}
\email[]{}

\author[0000-0002-5612-3427]{Jenny E. Greene}
\affiliation{Department of Astrophysical Sciences, Princeton University, 4 Ivy Lane, Princeton, NJ 08544, USA}
\email[]{}

\author[0000-0002-0106-7755]{Michael A. Strauss}
\affiliation{Department of Astrophysical Sciences, Princeton University, 4 Ivy Lane, Princeton, NJ 08544, USA}
\email[]{}

\author[0000-0003-1245-5232]{Richard Green}
\affiliation{Steward Observatory, University of Arizona, 933 N Cherry Ave, Tucson, AZ 85721, USA}
\email[]{}

\author[0000-0002-6221-1829]{Jianwei Lyu}
\affiliation{Steward Observatory, University of Arizona, 933 N Cherry Ave, Tucson, AZ 85721, USA}
\email[]{}


\author[0000-0002-6184-9097]{Jaclyn B. Champagne}
\affiliation{Steward Observatory, University of Arizona, 933 N Cherry Ave, Tucson, AZ 85721, USA}
\email[]{}

\author[0000-0003-4700-663X]{Andy D. Goulding}
\affiliation{Department of Astrophysical Sciences, Princeton University, 4 Ivy Lane, Princeton, NJ 08544, USA}
\email[]{}

\author[0000-0001-9840-4959]{Kohei Inayoshi}
\affiliation{Kavli Institute for Astronomy and Astrophysics, Peking University, Beijing 100871, China}
\email[]{}

\author[0000-0002-5768-738X]{Xiangyu Jin}
\affiliation{Steward Observatory, University of Arizona, 933 N Cherry Ave, Tucson, AZ 85721, USA}
\email[]{}

\author[0000-0002-9393-6507]{Gene C. K. Leung}
\affiliation{MIT Kavli Institute for Astrophysics and Space Research, 77 Massachusetts Ave., Cambridge, MA 02139, USA}
\email[]{}

\author[0000-0001-6251-649X]{Mingyu Li}
\affiliation{Department of Astronomy, Tsinghua University, Beijing 100084, China}
\email[]{}

\author[0000-0003-3762-7344]{Weizhe Liu \begin{CJK}{UTF8}{gbsn}(刘伟哲)\end{CJK}}
\affiliation{Steward Observatory, University of Arizona, 933 N. Cherry Ave., Tucson, AZ 85721, USA}
\email[]{}

\author[0000-0003-4247-0169]{Yichen Liu}
\affiliation{Steward Observatory, University of Arizona, 933 N Cherry Ave, Tucson, AZ 85721, USA}
\email{yichenliu@arizona.edu}  
\email[]{}

\author[0000-0001-7557-9713]{Junjie Mao}
\affiliation{Department of Astronomy, Tsinghua University, Beijing 100084, China}
\email[]{}

\author[0000-0003-4924-5941]{Maria Anne Pudoka}
\affiliation{Steward Observatory, University of Arizona, 933 N Cherry Ave, Tucson, AZ 85721, USA}
\email[]{}

\author[0000-0003-0747-1780]{Wei Leong Tee
}
\affiliation{Steward Observatory, University of Arizona, 933 N Cherry Ave, Tucson, AZ 85721, USA}
\email[]{}

\author[0000-0003-4877-1659]{Ben Wang}
\affiliation{Department of Astronomy, Tsinghua University, Beijing 100084, China}
\affiliation{Leiden Observatory, Leiden University, Leiden 2333 CA, Netherland}
\email[]{}

\author[0000-0002-7633-431X]{Feige Wang}
\affiliation{Department of Astronomy, University of Michigan, 1085 S. University Ave., Ann Arbor, MI 48109, USA}
\email[]{}

\author[0000-0003-0111-8249]{Yunjing Wu}
\affiliation{Department of Astronomy, Tsinghua University, Beijing 100084, China}
\email[]{}

\author[0000-0001-5287-4242]{Jinyi Yang}
\affiliation{Department of Astronomy, University of Michigan, 1085 S. University Ave., Ann Arbor, MI 48109, USA}
\email[]{}

\author[0000-0002-4321-3538]{Haowen Zhang}
\affiliation{Steward Observatory, University of Arizona, 933 N Cherry Ave, Tucson, AZ 85721, USA}
\email[]{}

\author[0000-0003-3307-7525]{Yongda Zhu}
\affiliation{Steward Observatory, University of Arizona, 933 N Cherry Ave, Tucson, AZ 85721, USA}
\email[]{}


%% file: 01_Introduction.tex
\section{Introduction}

The so-called ``little red dots" (LRDs) have been among the most significant discoveries of the early years of JWST operations \citep[e.g.,][]{Labbe2025,  Matthee2024, Lin2024, Akins2024}. These sources appear compact even in JWST imaging, and the majority of them (over 70–80\%) exhibit broad Balmer emission lines with FWHM $\gtrsim 1000$\,km\,s$^{-1}$, indicative of black holes (BHs) of $10^{6}$--$10^8\,M_\odot$ \citep[e.g.,][]{Greene2024, Kocevski2024,Zhang2025}.  They  
display V-shaped spectral energy distributions (SEDs), characterized by a blue UV continuum and a red optical continuum \citep{Kocevski2024, Setton2024, Hviding2025}. Moreover, a large fraction of these V-shaped objects presents Balmer absorption on top of the broad emission line, which is rarely seen in other type-1 AGNs \citep[e.g.,][]{Matthee2024, Lin2024, Wang2025, DEugenio2025}.

Debates about the nature of LRDs are ongoing, particularly concerning the origin of their V-shaped SEDs, the spectral shape inflection near the Balmer break, and the prevalence of Balmer absorption \citep[e.g.,][]{Setton2024, Lin2024}.   Early studies have shown that if the Balmer break arises from massive host galaxies, it would challenge our current understanding of structure formation \citep[e.g.,][]{Wang2024, Labbe2024}. Alternatively, if these features are attributed to reddened AGNs, it is difficult to account for the lack of strong rest-frame IR detections for most high-redshift LRDs, by JWST/MIRI, Spitzer, or ALMA \citep[e.g.,][]{Perez-Gonzalez2024, Williams2024, Ma2024, Xiao2025, Setton2025}. 
New models have started to explore the distribution of dust \citep{Li2025, Chen2025}, along with new constraints on the dust content in LRDs from multi-wavelength data \citep{Casey2024, Casey2025}.  \xj{In addition to their weak infrared and sub-millimeter emission, the weak X-ray \citep{Yue2024, Ananna2024, Sacchi2025} and radio emission \citep{Mazzolari2024, Perger2025}, further set LRDs apart from typical type-1 AGNs \citep{Kokubo2024, Maiolino2025}.}

The unusual continuum shape of LRDs has motivated multiple theoretical models \citep{Inayoshi2025b, Kido2025}. Recent studies have proposed that BHs embedded in dense gas with a high covering fraction can explain the presence of both the observed Balmer breaks and absorption features \citep{Inayoshi2024}. Such models have been applied to several high-redshift LRDs and have achieved reasonably good fits to their spectra \citep{Ji2025, Taylor2025, deGraaff2025, Naidu2025}. However, for sources with rest-frame IR detections, these models struggle to simultaneously reproduce the weak IR photometry and the shape of Balmer breaks. They typically require either an extremely high column density of hydrogen gas (e.g., $N_{\rm H} = 10^{26}\,\mathrm{cm}^{-2}$), exotic modified dust attenuation laws, or intrinsically red AGN SEDs \citep{deGraaff2025}. Even with these additional free parameters, challenges remain for fully explaining the IR broadband photometry \citep[e.g.,][]{deGraaff2025, Naidu2025, Taylor2025}.


Currently, only a limited number of high-redshift LRDs have been detected with JWST/MIRI, while the vast majority remain either not covered by or undetected in the mid-IR \citep[e.g.,][]{ Leung2024, Setton2025, Lin2025, Taylor2025}. Most JWST/NIRSpec spectroscopy of LRDs at $z>4$ has been conducted in the prism mode, and high- or medium-resolution grating observations of several objects reveal only their brightest features. In contrast, LRDs at low redshifts can be observed with a variety of facilities/instruments at multiple wavelengths, with high spectral and spatial resolution at relatively low cost, providing deep insights into the nature of LRDs as a population. Rest-frame IR and high-resolution grating spectra have been obtained for a small number of luminous LRDs at cosmic noon \citep[e.g.,][]{Juodzbalis2024, Wang2025, deGraaff2025}. Systematic searches for LRDs at $z<4$ using wide-field sky surveys are now underway and have already demonstrated successful spectroscopic follow-up with ground-based facilities \citep{RLin2025, Ma2025, LRD_EuclidCollaboration2025}. However, photometrically selected LRDs at cosmic noon are still relatively faint ($m_g\sim 24-25$ mag), making follow-up observations challenging with current ground-based facilities.

Identifying LRDs in the local Universe opens a valuable window into understanding the nature of this population.  Although the measurements of the evolution of LRD number density across cosmic time vary across different photometric and spectroscopic selections, \xjnew{recent observations and models show a decline in the LRD number density toward low redshift.} \citep{Ma2025, LRD_EuclidCollaboration2025, Loiacono2025, Inayoshi2025a}. This implies that local LRDs are rare and are linked to physical conditions that may be more common in the early Universe. Their discovery at low redshift would offer critical insights into the evolutionary pathways of LRDs and their connection to BH growth across cosmic time. Moreover, local LRDs may enable spatially resolved observations of themselves, their host galaxies, and their environments, revealing their structure, kinematics, and dust/gas content. These observations would not only be cost-effective but also may unravel details that remain challenging to obtain from their high-redshift counterparts, even with JWST.

In this paper, we present the search and discovery of three local LRDs at $z$=0.1–0.2 from the SDSS spectroscopic database, followed by high signal-to-noise (S/N) ratio ground-based spectroscopic observations. The paper is organized as follows. In \S\ref{sec:selection} we introduce the selection criteria we used to search the SDSS spectral database for local LRDs. We present the SDSS data and follow-up observations on the three selected local LRDs in \S\ref{sec:observation}. In \S\ref{sec:overview_spectral_properties} we describe their overall properties and spectral features, which are fully consistent with those of JWST-discovered high-redshift LRDs. In \S\ref{sec:emission_line_analysis}, we analyze their emission line properties and infer the physical conditions. We then conduct a detailed case study of one object in \S\ref{sec:J1025_case_study}. Finally, in \S\ref{sec:cartoon} we interpret the observed features to understand their physical nature. Throughout this work, we adopt the AB magnitude system for all photometric measurements. All wavelengths and line names are given in the vacuum frame. All equivalent widths (EWs) are reported in the rest frame. A flat $\Lambda$CDM cosmology is assumed, with $\rm H_0 = 70~km~s^{-1}~Mpc^{-1} $, $\Omega_{\Lambda,0} = 0.7$, and  $\Omega_{m,0}=0.3$.

%% file: 02_Selection.tex
\section{Selection and sample from SDSS}\label{sec:selection}

In the literature, a wide range of selection criteria have been used to define the LRD population \citep[e.g.,][]{Hainline2025, Hviding2025}. 
In this paper, we adopt a narrow definition of LRDs, referring specifically to   
objects that simultaneously satisfy the following criteria: (1) presence of broad Balmer emission lines; (2) compact morphology in the rest-frame optical, particularly in photometric bands dominated by \ha\ emission; and (3) a V-shaped SED, characterized by a blue UV continuum slope and red optical continuum slope. 
A detailed description of the selection of local LRDs will be presented in an upcoming paper. Here, we provide a summary of the procedure. 

\subsection{Selection}
We first built a library of high-redshift LRDs with well-characterized UV-to-optical spectra. We adopted the compilation of high-redshift LRDs from \cite{Setton2024}, which exhibit clear V-shaped SEDs with inflection points near the Balmer limit. Their JWST/NIRSpec PRISM spectra were obtained from the DAWN JWST Archive \footnote{\url{https://dawn-cph.github.io/dja/}}.

We then searched for local LRD candidates in the SDSS DR17\footnote{\url{https://www.sdss4.org/dr17/}} spectroscopic database \citep{Abazajian2009, Abdurrouf2022}.  We started with the GUVmatch catalog \citep{GUVcat_DOI}, which cross-matches GALEX far-UV (FUV) and near-UV photometry (NUV) with SDSS DR14 \textit{ugriz} photometry.  We shifted the high-redshift LRD PRISM spectra to $z = 0.0$--0.5, serving as templates for our local LRD search. This redshift range was chosen to ensure that the \ha\ emission line falls within the SDSS spectroscopic wavelength coverage.   We then re-projected the template spectra to GALEX and SDSS photometry to obtain the expected broadband photometry of local LRDs.  Using these synthetic colors, we searched for candidates in GUVmatch with similar SED shapes. Specifically, we computed the Euclidean distance in magnitude space between the observed photometry and each individual template. This distance was defined as the root mean square of the magnitude differences across the GALEX and SDSS bands, normalized by the $r$-band flux. Then, we selected sources with a distance of less than one magnitude from any of the templates. Finally, we imposed a compactness criterion by requiring the SDSS-reported Petrosian radii \citep{Petrosian1976} in both the \textit{g} and \textit{r} bands to be smaller than 1.8 arcsec (physical scale of 3.3~kpc at $z\sim0.1$).  This morphological criterion was relatively permissive, and variations in the threshold do not affect the selection results.

After the photometric selection, we retrieved the SDSS spectra of the selected candidates. We examined their emission line properties using the \texttt{SPZLINE} extension, which provides automated emission-line fits from the SDSS pipeline \citep{Bolton2012}.  We first required the EWs of [\ion{O}{3}] $\lambda$5008 to exceed 10 \AA, and the line ratio [\ion{O}{3}] $\lambda$5008/[\ion{O}{2}] $\lambda$3730 (O32) to be greater than 10. The first criterion effectively excludes stellar objects, quiescent galaxies, and post-starburst galaxies, which could exhibit strong Balmer breaks and are prevalent in the local Universe. The second criterion ensures that the selected galaxies have physical conditions similar to those of high-redshift galaxies \citep[e.g.,][]{Shapley2023, Sanders2023, Shapley2025}.

To further remove contaminants and confirm the presence of red optical continua, we inspected the continuum underlying emission lines in the \texttt{SPZLINE} extension. We required the continuum under \ha\ to be redder than that under [\ion{O}{3}] $\lambda$5008, i.e., $\frac{f_{{\rm cont, H\alpha}}}{f_{{\rm cont, [OIII]}}}>1$ in $f_\lambda$ space. We required the [\ion{O}{3}] $\lambda$5008 continuum to be redder than that under [\ion{O}{2}] $\lambda$3730.

\xj{After obtaining the initial parent sample based on the automated pipeline products, we visually inspected the candidates to exclude objects affected by pipeline uncertainties. For example, we removed cases where the more precise re-measured [\ion{O}{2}] $\lambda3730$ flux remained clearly stronger than [\ion{O}{3}] $\lambda5008$, or where emission lines had low S/N and were dominated by noise. This process resulted in a parent sample of 40 objects.}

Although the broad-line \ha\ criterion, which is one of the defining characteristics of LRDs, was not explicitly applied up to this stage, all selected targets already show clear broad \ha\ emission in their SDSS spectra. \xj{We note, however, that not all objects in the parent broad-line sample satisfy the LRD definition.} \xjnew{Many sources in the parent sample show very strong [\ion{N}{2}] or \ion{Mg}{2} emission accompanied by relatively red optical continua, in which case the spectral shapes are more naturally explained by type-1 AGNs with evolved stellar populations. In the end, we select three objects as the most representative LRDs.}

\subsection{Sample from SDSS}

\begin{figure*}[ht!]
    \centering
    \includegraphics[width=\linewidth]{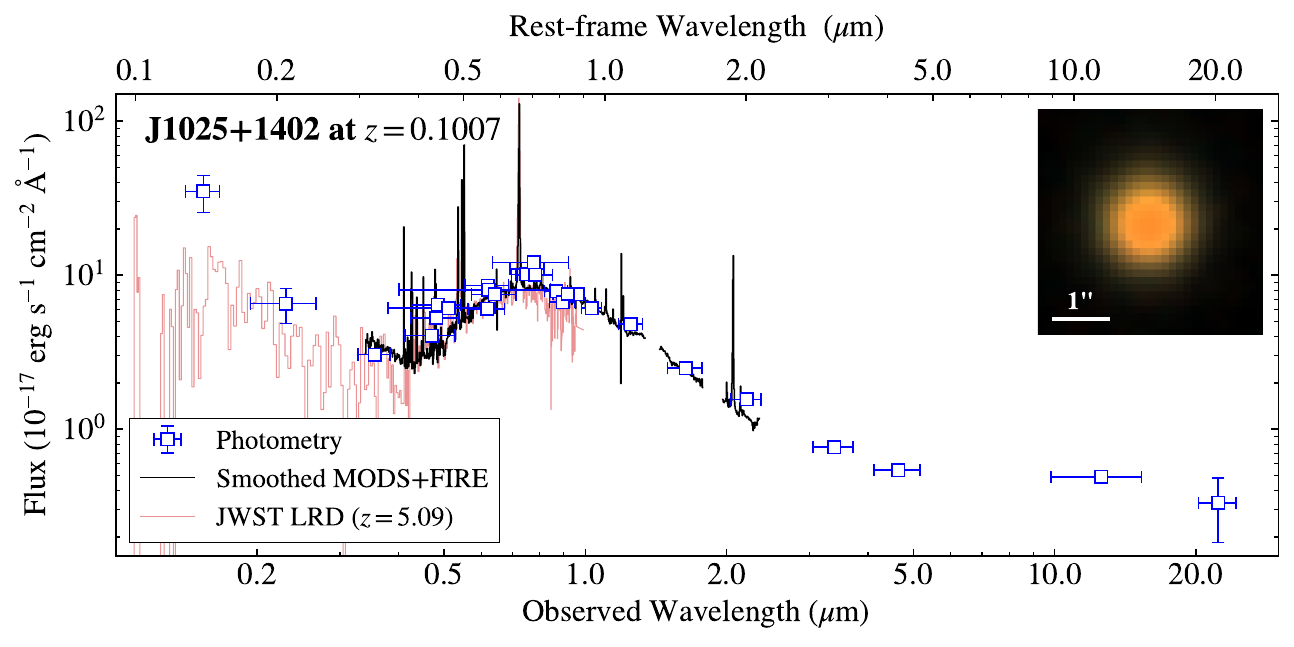}
     \includegraphics[width=0.495\linewidth]{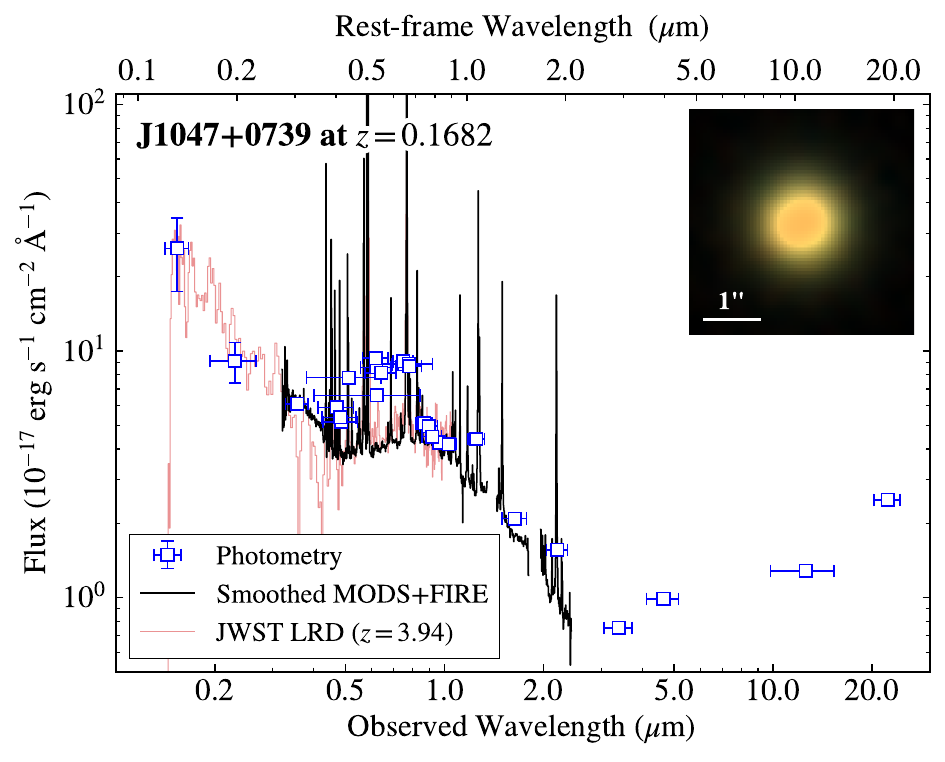}
      \includegraphics[width=0.495\linewidth]{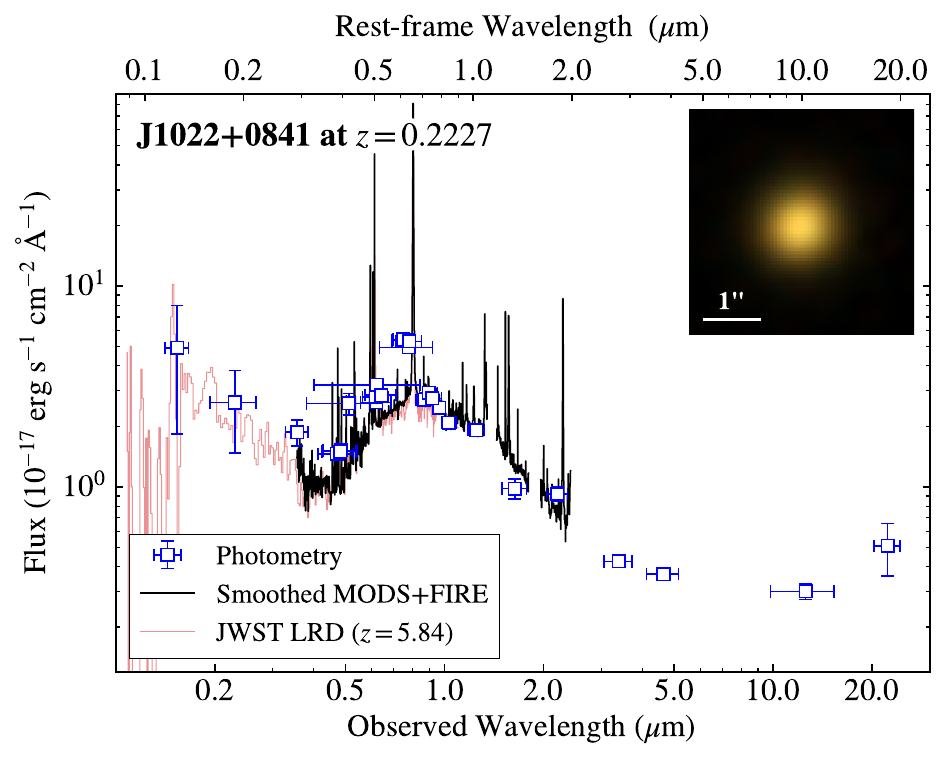}
    \caption{The images, multi-wavelength photometry, and LBT/MODS and Magellan/FIRE spectra of the three local LRDs.  The photometric data are shown as the blue squares. The LBT/MODS+ Magellan/FIRE spectra are shown as the black lines. The spectra are smoothed for display purposes only. The RGB thumbnails are composed of the $grz$ images from the Legacy Surveys DR10. The red lines show the best-matching LRD templates from the compilation by \cite{Setton2024}. These templates are JWST/NIRSpec PRISM spectra with their original redshifts labeled, but have been shifted to match the redshift of each local LRD.}
    \label{fig:SED}
\end{figure*}

In this paper, we report the discoveries and follow-up observations of three local LRDs: SDSS J102530.29+140207.3 (J1025+1402 hereafter), SDSS J104755.92+073951.2 (J1047+0739), and SDSS J102208.52+084156.1 (J1022+0841).  The full sample will be presented in a forthcoming paper. 

Figure \ref{fig:SED} shows the images, photometry, and spectra of the three selected local LRDs.  These objects closely match their corresponding LRD templates and exhibit unambiguous V-shaped UV-to-optical SEDs.

The local LRDs presented in this paper were originally selected as quasar candidates \citep{Richards2002}. The SDSS target selection flag for \obja\ and \objc\ is \texttt{QSO\_HIZ}, as their $riz$ colors meet the criteria for selecting high-redshift quasars, primarily at $z \gtrsim 3.5$. Both objects show rising flux toward and peaking in the $i$ band, followed by a decline toward the $z$ bands, resembling the spectral shape of a Lyman break with a blue UV continuum, characteristic of quasars at $z > 3.5$. \objb\ is flagged as both \texttt{QSO\_CAP} and \texttt{QSO\_HIZ}: the former because its $u-g$ color meets the selection criteria for quasars at $z < 2.2$, and the latter because it satisfies a relaxed $griz$ criterion for high-redshift quasars.  \xj{The misidentification} indicates that the three selected objects are outliers in both the $ugri$ and $griz$ color spaces, making them easily misclassified during color selection.

From the SDSS spectra, \obja\ shows clear blueshifted \ha\ absorption, while \objc\ exhibits tentative but unresolved \ha\ absorption. Both features are confirmed by our spectroscopic follow-up with higher spectral resolution (see \S\ref{sec:H_He_absorption} \xj{and Figure \ref{fig:H_He_lineprofile}}).  \obja\ and \objb\ were identified by \citet{Izotov2008} as metal-poor AGN in SDSS DR5, with 12 + $\log$(O/H) = 7.36 and 7.99, respectively. \citet{Simmonds2016} further noted that they are weak in X-rays, revealed by the non-detection of Chandra. \citet{Burke2021} found no significant variability in either optical broadband photometry or broad \ha\ luminosity.

\subsection{Number density of LRDs at $z<0.5$}

Our LRD selection at $z<0.5$ is incomplete. First of all, it requires that the object has an existing SDSS spectrum. \xj{The SDSS main galaxy survey is biased against LRDs because of their compact morphology, making them less likely to be prioritized for spectroscopic follow-up as galaxies. Their compactness is also one of the reasons that they are labeled as quasar candidates.} \xjnew{In addition}, although our [\ion{O}{3}] EW criteria effectively exclude stars, quiescent, and post-starburst galaxies at low redshifts, they may also omit potential LRD candidates with intrinsically weak [\ion{O}{3}] emission. Indeed, several high-redshift LRDs exhibit low [\ion{O}{3}] EWs, such as Abell2744-QSO1 (4.5 \AA; \citealt{Ji2025}), MoM-BH$^*$-1 (3 \AA; \citealt{Naidu2025}), and \textit{The Cliff} ($7.3$ \AA; \citealt{deGraaff2025}).

Therefore, we estimate a conservative lower limit on the number density of local LRDs at $z$=0--0.5, based on the effective SDSS survey area of 9,376 deg$^2$. The discovery of these three objects yields an estimated lower limit of approximately $5 \times 10^{-10}\,\mathrm{Mpc}^{-3}$ within this redshift range.   
Our estimated number density at $z$=0--0.5 is more than several orders of magnitude higher than the extrapolation of the log-normal distribution fitted to the LRD population at $z\sim$4--7 \citep{Inayoshi2025a}. However, this large discrepancy arises in the exponentially damped tail of the assumed distribution, where such divergence is expected due to the steep functional form and limited constraints at low redshifts. 
We find that a modest increase in the dispersion (by 10--20\%) of the log-normal model from \cite{Inayoshi2025a} yields a redshift evolution of the LRD abundance consistent not only with measurements at $z\approx$2--3 \citep{Ma2025}, but also with our observations at $z$=0--0.5.

\xj{The evolution of LRD number density across cosmic time remains uncertain, with different studies reporting varying trends. Photometric selection and spectroscopic follow-ups by \cite{Ma2025} indicate a substantial decline at $z<4$, with the number density decreasing by 1–2 dex from $z\sim4$ to $z\sim2$. In contrast, photometric selection in \cite{LRD_EuclidCollaboration2025} suggests a milder evolution, with LRD abundance rising to $z\approx 1.5-2.5$ before declining at lower redshift. \xjnew{Based on the discovery of three LRDs at $z\approx2-3$,} \cite{Loiacono2025} implies that LRD number density remains significant at cosmic noon, only 2–3 times lower than that of UV-selected quasars. Nevertheless, both \cite{Ma2025} and \cite{LRD_EuclidCollaboration2025} indicate a decreasing number density toward the local Universe. This trend, together with the discovery of the three local LRDs presented in this work, suggests that the BH accretion mode giving rise to LRDs, while rare in the local Universe, is not exclusively unique to the early Universe.}

%% file: 03_Observation.tex
\section{Data and Observations}\label{sec:observation}

\subsection{Photometry}
The three targets have been observed in multiple wide-field sky surveys. \xj{We compile photometric measurements from the following datasets, adopting PSF photometry where available since the targets are unresolved in the images}: FUV and NUV from GALEX DR6 \citep{Bianchi2020}; \textit{ugriz} PSF photometry from SDSS DR17 \citep{Abazajian2009, Abdurrouf2022};  $grizy$ PSF photometry from Pan-STARRS DR2 \citep{Flewelling2020}; $g,bp,rp$ mean photometry from GAIA DR3  \citep{Gaia2023}; $griz$ model photometry from Legacy Surveys DR10 \citep{Dey2019} where PSF models are used for the three objects;  $YJHK$ PSF photometry from UKIDSS DR11PLUS \citep{Lawrence2007}; and W1, W2, W3, W4 photometry from WISE  \citep{Wright2010}. The GALEX photometry is taken from the GUVmatch catalog. The WISE photometry has been incorporated into the Legacy Survey catalog based on the unWISE images \citep{Schlafly2019}. All other data are retrieved from Astro Data Lab\footnote{\url{https://datalab.noirlab.edu/}} \citep{ Fitzpatrick_SPIE, NIKUTTA2020100411}. 

\subsection{Spectroscopic Follow-up Observations and Data Reduction}

We obtained high S/N LBT/MODS and Magellan/FIRE spectroscopy to  
characterize the UV-to-near-IR spectra of the three local LRDs. We also obtained MMT/Binospec spectra of \obja\ and \objc\ to measure their high-resolution \ha\ profiles.

The LBT/MODS \citep{Pogge2010} is a pair of dual-channel spectrographs on the 2$\times$8.4\,m Large Binocular Telescope, with the blue channel covering 3300--6000\,\AA\ and the red channel 5000--10,000\,\AA.  We observed the three targets on February 24, 2025, using 1.0-arcsec slits in grating mode under an average seeing of $\sim$0.8 arcsec. This setup achieved a spectral resolution of $R\approx800$--1500 in the blue channel and $R\approx800$--1700 in the red channel. Observations of each target included 3$\times$600\,s exposures in both the blue and red channels, with the two LBT telescopes observing simultaneously. As a result, each target effectively has 6$\times$600\,s exposures in both the blue and red channels. 

The Magellan/FIRE \citep{Simcoe2013} is a near-IR echelle spectrograph covering 0.8--2.5\,$\micron$ on the 6.5\,m Magellan Baade Telescope. We observed \obja\ using a 0.75-arcsec slit on December 21, 2024, reaching $R=3000$--6000. The observation included 5$\times$900\,s exposures. We observed \objb\ and \objc\ using 1-arcsec slits on March 11, 2025, with 7$\times$900\,s exposures for \objb\ and 6$\times$900\,s exposures for \objc, reaching $R=2000$--4000. The seeing for these observations was $\sim0.5$--0.7 arcsec.

The MMT/Binospec observation of \obja\ was conducted in the IFU mode \citep{Fabricant2019, Fabricant2025} on December 7, 2024, using the 600 line/mm grating. The observation consisted of $3\times 600$\,s exposures and achieved a spectral resolution of $R \sim 5300$. The seeing during the observation was approximately 1.6 arcsec. For \objc, the observation was conducted in long-slit mode on January 23, 2024, using a 1-arcsec slit and the 600 line/mm grating, under a seeing of approximately 2.3 arcsec.
The observation included $6\times 600$\,s exposures, with a spectral resolution of $R \sim 4500$.  

The LBT/MODS and Magellan/FIRE data were reduced with \texttt{PypeIt} \citep{Prochaska2020}. We performed bias subtraction, flat-fielding, wavelength calibration, sky subtraction, spectral extraction, and telluric correction.  To correct for slit losses and calibrate the absolute flux of the LBT/MODS spectra, we first matched them to the SDSS spectra using a wavelength-dependent linear function. The spectra in SDSS DR17 have been processed with an updated flux calibration algorithm, yielding significant improvements at wavelengths below 6000\,\AA\footnote{\url{https://www.sdss4.org/dr17/algorithms/spectrophotometry/}}. We confirmed that the SDSS spectra closely follow the shape of the corresponding photometry. Then, we applied a constant scaling factor to align the semi-calibrated spectra with the SDSS $ugriz$ photometry. The Magellan/FIRE spectra were scaled to match the UKIDSS $Y$-band photometry using a constant factor. The MMT/Binospec data were reduced using the \texttt{Binospec} pipeline \citep{Kansky2019}. The IFU data were extracted with a rectangular aperture matching the seeing disk.  The Binospec spectra of \obja\ and \objc\ were flux-calibrated by scaling to match the calibrated LBT/MODS spectra after degrading them to the MODS resolution.
 
The SDSS spectra have a resolution of $R\sim1500$--2500. We convolve the SDSS spectra to match the spectral resolution of the MODS data. These SDSS spectra reach S/N$\approx$2.5 per \AA\ at 4000--5000 \AA\ and S/N$\approx$9 at 7000--8000 \AA\ for \obja. In stark contrast, the MODS-B and MODS-R spectra achieve S/N$\approx$13 and 81, respectively, improving the S/N by factors of 5 (blue) and 9 (red). For \objb, the S/N improves from 3.5 to 19 at 4000--5000 \AA, and from 6 to 60 at 7000--8000 \AA. For \objc, the S/N increases from 0.6 to 5.5 at 4000--5000 \AA, and from 2.6 to 29 at 7000--8000 \AA.

Figure \ref{fig:SED} shows the UV-to-near-IR spectra, multi-wavelength photometry, and images of the three targets. The LBT/MODS spectra confirm the presence of the V-shape of the UV-to-optical continua. The declining near-IR spectra in $f_\lambda$ space are consistent with the rest-frame IR photometry of bright high-redshift LRDs detected by MIRI \citep[e.g.,][]{Wang2025, deGraaff2025, Taylor2025, Setton2025}.

%% file: 04_OverviewProperties.tex
\section{Evidence as local LRDs}\label{sec:overview_spectral_properties}

\begin{table*}[t!]
\centering

\begin{tabular}{c|ccc}
\hline
Name & J102530.29+140207.3 & J104755.92+073951.2 & J102208.52+084156.1 \\
\hline
RA & 156.37622 & 161.98302 & 155.53552 \\
DEC & 14.03586 & 7.66423 & 8.69892 \\
$z$ & 0.1007 & 0.1682 & 0.2227 \\
$r$ (mag) & 19.003 & 18.846 & 20.011 \\
$\lambda L_{5100}$ ($10^{42}$ erg\,s$^{-1}$) & 7.57 $\pm$ 0.31 & 18.67 $\pm$ 0.53 & 17.78 $\pm$ 0.31 \\
$L_{2000\,\text{\AA}-20\,\mu\mathrm{m}}$ ($10^{43}$\,erg\,s$^{-1}$) & 4.95$^{+0.03}_{-0.02}$ & 29.93$\pm$0.01 & 17.72$^{+0.03}_{-0.04}$ \\
$L_{\rm 2-10keV}$ ($10^{40}$\,erg\,s$^{-1}$) & $<11$ & $8.1^{+3.8}_{-4.0}$ & -- \\
12 + $\log(\mathrm{O/H})$ & 7.58 $\pm$ 0.03 & 8.01 $\pm$ 0.02 & 7.43 $\pm$ 0.03 \\
$R_e$ (kpc) & $<0.85$ & 0.568 & $<1.6$ \\
$\beta_{\rm UV,\,2500\AA}$ &  $-1.37\pm0.01$ & $-0.97\pm0.02$ & $-1.57\pm0.02$ \\
$\beta_{\rm opt,\,4500\AA}$ & $3.22\pm0.01$ &  $0.31\pm0.01$ & $3.03\pm0.02$\\
Inflection point (\AA) & $3884\pm3$ &  $4168\pm4$ & $3704\pm2$ \\
\hline
\end{tabular}
\caption{Observed and derived properties of the three local LRDs. The systemic redshifts are determined by the center of the narrow [\ion{O}{3}] emission lines. The $r$-band magnitude is the observed magnitude from Legacy Survey DR10 \citep{Dey2019}.  The metallicities are derived using the direct $T_e$ method. The 2–10 keV luminosity of \obja\ is adopted from \cite{Simmonds2016}, \xj{and that of \objb\ is calculated using \textsc{WebPIMMS};} both assume a $\Gamma=1.8$ power-law spectrum absorbed by Galactic \ion{H}{1}. \xj{The half-light radii ($R_e$) of \obja\ and \objc\ are estimated from their unresolved morphologies in ground-based images, while $R_e$ of \objb\ is measured from its host galaxy in the HST image.} $\beta_{\rm UV,\,2500\AA}$ and $\beta_{\rm opt,\,4500\AA}$ denote the continuum slopes measured at rest-frame 2500\,\AA\ and 4500\,\AA, respectively. The inflection point corresponds to the transition where the continuum slope changes from negative to positive, marking the turnover of the UV–optical V-shaped continuum.}
\label{tab:basic_properties}
\end{table*}

\begin{table*}[htbp]
\centering
\begin{tabular}{c|ccccccc}
\hline
Name & $L_\mathrm{H\alpha, broad}$ & FWHM$^\star_\mathrm{H\alpha, broad}$ & $L_\mathrm{H\alpha, narrow}$ & FWHM$_\mathrm{H\alpha, narrow}$ & $\log M_\mathrm{BH}$ & $\log L_\mathrm{bol, H\alpha}$ $^b$ & $\lambda_\mathrm{Edd, H\alpha}$ $^b$ \\
 & ($10^{41}$ erg s$^{-1}$) & (km s$^{-1}$) & ($10^{41}$ erg s$^{-1}$) & (km s$^{-1}$) & ($M_\odot$) & (erg s$^{-1}$) &  \\
\hline
\obja & $8.20 \pm 0.24$ & $956 \pm 27$ & $1.98 \pm 0.02$ & $64 \pm 1$ & $6.52 \pm 0.03$ & $44.03 \pm 0.01$ & $0.21 \pm 0.01$ \\
\objb & $27.51 \pm 0.58$ & $952 \pm 25$ & $18.86 \pm 0.15$ & $54 \pm 1$ & $6.76 \pm 0.02$ & $44.55 \pm 0.01$ & $0.41 \pm 0.02$ \\
\objc & $51.99 \pm 1.28$ & $805 \pm 34$ & $2.84 \pm 0.72$ & $65 \pm 9$ & $6.74 \pm 0.04$ & $44.83 \pm 0.01$ & $0.81 \pm 0.07$ \\
\hline
\end{tabular}

\caption{\ha\ emission line and BH properties. FWHM$^\star_{\rm H\alpha,broad}$ refers to the effective FWHM  of the composite broad H$\alpha$ profile, modeled with multiple Gaussian components spanning intermediate ($\gtrsim 500$\,km\,s$^{-1}$) to very broad ($\sim 2000$\,km\,s$^{-1}$) line widths.   $^a$ \ha\ profile measurements for \obja\ and \objc\ are based on their Binospec spectra, with consistency confirmed against MODS-R spectra (see Appendix \ref{sec:appendix_emission_fitting}). $^b$ The bolometric luminosity ($L_{\rm bol,H\alpha}$) and Eddington ratio ($\lambda_{\rm Edd,H\alpha}$) are derived solely from $L_{\rm H\alpha,broad}$ and are provided for reference. However, we caution that these values should not be interpreted as the intrinsic bolometric luminosity and Eddington ratio.
}
\label{tab:Ha_properties}
\end{table*}

\begin{figure*}[htbp!]
    \includegraphics[width=\linewidth]{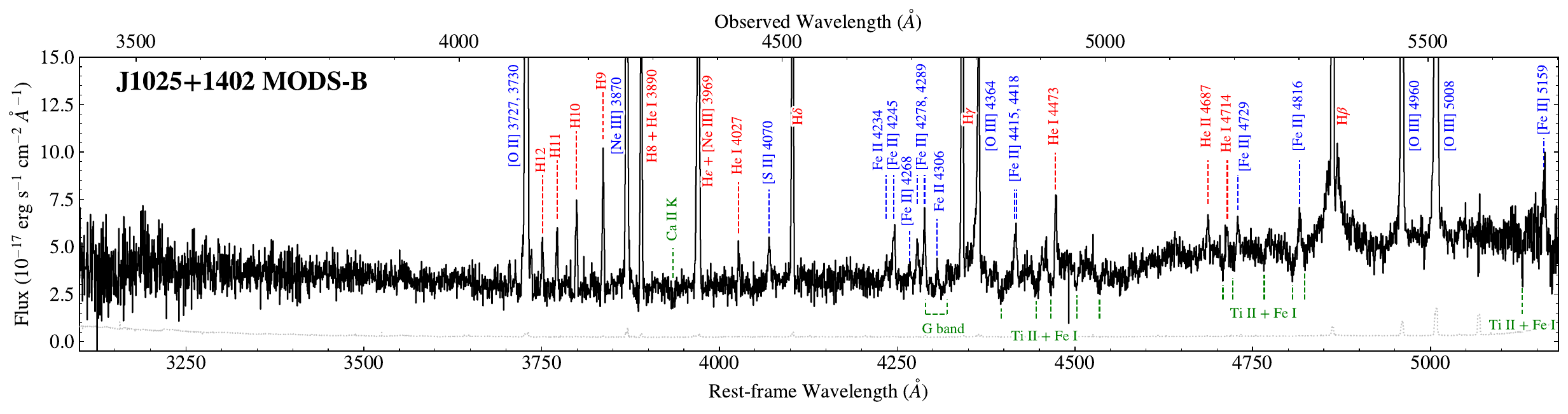}
    \includegraphics[width=\linewidth]{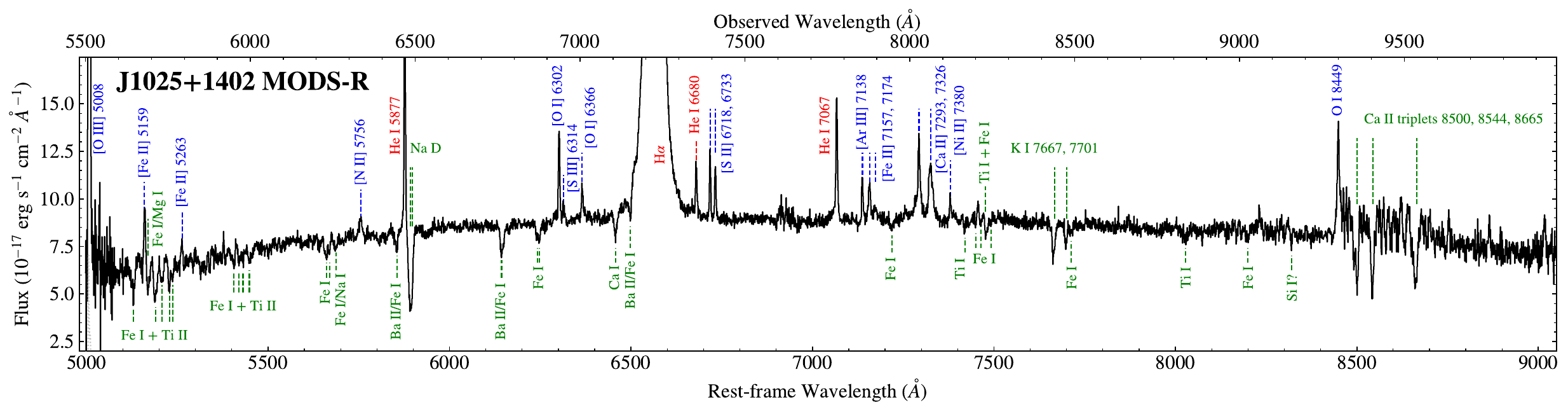}
    \includegraphics[width=\linewidth]{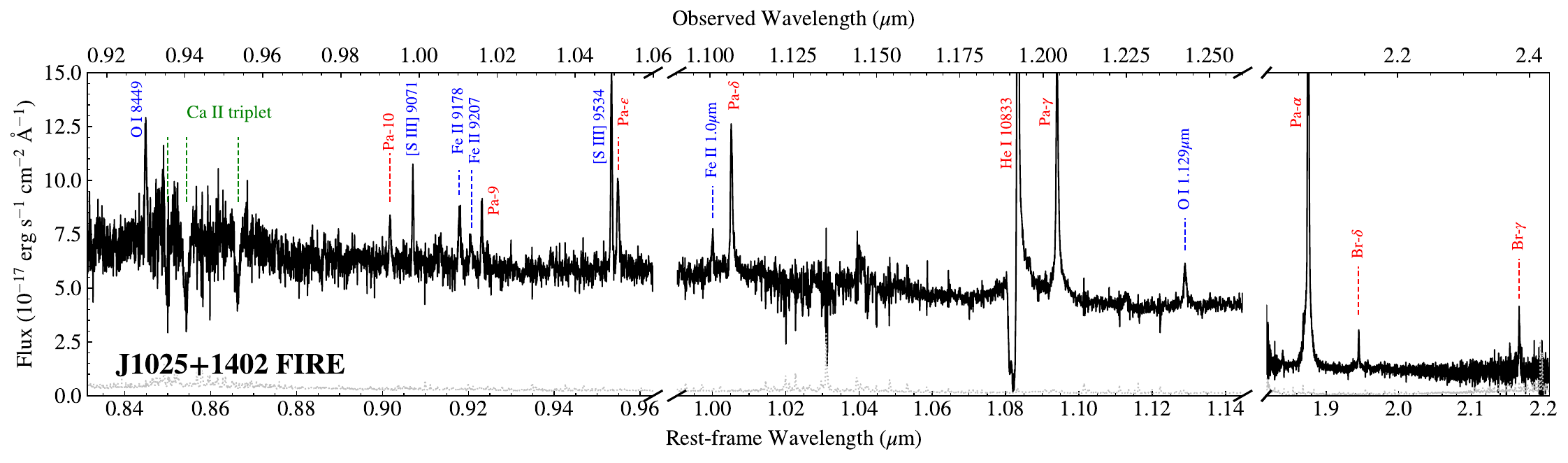}
    \caption{Zoomed-in view of the MODS blue channel (MODS-B), MODS red channel (MODS-R), and Magellan/FIRE spectra of \obja. Hydrogen and helium emission lines are marked with red dashed lines and labels, while other emission lines are indicated with blue dashed lines and labels. Absorption features are labeled as green dashed lines. The gray dotted lines represent the 1$\sigma$ flux uncertainties. In the MODS-R spectrum, these uncertainties are sufficiently low to fall below the plotted range. 
    }
    \label{fig:J1025_zoomin}
\end{figure*}

\begin{figure*}[htbp]
    \includegraphics[width=\linewidth]{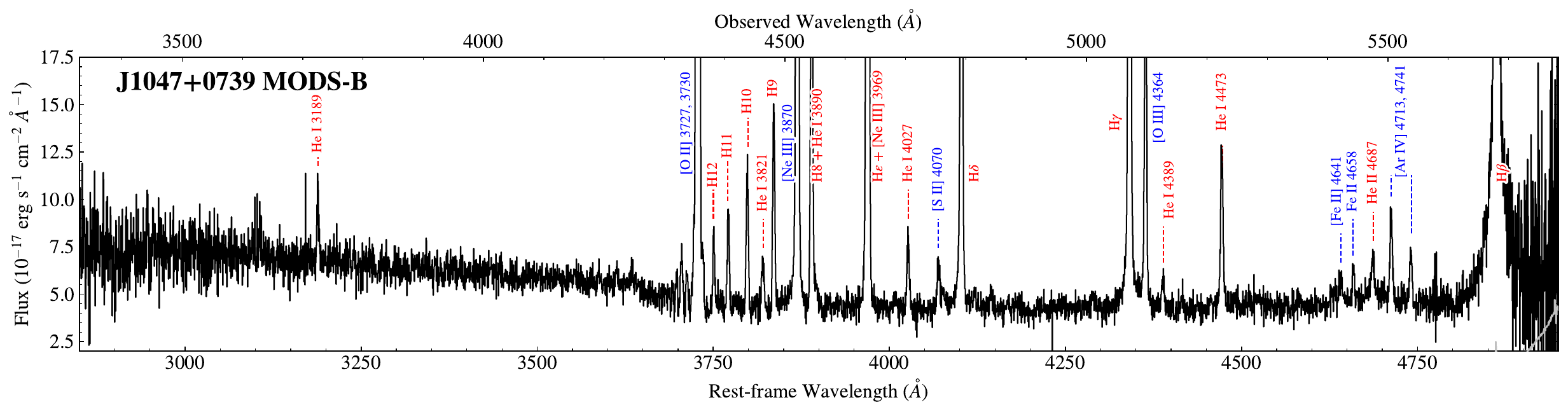}
    \includegraphics[width=\linewidth]{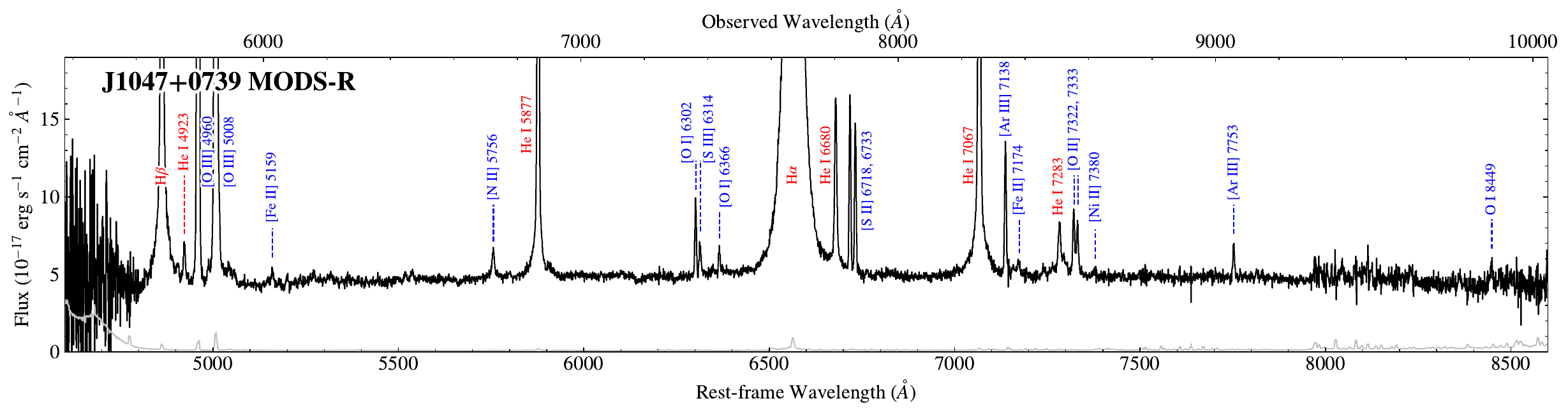}
    \includegraphics[width=\linewidth]{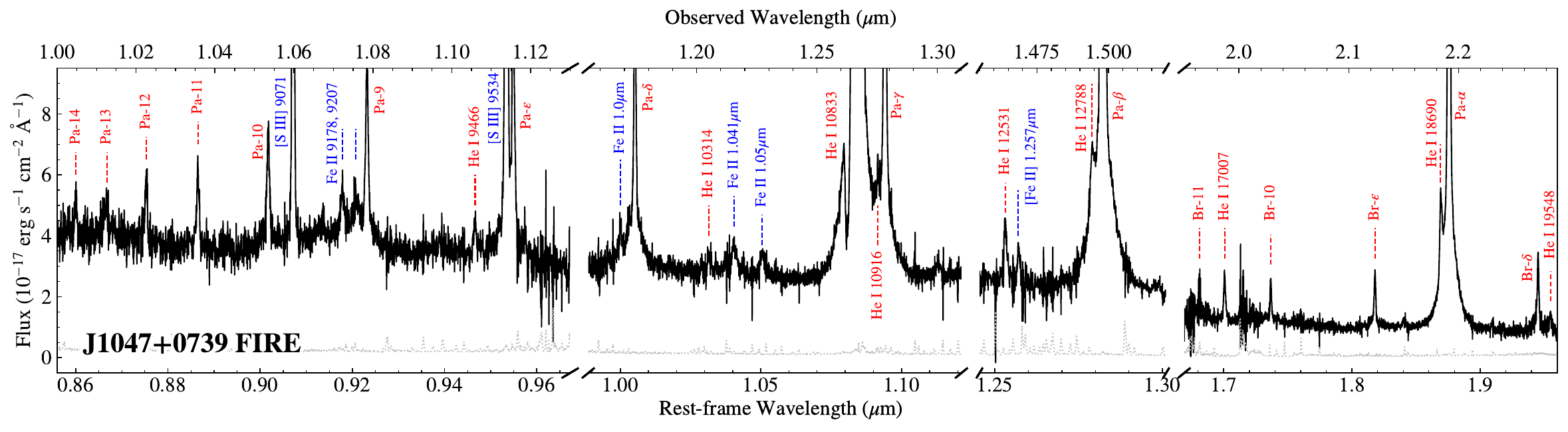}
    \caption{Zoomed-in view of the spectra of \objb, similar to Figure \ref{fig:J1025_zoomin}}
    \label{fig:J1047_zoomin}
\end{figure*}

\begin{figure*}[htbp]
    \includegraphics[width=\linewidth]{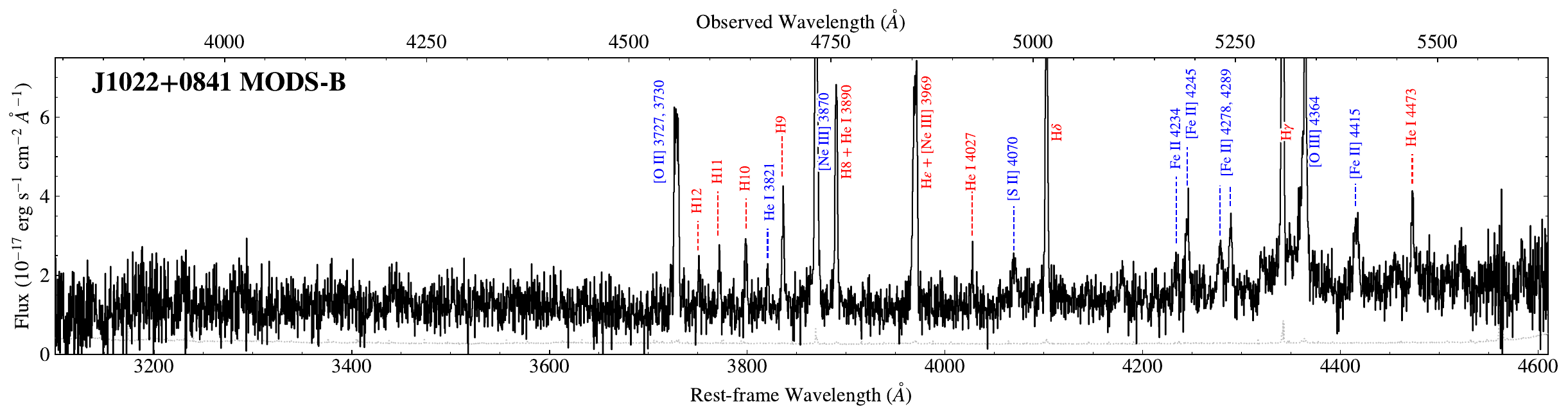}
    \includegraphics[width=\linewidth]{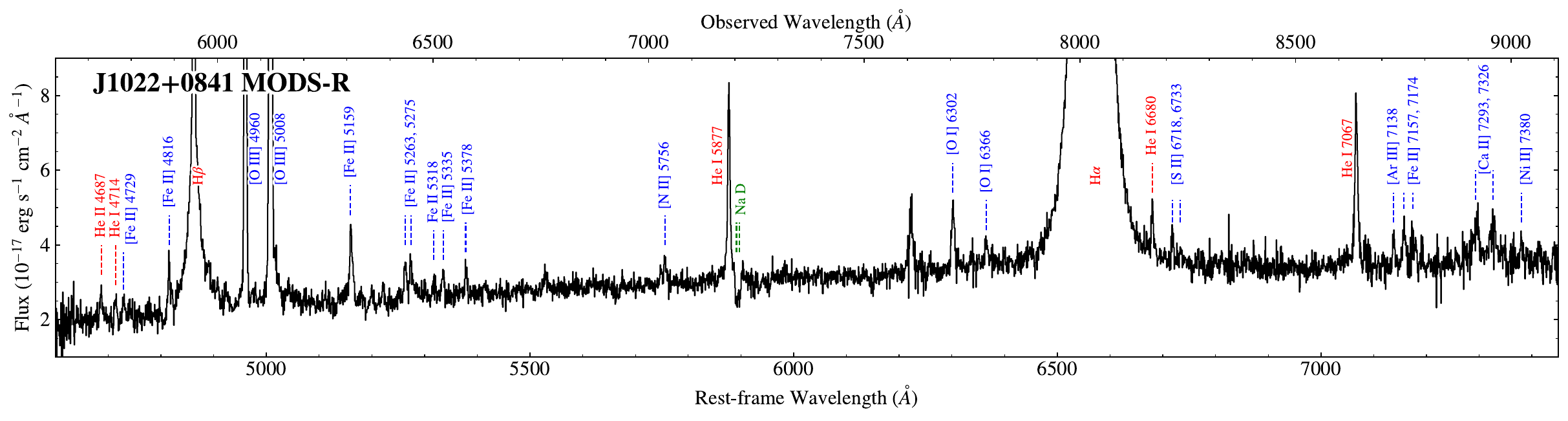}
    \includegraphics[width=\linewidth]{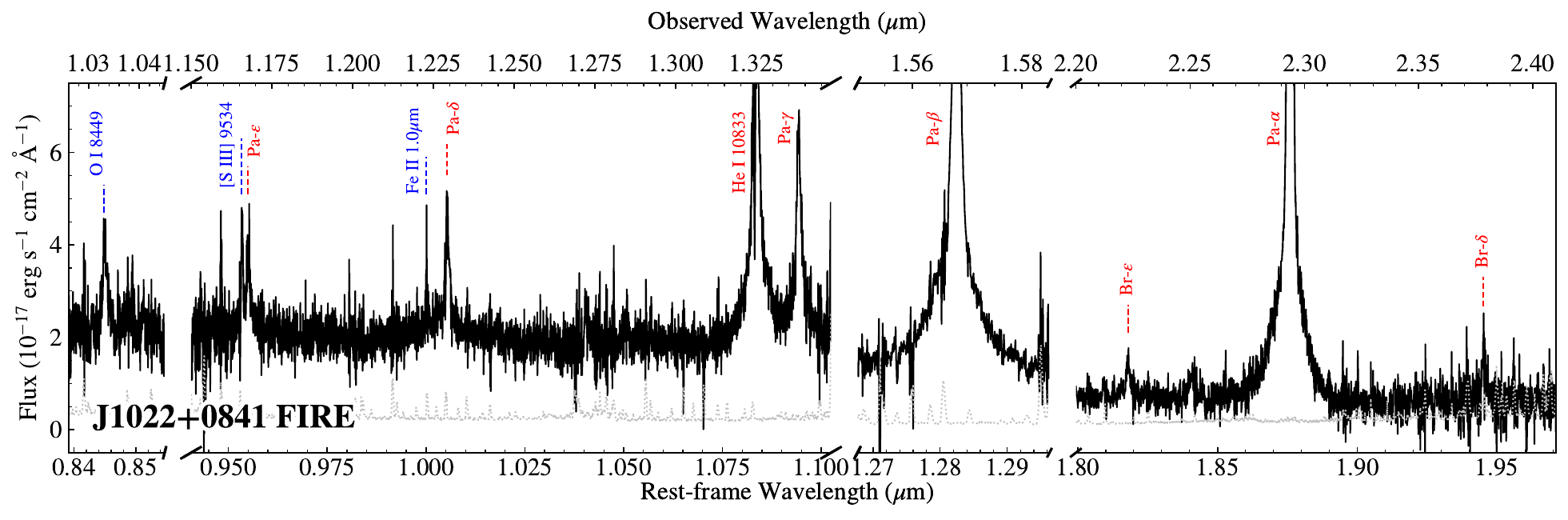}
    \caption{Zoomed-in view of \objc, similar to Figure \ref{fig:J1025_zoomin}. }
    \label{fig:J1022_zoomin}
\end{figure*}

In this section, we present the observed characteristics of three local objects that closely resemble high-redshift LRDs. The section is organized as follows: morphology and continuum luminosities are discussed in \S\ref{sec:morph_lum}; broad \ha\ emission line and BH properties in \S\ref{sec:broad_ha_mbh}; Balmer and \ion{He}{1} absorption features in \S\ref{sec:H_He_absorption}; variability properties in \S\ref{sec:variability}; X-ray and radio properties in \S\ref{sec:xray}; and a summary of the observational evidence as LRDs is provided in \S\ref{sec:evidence_as_LRD}. We present their zoomed-in spectra in Figures \ref{fig:J1025_zoomin}, \ref{fig:J1047_zoomin}, and \ref{fig:J1022_zoomin}, and summarize their properties in Table \ref{tab:basic_properties} and  \ref{tab:Ha_properties}.

\subsection{Morphology and Continuum}\label{sec:morph_lum}
 
All three objects exhibit very compact morphologies across the optical bands. They are not spatially resolved in LBT/MODS direct imaging with a seeing of $\sim$0.9\arcsec.   This implies that the diameter of \obja\ is comparable to or smaller than 1.7\,pkpc, while those of \objb\ and \objc\ are constrained to be smaller than 2.6\,pkpc and 3.2\,pkpc, respectively. 

\objb\ was observed with the HST/ACS FR782N ramp filter narrow-band imaging that covers both the \ha\ line and the adjacent continuum (GO-15617; PI: F. Bauer), achieving a spatial resolution of 206 pc. The continuum map reveals a bright central point source contributing about 85\% of the light, alongside a compact yet clumpy host galaxy with a half-light radius of 568 pc \xj{(see Appendix \ref{sec:appendix_J1047_morph})}. The \ha\ line map is dominated by the central point source with minimal host contribution. The spatially resolved map shows that, in \objb, AGN emission dominates both the continuum and the \ha\ line, in agreement with our definition of LRDs, although with a non-negligible contribution of the stellar light from the host galaxy to the continuum. The detailed morphological analysis is presented in Appendix \ref{sec:appendix_J1047_morph}. Future high-resolution imaging with HST, JWST, or ground-based AO will further constrain the properties of the physical sizes of the local LRDs and their host galaxy properties.

All three objects exhibit continuum shapes characterized by pronounced V-shapes and declining IR slopes. We determine the inflection points by fitting a two-component power law to a 300\,\AA\ segment of the continuum centered on the minimum within the rest-frame 3000–4500\,\AA\ range. The inflections occur near the Balmer limit. \xj{We measure the UV continuum slopes at rest-frame 2500 \AA, $\beta_{\rm UV,\,2500\AA}$, by fitting a power law to the GALEX/NUV photometry and MODS spectra below 3000 \AA. All the three objects have $\beta_{\rm UV,\,2500\AA}\lesssim-1$. The optical continuum slopes ($\beta_{\rm opt,\,4500\AA}$) are measured from the MODS spectra around rest-frame 4500\,\AA, after masking emission lines and potential absorption features. \obja\ and \objc\ exhibit steep $\beta_{\rm opt,\,4500\AA}>3$, while \objb\ shows a flatter, yet still red, optical continuum slope ($\beta_{\rm opt,\,4500\AA}\approx0.3$). } \objb\ also exhibits a potential Balmer jump, presumably due to the presence of a host galaxy. The continuum luminosity of the three objects at rest-frame 5100 \AA\ ($\lambda L_{5100}$) is around $10^{43}$\,erg\,s$^{-1}$.  For all three objects, the WISE photometry at rest-frame 3--20\,\micron\ reveals the presence of weak dust emission, at a wavelength range beyond what is possible with JWST observations of high-redshift LRDs. We integrate the continua from rest-frame 2000\,\AA\ to 20\,\micron\ to estimate the total continuum luminosity over this wavelength range, yielding $L_{\rm 2000\text{\AA}-20\micron} \sim 5\times10^{43}$--$3\times 10^{44}$\,erg\,s$^{-1}$ for all three objects.

\subsection{Broad \ha\ Emission Lines and BH properties}\label{sec:broad_ha_mbh}
 
The three objects show clear broad profiles in \ha, \hb, Pa$\alpha$ to Pa$\delta$, and several \ion{He}{1} lines (e.g., \ion{He}{1} $\lambda$7067, $\lambda$10833). \obja\ and \objc\ also exhibit broad infrared \ion{O}{1} \xj{$\lambda$1.29 \micron} emission lines, consistent with the line broadening originating from gas motions in the BLRs around BHs. The methodology for the line measurements is described in Appendix~\ref{sec:appendix_emission_fitting}. For the \ha\ lines, the narrow component is modeled with a single Gaussian, while the broad component is fitted with two to three broad Gaussians. The [\ion{N}{2}] $\lambda\lambda$6550, 6585 doublet is modeled with two narrow Gaussians sharing the same FWHM as the narrow \ha\ component and a fixed line ratio of 3.  For \obja\ and \objc, we adopt the MMT/Binospec \ha\ profile measurements as fiducial and confirm their consistency with the MODS spectra. The resulting \ha\ measurements are summarized in Table~\ref{tab:Ha_properties}. \xj{The effective FWHM of the broad \ha\ profiles (FWHM$^\star_{\rm H\alpha, broad}$), used to estimate BH masses, is derived from the summed profile of all broad components.}

All three objects exhibit broad H$\alpha$ emission with luminosities around $10^{42}$\,erg\,s$^{-1}$, and their individual broad components exhibit FWHMs ranging from $>500$\,km\,s$^{-1}$ to 2000\,km\,s$^{-1}$ 
, with the effective FWHMs around $\sim$1000\,km\,s$^{-1}$.  We estimate their BH masses using the broad H$\alpha$ luminosity ($L_{\rm H\alpha,broad}$), the effective broad \ha\ FWHM (FWHM$^\star_{\rm H\alpha,broad}$), and the empirical relation from \citet{Reines2015}. We find BH masses around $10^6$--$10^7\,M_\odot$.  This scaling relation for BH mass estimates is widely used in high-redshift LRD studies \citep[e.g.,][]{Matthee2024, Lin2024, Lin2025, Taylor2025a}. \xj{If estimated using the \citet{Greene2005} relation, the $M_{\rm BH}$ values would be 0.15–0.33 dex lower.}
However, these estimators might not be fully applicable to LRDs, 
if their physical nature is different from classical AGNs, particularly if their line profiles and continua arise from distinct physical mechanisms (see \S\ref{sec:cartoon}).  The measured $L_{\rm H\alpha, broad}$ values in these objects are 3–7 times higher than the $L_{\rm H\alpha}$ predicted by the $L_{5100}$–$L_{\rm H\alpha}$ relation in \citet{Greene2005}, where $L_{\rm H\alpha}$ refers to the total (broad + narrow) H$\alpha$ luminosity from AGNs. \xj{If using $L_{\rm 5100}$ and FWHM$_{\rm H\alpha, broad}$ to estimate $M_{\rm BH}$ following the relations in \cite{Greene2005}, the resulting values are 0.5–0.7 dex lower than those listed in Table~\ref{tab:Ha_properties}.} It remains unclear whether $L_{5100}$ or $L_{\rm H\alpha, broad}$ reliably traces the BLR size in these systems.  \xj{On the other hand, if the broadening of \ha\ lines is partly due to electron scattering in dense ionized gas rather than purely from BLR gas motion, the derived $M_{\rm BH}$ could be overestimated by 1–2 dex \citep{Naidu2025, Rusakov2025}. However, whether this scenario applies to the general LRD population remains uncertain \citep{Juodzbalis2025, Brazzini2025}. }

\xj{We note that no dust attenuation correction has been applied in the estimates above. The intrinsic Balmer decrements in BLRs remain uncertain because of the complex radiative transfer effects in high-density gas clouds. The broad \ha/\hb\ line ratios in the three objects are 8-34 (see \S\ref{sec:Balmer_Paschen_decrements} for a more detailed analysis). Assuming Case B recombination, although it may not hold in BLRs, such large ratios correspond to $A_V \approx 3-7$ under an SMC attenuation curve \citep{Pei1992}. If this attenuation correction is applied,  the intrinsic \ha\ luminosity could be a factor of $4-31$ higher than the directly measured values, implying that the BH masses would increase by $\sim 0.3-0.7$ dex. However, if energy balance holds, such a high level of dust attenuation is in tension with the weak IR emission observed by WISE. }

\xj{For reference and comparison with literature LRDs, we also list bolometric luminosities derived solely from \ha\ following \cite{Stern2012}, along with the corresponding Eddington ratios, in Table \ref{tab:Ha_properties} ($L_{\rm bol,H\alpha}$ and $\lambda_{\rm Edd,H\alpha}$). These values are generally consistent with high-redshift LRDs, which span a wide range of $L_{\rm bol,H\alpha}$ from $\sim 10^{44}~{\rm erg~s^{-1}}$ to $\sim10^{45}~{\rm erg~s^{-1}}$, and $\lambda_{\rm Edd,H\alpha}$ from $\gtrsim0.01$ to $\gtrsim1$ \citep[e.g.,][]{Maiolino2024, Greene2024, Lin2024, Juodzbalis2025}. We emphasize, however, that these $L_{\rm H\alpha,broad}$-derived estimates may not accurately reflect the intrinsic bolometric luminosities and Eddington ratios due to the uncertain physical processes in such systems and the systematics discussed above \citep{Greene2025}.}

\subsection{Balmer and \ion{He}{1} Absorption}\label{sec:H_He_absorption}

\begin{figure*}[htbp]
    \centering
    \includegraphics[width=\linewidth]{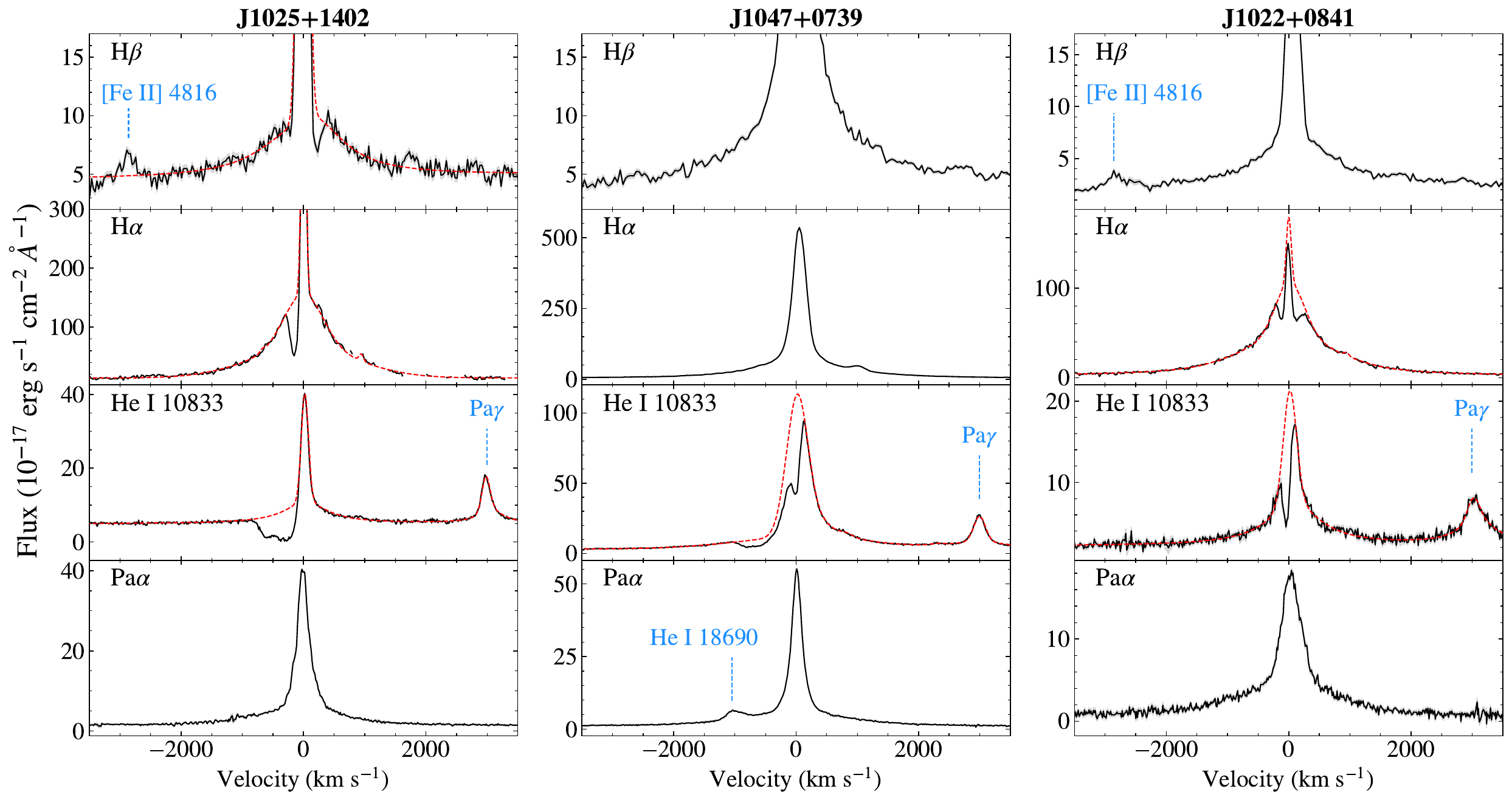}
    \caption{The \hb, \ha, \hei,  and Pa$\alpha$ line profiles of the three local LRDs \xj{studied in this work}. These line profiles are from the MODS and FIRE observations, except for the H$\alpha$ profile of \obja\ and \objc, which are from the Binospec observations. For emission lines with absorption, the red dashed lines depict the intrinsic profiles without absorption.}
    \label{fig:H_He_lineprofile}
\end{figure*}

Figure \ref{fig:H_He_lineprofile} presents the line profiles of \hb, \ha, \ion{He}{1}, and Pa$\alpha$ for the three local LRDs. All three objects exhibit absorption lines in their Balmer series and/or \ion{He}{1} $\lambda$10833 lines. For \ha\ of \obja\ and \objc, we model the absorber as Gaussian components superimposed on the intrinsic broad \ha\ emission. We then simultaneously fit the absorbers, the narrow \ha, broad \ha, and the [\ion{N}{2}] $\lambda\lambda$6550, 6585 doublet. For \hb\ of \obja\ and \ion{He}{1} $\lambda$10833 of all three objects, absorbed regions are masked, and the intrinsic emission line profile is modeled using a narrow Gaussian combined with multiple broad Gaussians.  The detailed fitting methodology is described in Appendix~\ref{sec:appendix_emission_fitting}.  The measured absorption properties are presented in Table \ref{tab:absorber_fit}.

\obja\ exhibits a redshifted \hb\ absorption feature superimposed on the broad \hb\ emission and continuum, with $\Delta v = 73$\,km\,s$^{-1}$ and $\mathrm{EW} = 4.16 \pm 0.43$\,\AA.  A similar redshifted \hb\ absorber is also seen in the triply imaged $z=7.04$ LRD in the Abell 2744 field \citep{Ji2025}.  \obja\ and \objb\ both have \ha\ absorption, superimposed on the broad \ha\ emission. The \ha\ absorber of \obja\ has an EW of $2.54\pm0.07$\,\AA, and is blueshifted by $\Delta v=-134$\,km\,s$^{-1}$ relative to the center of the broad \ha\ component. The H$\alpha$ absorber in \objc\ can be described with two  Gaussian components,  one blueshifted by $\Delta v=-95$\,km\,s$^{-1}$ and one redshifted by $\Delta v=102$\,km\,s$^{-1}$. Their EWs are $0.91\pm0.24$\,\AA\ and  $2.04\pm0.60$\,\AA, respectively,  yielding a total EW of  $2.96\pm0.64$\,\AA. An alternative solution is one Gaussian absorber with $\Delta v=11$\,km\,s$^{-1}$ and EW of $4.16\pm0.33$\,\AA, \xj{but we find that this solution leaves more residuals as commented in  Appendix  \ref{sec:appendix_emission_fitting}}.  \objb\ does not show clear Balmer absorption.

The lack of \hb\ absorbers corresponding to \ha\ absorbers in \obja\ and \objc\ might be attributed to the S/N of the \hb\ spectra. However, a puzzling aspect of \obja\ is that its redshifted \hb\ absorber does not have corresponding \ha\ absorption. This suggests that the two lines might have distinct origins.

For reference, we also present the Pa$\alpha$ line profile in Figure \ref{fig:H_He_lineprofile}. \xj{No absorption features associated with Pa$\alpha$ are detected, consistent with the findings of \cite{Juodzbalis2024, Juodzbalis2025}.}  This suggests that the Balmer absorbers originate from the neutral hydrogen elevated to the $n=2$ state but not significantly populated at $n=3$.

\begin{figure}
    \centering
    \includegraphics[width=\linewidth]{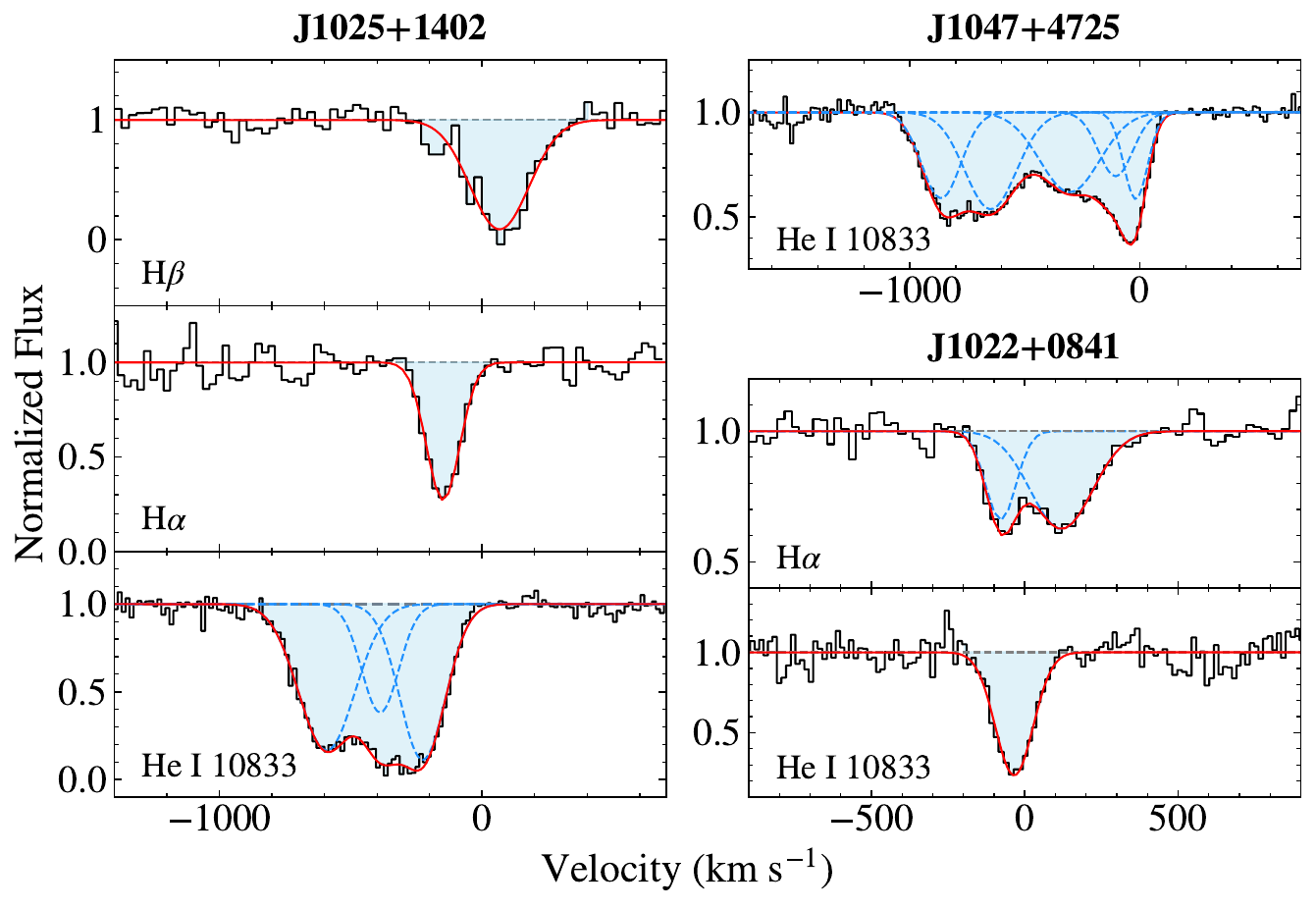}
    \caption{Normalized Balmer and \ion{He}{1}~$\lambda10833$ absorption line profiles. For multi-component absorption profiles, the individual components are shown as the blue dashed lines.
    }
    \label{fig:HaHeI_absorber_compare}
\end{figure}

All three targets show prominent blueshifted \hei\ absorption. The \hei\ absorber of \obja\ and \objb\  has multiple components with distinct kinematics (Figure \ref{fig:HaHeI_absorber_compare}). The line profile of the \obja\ \hei\ absorber, as shown in Figure \ref{fig:H_He_lineprofile}, can be described using three Gaussian components, blueshifted with $\Delta v=-613, -410$, and $-248$\,km\,s$^{-1}$, respectively. This profile may alternatively result from a mixture of saturated \ion{He}{1} absorbers. The line profile of the \objb\ \hei\ absorber can be described using five Gaussian components with $\Delta v=-1646, -1431, -1088, -889$, and $-801$\,km\,s$^{-1}$, respectively.  For \objc, the \hei\ absorber is well described by a single Gaussian component with $\Delta v = -55$\,km\,s$^{-1}$. They highlight the prevalence of \ion{He}{1} absorption in local LRDs. {The \ion{He}{1} line is a resonant transition from the metastable 2$^3$S state to the 2$^3$P state. The 2$^3$S state can be populated either through collisional excitation in high-density gas or via recombination from He$^+$. \xjnew{It remains unclear whether the \ion{He}{1} absorption in LRDs originates in the same dense clouds responsible for the Balmer absorption \citep{Juodzbalis2024}, where neutral He is collisionally excited into the 2$^3$S state, or instead in lower-density \xjnew{partially ionized} outflows \citep{Wang2025}, where He is first ionized to He$^+$ and then recombines into the 2$^3$S state.} Although strong \ion{He}{1} recombination lines ($\lambda$4473, $\lambda$5877, $\lambda$7067) with 2$^3$P as the lower state are detected in all three objects, the origin remains challenging to robustly constrain, as the emitting and absorbing gas clouds may not spatially co-exist. The complex kinematics of \ion{He}{1} absorbers (Figure~\ref{fig:HaHeI_absorber_compare}) may also be attributed to mixed origins.   
}

\begin{table*}[]
    \centering
    \begin{tabular}{c|ccc|ccc|ccc}
    \hline
      Name & \multicolumn{3}{c|}{\hb} & \multicolumn{3}{c|}{\ha} & \multicolumn{3}{c}{\hei} \\
      \hline
          & $\Delta v$ & FWHM & EW & $\Delta v$ & FWHM & EW & $\Delta v$ & FWHM & EW    \\
         & (km\,s$^{-1}$) & (km\,s$^{-1}$) & (\AA) & (km\,s$^{-1}$) & (km\,s$^{-1}$) & (\AA) & (km\,s$^{-1}$) & (km\,s$^{-1}$) & (\AA) \\
        \hline 

        \multirow{4}{*}{\obja}  & \multirow{4}{*}{73} & \multirow{4}{*}{$232\pm25$} & \multirow{4}{*}{$4.16\pm0.43$} & \multirow{4}{*}{$-134$} & \multirow{4}{*}{$138\pm5$} & \multirow{4}{*}{$2.54\pm0.07$} & $-613$ & $222\pm17$ & $8.05\pm0.71$ \\
          &  &   &  &   &    &   & $-410$ & $117\pm31$ & $3.93\pm1.67$ \\
           &   &   &  &   &   &    & $-248$ & $167\pm20$ & $6.97\pm1.23$ \\
        &   &   &  &   &   &   &  &  total & $18.95\pm2.19$ \\
        \hline 
     \multirow{6}{*}{\objb} &  \multicolumn{3}{c|}{\multirow{6}{*}{No absorber}}   &  \multicolumn{3}{c|}{\multirow{6}{*}{No absorber}}   & $-1646$ & $159\pm17$ & $3.15\pm0.44$ \\
    &   &   &    &    &    &    & $-1431$ & $234\pm41$ & $4.69\pm0.80$ \\
    &    &   &    &    &    &    & $-1088$ & $278\pm58$ & $4.47\pm0.93$ \\
  &    &   &    &    &    &    & $-889$ & $128\pm29$ & $2.06\pm0.61$ \\
  &    &   &    &   &    &   & $-801$ & $30\pm14$ & $1.96\pm0.29$ \\ 
 &    &    &    &    &    &    &  &  total & $16.33\pm1.47$ \\
    \hline
     \multirow{3}{*}{\objc} &  \multicolumn{3}{c|}{\multirow{3}{*}{No absorber}} & $-95$ & $94\pm19$ & $0.91\pm0.24$ & \multirow{3}{*}{$-55$} & \multirow{3}{*}{$85\pm4$} & \multirow{3}{*}{$4.25\pm0.11$}\\
     &   &   &   & $102$ & $227\pm45$ & $2.04\pm0.60$ \\
    &   &   &   &   &  total & $2.96\pm0.64$ \\
    \hline
    \end{tabular}
    \caption{Properties of the Balmer and \ion{He}{1} absorbers in the three local LRDs. For multi-component profiles, we list the parameters of individual components along with the summed EWs. 
    }
    \label{tab:absorber_fit}
\end{table*}

\subsection{Variability}\label{sec:variability}

We examine the variability of all three objects in the optical ($grizy$) and infrared (WISE W1 and W2) bands. For the optical $grizy$ bands, we retrieve multi-epoch PSF-modeled photometry from the Pan-STARRS DR2 detection catalog. Over 1500 days ($\sim$1300 days in the rest frame), we find no significant variability, with observed variation of 0.1–0.2 mag consistent with systematic uncertainties as well as the scatter in non-variable sources \citep{Simm2015}. Following \cite{Lyu2019}, we construct the WISE W1 ($\sim$3.4\,\micron) and W2 ($\sim$4.6\,\micron) band light curves from the WISE and NEOWISE \citep{Mainzer2011} missions.  In the W1 and W2 bands, we find variations of 0.25--0.5 mag over 5000 days ($\sim$4000 days in the rest frame), also consistent with the typical scatter expected for non-variable sources. We thus conclude that no significant variability is detected in either the rest-frame optical or infrared bands, consistent with the lack of significant variability seen in high-redshift LRDs \citep{Kokubo2024, ZZhang2025, Lin2025, Tee2025}.

\cite{Burke2021} reported no significant variability in the broad \ha\ luminosity of \obja\ and \objb\ over a continuous 15-year period. The \ha\ luminosity range of \obja\ in \cite{Burke2021} is approximately $(4$–$7)\times10^{41}$\,erg\,s$^{-1}$, and for \objb\ it is $(2$–$3)\times10^{42}$\,erg\,s$^{-1}$, both well consistent with our measurements. They also presented V-band light curves from the Catalina Real-Time Transient Survey \cite[CRTS; ][]{Drake2009} and $r$-band light curves from the Zwicky Transient Facility \citep[ZTF; ][]{Masci2019} for both objects. Although the limited quality of the light curves precluded definitive conclusions, they found potential low-level variability consistent with the stochastic damped random walk model of AGN variability \citep{Kelly2009, MacLeod2010}.

\subsection{X-ray and Radio}\label{sec:xray}

\citet{Simmonds2016} reported $<4.6$ net counts for \obja\ in the 0.5–7.0\,keV band with Chandra, based on a net exposure time of 4937\,s, placing an upper limit\footnote{Assuming a $\Gamma=1.8$ power law spectrum absorbed by Galactic $N_{\rm H}$.} on the apparent X-ray luminosity of $L_{2-10\,\rm keV} < 1.1 \times 10^{41}\,{\rm erg\,s^{-1}}$. For \objb, they reported $3.0^{+2.9}_{-1.6}$ net counts in 4782\,s. We further examined the Chandra Source Catalog Release 2.1 \citep{Evans2024}, which includes more recent observations of \objb, and found $5.60^{+2.80}_{-2.67}$ net counts in the 0.5–7 keV band over a 53\,ks exposure. This corresponds to an apparent X-ray luminosity of $L_{2-10\,\mathrm{keV}} \approx (8.1^{+3.8}_{-4.0})\times 10^{40}\,{\rm erg\,s^{-1}}$
, although the detection is only significant at $\lesssim 2\sigma$ \footnote{We use \href{https://cxc.harvard.edu/toolkit/pimms.jsp}{WebPIMMS} to convert counts into fluxes. }. For \objc, no Chandra observations are currently available.

The comparison of the  
X-ray emission of \obja\ and \objb\ with that of typical type-1 AGNs is detailed in \citet{Simmonds2016}. 
The upper limits or $<2\sigma$ X-ray detections suggest that \obja\ and \objb\ are X-ray weak, exhibiting $L_{\rm 2-10\,keV}/L_{\rm H\alpha}$ ratios at least 1–2 dex below those of typical type-1 AGNs with similar broad \ha\ luminosities \citep{Greene2004, Panessa2006, Desroches2009}. This X-ray weakness is consistent with that seen in high-redshift LRDs \citep{Yue2024, Ananna2024}.

We examine the radio data from the VLA Sky Survey \citep{Lacy2020}, the TIFR GMRT Sky Survey \citep{Intema2017}, and the VLA FIRST Survey \citep{Becker1995}. We find no radio detections at 153 MHz, 1.4 GHz, and 3 GHz for any of the three sources.
The corresponding $5\sigma$ upper limits for \obja, \objb, and \objc\ are: 13.8, 12.5, and 15.1 mJy at 153 MHz; 0.7, 0.7, and 0.8 mJy at 1.4 GHz; and 0.4, 0.5, and 0.4 mJy at 3 GHz, respectively. 
The $5\sigma$ upper limits of radio-loudness ($\mathcal{R} \equiv L_\mathrm{3GHz}/L_\mathrm{5100}$, \citealt{Sikora2007}) for the three objects are estimated to be 8, 12, and 20, respectively. The current radio data are not deep enough to determine whether these sources are radio-quiet ($\mathcal{R}<1$).

\subsection{Summary of Evidences as LRDs}\label{sec:evidence_as_LRD}

As discussed in this section and in the previous ones, the three objects at $z = 0.1$–$0.2$ satisfy all the defining criteria for LRDs. We summarize the similarity between the observed properties of the three local sources to high-redshift LRDs as follows:

\begin{itemize}
    \item compact morphology;

    \item V-shaped continuum (blue UV continuum slope, red optical continuum slope) and decreasing IR continuum at $\lesssim0.7-2$\,\micron;

    \item  broad Balmer emission lines, consistent with $\sim10^6$--$10^7M_\odot$ BHs;

    \item weak X-ray emission; \xj{no radio detection in the current data;}

    \item no significant (or low-level) variability;

    \item high occurrence rate of Balmer and \ion{He}{1} $\lambda$10833 absorption.

\end{itemize}

Their broad H$\alpha$ luminosities, $L_{\rm H\alpha, broad}\sim10^{42}$\,erg\,s$^{-1}$, fall within the range observed in high-redshift LRDs \citep[e.g.,][]{Matthee2024, Maiolino2024, Lin2024, Greene2024, Zhang2025, Hviding2025}. If redshifted to $z = 6$, their expected apparent magnitudes in JWST/NIRCam F444W would be 25.7--26.8\,mag.   
These values lie within the magnitude range observed for the diverse LRDs discovered at $z > 5$ \citep[e.g.,][]{Greene2024, Kocevski2024, Akins2024, Kocevski2025}. 

\section{Emission Line Analysis }\label{sec:emission_line_analysis}

In this section, we discuss in detail the properties of the emission lines detected in the local LRDs to constrain their origins, gas density, metallicity, and ionizing sources. 

\subsection{Balmer Lines and Paschen Line Decrements. 
}\label{sec:Balmer_Paschen_decrements}

\begin{figure*}[htbp]
    \centering
    \includegraphics[width=0.32\linewidth]{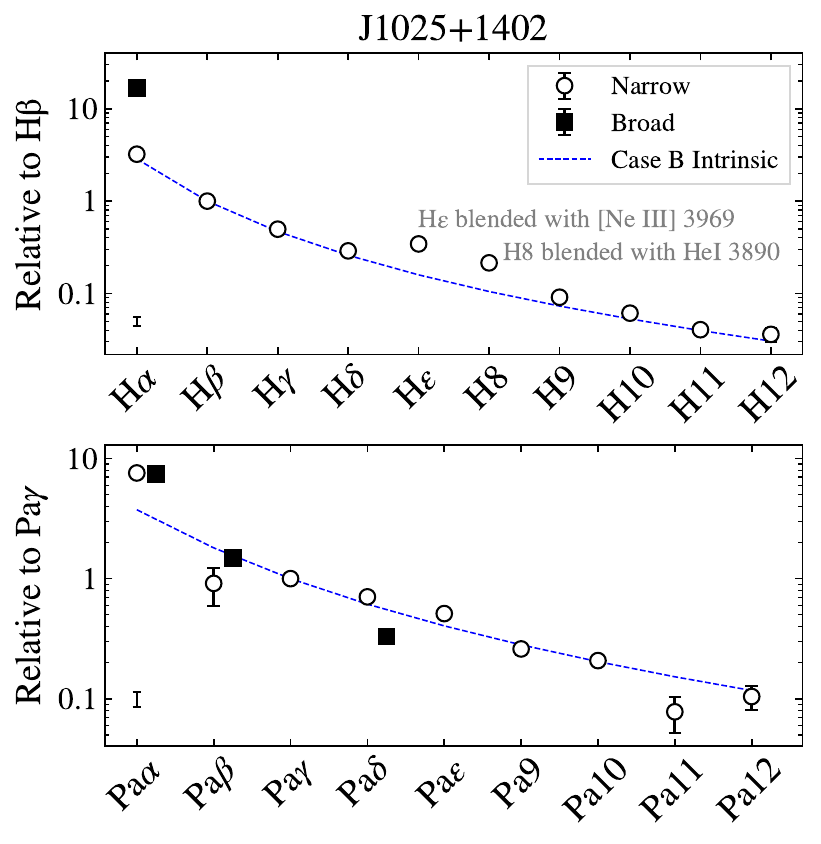}
    \includegraphics[width=0.32\linewidth]{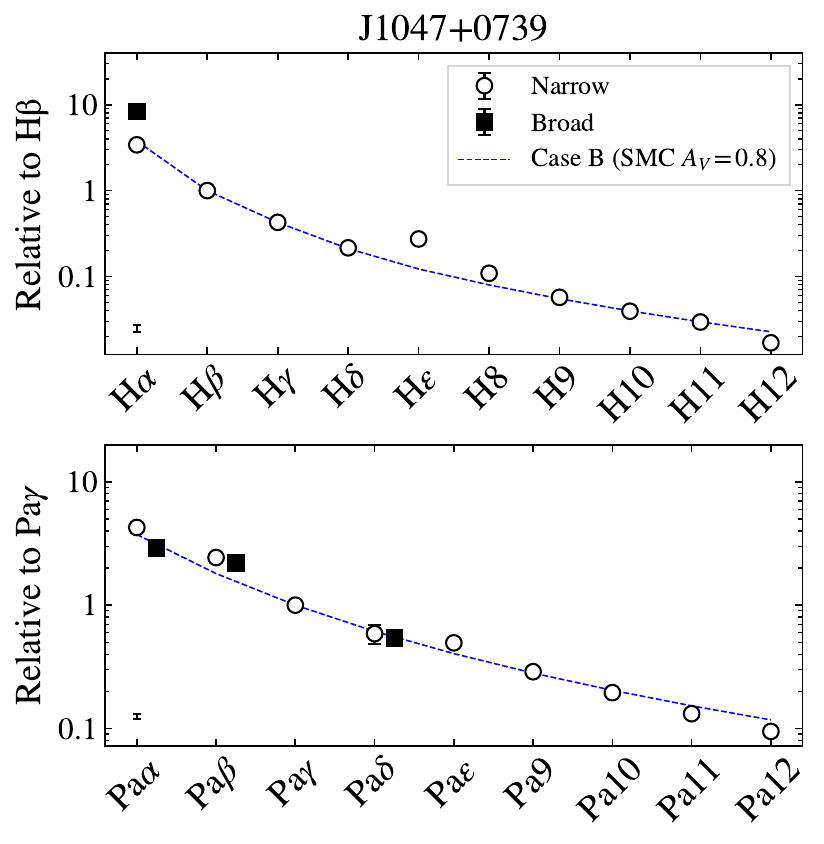}
    \includegraphics[width=0.32\linewidth]{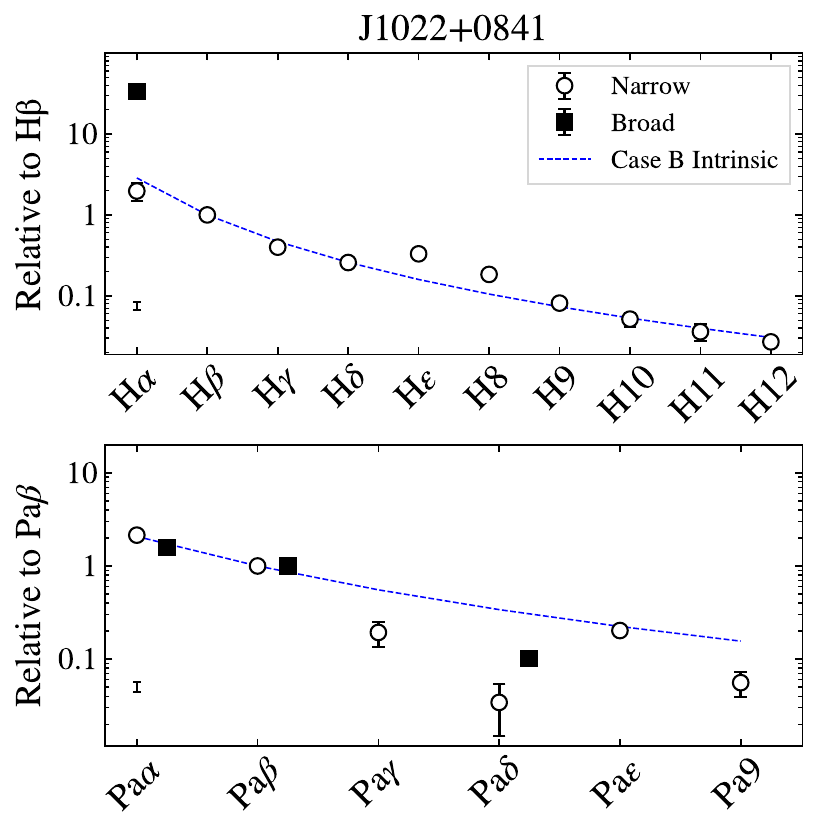}
    \caption{The Balmer and Paschen decrements of the three local LRDs. The decrements of the broad and narrow components are shown as black squares and white circles, respectively. The intrinsic line decrements under Case B conditions ($n_e = 100\,\mathrm{cm^{-3}}\,, T_{\rm e} = 10^4\,\mathrm{K}$) are indicated by the blue dashed lines for \obja\ and \objc. For \objb, the Case B predictions with the SMC dust attenuation law of $A_V = 0.8$ are shown.  The median error bars are indicated in the lower left corner of each panel. For the Paschen series in \objc, we normalize to Pa$\beta$ due to the relatively large uncertainty in Pa$\gamma$. 
    }
    \label{fig:H_decrement}
\end{figure*}

We utilize the rich series of Balmer and Paschen lines to measure the Balmer and Paschen decrements. For high-order hydrogen lines with insufficient S/N to detect reliable broad components, we fit them with a single Gaussian representing the narrow component, assuming that the broad components are buried in the noise.  As shown in Figure \ref{fig:H_decrement}, the narrow Balmer lines from \ha\ to H12 in \obja\ and \objc\ closely match the intrinsic Case B recombination predictions \citep[$n = 100\,\mathrm{cm^{-3}}$, $T = 10^4\,\mathrm{K}$; ][]{Brocklehurst1971,  Hummer1987, Osterbrock2006}.  The narrow Balmer lines in \objb\ are consistent with the SMC dust attenuation law \citep{Pei1992} with modest attenuation of $A_V = 0.8$.  The exception is H$\epsilon$ and H8, which are blended with [\ion{Ne}{3}] $\lambda$3969 and \ion{He}{1} $\lambda$3890, respectively. The agreement indicates that the narrow hydrogen lines originate from regions with minimal dust attenuation. However, we caution that the decomposition of narrow and broad components is degenerate, and the strength of the potential underlying absorption remains uncertain.

The narrow Paschen lines generally follow the Case B predictions, similar to the Balmer series, although \objc\ exhibits significant scatter due to limited S/N. The P$\alpha$ line in \obja\ deviates from the Case B prediction, but its flux measurement may be contaminated by nearby \ion{He}{1} emission lines adjacent to the Paschen transitions. This issue may also affect the measurement in \objc. In contrast, the \ion{He}{1} emission is particularly strong in \objb,  allowing both the \ion{He}{1} and Paschen lines to be reliably fitted.

The broad H$\alpha$/H$\beta$ line ratios are $16.9 \pm 1.8$ for \obja, $8.3 \pm 0.9$ for \objb, and $33.5 \pm 3.5$ for \objc. Such large decrements may be attributed to either radiative transfer effects in high-density gas (e.g., collisional excitation and de-excitation, resonance scattering, etc.) or substantial dust attenuation.  Notably, the measured decrements of the broad Balmer lines are well consistent with those observed in high-redshift LRDs \citep{Brooks2025, D'Eugenio2025_z7p04}. These values are comparable to those expected for the SMC  attenuation law with $A_V \approx 3-7$ under Case B conditions.  However, one caveat in attributing the broad-line decrements to dust is the survival of dust grains in the BLRs, which likely lies within the sublimation zone \citep[e.g.,][]{Martin1980, Kishimoto2007, Gaskell2018, Baskin2018}. For comparison, type-1 AGNs in the SDSS typically show broad H$\alpha$/H$\beta$ ratios around 3.7, with the most extreme cases reaching $\sim$5 \citep{Ilic2012}. The much larger broad Balmer decrements observed in LRDs suggest that their BLRs may have extreme conditions with unusually dense gas or atypical dust properties,  which differ substantially from those in typical type-1 AGNs \citep[e.g.,][]{Inayoshi2022}.

\subsection{BPT Diagram and Metallicity}
\label{subsec:BPT_metallicity}

\begin{figure*}[htbp]
    \centering
    \includegraphics[width=\linewidth]{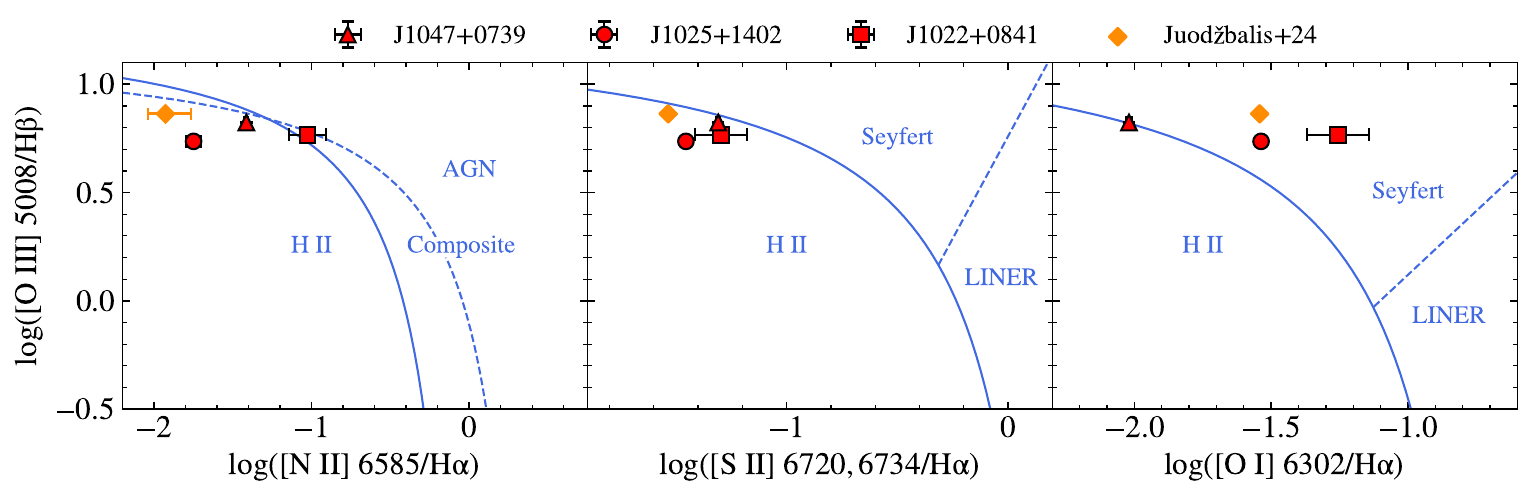}
    \caption{The BPT diagrams for the three local LRDs. The classification boundaries for different ionization mechanisms \citep{Kewley2006} are shown as blue solid and dashed lines. The corresponding regions are labeled. We also mark the cosmic-noon LRD in \cite{Juodzbalis2024} for reference.}
    \label{fig:BPT}
\end{figure*}

We estimate the BPT diagnostics \citep{Baldwin1981} and the gas-phase metallicity of the narrow line regions (NLRs). To measure the fluxes of forbidden lines, we fit Gaussian profiles. Since the [\ion{O}{3}] $\lambda4364$ and $\lambda5008$ lines are blended with [\ion{Fe}{2}] emission, their fluxes are measured from spectra where the [\ion{Fe}{2}] contribution has been subtracted using the best-fit models (see \S\ref{sec:fe_origin}).

We show the positions of the three local LRDs on the [\ion{N}{2}], [\ion{S}{2}], and [\ion{O}{1}] BPT diagrams in Figure~\ref{fig:BPT}. In the [\ion{N}{2}] and [\ion{S}{2}] diagrams, all three LRDs lie near the boundary between \ion{H}{2} regions and AGN ionization. In the [\ion{O}{1}] diagram, \obja\ and \objc\ fall within the Seyfert region, while \objb\ lies near the edge of the \ion{H}{2} region. Their positions on the [\ion{N}{2}] and [\ion{S}{2}] diagrams also overlap with those of low-metallicity AGNs \citep{Kewley2013, Simmonds2016}. Notably, the locations of these local LRDs closely match that of the cosmic-noon LRD in \citet{Juodzbalis2024}. This suggests that the narrow emission lines in both local and high-redshift LRDs are excited by similar mechanisms associated with AGN activity.  

\begin{figure*}[htbp]
    \centering
    \includegraphics[width=\linewidth]{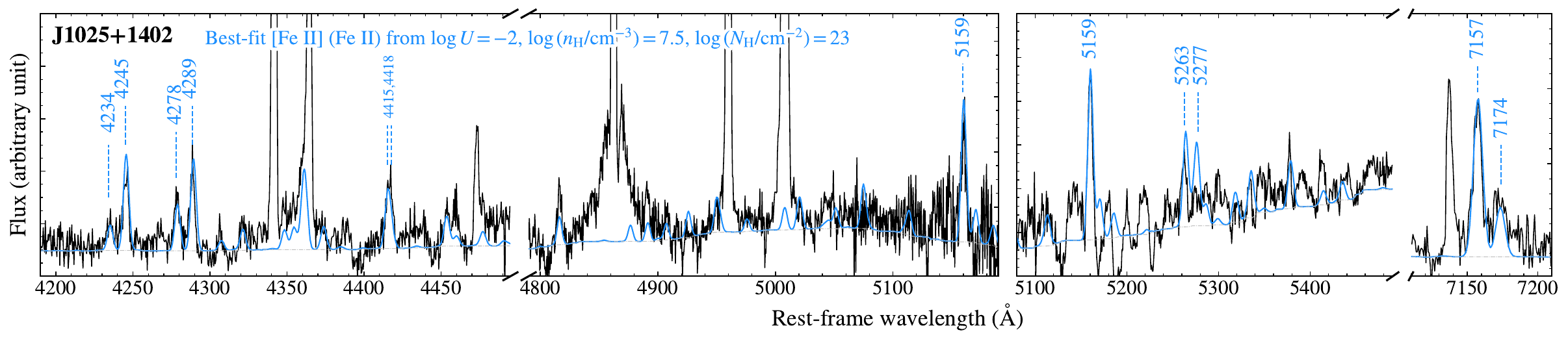}
    \caption{The narrow [\ion{Fe}{2}] and \ion{Fe}{2} emission lines in \obja\ and the best-fit \textsc{cloudy} model. All labeled lines, except \ion{Fe}{2} $\lambda$4234 and \ion{Fe}{2} $\lambda$5277, are forbidden [\ion{Fe}{2}] transitions. The left panel shows the MODS-B spectrum, and the right panel shows the MODS-R. The observed [\ion{Fe}{2}] lines used for the modeling are labeled.  The gray dashed line indicates the continuum level. The best-fit [\ion{Fe}{2}] (\ion{Fe}{2}) spectrum is shown as the blue line. 
    }
    \label{fig:J1025_bestfit_FeII_model}
\end{figure*}

We determine the metallicities of the NLRs using the direct electron temperature ($T_e$) method. The $T_e$ of the O$^{++}$ zone is derived from the  ratio between [\ion{O}{3}]$\lambda4364$ and [\ion{O}{3}]$\lambda5008$, assuming $n_e=250$\,cm$^{-3}$.  The estimated $T_e$(O$^{++}$) is $21386^{+517}_{-589}$ K for \obja, $14095^{+180}_{-171}$ K for \objb, and $23908^{+827}_{-767}$ K for \objc. The $T_e$ of the O$^{+}$ zone is estimated using the empirical $T_e$(O$^+$)–$T_e$(O$^{++}$) relation from \citet{Izotov2006}. We apply a dust attenuation correction of $A_V=0.8$ to \objb. 
As summarized in Table~\ref{tab:basic_properties}, all three objects exhibit metal-poor NLRs with 12 + $\log(\mathrm{O/H})$ ranging from 7.43 to 8.01. The values are broadly consistent with those reported by \cite{Izotov2008} for \obja\ and \objb.

We note that although the [\ion{O}{2}]$\lambda\lambda7322,7333$ lines are detected in \objb, they are significantly affected by recombination excitation enhanced by the hard ionizing radiation field  \citep{Liu2000, Tan2024}. Therefore, we do not use them for $T_e$(O$^+$) diagnostics. The $T_e$(O$^+$) value directly derived from the [\ion{O}{2}]$\lambda\lambda7322,7333$ and [\ion{O}{2}]~$\lambda\lambda3727,3730$ line ratios exceeds $\sim 3 \times 10^4$ K, more than twice the values of $T_e$(O$^{++}$) \xj{and the empirical $T_e$(O$^+$)}.  \xj{This discrepancy suggests that [\ion{O}{2}]$\lambda\lambda7322,7333$ is enhanced by recombination from the hard ionizing radiation field, and further supports an AGN origin for the narrow emission lines.}

Based on the Balmer decrements, BPT diagram, and direct $T_e$-based metallicity, we conclude that the narrow lines in local LRDs originate from metal-poor NLRs powered by AGNs with minimal dust content.

\subsection{[\ion{Fe}{2}] (\ion{Fe}{2}) Emission Analysis}\label{sec:fe_origin}

We observed a series of narrow [\ion{Fe}{2}] and \ion{Fe}{2} emission lines in \obja\ and \objc\ (see Figures \ref{fig:J1025_bestfit_FeII_model} and \ref{fig:J1022_bestfit_FeII_model}), most of which are forbidden transitions. These lines stand in stark contrast to the broad Fe emission lines typically observed in type-1 AGNs \citep[e.g., I Zw 1;][]{VMP2004}.
Though narrow optical [\ion{Fe}{2}] (\ion{Fe}{2}) lines have been reported in type-1 AGNs, they are blended with broad Fe lines and exhibit transitions and line ratios different from those in our sample; these narrow Fe lines are absent in type-2 AGNs \citep{Dong2010}.  To constrain the physical conditions such as temperature and density, we employ \textsc{cloudy} for photoionization modeling \citep[C23,][]{Gunasekera2023}, with the latest \ion{Fe}{2} atomic database \citep{Smyth2019}. 
We consider scenarios in which the gas is photoionized by AGN radiation. The modeling methodology is detailed in Appendix \ref{sec:appendix_feii}.

The AGN photoionization models yield best-fit parameters for both \obja\ and \objc\ of $\log (n_{\rm H}/{\rm cm^{-3}}) = 7.5$, $\log (N_{\rm H}/{\rm cm^{-2}}) = 23$, and $\log U = -2$. The relative [\ion{Fe}{2}] (\ion{Fe}{2}) line ratios are not strongly sensitive to metallicity, so we adopt a fiducial value of $Z_\odot$ without attempting to constrain it. The average temperature of the [\ion{Fe}{2}]-emitting gas\footnote{For simplicity, we hereafter refer to the gas as [\ion{Fe}{2}]-emitting, as most emission lines are forbidden [\ion{Fe}{2}] transitions, although a few \ion{Fe}{2} lines are also present.} is approximately $10^4$ K, ranging from $4 \times 10^3$ K in the inner region to $2 \times 10^4$ K on the surface. As shown in Figure~\ref{fig:J1025_bestfit_FeII_model}, the best-fit model reproduces the observed [\ion{Fe}{2}] (\ion{Fe}{2}) line ratios of \obja\ reasonably well, although some lines (e.g., \ion{Fe}{2}~$\lambda$5277) remain imperfectly matched. A similar level of agreement is achieved for \objc\ using the same best-fit parameters (Figure \ref{fig:J1022_bestfit_FeII_model}).

The inferred density lies between typical values for the BLRs and NLRs. The FWHMs of the [\ion{Fe}{2}] (\ion{Fe}{2}) lines are also intermediate ($\approx100-200$\,km\,s$^{-1}$ for \obja\ and $\approx200-300$\,km\,s$^{-1}$ for \objc), broader than the narrow \ha\ and [\ion{O}{3}] (\ha\ $\sim 65$ km\,s$^{-1}$ in Binospec, [\ion{O}{3}] not resolved in MODS). This suggests that the [\ion{Fe}{2}] (\ion{Fe}{2}) emission may arise from dense gas located outside the BLRs but inside the NLRs.   We emphasize that our models are qualitative. Fe emission lines in AGNs are inherently complex and can be influenced by a variety of physical processes, making comprehensive theoretical modeling challenging \citep[e.g.,][]{Sarkar2021,Pandey2025}. Further refinement is required to fully understand the origin and nature of [\ion{Fe}{2}] (\ion{Fe}{2}) emission in the LRD population.

%% file: 05_J1025.tex
\section{Case study: \obja,  ``The Egg"}\label{sec:J1025_case_study}

\begin{figure*}[htbp]
    \centering
    \includegraphics[width=1
    \linewidth]{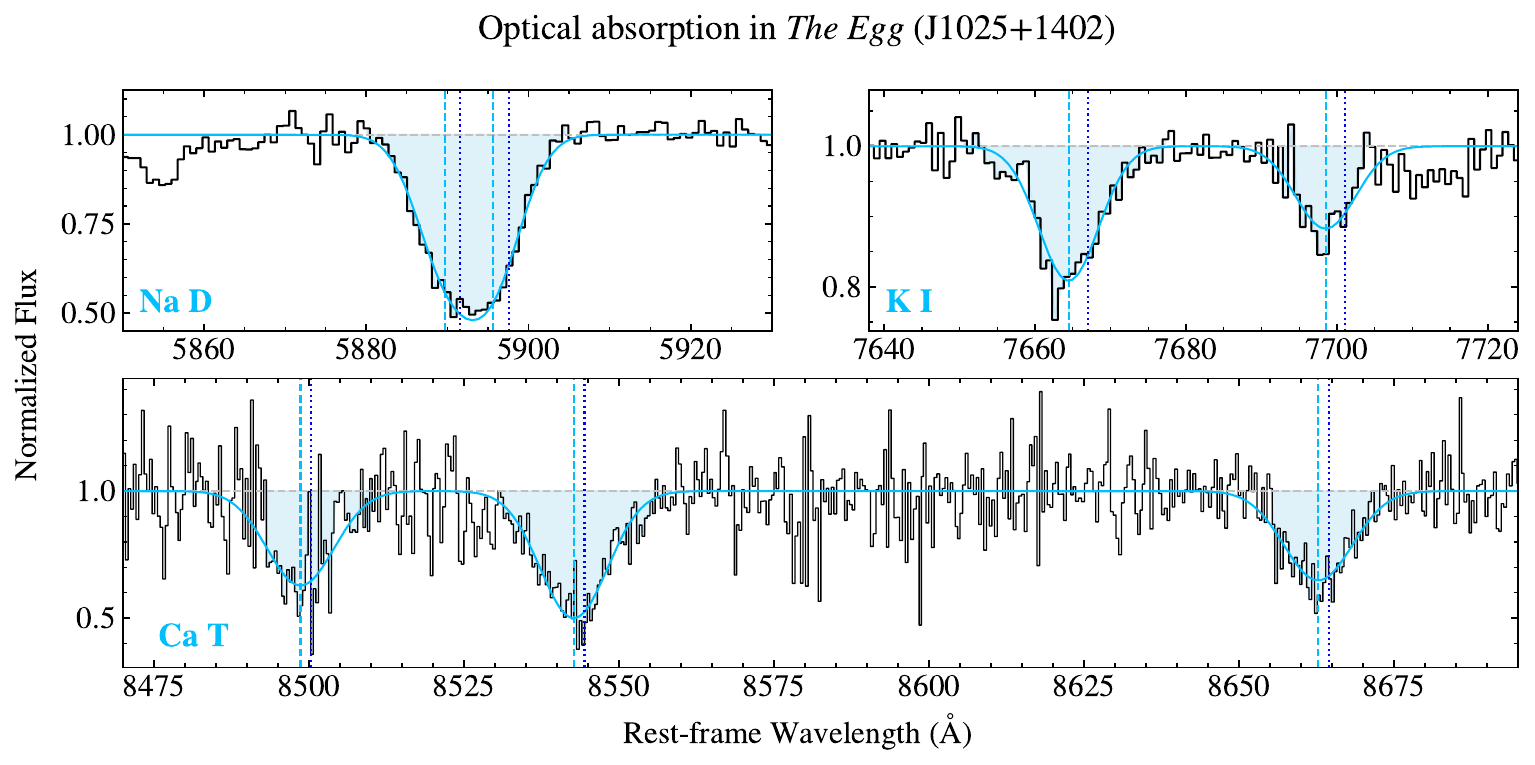}
    \caption{The Na D, \ion{K}{1}, and Ca T absorption in the optical spectrum of \egg\ (\obja). The normalized profiles of Na D, \ion{K}{1} are from the MODS spectra, and the Ca T is from the FIRE. The Na D profile is fitted with two blended Gaussian components, while each of the remaining absorbers is fitted with a single Gaussian. The light blue dashed lines indicate the line centers of metal absorbers, while the blue dotted lines mark their rest-frame wavelengths at the NLR redshift, as defined by [\ion{O}{3}] $\lambda5008$.
    }
    \label{fig:J1025_metal_absorber}
\end{figure*}

In this section, we connect the observed spectral characteristics to the underlying physical picture of LRDs, using \obja\ as a case study.  \obja\ is the most extreme object in our sample, distinguished by its prominently V-shaped continuum and high-EW absorption features in the optical spectrum. These characteristics provide important insights into the nature of LRDs. We extend the insights gained from \obja\ to the broader LRD population in \S\ref{sec:cartoon}. Although the extreme metal absorbers seen in \obja\ are not observed in the other two objects, we attribute this diversity to the effects of geometry, viewing angle and metallicity (see \S\ref{sec:cartoon}).

Hereafter, we nickname \obja\ as the \emph{Egg}, inspired by its Ca-rich envelope, a temperature comparable to that of yellow supergiant stars, and its distinctive geometry. These features will be discussed in more detail below. An independent discovery and analysis of this object is also presented in \citep{Ji2025_lord}.

\subsection{Strong Metal Absorption: Evidence for a Cool Gas Envelope}\label{sec:j1025_absorber}

\subsubsection{Na D, \ion{K}{1}, and Ca T Absorption}

\egg\ exhibits exceptionally high-EW Na D $\lambda\lambda$5892, 5898, \ion{K}{1} $\lambda\lambda$7667, 7701,  and Ca triplets ($\lambda$8500, $\lambda$8544, $\lambda$8665; Ca T) absorption features in its optical continuum. We show their profiles in Figure \ref{fig:J1025_metal_absorber} and summarize their properties in Table~\ref{tab:J1025_metal_abs}.

\begin{table}[h!]
    \centering
    \begin{tabular}{c|ccc}
    \hline
    Absorber & EW & $\Delta v$ & FWHM \\
    & (\AA) & (km\,s$^{-1}$) & (km\,s$^{-1}$) \\ 
    \hline
       Na D doublets  & $6.6\pm0.4$  & $-98\pm16$ & $337\pm20$ \\
       \ion{K}{1}  doublets &  $3.1\pm0.1$ & $-101\pm9$ & $296\pm20$ \\
       \ion{Ca}{2} triplets & $16.7\pm0.5$ & $-61\pm7$ &$441\pm18$\\
       \hline
    \end{tabular}
    \caption{The properties of metal absorbers in \egg\ (\obja). The EWs are rest-frame equivalent widths summed over all absorption lines for each ionic species. 
    }
    \label{tab:J1025_metal_abs}
\end{table}

The Na D absorption in \egg\ has an EW of $6.6 \pm 0.4$\,\AA\ and \ion{K}{1} absorption shows an EW of $3.1 \pm 0.1$\,\AA. These two neutral alkali metals can be easily ionized by photons with low ionization potentials (5.14 eV for Na and 4.34 eV for K). This implies that the absorbing gas must be both cool and dense to preserve its neutral state. 
One possible origin for these absorbers is the cold, self-shielded intervening gas in the interstellar medium (ISM), with a minor contribution from late-type stars. However,  the observed Na D EW in \egg\ exceeds those typically found in such environments. The Na D EW in \egg\ is higher than that observed in most galaxies and AGNs \citep[typically $\lesssim 2$\AA;][]{Chen2010, Cazzoli2016,Rupke2021, Sun2024}, and is only comparable to a few extreme IR-luminous LINERs \citep{Rupke2005}.

Likewise, the \ion{K}{1} EW is orders of magnitude larger than those found in the Galactic interstellar medium or in circumstellar gas clouds around young stellar objects, where values typically range from a few up to $\sim$200\,m\AA\ \citep{Hobbs1975, Welty2001, Pascucci2015}. 

\begin{figure}[htbp]
    \centering
    \includegraphics[width=\linewidth]{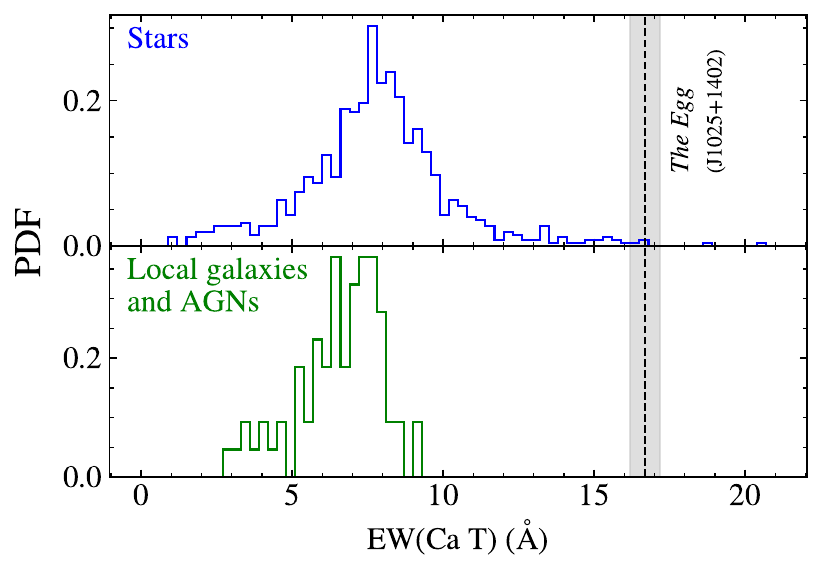}
    \caption{The Ca T absorption EW of \egg\ (\obja) compared to those typically observed in stars, galaxies, and AGN host galaxies in the local Universe. The stellar Ca T absorption EWs are compiled from \cite{Mallik11997, Cenarro2001}, and the Ca T absorption EWs of galaxies and AGNs are compiled from \cite{GR2005}. The Ca T absorption EW and its uncertainty are shown as the black dashed line and gray shaded region.}
    \label{fig:J1025_CaT_comparison}
\end{figure}

The Ca T absorption at 8500, 8544, and 8665 \AA\ in \egg\ has an EW of $16.7 \pm 0.5$\,\AA. In Figure \ref{fig:J1025_CaT_comparison}, we compare this value to typical Ca T EWs observed in stars, star-forming galaxies, and AGNs. The Ca T EW in \egg\ exceeds those of most stars and all known galaxies and AGNs, where Ca T EWs are generally below 10\,\AA.

Only a few individual stars, such as yellow hypergiants or supergiants \citep{Mallik11997, Cenarro2001}, can exhibit comparable Ca T EWs. However, reproducing the observed Ca T EWs in the integrated galaxy light would require an unusually exotic stellar population composed solely of these stars.

All the evidence above points to an origin that is neither stellar nor associated with the ISM. 
The distinct physical characteristics of these absorbers argue against a common origin within a single cloud. We discuss another possible origin in \S\ref{sec: absorber origin}.

\subsubsection{Other potential absorption lines}

\begin{figure*}[htbp]
    \includegraphics[width=\textwidth]{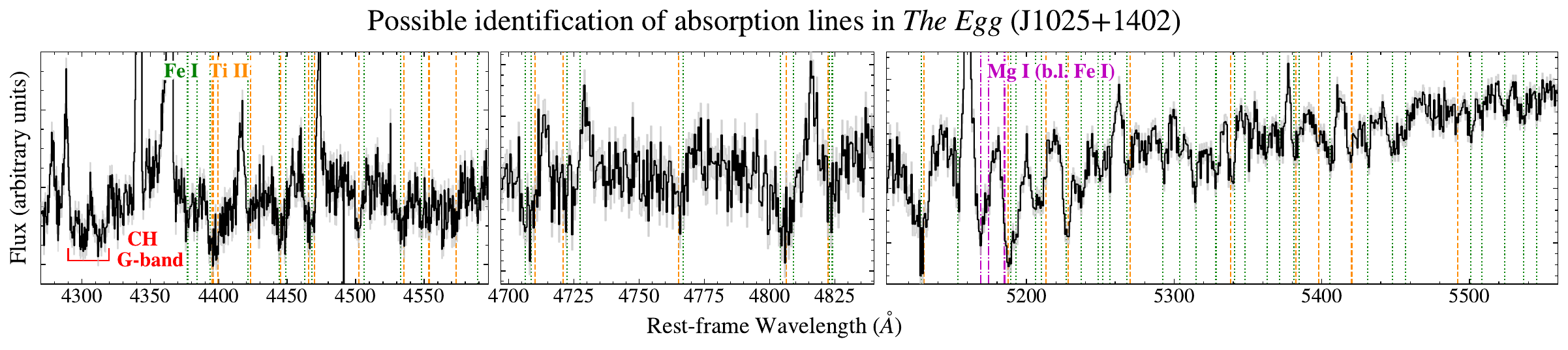}
    \includegraphics[width=\textwidth]{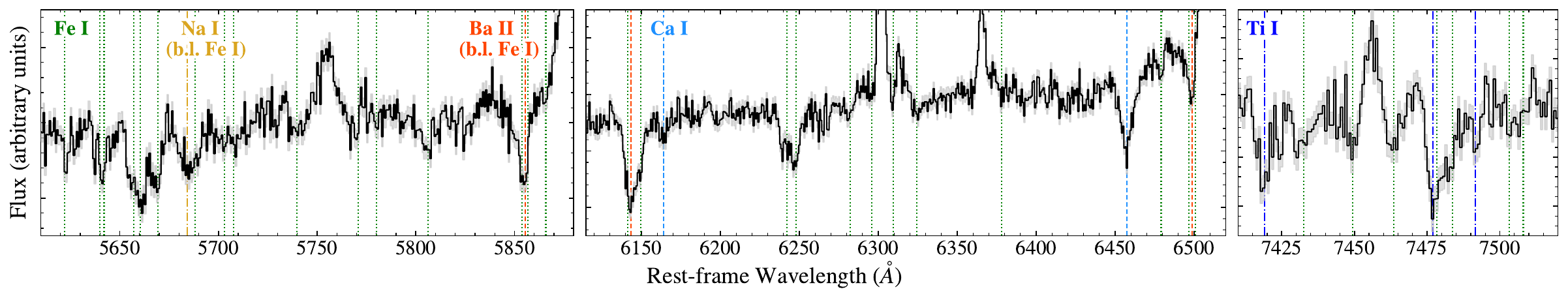}
    \caption{Possible identification of absorption lines in \egg\ (\obja). The CH G-band absorption is highlighted in red; the \ion{Fe}{1} absorption series is marked with green dashed lines; the \ion{Ti}{2} series with orange dashed lines. The absorption lines of \ion{Mg}{1}, \ion{Na}{1}, and \ion{Ba}{2}, blended with \ion{Fe}{1}, are shown in magenta, gold, and orange-red, respectively. The \ion{Ca}{1} and \ion{Ti}{1} lines are labeled with light blue dashed lines and blue dash-dotted lines, respectively.  }
    \label{fig:J1025_absorption_panel}
\end{figure*}

\xj{
We identify additional weak absorption features in the optical spectrum of \egg. We cross-matched the visual spectral troughs against three references: the stellar absorption-line list of \cite{Struve1934}, the solar spectrum line list of \cite{Moore1966}, and the theoretical line catalog from the \textsc{vald} stellar database \citep{Piskunov1995,Ryabchikova2015}. As shown in Figure \ref{fig:J1025_absorption_panel}, these spectral troughs likely correspond to low-ionization absorbers of \ion{Fe}{1}, \ion{Ti}{2}, \ion{Mg}{1}, \ion{Na}{1}, \ion{Ba}{2}, \ion{Ca}{1}, and \ion{Ti}{1}. Some of the absorption features are also reported in \cite{Ji2025_lord}. These species all have low ionization potentials: 7.9 eV for \ion{Fe}{1}, 13.6 eV for \ion{Ti}{2}, 7.6 eV for \ion{Mg}{1}, 5.1 eV for \ion{Na}{1}, 10.0 eV for \ion{Ba}{2}, 6.1 eV for \ion{Ca}{1}, and 6.8 eV for \ion{Ti}{1}. These absorbers are commonly observed in cool G-to-K stars \citep[e.g.,][]{Heiter2021}. Additionally, CH G-band absorption is present, which was also highlighted in \cite{Ji2025_lord}. This feature is characteristic of G–K type stars and traces cool, neutral regions that favor carbon-hydrogen molecular formation.
}

\subsubsection{The Origin: a Cool Gas Envelope around BHs}\label{sec: absorber origin}

\begin{figure*}[t]
    \centering
    \includegraphics[width=\linewidth]{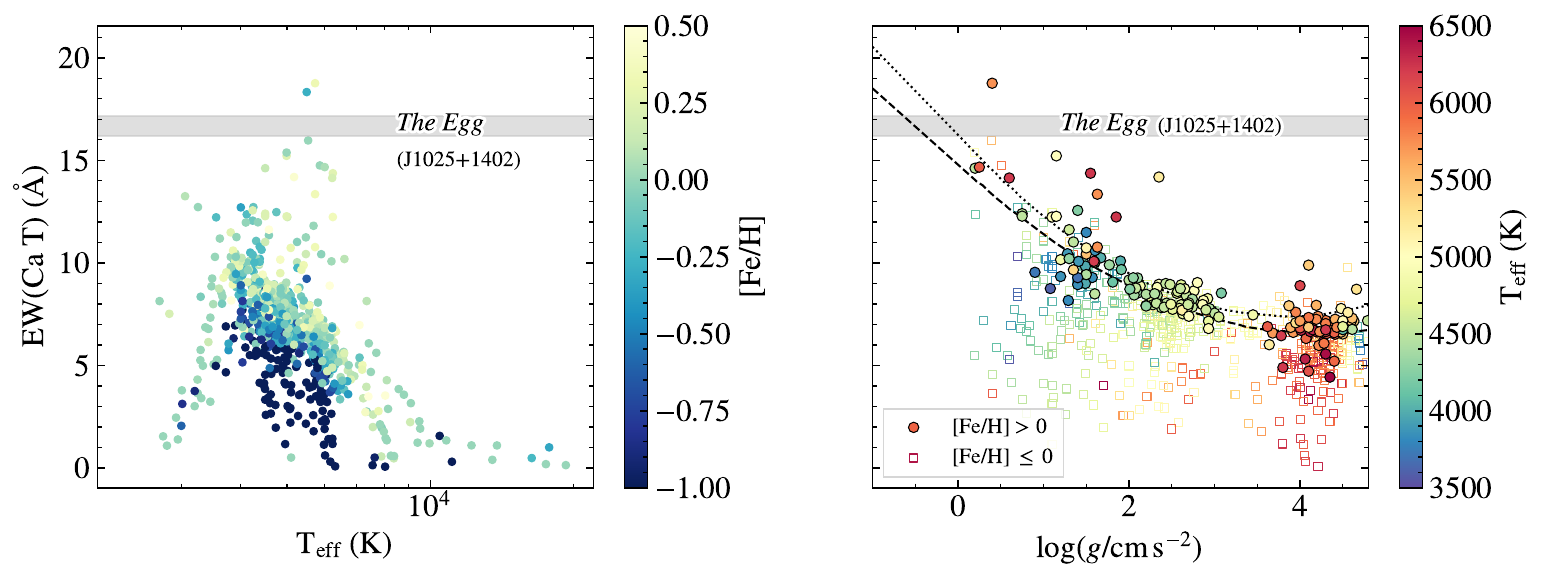}
    \caption{The \caT\ absorption of \egg\ (\obja) in the context of stellar \caT\ absorption.  \textit{Left}: EWs of stellar Ca T absorption as a function of the effective temperature ($T_{\rm eff}$). The data points are color-coded by the metallicity of the stellar atmosphere. The range of \egg's Ca T absorption within $\pm 1\sigma$ is indicated by the gray-shaded region. \textit{Right:} EWs of stellar Ca T absorption as a function of the surface gravity ($\log g$). The data points are color-coded with $T_{\rm eff}$. Metal-rich stars ([Fe/H]$>0$) are shown as the filled circles, and the metal-poor ones ([Fe/H]$\leq0$) are shown as the open squares. The empirical relations between $\log g$ and Ca T EW from \cite{Cenarro2002} are shown: the dashed line corresponds to the relation for stars with $-0.25 < {\rm [Fe/H]} < 0.25$, and the dotted line represents that for stars with ${\rm [Fe/H]} > 0.25$.
    }
    \label{fig:J1025_CaT_stellar}
\end{figure*}

\xj{Motivated by the unusually high-EW Na D, \ion{K}{1}, and Ca T absorption, along with the simultaneous presence of other cool-star-like low-ionization absorptions in \egg, we propose that these absorption lines originate from an atmosphere-like gas envelope surrounding the central BH.} The microphysics of line formation from stellar atmospheres may likewise apply to this atmosphere-like gas envelope. A comparison with stellar spectra, therefore, enables us to infer the physical conditions of the absorbing gas.

In the following, we investigate the EW of Ca T absorption as a key tracer.
Figure \ref{fig:J1025_CaT_stellar} places \egg's CaT absorption in the context of stellar atmospheres. As shown in the left panel of Figure~\ref{fig:J1025_CaT_stellar}, within the stellar catalog presented here \citep{Mallik11997, Cenarro2001}, only two metal-rich stars 
exhibit Ca T EWs greater than that of \egg. These two stars are classified as a G0 Ia yellow supergiant and a G4 0–Ia yellow hypergiant, with effective temperature ($T_{\rm eff}$) of 5500 and 5727 K, surface gravities ($\log g$) of 0.0 and 0.4, and metallicities [Fe/H] of $-0.28$ and $0.32$, respectively.
 
The yellow supergiant and hypergiant stars exhibit the strongest Ca T absorption due to their low surface gravity and their effective temperatures of $T_{\rm eff}\sim5000$~K. This temperature allows Ca in the atmosphere to exist mainly in the singly-ionized state. In stellar atmospheres, the EWs of Ca~T absorption strongly anti-correlate with surface gravity \cite[e.g.,][]{Mallik11997,Cenarro2002}.  
This is due to the gas opacity changing with the photosphere density, which generally increases with $\log g$ in stellar atmosphere models.  
As $\log g$ decreases, the photosphere density decreases, leading to a diminished continuum absorption opacity (the main source being H$^-$ ions at $T\sim5000{\rm~K}$ and at wavelengths around $8000$--$9000$\,\AA), while the line opacity of \ion{Ca}{2} changes only mildly \citep{Jorgensen1992}. This enhances the contrast between the line and continuum opacities, increasing the EWs of the Ca T absorption.  Thus, the Ca~T EWs are essentially atmospheric density indicators at a given effective temperature.


We interpret the Ca T absorption in \egg\ as originating from an atmosphere-like gas envelope surrounding the BH, with a structure and opacity analogous to that of cool stellar atmospheres. We apply stellar Ca T diagnostics to estimate $\log g$ to first order. The right panel of Figure \ref{fig:J1025_CaT_stellar} demonstrates the stellar Ca T EWs as a function of $\log g$. The Ca T EW–$\log g$ relation also depends on stellar metallicity. Extrapolating the empirical Ca T EW-$\log g$ relation for metal-rich stars with [Fe/H]$>$0.25 to match the observed Ca T EW of \egg\ yields an estimated $\log g$ of approximately $-0.2$. If we assume a slightly lower metallicity (${\rm [Fe/H]}\approx-0.25$ to $0.25$), the inferred $\log g$ would be about $-0.4$. The value may be even lower given the low metallicity we have inferred from the NLR (\S\ref{subsec:BPT_metallicity}).

In conclusion, the Ca T absorption in \egg\ indicates the presence of a metal-rich cool gas envelope with low surface gravity ($\log g < 0$). Comparison with known supergiant and hypergiant stars suggests that the $T_{\rm eff}$ of such an atmosphere is likely around 5000--6000 K. However, we emphasize that this is a first-order approximation with simplified assumptions; the actual physical conditions of the absorbing gas may differ substantially from those in stellar atmospheres. In particular, the mapping between $\log g$ and the stellar atmosphere density (and hence opacity) is based on hydrostatic equilibrium, which is not necessarily true for the envelope of \egg. Nevertheless, as we argue above, the line formation process mainly depends on the atmospheric temperature and density. The inferred value of $\log g$ above may not reflect the dynamical condition of the envelope, but does suggest that its atmosphere density may resemble that of very low-surface gravity stars. Standard stellar photosphere relations imply that $\log g=0$ at $T_{\rm eff}=5000$~K corresponds to a photosphere density of $3\times10^{-10}{\rm~g~cm^{-3}}$ \citep[e.g.,][their Equations~(4.56)(4.65)]{Hansen2004}. Thus, we expect the photosphere density of \egg\ to be on the order of or lower than $10^{-10}{\rm~g~cm^{-3}}$.

\xj{We note that these metal absorbers are offset by $-100$ to $-60$ km s$^{-1}$ from the NLRs, with FWHMs of approximately 300–440 km s$^{-1}$ (Table \ref{tab:J1025_metal_abs}). It is unclear whether the offset arises  from a velocity shift of the NLRs relative to the systemic redshift of the AGNs, or from the expansion of the gas envelope. Velocity offsets of tens to several hundred km s$^{-1}$  between NLRs and AGN systemic redshifts are common in type-1 AGNs, driven by various dynamical and radiative processes \citep{Bae2014, Shen2016}. Meanwhile, the cool gas envelope itself is turbulent and dynamically complex.} The FWHMs are significantly broader than those typically observed in stellar atmospheres. For instance, Ca T absorption lines in stars are intrinsically narrow, with FWHMs reaching only up to $\sim$70 km\,s$^{-1}$ in supergiant stars \citep{Filippenko2003}. A more detailed modeling of the gas properties,  \xj{dynamical state, and the stability of the envelope under the BH accretion activity} is necessary for a comprehensive understanding.

We have limited our analysis to the Ca T lines; similar atmospheric conditions are presumably responsible for the other absorption lines identified here. However, we do not perform the same analysis on them due to our unawareness of dedicated stellar datasets on these lines. Future characterization of observed stellar spectral lines, combined with analysis of theoretical spectral libraries, can directly test our scenario.

\subsection{Optical to IR Continuum: Evidence of Thermalized Emission}\label{sec:opt_ir_continua}

\begin{figure*}[t]
    \centering
    \includegraphics[width=1\linewidth]{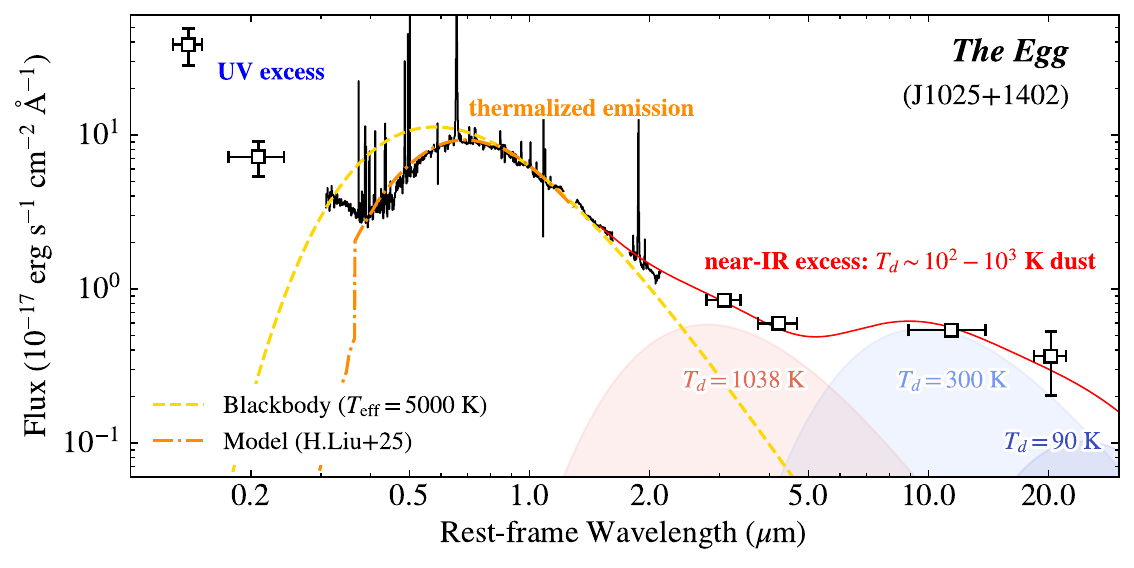}
    \caption{The SED of \egg\ (\obja) in comparison with a 5000 K blackbody spectrum and the theoretical model from \cite{Liu2025}. The dust components are indicated by the shaded regions in the near-IR wavelength range. }
    \label{fig:J1025_BB_comparison}
\end{figure*}

As discussed in \S\ref{sec:j1025_absorber}, we find evidence for cool gas envelopes that may be analogous to (supergiant or hypergiant) stellar atmospheres at $\sim$5000 K. In this section, we investigate the potential application of atmosphere-like gas envelopes to explain the observed continuum shape of the local LRDs. A similar comparison conducted by \citet{deGraaff2025} also noted a resemblance to supergiant stars, though with notable discrepancies.  In Figure~\ref{fig:J1025_BB_comparison}, we compare the spectrum of \egg\ to a 5000 K blackbody. The \xj{$\sim8000$\,\AA--1.5\,\micron\  regime of} \egg's spectrum can be reasonably well described by the Rayleigh-Jeans tail of a 5000 K blackbody. However, the $<8000$\,\AA\ region shows a significant flux deficit relative to the blue side of the blackbody spectrum, even without subtracting the extrapolation from the UV continuum as a separate component. Varying the blackbody temperature does not yield a better match, as the optical-to-near-IR continuum of \egg\ exhibits stronger curvature than a standard blackbody. 
Such a pseudo-blackbody continuum shape signifies thermalized emission, but may involve intricate physical mechanisms. 

Theoretical models have proposed that envelope-like structures can form around BHs under super-Eddington accretion and emit thermal radiation  
\citep{Meier1982a, Meier1982b, Zhou2019, Kido2025, Liu2025}. To investigate such scenarios, we compare the observed continuum of \egg\ with the models from \cite{Liu2025}, which assume an atmosphere formed by quasi-spherical super-Eddington accretion flows. As shown in Figure \ref{fig:J1025_BB_comparison}, one such model from \cite{Liu2025} provides a good match to the optical continuum shape of \egg. This model has a total luminosity $L=2.1\times10^{43}{\rm~erg~s^{-1}}$ and a density of $\rho_{\rm ph}=3.5\times10^{-12}{\rm~g~cm^{-3}}$ at the photosphere \footnote{$\rho_{\rm ref}=1.2\times10^{-12}{\rm~g~cm^{-3}}$ in their notation.}; a star at $T_{\rm eff}=5000{\rm~K}$ would have this photosphere density at $\log (g/\mathrm{cm\,s^{-2}})\sim-2.8$ \citep{Hansen2004}.  
The photosphere (defined by the Rosseland-mean optical depth) is located at 510 AU with $T_{\rm eff}= 4761$~K. The frequency-dependent gas optical depth effects suppress the UV-to-optical flux, making the predicted continuum deviate from a blackbody \footnote{We invoke no dust reddening to the model curve; we have checked that adding a modest level of dust extinction ($A_V\lesssim1$~mag using the SMC dust extinction law) would further improve the agreement.}. Essentially, the optical spectrum is ``reddened'' by the atmospheric gas; otherwise, we would have to invoke a very steep dust extinction law to explain the optical redness while retaining the declining IR slope (see Appendix~\ref{sec:appendix_sed_fit}). However, we caution that the model of \cite{Liu2025} assumes super-Eddington accretion, while current observations do not provide direct evidence of it. The integrated $L_{2000\,\text{\AA}–20\,\mu\mathrm{m}}$ corresponds to about 10\% of the Eddington luminosity, based on the $M_{\rm BH}$ listed in Table~\ref{tab:Ha_properties}. The true Eddington ratio is unconstrained, as both $M_{\rm BH}$ and the intrinsic bolometric luminosity are uncertain. In addition to the theoretical model, we present an alternative empirical parameterization of the pseudo-blackbody continuum shape in Appendix~\ref{sec:appendix_sed_fit}.

The origin of the blue UV continuum  \xj{($\lambda\lesssim4000$\,\AA)} remains uncertain.  
Several observations suggest a stellar origin, \xj{rather than an AGN one}, from young host galaxies, based on the lack of UV variability, the absence of high-ionization UV lines, and the extended, complex UV morphologies in a number of high-redshift LRDs \citep{Rinaldi2024, Akins2025, Tee2025, Naidu2025, Zhuang2025}. However, these features do not definitively rule out an AGN origin, either directly from the accretion disk or scattered by electrons/dust \citep{Lambrides2024, Li2025, Inayoshi2025b}. This is especially the case when considering the recent detection of AGN-powered high-ionization UV lines in a subset of LRDs \citep{Tang2025}. Recently, \cite{Rinaldi2025} found that a V-shaped continuum, similar to those observed in LRDs, originates from the nuclei of a cosmic-noon IR-bright galaxy. \xj{This finding further suggests that AGN emission could contribute to the blue UV continuum, although a circumnuclear starburst on unresolved scales could still play a role.} At present, we lack spectral information at rest-frame wavelengths below 2000\,\AA\ to map diagnostic UV lines that could distinguish between AGN and star-forming activity \citep[e.g., \ion{C}{4}, \ion{He}{2}, \ion{N}{5}; ][]{Feltre2016, Gutkin2016, Mingozzi2022}. As such, the origin of this component remains inconclusive given the current data. Deep UV spectroscopy or polarization measurements, both of which could trace AGN light, are needed to clarify its nature.

\subsection{The Dust Emission}\label{sec:dust}

The WISE photometry of local LRDs suggests that the dust content in these objects, while not as strong as implied by a standard AGN dust torus,  is also not negligible. A typical AGN dust torus, when scaled to LRD luminosities, would show rapidly increasing emission beyond 1–2\,\micron, peaking around 10\,\micron, which is clearly not the case for LRDs \citep{Stalevski2016, BWang2025}.
 All three local LRDs show flattening or an increase in flux beyond rest-frame 10\,\micron\ (W3 and W4 bands; Figure~\ref{fig:SED}).
  At $z > 3$, this wavelength range is beyond JWST coverage, and it is either only partially covered by Spitzer and Herschel or falls into gaps between their coverage, leaving it largely unconstrained for high-redshift LRDs \citep{Akins2024, Casey2024, Casey2025, Setton2025}.

Qualitatively, the IR excess from the WISE photometry in these local LRDs (Figure~\ref{fig:SED}) reveals at least dust components with temperatures approximately between $10^2$ and $10^3$ K. The emission at rest-frame 3\,\micron\ corresponds to the peak of a $\sim1000$ K blackbody, and the emission at 10\,\micron\ matches a $\sim300$ K blackbody.  As shown in Figure~\ref{fig:J1025_BB_comparison},  if assuming templates from \citet{Lyu2021} that include dust components at $T_d = 2100$, 1038, 300, and 90 K (see Appendix~\S\ref{sec:appendix_empirical_fit}), the IR emission of \egg\ can be well reproduced by significant dust components at 1038 K and 300 K. However, the derived dust properties depend sensitively on the chosen templates.  We discuss alternative dust templates based on Haro 11 \citep{Lyu2016} and the degeneracies in Appendix~\ref{sec:appendix_haro11_dust}.  
In the absence of continuous near- to mid-infrared coverage and far-infrared constraints, it is challenging to robustly constrain the dust structures and grain properties in these systems.

Recent IR and sub-mm observations have placed \xjnew{first} constraints on the dust content of high-redshift LRDs \citep{Casey2024, Casey2025, Setton2025, Xiao2025, Li2025, Chen2025}. The weak IR emission observed in the three local LRDs in this work is consistent with both the detections and upper limits reported for high-redshift LRDs. Deep ALMA non-detections and stacking analyses indicate modest dust masses ($\lesssim 10^6 M_\odot$) and peak temperatures of 100–300 K \citep{Casey2025, Setton2025} for high-redshift LRDs. These inferred temperatures agree with the dust components detected in the three local LRDs at rest-frame wavelengths $>10$\,\micron.  For dust attenuation, \cite{Chen2025} placed stringent upper limits on high-redshift LRDs ($A_V \lesssim 1–2$), whereas other studies have inferred non-negligible values ($A_V \sim 2$) from broad-line Balmer decrements or continuum fitting \citep{Ji2025, D'Eugenio2025_z7p04, Taylor2025}, underscoring a tension in current estimates.  We do not attempt to precisely quantify the attenuation on the continuum or broad lines in this work. Proper modeling of dust attenuation in such atypical environments requires careful treatment of multiple factors, including the geometry and distribution of dust needed to maintain nearly dust-free NLRs, uncertainties in broad-line Balmer decrements, the survivability of dust grains in BLRs, and the validity of energy balance along the line of sight. We also emphasize the need for longer-wavelength follow-up observations of local LRDs, such as with  ALMA. These observations can yield more stringent constraints on the dust content than those currently available for their high-redshift counterparts, pending the arrival of next-generation IR facilities \citep[e.g., PRIMA,][]{Moullet2023}.

\section{A conceptual picture of LRDs}\label{sec:cartoon}

\begin{figure*}[htbp]
    \includegraphics[width=\linewidth]{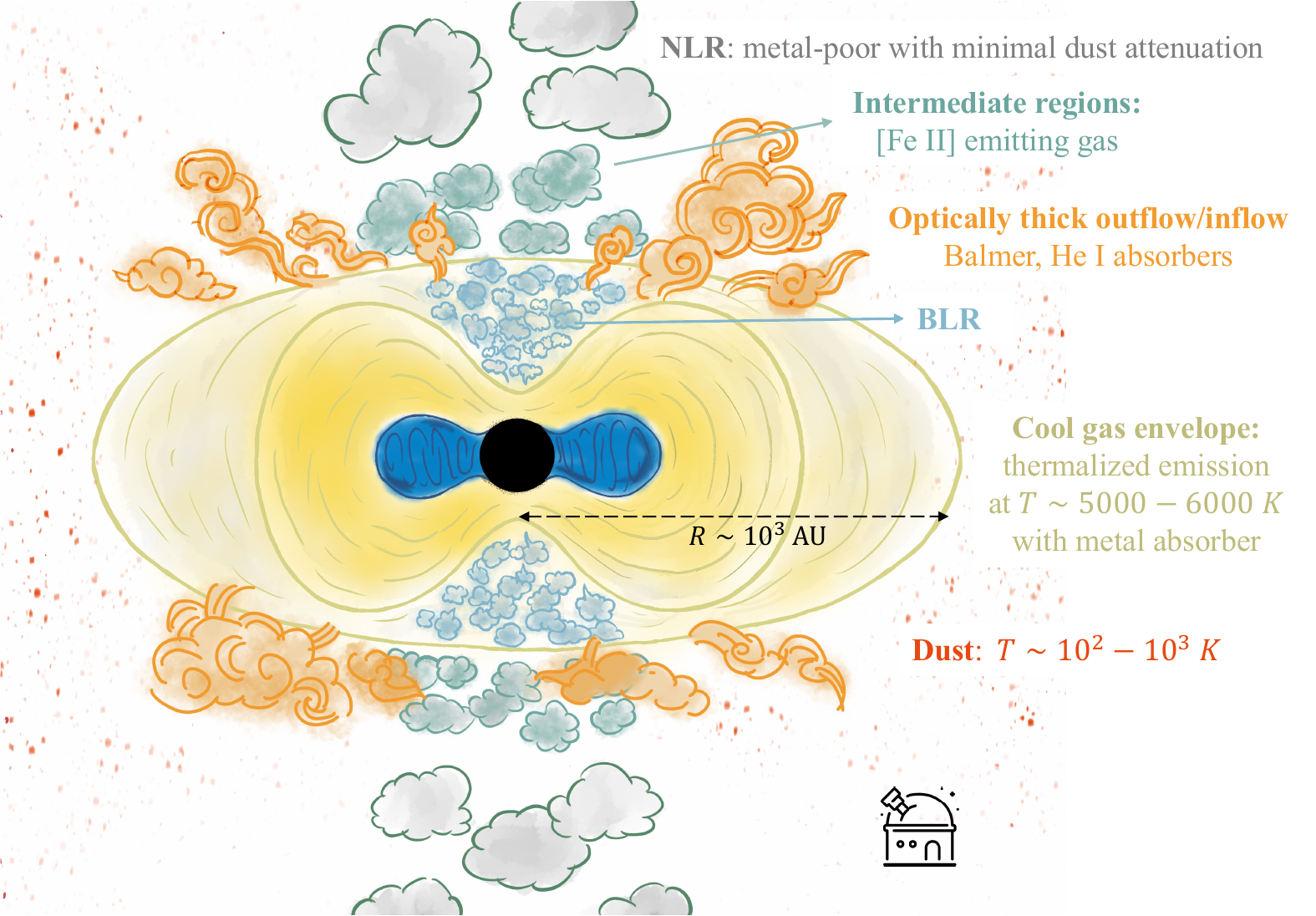}
    \caption{Schematic illustration of the structural components of LRDs. The viewing angle, indicated by the telescope icon, is randomly selected here for illustrative purposes; its impact on observed properties is discussed in the main text.}
    \label{fig:cartoon}
\end{figure*}

\begin{table*}[htbp]
    \centering
    \begin{tabular}{p{0.175\linewidth}|p{0.425\linewidth}|p{0.35\linewidth}}
    \hline\hline
    Physical Process & Observational Evidence & Caveat and prospects\\
    \hline\hline
    BH accretion & Broad emission lines (e.g., Balmer and Paschen series, infrared \ion{O}{1}, infrared \ion{Fe}{2}) with FWHM $\sim$ 1000\,km\,s$^{-1}$ & Optically thick gas envelope prevents direct observations of BH accretion process \xj{(e.g., emission directly from the accretion disks)}. \\

      \hline
    AGN Emission Line Regions & Narrow and broad line regions consistent with AGN ionization; [\ion{Fe}{2}] and \ion{Fe}{2} emission lines with FWHM and inferred density between broad and narrow components & Modeling of [\ion{Fe}{2}] and \ion{Fe}{2} is complex and potentially degenerate \\
    
    \hline 
    Cool gas envelope & Cool-star like metal absorption features, especially extremely high-EW Na D, \ion{K}{1}, and Ca T absorption lines in \egg; the thermalized emission in good agreement with theoretical atmosphere models \citep{Liu2025}  & \xj{Envelopes are dynamically complex, requiring models that capture absorption-line FWHMs and velocities, as well as detailed prescriptions of opacity, cooling, and circum-BH environments.} \\

    \hline 
    Non-spherical geometry and viewing angle & Optically thin channels within the envelope are required for AGN ionizing photons to escape and illuminate the BLRs, intermediate [\ion{Fe}{2}] regions, and NLRs; may also explain the diversity in continuum shapes and the presence or absence of high-EW metal absorbers.   & The impact of viewing angle is difficult to quantify without a statistically large sample. \\
    
    \hline
        Outflowing/inflowing optically-thick gas & Blueshifted Balmer and \ion{He}{1} absorption lines & Their correlation with other properties of LRDs (i.e., Balmer break strength, continuum shape) remains uncertain; the origin of \ion{He}{1} emission (ionized gas versus dense neutral gas) requires further study. \\
    \hline
    Dust emission & near- to mid-IR excess in WISE photometry (rest-frame 3-20\,\micron) & Constraints on dust components depend on model assumptions; cannot fully characterize dust properties from the AGN to the ISM  due to the lack of longer-wavelength data. \\

    \hline\hline
    \end{tabular}
    \caption{Summary of physical processes potentially contributing to the observed features, along with associated limitations and future prospects.}
    \label{tab:phys_process_summary}
\end{table*}

Based on the observational evidence and physical interpretations discussed above, we construct a conceptual picture of LRDs, as shown in Figure \ref{fig:cartoon}. Table \ref{tab:phys_process_summary} summarizes the proposed physical processes, their observational signatures, and the associated caveats in our interpretation. 
 
We first conduct an order-of-magnitude estimate of the characteristic physical scales that support the proposed framework. Assuming that the observed emission line broadening is gravitational in origin, the enclosed mass is given by $M \approx v^2 r / G$. For a broad emission line with FWHM of 1000\,km\,s$^{-1}$, and adopting a low surface gravity atmosphere with $\log (g/\mathrm{cm\,s^{-2}}) = 0$ as implied by the strong Ca T absorption, we derive
\begin{equation}
M \approx \frac{v^4}{Gg} \approx 8 \times 10^5\, M_\odot,
\qquad
r \approx \frac{v^2}{g} \approx 700\,\mathrm{AU}.
\end{equation}
In the calculation above, we use the broad emission lines as velocity indicators, rather than the absorption line widths, and place the BLRs near the edge of the photosphere. The kinematics of the BLR clouds are assumed to be primarily governed by gravity, allowing for a reasonably simplified estimate based on Kepler's laws. In contrast, the broadening of metal absorption lines in the atmosphere is more complex, influenced by intricate  
processes such as anisotropically distributed turbulence, making their kinematics more difficult to interpret.

The total luminosity of a blackbody radiation with a temperature of 5000 K (Figure \ref{fig:J1025_BB_comparison}) and a radius of 700 AU is about $5\times 10^{43}$ erg s$^{-1}$. This estimated luminosity is consistent within an order of magnitude with the observed luminosities ($L_{2000\text{\AA} - 20\,\mu{\rm m}}$, Table \ref{tab:basic_properties}) and the luminosities of the theoretical models (Figure~\ref{fig:J1025_BB_comparison} and \ref{fig:sed_bb_two}) for the three local LRDs.  For comparison, the corresponding Eddington luminosity for a 8$\times 10^5\, M_{\odot}$ BH is 10$^{44}$ erg s$^{-1}$.
The BH mass estimated here is up to an order of magnitude smaller than the H$\alpha$-based BH masses listed in Table \ref{tab:basic_properties}.   Given the approximate nature of these estimates, the overall agreement between the observed properties and those inferred from this conceptual model supports the feasibility of an atmosphere-enshrouded BH model. The individual components of our conceptual framework are described in detail below.

\begin{itemize}
    \item \textbf{BLRs and NLRs}. The broad Balmer and Paschen lines, along with the infrared \ion{O}{1}$\lambda1.29$\,\micron\ emission lines (in \egg, see Figure \ref{fig:J1025_zoomin}), indicate that the observed line broadening arises from gas motions within the BLRs surrounding the BHs. Analysis of the Balmer decrements, BPT diagrams, and metallicity suggests that the narrow emission lines originate from AGN-powered NLRs with low metallicity and minimal dust. \xj{The high decrements of broad Balmer lines (\ha/\hb $\gtrsim 10$) imply either high-density gas with complex radiative transfer effects or atypical dust properties in BLRs.
    }

    \item \textbf{[\ion{Fe}{2}]-emitting gas intermediate between the BLRs and NLRs}. The AGN-excited [\ion{Fe}{2}] (or \ion{Fe}{2})  emission lines indicate gas with a density of approximately $n_{\rm H} \approx 10^{7.5}$ cm$^{-3}$, intermediate between the BLRs and NLRs. The FWHM of the [\ion{Fe}{2}] lines ( $\sim$100-300\,km\,s$^{-1}$) is likewise between that of the broad ($\sim1000$\,km\,s$^{-1}$) and narrow lines ($\sim65$\,km\,s$^{-1}$). \xj{Determining whether these regions can emit other lines requires deeper, higher-resolution spectroscopy to accurately decompose broad lines, such as those in the Balmer and Paschen series, or to detect fainter features.} The absence of these [\ion{Fe}{2}] lines in \objb\ may be attributed to a lower covering fraction of the intermediate emitting regions.

    \item \textbf{A cool gas envelope with $T\sim$ 5000--6000\, K and photosphere density of $\rho_{\rm ph}\lesssim10^{-10}{\rm~g~cm^{-3}}$ ($\log(g/{\rm cm\,s^{-2})} \lesssim 0$)}.  The optical–IR continua of the three local LRDs exhibit pseudo-blackbody shapes indicative of thermalized emission and are in good agreement with theoretical atmosphere models \citep{Liu2025}. The presence of extremely high-EW metal absorbers (Na D, \ion{K}{1}, Ca T) in \egg, along with other potential G-to-K star-like low-ionization absorbers (Figure~\ref{fig:J1025_absorption_panel}), further indicates a cool, optically thick gas envelope surrounding the BH.  Analysis of the Ca T EW, by analogy with stellar atmospheres, suggests an atmospheric temperature of approximately 5000–6000 K and surface gravity of $\log(g/\mathrm{cm\,s^{-2}}) \lesssim 0$ at the photosphere, corresponding to a density of $\rho_{\rm ph} \lesssim 10^{-10}\,\mathrm{g\,cm^{-3}}$. The size of this envelope is on the order of $10^3$ AU as inferred from the optical-IR luminosity and the effective temperature. A consequence of an optically thick cool gas envelope is that the BH accretion process, i.e., the emission from the accretion disk itself, is not directly observable. \xj{The origin of the UV emission remains inconclusive with the current data, and we do not attempt to address it within this framework. It could originate either from stars in the host galaxy or from AGN light scattered by the surrounding gas. We leave a more thorough investigation of its nature to future work and observations.}

      \item \textbf{Non-spherical geometry and role of the viewing angle.} The observed AGN-powered emission lines imply that the BLRs, intermediate regions, and NLRs cannot be located entirely outside the gas envelope. Otherwise, the $T\sim5000$–6000 K central source would not produce enough ionizing photons responsible for the recombination lines or the photo-ionization conditions required for forbidden lines (e.g., [\ion{Fe}{2}], [\ion{N}{2}], \ion{O}{1}). Therefore, there should be a relatively optically thin channel that allows high-ionization photons to escape from the disk and gas envelope.  We attribute such an optically thin channel along the direction of the line-emitting regions. This non-spherical geometry allows emission lines to form while preserving the envelope structure.

      The viewing angle in this non-spherical geometry can explain the diversity in continuum shapes and the presence or absence of metal absorption features among different objects. Changes in viewing angle can modulate the relative contributions of thermalized emission and AGN photoionized radiation. It also determines whether the line of sight intersects the metal-absorbing layers within the cool gas envelope.  For example, the line of sight toward \egg\ may intersect the optically thick part of the envelope with high metal-line opacity. This orientation results in a significant contribution to the continuum from the thermalized emission, making the metal absorbers observable. \objc\ could be oriented closer to the optically thin region, where the optical depth of metal lines is lower. \objb\ could lie even closer to the optically thin region, leading to a reduced contribution from the thermalized emission to the total optical continuum.  
      However, we emphasize that this is a speculative scenario, and a larger statistical sample is needed to better understand the roles of geometry and viewing angle. Differences in the metallicities of the cool gas envelope would also contribute to the diversity of observed metal absorption features.

 \item \textbf{Optically thick outflows/inflows}. In most high-redshift LRDs, the Balmer absorption lines are blueshifted, with only a few cases showing redshifted \ha\ absorption with minimal velocity offsets \citep{Lin2024, Ji2025, DEugenio2025}. Up to now, all detected \ion{He}{1} absorbers in LRDs are blueshifted \citep{Juodzbalis2024, Wang2025, Loiacono2025}. The Balmer absorbers are superimposed on broad Balmer emission lines, and the \ion{He}{1} absorbers are superimposed on both the broad emission and the underlying continuum. We thus attribute these absorbers to optically thick outflowing gas located either within or outside the BLRs. Balmer absorption arises from high-density neutral gas. \ion{He}{1} absorbers may originate from either high-density neutral gas clouds \xjnew{responsible for the Balmer absorption} or \xjnew{partially ionized} gas clouds with lower density. The rare instances of redshifted Balmer absorption with small velocity shifts, \xj{like those seen in \egg\ and \objc,} may indicate inflows, which could represent accreting material or gas cooling and moving back toward the nucleus. \xj{The relationship between these absorptions and other properties of LRDs (e.g., Balmer break strength,  continuum shape) is not yet established. However, since their presence can be attributed to high-density hydrogen ($n_{\rm H} > 10^9$ cm$^{-3}$; \citealt{Inayoshi2024}), there may be plausible connections that merit further investigation.}

    \item \textbf{Dust at $10^2$--$10^3$ K}. The WISE photometry of the three local LRDs is consistent with dust components with temperatures ranging from $10^2$ to $10^3$ K. These dust components, cooler than typical AGN-heated hot dust but warmer than ISM dust, are likely located outside the $\sim$5000 K atmosphere yet still within the AGN vicinity. However, we note that the analysis of the dust components is highly uncertain, model-dependent, and subject to degeneracies. Due to the lack of observational constraints, the exact dust geometry and grain properties remain unknown. Future modeling efforts with improved IR observations are needed.
\end{itemize}

In the proposed scenario, the BH resides in an environment distinct from that of typical type-1 AGNs. As discussed in \S\ref{sec:overview_spectral_properties}, it remains unclear whether the empirical BH mass estimators calibrated for type-1 AGNs are valid for LRDs, given their potentially different physical conditions. Assuming a BH mass of $M_{\rm BH} \sim 10^6$--$10^7\,M_\odot$, the corresponding Eddington luminosity is $10^{44}$--$10^{45}$\,erg\,s$^{-1}$. For the three objects, if we take the total integrated luminosity $L_{2000\text{\AA}–20\,\mu{\rm m}}$, assuming that the UV continuum is AGN light, the Eddington ratio would be $\sim 0.1–0.3$.  The current observations do not provide sufficient evidence to determine whether LRDs are in a super-Eddington accretion mode, as proposed by some theoretical scenarios \citep[e.g.,][]{Inayoshi2024a, Pacucci2024, Madau2025}. The accretion rate of the central BH and its disk can significantly influence the geometry of both the disk and the surrounding gas envelope \citep[e.g.,][]{Dotan2011}. We leave a detailed exploration of this to future work.

%% file: 06_Summary.tex
\section{Summary}\label{sec:summary}

We searched the SDSS spectroscopic database for local LRDs using high-redshift LRD templates and identified three LRDs at $z = 0.1$–$0.2$. \xj{Their discovery suggests a conservative lower limit for LRD number density at $z<0.5$ of $\sim 5\times10^{-10}\,\mathrm{Mpc}^{-3}$.} We followed up these sources with MMT/Binospec, LBT/MODS, and Magellan/FIRE spectroscopy, which reveal detailed spectral features that probe the physical nature of these local LRDs. Our findings are summarized below.

\begin{itemize}
    \item The three local objects closely resemble the high-redshift LRDs. They display compact morphologies ($R_e\lesssim$0.85-1.6\,pkpc). HST narrow-band imaging of \objb\ reveals that both the continuum and the \ha\ line emission are AGN-dominated, with the compact yet clumpy host galaxy ($R_e\approx568$\,pc) contributing a minor fraction to the continuum. All three objects exhibit clear V-shaped UV-to-optical continua with inflections near the Balmer break, and a decreasing IR continuum from $\sim7000$\,\AA\ to 2\,\micron. All of them show broad \ha\ emission with FWHMs$\sim 1000$\,km\,s$^{-1}$, along with broad Paschen and \ion{He}{1} lines. All of them exhibit blueshifted \ion{He}{1} $\lambda$10833 absorbers; two of them show \ha\ absorbers, and one even shows a redshifted \hb\ absorber without an \ha\ counterpart. None of the objects shows significant optical or IR variability. Two were targeted in hard X-rays but were either not detected or detected with low statistical significance.

    \item The Balmer and Paschen decrements, covering from \ha\ to H12 and from Pa$\alpha$ to Pa12 (or Pa9 for one object), indicate negligible dust attenuation in the NLRs, \xj{consistent with Case B recombination.} The broad \ha/\hb\ ratios are 8-34 in the three objects, higher than those observed in most local AGNs, indicating either unusually high-density gas or atypical dust properties in the BLRs. 
    
    \item The BPT diagram shows that the narrow emission lines are photoionized by AGNs, \xj{with the three LRDs occupying positions consistent with those reported by \citet{Juodzbalis2024}.} Their NLRs are metal-poor, with 12 + $\log$(O/H) $\approx$ 7.43–8.01.

    \item  Two of the objects exhibit prominent narrow [\ion{Fe}{2}] (or \ion{Fe}{2}) emission lines that are distinct from those typically seen in type-1 AGNs. Photoionization modeling suggests that the [\ion{Fe}{2}]-emitting gas is AGN-powered, with physical conditions characterized by $\log(n_{\rm H}/{\rm cm^{-3}}) = 7.5$, $\log(N_{\rm H}/{\rm cm^{-2}}) = 23$, $\log U = -2$, and an average temperature of $10^4$ K. Based on the inferred gas density and FWHMs, we propose that the [\ion{Fe}{2}]-emitting region lies at an intermediate scale between the BLRs and NLRs.

    \item  One of the local LRDs, \obja\ (\egg) exhibits exceptionally high EWs in Na D, \ion{K}{1}, and Ca T absorption lines, with  FWHMs of about 300–440 km\,s$^{-1}$ and no significant velocity shifts. Such strong metal absorbers are rarely seen in the ISM, galaxies, AGNs, or stars.  
    Its Ca T EW is matched only by that of two supergiant and hypergiant stars. \egg\ also exhibits potential low-ionization absorbers, such as \ion{Fe}{1}, \ion{Ti}{2}, \ion{Mg}{1}, \ion{Ba}{2}, etc., characteristic of G- to K-type stars. These extreme properties suggest a non-stellar and non-ISM origin. We interpret these absorption features in \egg\ as arising from a cool gas envelope around the BH, analogous to stellar atmospheres. Based on its Ca T EW, we infer $T_{\rm eff} \approx 5000$–$6000\,\mathrm{K}$ and $\log(g/\mathrm{cm\,s^{-2}}) < 0$ at the photosphere, corresponding to a density of $\rho_{\rm ph}\lesssim10^{-10}$\,g\,cm$^{-3}$.

    \item The optical-IR continua of these local LRDs exhibit pseudo-blackbody shapes. The shapes are consistent with theoretical models featuring atmospheres surrounding BHs \citep{Liu2025}, with characteristic temperatures of $\sim5000$ K. The WISE photometry indicates modest dust emission at temperatures of $10^2$–$10^3$ K. 

    \item  By combining the observational evidence, we propose a conceptual model for the structure of LRDs. In this scenario, BHs are embedded within a cool gas envelope that emits thermalized radiation, giving rise to the observed pseudo-blackbody continuum and metal absorption features. The Balmer and \ion{He}{1} absorbers are attributed to optically thick outflows launched from the envelope. The overall AGN structure consists of BLRs, an intermediate region that contains [\ion{Fe}{2}]-emitting gas, and NLRs with minimal dust attenuation. The geometry of these structures, along with the viewing angle, may account for the diverse observed features across different objects \xj{(i.e., the presence or absence of cool star–like absorption)}.

\end{itemize}

Local LRDs offer a valuable opportunity to investigate the physical nature of this population. The conceptual framework presented in \S\ref{sec:cartoon} provides one possible scenario to explain the observed features. However, open questions remain. Do LRDs, and the scenario illustrated in Figure~\ref{fig:cartoon}, represent a specific evolutionary phase of SMBHs? What is the accretion mode of these objects (super-Eddington or sub-Eddington)? Are they connected to conventional type-1 and type-2 AGNs? How do the physical conditions that give rise to LRDs evolve with redshift, and why are they so rare in the local Universe?  How do the BH and host galaxy properties of LRDs change over cosmic time? While answering these questions is beyond the scope of this paper, they are critical for understanding the broader context of BHs and merit future investigation.

Ongoing wide-field surveys, such as DESI and Euclid, along with upcoming missions such as the Prime Focus Spectrograph (PFS), LSST, and Roman Space Telescope, will open new windows for identifying LRDs at low redshift. These efforts will enable studies of their cosmic evolution and contributions to the history of SMBH growth. Large statistical samples will be essential for establishing their demographics. Future observations with high-resolution facilities such as HST and JWST will provide critical insights to test the proposed scenarios and directly probe the properties of their host galaxies.

%% file: 99_Acknowledgement.tex
We thank the anonymous referee for providing constructive comments. We thank Zuyi Chen, Anna de Graaf, Christina Eilers, Carl Fields, Yifei Jin, Raphael Hviding, Rob Kennicutt, Serena Kim, Rohan Naidu, George Rieke, Pierluigi Rinaldi, Hans-Walter Rix, David Setton, Jinyi Shangguan, Nathan Smith, Yang Sun, and Marianne Vestergaard for valuable discussions and comments. We thank Joe Shields and Joannah Hinz for approving LBT Directory Discretionary Time, Jenny Powers for her expert help with LBT observations, and Olga Kuhn for her help with the data reduction.  We thank Ben Weiner and Sean Moran for their assistance with the Binospec data reduction. We thank Cass Fan for help with the illustration.  X.L, X.F and F.W acknowledge support from the NSF grant AST-2308258. K.I acknowledges support from the National Natural Science Foundation of China (12573015, 1251101148, 12233001), 
the Beijing Natural Science Foundation (IS25003), and the China Manned Space Program (CMS-CSST-2025-A09).

Observations reported here were obtained at the MMT Observatory, a joint facility of the University of Arizona and the Smithsonian Institution. {The LBT is an international collaboration among institutions in the United States, Italy and Germany. LBT Corporation partners are: The University of Arizona on behalf of the Arizona university system; Istituto Nazionale di Astrofisica, Italy; LBT Beteiligungsgesellschaft, Germany, representing the Max-Planck Society, the Astrophysical Institute Potsdam, and Heidelberg University; The Ohio State University, and The Research Corporation, on behalf of The University of Notre Dame, University of Minnesota and University of Virginia.} This paper used data obtained with the MODS spectrographs built with
funding from NSF grant AST-9987045 and the NSF Telescope System
Instrumentation Program (TSIP), with additional funds from the Ohio
Board of Regents and the Ohio State University Office of Research.  

This paper includes data gathered with the 6.5 meter Magellan Telescopes located at Las Campanas Observatory, Chile.

This research uses services or data provided by the Astro Data Lab, which is part of the Community Science and Data Center (CSDC) Program of NSF NOIRLab. NOIRLab is operated by the Association of Universities for Research in Astronomy (AURA), Inc. under a cooperative agreement with the U.S. National Science Foundation.

This research is based on observations made with the Galaxy Evolution Explorer, obtained from the MAST data archive at the Space Telescope Science Institute, which is operated by the Association of Universities for Research in Astronomy, Inc., under NASA contract NAS 5–26555.

This work has made use of data from the European Space Agency (ESA) mission
{\it Gaia} (\url{https://www.cosmos.esa.int/gaia}), processed by the {\it Gaia}
Data Processing and Analysis Consortium (DPAC,
\url{https://www.cosmos.esa.int/web/gaia/dpac/consortium}). Funding for the DPAC
has been provided by national institutions, in particular the institutions
participating in the {\it Gaia} Multilateral Agreement.

This work is based in part on data obtained as part of the UKIRT Infrared Deep Sky Survey.

The Pan-STARRS1 Surveys (PS1) and the PS1 public science archive have been made possible through contributions by the Institute for Astronomy, the University of Hawaii, the Pan-STARRS Project Office, the Max-Planck Society and its participating institutes, the Max Planck Institute for Astronomy, Heidelberg and the Max Planck Institute for Extraterrestrial Physics, Garching, The Johns Hopkins University, Durham University, the University of Edinburgh, the Queen's University Belfast, the Harvard-Smithsonian Center for Astrophysics, the Las Cumbres Observatory Global Telescope Network Incorporated, the National Central University of Taiwan, the Space Telescope Science Institute, the National Aeronautics and Space Administration under Grant No. NNX08AR22G issued through the Planetary Science Division of the NASA Science Mission Directorate, the National Science Foundation Grant No. AST-1238877, the University of Maryland, Eotvos Lorand University (ELTE), the Los Alamos National Laboratory, and the Gordon and Betty Moore Foundation.

The Legacy Surveys consist of three individual and complementary projects: the Dark Energy Camera Legacy Survey (DECaLS; Proposal ID \#2014B-0404; PIs: David Schlegel and Arjun Dey), the Beijing-Arizona Sky Survey (BASS; NOAO Prop. ID \#2015A-0801; PIs: Zhou Xu and Xiaohui Fan), and the Mayall \textit{z}-band Legacy Survey (MzLS; Prop. ID \#2016A-0453; PI: Arjun Dey). DECaLS, BASS, and MzLS together include data obtained, respectively, at the Blanco telescope, Cerro Tololo Inter-American Observatory, NSF’s NOIRLab; the Bok telescope, Steward Observatory, University of Arizona; and the Mayall telescope, Kitt Peak National Observatory, NOIRLab. Pipeline processing and analyses of the data were supported by NOIRLab and the Lawrence Berkeley National Laboratory (LBNL). The Legacy Surveys project is honored to be permitted to conduct astronomical research on Iolkam Du’ag (Kitt Peak), a mountain with particular significance to the Tohono O’odham Nation.

NOIRLab is operated by the Association of Universities for Research in Astronomy (AURA) under a cooperative agreement with the National Science Foundation. LBNL is managed by the Regents of the University of California under contract to the U.S. Department of Energy.

This project used data obtained with the Dark Energy Camera (DECam), which was constructed by the Dark Energy Survey (DES) collaboration. Funding for the DES Projects has been provided by the U.S. Department of Energy, the U.S. National Science Foundation, the Ministry of Science and Education of Spain, the Science and Technology Facilities Council of the United Kingdom, the Higher Education Funding Council for England, the National Center for Supercomputing Applications at the University of Illinois at Urbana-Champaign, the Kavli Institute of Cosmological Physics at the University of Chicago, the Center for Cosmology and Astro-Particle Physics at the Ohio State University, the Mitchell Institute for Fundamental Physics and Astronomy at Texas A\&M University, Financiadora de Estudos e Projetos, Fundação Carlos Chagas Filho de Amparo à Pesquisa do Estado do Rio de Janeiro, Conselho Nacional de Desenvolvimento Científico e Tecnológico, and the Ministério da Ciência, Tecnologia e Inovação, the Deutsche Forschungsgemeinschaft, and the Collaborating Institutions in the Dark Energy Survey. The Collaborating Institutions are Argonne National Laboratory, the University of California at Santa Cruz, the University of Cambridge, Centro de Investigaciones Energéticas, Medioambientales y Tecnológicas-Madrid, the University of Chicago, University College London, the DES-Brazil Consortium, the University of Edinburgh, the Eidgenössische Technische Hochschule (ETH) Zürich, Fermi National Accelerator Laboratory, the University of Illinois at Urbana-Champaign, the Institut de Ciències de l’Espai (IEEC/CSIC), the Institut de Física d’Altes Energies, Lawrence Berkeley National Laboratory, the Ludwig-Maximilians-Universität München and the associated Excellence Cluster Universe, the University of Michigan, NSF’s NOIRLab, the University of Nottingham, the Ohio State University, the University of Pennsylvania, the University of Portsmouth, SLAC National Accelerator Laboratory, Stanford University, the University of Sussex, and Texas A\&M University.

BASS is a key project of the Telescope Access Program (TAP), which has been funded by the National Astronomical Observatories of China, the Chinese Academy of Sciences (the Strategic Priority Research Program ``The Emergence of Cosmological Structures'', Grant \#XDB09000000), and the Special Fund for Astronomy from the Ministry of Finance. BASS is also supported by the External Cooperation Program of Chinese Academy of Sciences (Grant \#114A11KYSB20160057), and the Chinese National Natural Science Foundation (Grant \#12120101003, \#11433005).

The Legacy Survey team makes use of data products from the Near-Earth Object Wide-field Infrared Survey Explorer (NEOWISE), which is a project of the Jet Propulsion Laboratory/California Institute of Technology. NEOWISE is funded by the National Aeronautics and Space Administration.

The Legacy Surveys imaging of the DESI footprint is supported by the Director, Office of Science, Office of High Energy Physics of the U.S. Department of Energy under Contract No. DE-AC02-05CH1123, by the National Energy Research Scientific Computing Center, a DOE Office of Science User Facility under the same contract, and by the U.S. National Science Foundation, Division of Astronomical Sciences under Contract No. AST-0950945 to NOAO.

Funding for the Sloan Digital Sky 
Survey IV has been provided by the 
Alfred P. Sloan Foundation, the U.S. 
Department of Energy Office of 
Science, and the Participating 
Institutions. 

SDSS-IV acknowledges support and 
resources from the Center for High 
Performance Computing  at the 
University of Utah. The SDSS 
website is www.sdss4.org.

SDSS-IV is managed by the 
Astrophysical Research Consortium 
for the Participating Institutions 
of the SDSS Collaboration including 
the Brazilian Participation Group, 
the Carnegie Institution for Science, 
Carnegie Mellon University, Center for 
Astrophysics | Harvard \& 
Smithsonian, the Chilean Participation 
Group, the French Participation Group, 
Instituto de Astrof\'isica de 
Canarias, The Johns Hopkins 
University, Kavli Institute for the 
Physics and Mathematics of the 
Universe (IPMU) / University of 
Tokyo, the Korean Participation Group, 
Lawrence Berkeley National Laboratory, 
Leibniz Institut f\"ur Astrophysik 
Potsdam (AIP),  Max-Planck-Institut 
f\"ur Astronomie (MPIA Heidelberg), 
Max-Planck-Institut f\"ur 
Astrophysik (MPA Garching), 
Max-Planck-Institut f\"ur 
Extraterrestrische Physik (MPE), 
National Astronomical Observatories of 
China, New Mexico State University, 
New York University, University of 
Notre Dame, Observat\'ario 
Nacional / MCTI, The Ohio State 
University, Pennsylvania State 
University, Shanghai 
Astronomical Observatory, United 
Kingdom Participation Group, 
Universidad Nacional Aut\'onoma 
de M\'exico, University of Arizona, 
University of Colorado Boulder, 
University of Oxford, University of 
Portsmouth, University of Utah, 
University of Virginia, University 
of Washington, University of 
Wisconsin, Vanderbilt University, 
and Yale University.

\xj{This research is based on observations made with the NASA/ESA Hubble Space Telescope obtained from the Space Telescope Science Institute, which is operated by the Association of Universities for Research in Astronomy, Inc., under NASA contract NAS 5–26555. These observations are associated with program 15617. The HST data presented in this article were obtained from the Mikulski Archive for Space Telescopes (MAST) at the Space Telescope Science Institute. The specific observations analyzed can be accessed via \dataset[doi: 10.17909/w7w7-r445]{http://dx.doi.org/10.17909/w7w7-r445}.}

This work has made use of the VALD database, operated at Uppsala University, the Institute of Astronomy RAS in Moscow, and the University of Vienna.

%% file: 99_Appendix.tex
\counterwithin{figure}{section}
\counterwithin{table}{section}

\section{HST FR782N imaging of \objb}\label{sec:appendix_J1047_morph}

As described in \S\ref{sec:morph_lum}, HST program GO-15617 (PI: F. Bauer) obtained  high resolution ACS FR782N ramp filter images on \objb. To cover the continuum, broad \ha, and narrow \ha\ lines, the observations were taken using the FR782N ramp filter with four configurations centered at  7474\,\AA\, 7583\,\AA\, 7669\,\AA, and 7755\,\AA, respectively. 
The total integration time of each configuration is $\sim 5000$\,s. 
In our analysis, we focus on the narrow-band images centered at 7474\,\AA\ and 7669\,\AA. The 7474\,\AA\ narrow-band image serves as the continuum map at rest-frame 6397\,\AA. We subtract it from the 7669\,\AA\ narrow-band image to obtain the pure \ha\ line map without continuum contribution.  The \ha\ line map covers the entire narrow \ha\ component and 91\% of the broad component.   

\begin{figure*}[htbp]
    \includegraphics[width=\textwidth]{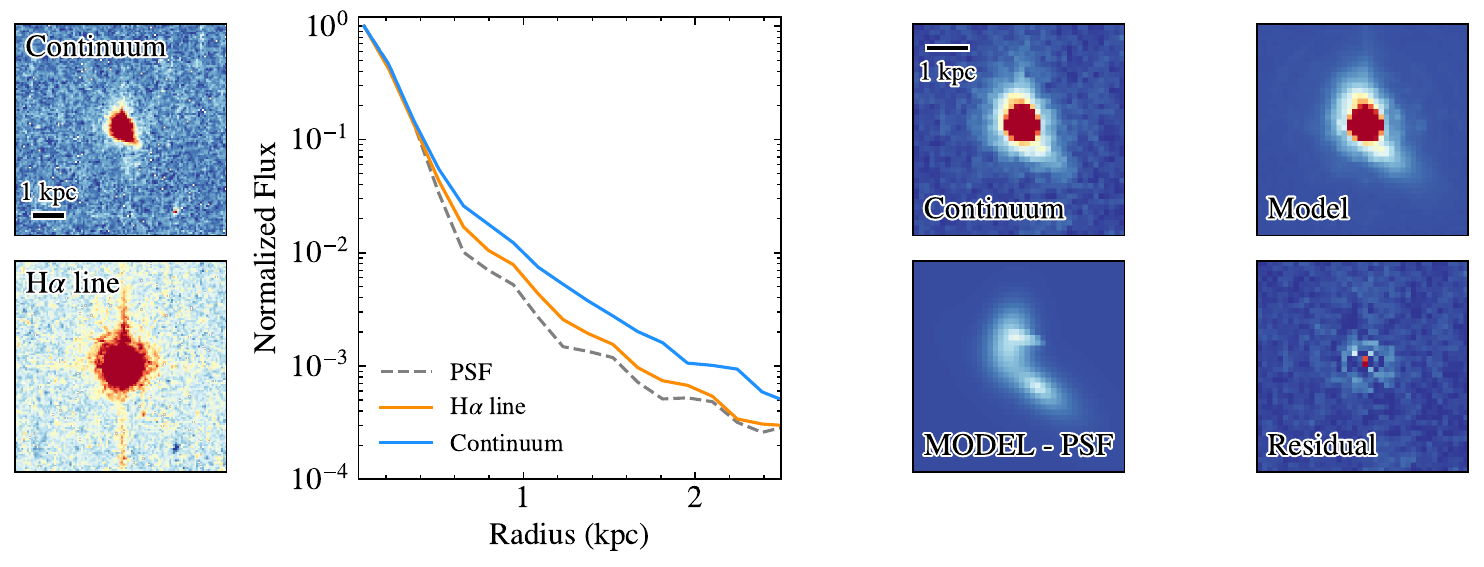}
    \caption{Morphology of the continuum at rest-frame 6397\,\AA\ and \ha\ line maps of \objb. \textit{Left panel}: the continuum map, \ha\ line map, and the radial profiles compared with that of the PSF. \textit{Right panel}: The best-fit model of the continuum, the PSF-subtracted component representing the galaxy light, and the residual. The four figures are shown in the same color scale. 
    \label{fig:J1047_morph}
    }
\end{figure*}

We compare the profile of the continuum and \ha\ line map with that of the PSF. The PSF is constructed using a star within the HST field-of-view 16\arcsec\ away from the object. As shown in the left panel of Figure \ref{fig:J1047_morph}, the continuum map has a bright central point source and an extended asymmetric structure, which we assume to be stellar light from the LRD host galaxy. Its radial profile is consistent with a PSF profile within $\lesssim 0.5$ kpc, while showing a deviation at $\gtrsim 0.5$ kpc. The \ha\ line map exhibits a point-source-like morphology with clear diffraction spikes. Its radial profile closely matches that of the PSF. It indicates that the \ha\ line is primarily contributed by the AGN.

We model the morphology of the continuum map using \textsc{galfit} \citep{Peng2002}. \xj{We assume a PSF and four S\'ersic components, with the number of S\'ersic components determined through several trials; four S\'ersic yielded the optimal fit.} The right panel of Figure \ref{fig:J1047_morph} presents the best-fit model and the PSF-subtracted components, which represent the host galaxy. The host galaxy has compact, yet clumpy morphology, with a half-light radius of 568 pc. The luminosity of the central PSF is 5.5 times that of the host galaxy, indicating that the AGN contributes about 85\% of the total continuum light in the optical. 

We note that the compact host galaxy of \objb\ detected by HST at $z = 0.17$ will be very challenging to resolve and detect for an LRD at high redshift, even at JWST resolution. This highlights the need for spatially resolved studies of local LRDs. 

\section{Emission line fitting}\label{sec:appendix_emission_fitting}

\begin{figure*}[htbp!]
    \includegraphics[width=0.757\textwidth]{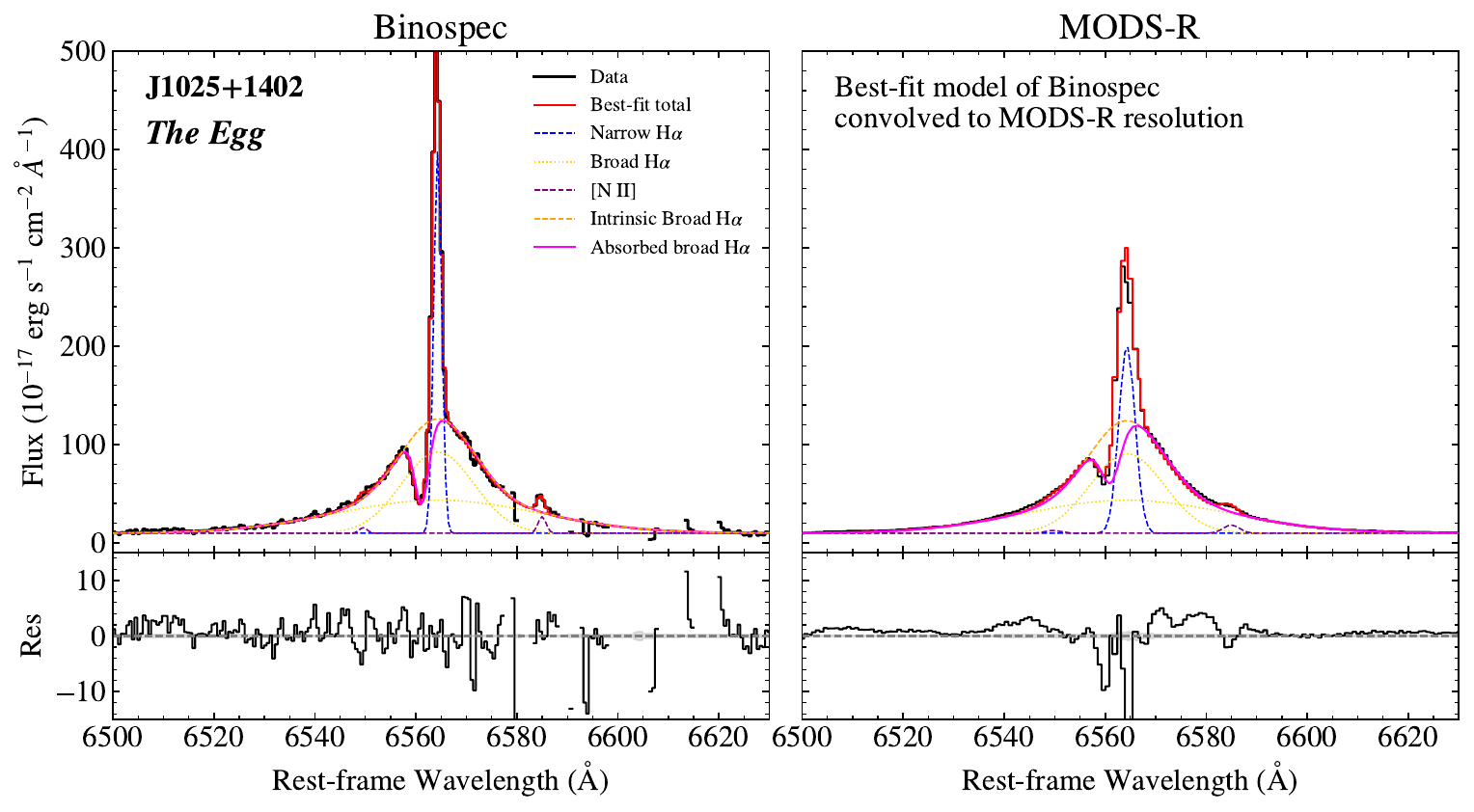}
    \includegraphics[width=\textwidth]{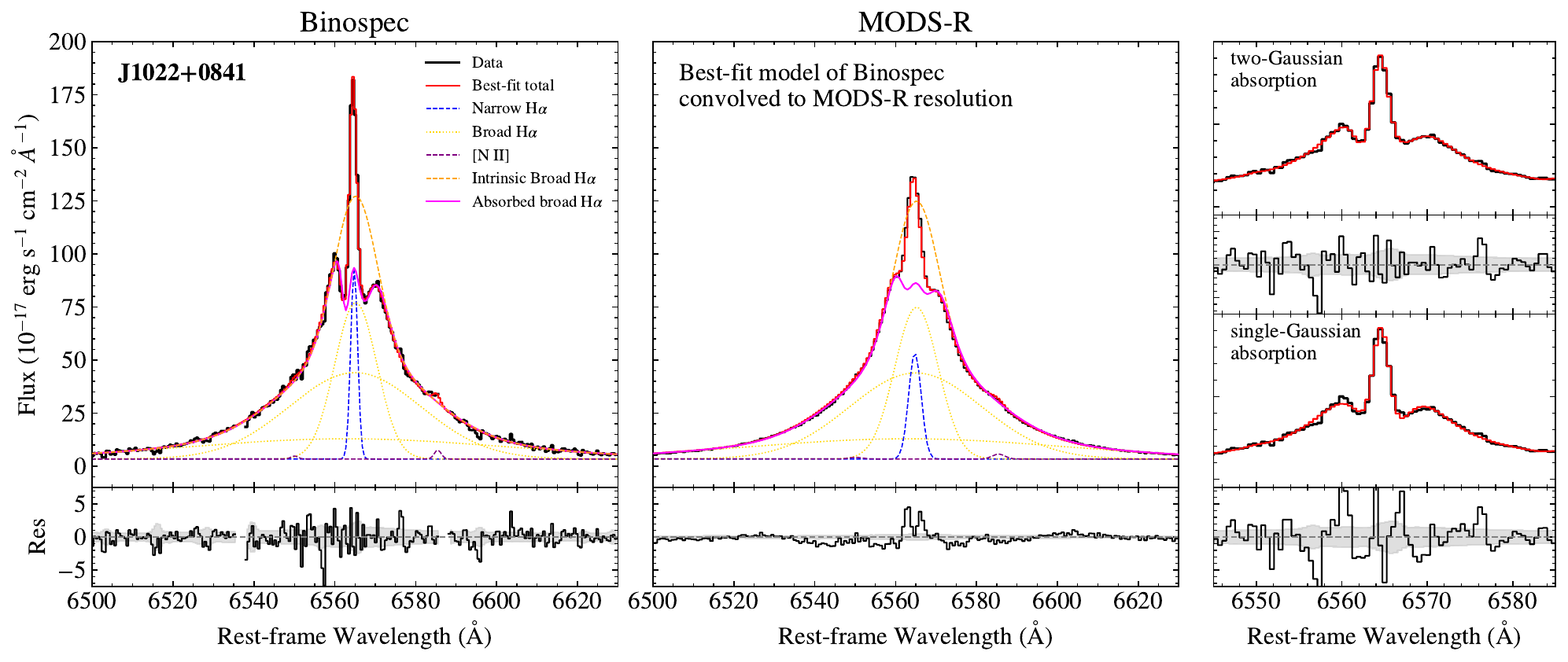}
    \caption{\textit{Left:} Best-fit line profiles for \egg\ and \objc\ from Binospec spectra, with absorption components overlaid on the emission profiles.  \textit{Middle:} Best-fit Binospec models convolved to match the resolution of MODS-R. The convolved models are in good agreement with the observed MODS-R spectra. \textit{Right:} Two-Gaussian and single-Gaussian absorption models for \objc. The single-Gaussian absorption model does not sufficiently reproduce the profile wings. For each fit, the residuals (data – model) are shown at the bottom, with the gray shaded region indicating the data uncertainty.
    \label{fig:Ha_fitting}
    }
\end{figure*}

For the \ha\ emission lines of \egg\ and \objc, we use observations from LBT/MODS and MMT/Binospec. Given the higher resolution of Binospec ($R \approx 4400$–5300, compared to $R \approx 1200$–1300 for MODS), we adopt the Binospec spectra for the fiducial fits and degrade the best-fit models to the MODS resolution for consistency checks. In the Binospec spectra, the absorption and emission components are fitted simultaneously, with a linear polynomial included to account for the local continuum. The [\ion{N}{2}] $\lambda\lambda$6550, 6585 doublet is modeled jointly with \ha\ using two narrow Gaussians, with their FWHMs tied to that of the narrow \ha\ component and their flux ratio fixed to 3.  As shown in Figure~\ref{fig:Ha_fitting}, the best-fit models reproduce the Binospec spectra well. When degraded to the resolution of MODS, the models remain in good agreement with the MODS \ha\ profiles. Small discrepancies are present in the broad wings of \egg, arising from the fact that the instrumental line-spread functions might not be perfectly Gaussian. We verify that such discrepancies introduce only a $\sim$5\% difference in the broad \ha\ luminosity when fitting the MODS spectrum alone, and thus have a negligible impact on the derived physical properties (e.g., $M_{\rm BH}$). For \objc, we test two different fitting models: one assuming a single Gaussian absorber superimposed on the broad emission lines, and the other using two Gaussian absorbers. As shown in Figure~\ref{fig:Ha_fitting}, the two-Gaussian model provides a better fit, more accurately capturing the wing of the absorption profile. \xj{The fitting results using the two-Gaussian model are reported in Table \ref{tab:absorber_fit}. For the single-Gaussian model, we obtain $\Delta v = 11$ km s$^{-1}$, FWHM=$273 \pm 26$ km s$^{-1}$, and EW=$4.16 \pm 0.33$\,\AA.}  The normalized \ha\ absorption profiles shown in Figure \ref{fig:HaHeI_absorber_compare} are derived by dividing the absorbed broad \ha\ emission by the intrinsic broad \ha\ line profile.

For strong emission lines exhibiting visible broad components (e.g., \hb, Pa$\alpha$) in the LBT/MODS and Magellan/FIRE spectra, each line is modeled with a single narrow Gaussian plus one or more broad Gaussian or Lorentzian profiles to capture the composite broad features. A linear polynomial is added to account for the local continuum. Specifically, broad Paschen lines are modeled with Lorentzian profiles, leveraging the high-resolution FIRE spectra to resolve line wings.  In \objb, the Paschen lines and adjacent IR \ion{He}{1} lines are modeled simultaneously, with the \ion{He}{1} emission represented by broad and narrow Gaussian components. Note that for each line profile fit, the parameters of the narrow Gaussian component are treated independently, as the spectral resolution varies across wavelengths and instruments. For example, [\ion{O}{3}] $\lambda$5008 in the MODS spectra is unresolved, with its observed FWHM determined by the wavelength-dependent line spread function. Therefore, we do not use its fit to constrain the narrow components of \hb\ or Pa$\alpha$. The measurements of key emission lines are summarized in Table~\ref{tab:measurement_other_lines}.

\begin{table*}[h!]
\centering
\begin{tabular}{l|ccc}
\hline
 & \obja & \objb & \objc \\
\hline
$L_{\rm He~II~4687}$ ($10^{39}$ erg s$^{-1}$) & $ 1.59\pm 0.22$ & $8.42\pm0.91$ & $<14.75$ \\
${\rm FWHM}_{\rm He~II~4687}$ (km s$^{-1}$) &  -- & $239\pm36$ & -- \\

\hline
$L_{\mathrm{[O~III]~5008}}$ ($10^{40}$ erg s$^{-1}$) & $33.48 \pm 0.28$ & $392.41 \pm 3.35$ & $84.18 \pm 0.45$ \\
$L_{\mathrm{[O~III]~4364}}$ ($10^{40}$ erg s$^{-1}$) & $1.25 \pm 0.05$ & $5.95 \pm 0.16$ & $3.68 \pm 0.16$ \\
$L_{\mathrm{H\beta, narrow}}$ ($10^{40}$ erg s$^{-1}$) & $5.80 \pm 0.05$ & $58.16 \pm 0.49$ & $14.40 \pm 0.20$ \\
$L_{\mathrm{H\beta, broad}}$ ($10^{40}$ erg s$^{-1}$) & $5.38 \pm 0.33$ & $28.24 \pm 1.52$ & $15.50 \pm 0.48$ \\
FWHM$_{\mathrm{H\beta, broad}}$ (km s$^{-1}$) & $1402 \pm 35$ & $1236 \pm 81$ & $1511 \pm 43$ \\
\hline
$L_{\mathrm{Pa\alpha, broad}}$ ($10^{40}$ erg s$^{-1}$) & $21.19 \pm 0.24$ & $82.82 \pm 0.55$ & $117.21 \pm 2.86$ \\
FWHM$_{\mathrm{Pa\alpha, broad}}$ (km s$^{-1}$) & $819 \pm 24$ & $1531 \pm 19$ & $1274 \pm 70$ \\
$L_{\mathrm{Pa\alpha, narrow}}$ ($10^{40}$ erg s$^{-1}$) & $10.96 \pm 0.20$ & $34.53 \pm 0.46$ & $37.50 \pm 1.36$ \\
FWHM$_{\mathrm{Pa\alpha, narrow}}$ (km s$^{-1}$) & $183 \pm 3$ & $287 \pm 4$ & $307 \pm 7$ \\
\hline
$L_{\mathrm{Pa\beta, broad}}$ ($10^{40}$ erg s$^{-1}$) & $4.20 \pm 0.39$ & $62.83 \pm 0.87$ & $74.43 \pm 2.66$ \\
FWHM$_{\mathrm{Pa\beta, broad}}$ (km s$^{-1}$) & $317 \pm 37$ $^a$ & $1548 \pm 36$ & $1757 \pm 73$ \\
$L_{\mathrm{Pa\beta, narrow}}$ ($10^{40}$ erg s$^{-1}$) & $1.32 \pm 0.46$ & $21.88 \pm 0.60$ & $17.50 \pm 0.55$ \\
FWHM$_{\mathrm{Pa\beta, narrow}}$ (km s$^{-1}$) & -- $^a$ & $288 \pm 8$ & $316 \pm 6$ \\
\hline
$L_{\mathrm{Pa\gamma, broad}}$ ($10^{40}$ erg s$^{-1}$) & $2.84 \pm 0.12$ & $28.66 \pm 0.40$ & $19.65 \pm 1.09$ \\
FWHM$_{\mathrm{Pa\gamma, broad}}$ (km s$^{-1}$) & $448 \pm 30$ & $1196 \pm 32$ & $677 \pm 100$ \\
$L_{\mathrm{Pa\gamma, narrow}}$ ($10^{40}$ erg s$^{-1}$) & $1.45 \pm 0.08$ & $13.04 \pm 0.18$ & $3.38 \pm 1.03$ \\
FWHM$_{\mathrm{Pa\gamma, narrow}}$ (km s$^{-1}$) & $95 \pm 7$ & $189 \pm 3$ & $190 \pm 36$ \\
\hline
$L_{\mathrm{Pa\delta, broad}}$ ($10^{40}$ erg s$^{-1}$) & $0.93 \pm 0.37$ & $15.44 \pm 0.61$ & $7.55 \pm 0.66$ \\
FWHM$_{\mathrm{Pa\delta, broad}}$ (km s$^{-1}$) & $890 \pm 200$ & $1512 \pm 110$ & $375 \pm 50$ \\
$L_{\mathrm{Pa\delta, narrow}}$ ($10^{40}$ erg s$^{-1}$) & $1.02 \pm 0.06$ & $4.78 \pm 0.86$ & $0.60 \pm 0.34$ \\
FWHM$_{\mathrm{Pa\delta, narrow}}$ (km s$^{-1}$) & $126 \pm 9$ & $269 \pm 32$ & -- \\
\hline
\end{tabular}
\caption{The measurements of \ion{He}{2}, [\ion{O}{3}], H$\beta$, and Paschen lines. The \ion{He}{2}, [\ion{O}{3}], and H$\beta$ lines measurements are based on the LBT/MODS spectra. The Paschen lines measurements are based on the Magellan/FIRE spectra. The [\ion{O}{3}] 4364, \hb, and [\ion{O}{3}] 5008 are not resolved in the MODS spectra. $^a$ The Pa$\beta$ emission of \obja\ lies within strong telluric absorption regions, and the spectral fitting is subject to significant systematics. }
\label{tab:measurement_other_lines}
\end{table*}

For \hb\ line of \egg, we mask the absorbed region and fit the intrinsic emission profile. Similarly, the intrinsic \ion{He}{1} $\lambda$10833 emission is modeled by excluding regions affected by absorption. The unaffected line wings are used to constrain and recover the symmetric emission profile. The emission is modeled with one or two broad Gaussian components plus a narrow component. In Figure~\ref{fig:HaHeI_absorber_compare}, the \hb\ profile in \egg\ and the \ion{He}{1} $\lambda$10833 profiles in all three objects are normalized by the total unabsorbed profiles, including both the broad emission lines and continua. Normalizing \hb\ using only the broad emission line is tested, but results in an absorption peak below zero. We then model the normalized \hb\ absorber with a single Gaussian and the normalized \ion{He}{1} $\lambda$10833 absorber with multiple Gaussians. Table~\ref{tab:absorber_fit} summarizes the best-fit parameters for each absorber.

\section{[\ion{Fe}{2}] modeling}\label{sec:appendix_feii}

To model AGN photoionization scenarios for [\ion{Fe}{2}] emission, we adopt AGN SEDs from \citet{Jin2012}. While the chosen SED corresponds to an Eddington ratio of 0.17, our \textsc{cloudy} modeling is intended to be qualitative and is not highly sensitive to this assumed value. 
How the Eddington ratio impacts the AGN SED, and consequently the resulting [\ion{Fe}{2}] spectrum, is beyond the scope of this paper.  Our \textsc{cloudy} models span a grid of hydrogen densities from $\log (n_{\rm H}/{\rm cm^{-3}}) = 4$ to $15$ (0.5 dex step), column densities from $\log (N_{\rm H}/{\rm cm^{-2}}) = 18$ to $27$ (0.5 dex step), ionization parameters from $\log U = -4$ to $0$ (1 dex step), and metallicities from $Z/Z_\odot = -1$ to $2$ (1 dex step). We run a suite of \textsc{cloudy} models over this parameter space, assuming a plane-parallel geometry, and obtain the predicted [\ion{Fe}{2}] line spectra.

To determine the best-fit models, we compare the observed emission line fluxes with those predicted by the modeled [\ion{Fe}{2}] spectrum. Specifically, we use the fluxes of  \ion{Fe}{2} $\lambda$4234, [\ion{Fe}{2}] $\lambda$4245, $\lambda$4278, $\lambda$4289, $\lambda$4416 (blend of 4415 and 4418), $\lambda$5159, $\lambda$5263, $\lambda$7157, and $\lambda$7174. Additionally, since \ion{Fe}{2} $\lambda$5171 is highly sensitive to the gas conditions but is not detected, we include its 3$\sigma$ upper limit as an extra constraint. All line fluxes, both observed and modeled, are normalized to that of [\ion{Fe}{2}] $\lambda$4245, and we compute the $\chi^2$ value for each model grid.

For comparison, we also explored models using a 5000\,K blackbody as the ionizing source. The temperature is chosen based on our analysis in \S\ref{sec:J1025_case_study} and \S\ref{sec:cartoon}. These models fail to reproduce the observed [\ion{Fe}{2}] line ratios, with significant discrepancies particularly in \ion{Fe}{2}~$\lambda$4234,  [\ion{Fe}{2}]$\lambda$4245, $\lambda$4278, $\lambda$4289, and $\lambda$4416.

\begin{figure*}[htbp!]
    \centering
    \includegraphics[width=\linewidth]{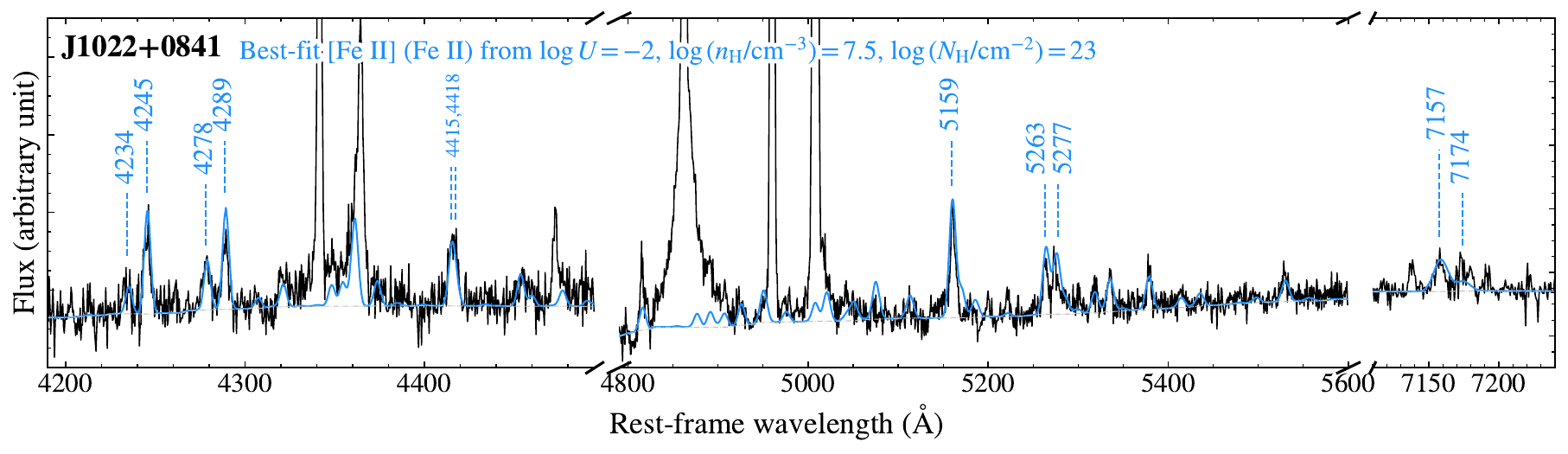}
    \caption{Same as Figure~\ref{fig:J1025_bestfit_FeII_model}, but for \objc. The best-fit parameters are identical to those for \obja.}
    \label{fig:J1022_bestfit_FeII_model}
\end{figure*}

In Figure~\ref{fig:J1022_bestfit_FeII_model}, we present the best-fit [\ion{Fe}{2}] model for \objc. Its best-fit parameters are identical to those used for \obja.

\section{The parameterization of the continua}\label{sec:appendix_sed_fit}

\subsection{Theoretical atmosphere model + dust templates}\label{sec:appendix_dust_modeling}

We begin with the best-matched theoretical models from \cite{Liu2025} and model the dust emission using the dust templates from \citet{Lyu2021}. As demonstrated in \cite{Lyu2017} and \cite{Lyu2021}, typical AGN-heated dust emission SED can be well can be well characterized by the combination of four major dust components: blackbodies at 2100 K, 1038 K, and 300 K, and a graybody at 90 K.  We extrapolate the theoretical model spectra by assuming a blackbody tail in the near-IR, consistent with the model’s effective temperature. We then fit the amplitudes of the four dust components to reproduce the overall continuum and WISE photometry at wavelengths longer than 1.5\,\micron.

\begin{figure*}    \includegraphics[width=0.49\textwidth]{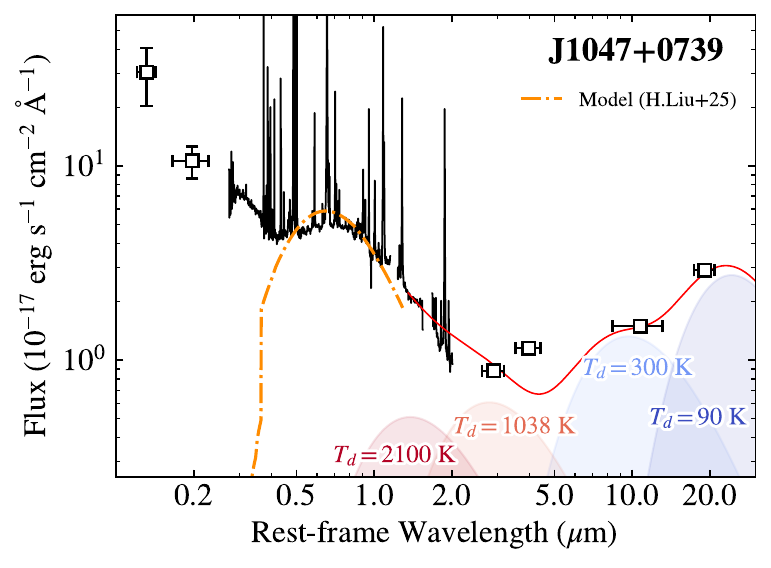}
\includegraphics[width=0.5\textwidth]{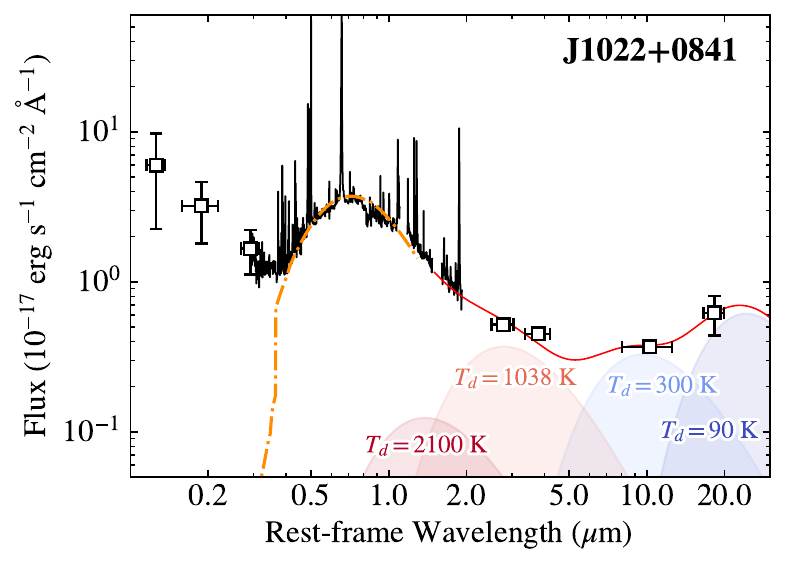}
\caption{The SEDs of \objb\ and \objc\ and their respective best-matched models from \cite{Liu2025}. The dust components, assuming templates from \cite{Lyu2021}, are shown as the color-shaded regions.
\label{fig:sed_bb_two}
}
\end{figure*}

Figure~\ref{fig:J1025_BB_comparison} demonstrates that this model successfully reproduces the near-infrared photometry of \egg.  Figure~\ref{fig:sed_bb_two} shows the SEDs of \objb\ and \objc\ compared with and the models from \cite{Liu2025} that visually match the data. The intrinsic luminosity of the \cite{Liu2025} model is $4.1 \times 10^{43}$\,erg\,s$^{-1}$ for \objb, with a reference density of $\rho_{\rm ref} = 1.1 \times 10^{-12}$\,g\,cm$^{-3}$ (which is comparable to the photosphere density; see \cite{Liu2025} for exact definition). These model parameters give an effective temperature of $T_{\rm eff}=4816$~K. For \objc, the model luminosity is $5.1 \times 10^{43}$\,erg\,s$^{-1}$ with $\rho_{\rm ref} = 1.9 \times 10^{-12}$\,g\,cm$^{-3}$, and the resulting $T_{\rm eff}=4614$~K. 
We note that the SED of \objb\ is less well reproduced, \xj{possibly} due to the non-negligible galaxy light. 
\xj{Indeed,} the HST narrow-band imaging (see Appendix \ref{sec:appendix_J1047_morph}) reveals that, although the AGN dominates the continuum, the host galaxy contributes about 15\% of the continuum light around the \ha\ wavelengths.  The contribution from the host galaxy is likely minor in \egg\ and \objc, both of which exhibit more prominent V-shaped SEDs with redder optical continua. This suggests stronger AGN dominance in the optical continua. Nevertheless, the comparison between the model and observed spectra aims to demonstrate the plausibility of the theoretical models rather than to constrain the physical properties of these objects. 

As shown in Figure~\ref{fig:sed_bb_two}, the rest-frame near-IR emission of \objb\ and \objc\ can be described by dust components with temperatures ranging from 2100 K to 90 K. However, the inferred amplitudes are highly sensitive to the assumed optical-to-IR continuum shape and the choice of dust templates. We discuss alternative approaches and associated degeneracies below.

\subsection{An empirical parameterization of the continuum}\label{sec:appendix_empirical_fit}

We attempt to empirically parameterize the optical-IR continuum shapes of local LRDs with a pseudo-blackbody spectrum.  This parameterization is purely descriptive, intended to capture the observed continuum shape but not to provide a physical interpretation. We adopt the functional form of a modified Calzetti law \citep{Noll2009}, but it should not be interpreted as indicating any particular amount of dust attenuation.
  The parameterization can be expressed as:
\begin{equation}\label{eq:modified_bb}
\begin{split}
    f_{\rm MB}(\lambda) =\ & f_{\rm UV} \left( \frac{\lambda}{3000\,\text{\AA}} \right)^{\alpha} 
     + f_{\rm BB} \left[ e^{-0.4\, \tilde{A}_\lambda(\tilde{A}_V, \delta)} \cdot BB(\lambda, T) \right] + \sum_{i} f_{d,i}  B_{d}(\lambda, T_i)
\end{split}
\end{equation}
where the free parameters are $f_{\rm UV}$, $\alpha$, $\tilde{A}_V$, $\delta$,  $f_{\rm BB}$,  $T$, and $f_{d,i}$. $f_{\rm UV}$ and $\alpha$ are the amplitude and power-law slope of the UV continuum. It represents a distinct component, which could be either due to the AGN light or the host galaxy. $\tilde{A}_\lambda(\tilde{A}_V, \delta)$ denotes the form of the modified Calzetti law \citep{Noll2009}, where $\tilde{A}_\lambda$ represents a generalized attenuation curve (not necessarily tied to physical dust properties), and $\delta$ is the power-law modification parameter, normalized at 5500\,\AA. For \objb, we introduce an additional free parameter to model the Balmer jump as a discontinuity in the power-law at 3645\,\AA.  $BB(\lambda, T)$ is the blackbody spectrum at temperature $T$, and $f_{\rm BB}$ is its amplitude.   For the dust emission, we adopt the dust templates from \citet{Lyu2021}. $f_{d,i}$ and $B_{d}(\lambda, T_i)$ are the amplitude and blackbody or gray body spectrum of the $i$-th dust component.

\begin{figure*}[htbp]
    \centering
    \includegraphics[width=0.325\linewidth]{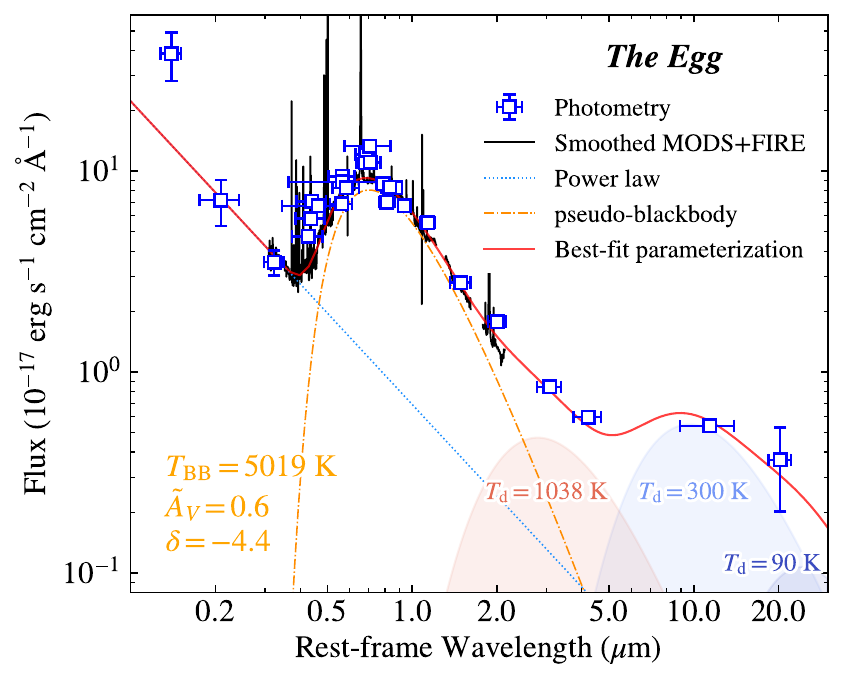}
    \includegraphics[width=0.325\linewidth]{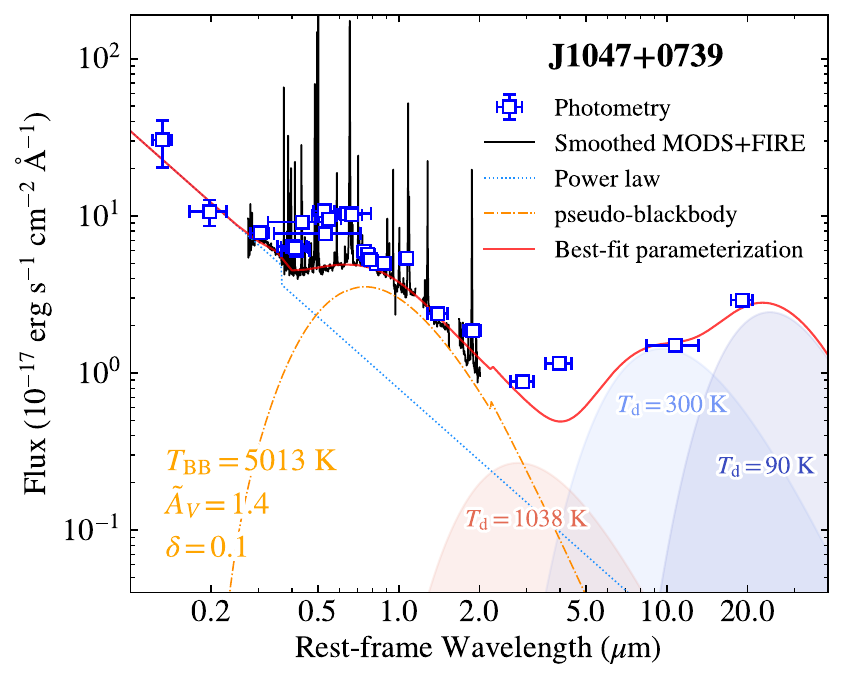}
    \includegraphics[width=0.325\linewidth]{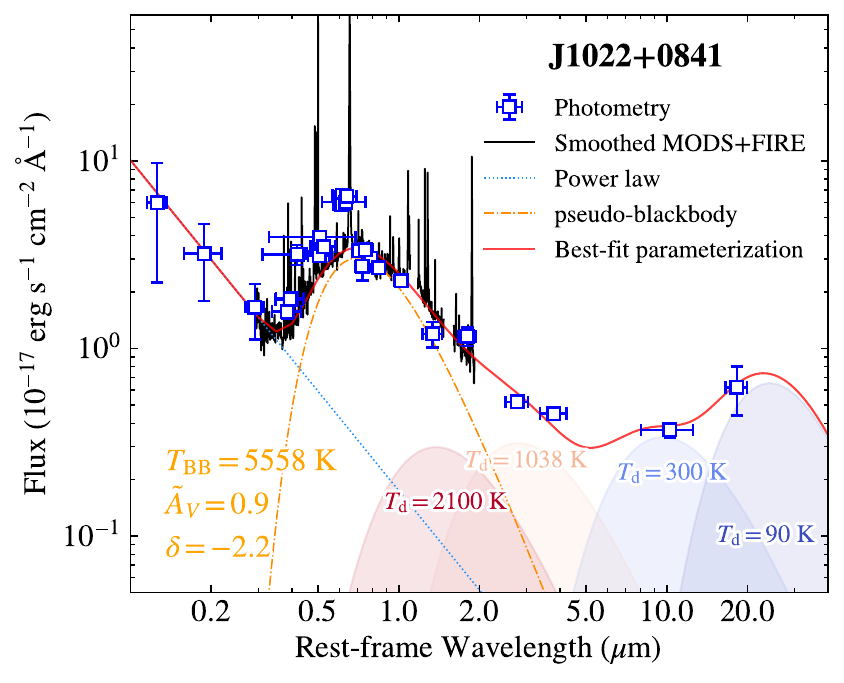}
    
    \caption{The best-fit \xj{empirical} parameterization.  The best-fit model (Equation \ref{eq:modified_bb}) is shown as the red line, which includes a power-law component shown as the blue dotted line, a pseudo-blackbody component shown as the orange dashed-dotted line, and dust components shown as the color-shaded regions. The discontinuity around 2\,\micron\ in \objb\ arises from the range limitation of the modified Calzetti law, but it does not affect the result. The best-fit parameters are labeled; we emphasize that $\tilde{A}_V$ and $\delta$ are purely descriptive of the continuum shape and are not intended to represent physical dust attenuation.
    }
     \label{fig:continuum_template_mb}
\end{figure*}

The best-fit continua of the three local LRDs are shown in Figure \ref{fig:continuum_template_mb}. The corresponding best-fit parameters and component luminosities are summarized in Table~\ref{tab:bestfit_parameter}. We do not include associated uncertainties given the descriptive nature of this parameterization.  
All three objects exhibit temperatures around 5000--6000 K and very negative values of $\delta$. 
The extreme $\delta$ produces a very steep modification to the original blackbody.  
Such a negative $\delta$ strongly suppresses the UV-to-optical flux of the blackbody spectrum while leaving the IR spectrum largely unaffected.  As discussed in \S\ref{sec:opt_ir_continua},  physical mechanisms, such as metal cooling and gas opacity, may contribute to this modification.  In particular, the contribution of the UV power-law component to the optical continuum in \objb\ is non-negligible. At the \ha\ wavelength, the power-law component is only a factor of 2.3 fainter than the pseudo-blackbody. This is broadly consistent with measurements from its HST narrow-band image, which show the AGN luminosity to be 5.5 times that of the host galaxy. In the other two objects, the power-law components at optical wavelengths are orders of magnitude fainter than the pseudo-blackbody components.

\begin{table}[htbp]
    \centering
    \begin{tabular}{c|ccc}
    \hline
        Parameter & \obja\ (\egg) & \objb & \objc \\
    \hline 
     $\alpha$    &   $-1.51$ & $-1.52$ & $-1.76$ \\
     $f_{\rm UV}$ ($10^{-17}$\,erg\,s$^{-1}$\,\AA$^{-1}$) &  $4.25$ & $4.92$ & $1.45$ \\
     $\tilde{A}_V$ &  $0.58$  & $1.81$  & $0.87$\\
     $\delta$ & $-4.37$ & $-0.11$ & $-1.65$ \\
     $T_{\rm BB}$ (K) &  $5019$ & $5013$ &  $5558$ \\
     $L_{\rm pseudo-BB}$  ($10^{43}$\,erg\,s$^{-1}$) &  $1.89$ & $3.58$ & $4.17$\\
      $L_{\rm dust, 2000K}$ ($10^{43}$\,erg\,s$^{-1}$) &  -- & -- & $0.98$\\
     $L_{\rm dust, 1038K}$ ($10^{43}$\,erg\,s$^{-1}$)  &  $0.55$ & $0.95$ & $2.07$ \\
     $L_{\rm dust, 300K}$ ($10^{43}$\,erg\,s$^{-1}$)  & $2.21$ & $17.5$ & $76.90$\\
     $L_{\rm dust, 90K}$ ($10^{43}$\,erg\,s$^{-1}$) &  $0.84$ & $57.9$ & $28.94$\\
     \hline
    \end{tabular}
    \caption{The best-fit continuum parameters for the three local LRDs are derived using the parametrization in Equation~\ref{eq:modified_bb}. The normalization amplitude in Equation~\ref{eq:modified_bb} is converted into the luminosity of each component. $T_{\rm BB}$ is the effective temperature of the pseudo-blackbody spectrum, and $L_{\rm pseudo-BB}$ is the total luminosity of the pseudo-blackbody component. $L_{\rm dust, 2000K}$, $L_{\rm dust, 1038K}$, $L_{\rm dust, 300K}$, and $L_{\rm dust, 90K}$ denote the luminosities of dust emission at 2000 K, 1038 K, 300 K, and 90 K, respectively, based on the templates in  \cite{Lyu2021}.
    \label{tab:bestfit_parameter}
    }
\end{table}

In \egg, the two most prominent dust components are the 1038 K and 300 K blackbody components, which dominate the SED at wavelengths of 3-10\,\micron\ and 10-20\,\micron, respectively. They drive the rising flux observed in the four WISE bands. The contributions from the remaining two dust components are minor.

\subsection{The Haro 11 dust template}\label{sec:appendix_haro11_dust}
\begin{figure}[htbp]
    \centering
    \includegraphics[width=0.325\linewidth]{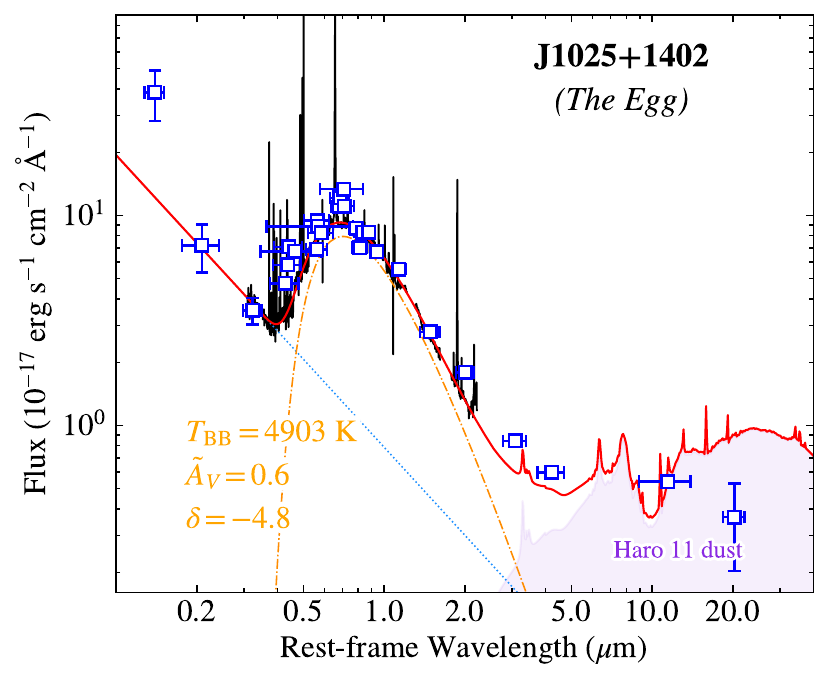}
    \includegraphics[width=0.325\linewidth]{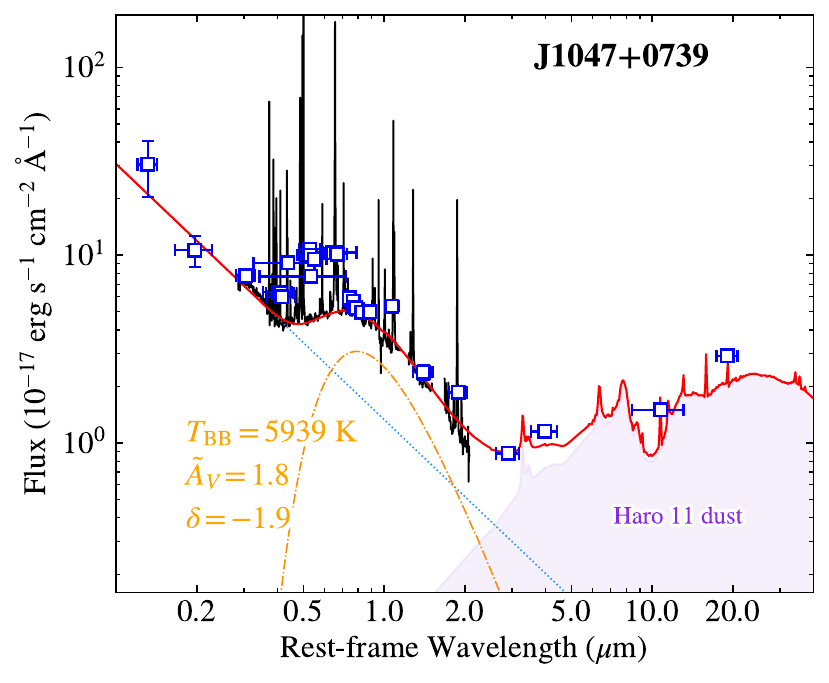}
    \includegraphics[width=0.325\linewidth]{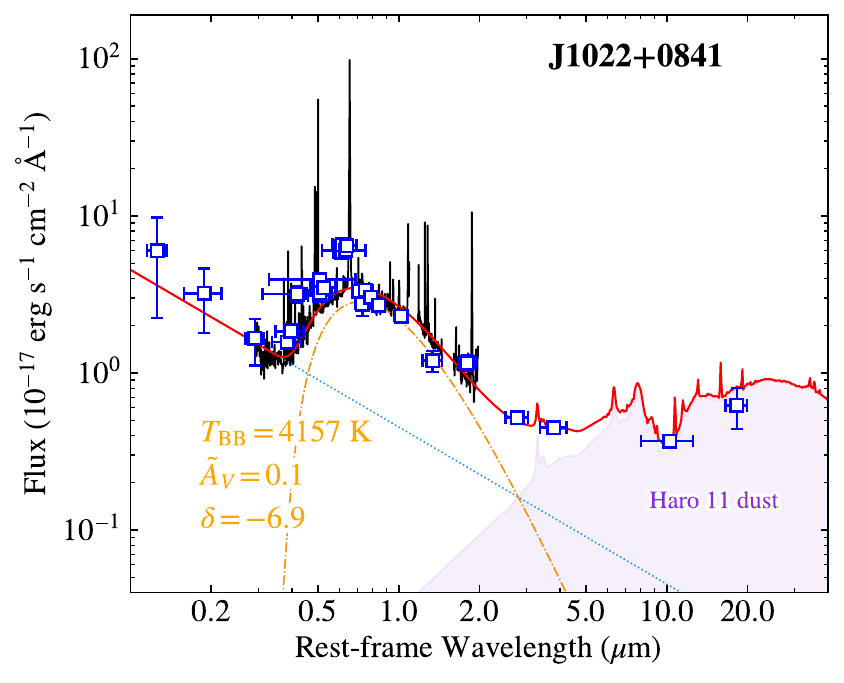}
    \caption{Similar to Figure \ref{fig:continuum_template_mb}, but using the Haro 11 dust template. }
    \label{fig:best_parameterization_haro11}
\end{figure}

The inferred dust components are highly sensitive to the choice of fitting templates. In \objb, the W2 photometry at a rest-frame wavelength of 4\,\micron\ is not well reproduced, as simple blackbody or graybody models fail to capture features such as PAH emission and silicate absorption. In \objc, a dust component with $T_d \approx 2100$\,K appears, but this primarily reflects the influence of the four dust templates adopted from \citet{Lyu2021}. To explore alternative interpretations, we replace the dust component in Equation~\ref{eq:modified_bb} with the Haro 11 dust template from \citet{Lyu2016} and fit the continua of the three local LRDs, as shown in Figure~\ref{fig:best_parameterization_haro11}. Notably, the dust emission in \objc\ is well reproduced by the Haro 11 template, suggesting that its dust may reside in the ISM and be heated by AGN-driven feedback or massive star clusters in the host galaxy. This degeneracy highlights the need for mid- to far-infrared observations to robustly constrain the dust content and its origin in these systems.

%% file: main.bib
@ARTICLE{Setton2024,
       author = {{Setton}, David J. and {Greene}, Jenny E. and {de Graaff}, Anna and {Ma}, Yilun and {Leja}, Joel and {Matthee}, Jorryt and {Bezanson}, Rachel and {Boogaard}, Leindert A. and {Cleri}, Nikko J. and {Katz}, Harley and {Labbe}, Ivo and {Maseda}, Michael V. and {McConachie}, Ian and {Miller}, Tim B. and {Price}, Sedona H. and {Suess}, Katherine A. and {van Dokkum}, Pieter and {Wang}, Bingjie and {Weibel}, Andrea and {Whitaker}, Katherine E. and {Williams}, Christina C.},
        title = "{Little Red Dots at an Inflection Point: Ubiquitous ``V-Shaped'' Turnover Consistently Occurs at the Balmer Limit}",
      journal = {arXiv e-prints},
         year = 2024,
        month = nov,
          eid = {arXiv:2411.03424},
        pages = {arXiv:2411.03424},
          doi = {10.48550/arXiv.2411.03424},
archivePrefix = {arXiv},
       eprint = {2411.03424},
 primaryClass = {astro-ph.GA},
       adsurl = {https://ui.adsabs.harvard.edu/abs/2024arXiv241103424S}
}

@ARTICLE{Cenarro2001,
       author = {{Cenarro}, A.~J. and {Cardiel}, N. and {Gorgas}, J. and {Peletier}, R.~F. and {Vazdekis}, A. and {Prada}, F.},
        title = "{Empirical calibration of the near-infrared Ca ii triplet - I. The stellar library and index definition}",
      journal = {\mnras},
         year = 2001,
        month = sep,
       volume = {326},
       number = {3},
        pages = {959-980},
          doi = {10.1046/j.1365-8711.2001.04688.x},
archivePrefix = {arXiv},
       eprint = {astro-ph/0109157},
 primaryClass = {astro-ph},
       adsurl = {https://ui.adsabs.harvard.edu/abs/2001MNRAS.326..959C}
}

@ARTICLE{Cenarro2002,
       author = {{Cenarro}, A.~J. and {Gorgas}, J. and {Cardiel}, N. and {Vazdekis}, A. and {Peletier}, R.~F.},
        title = "{Empirical calibration of the near-infrared Ca II triplet - III. Fitting functions}",
      journal = {\mnras},
         year = 2002,
        month = feb,
       volume = {329},
       number = {4},
        pages = {863-876},
          doi = {10.1046/j.1365-8711.2002.05029.x},
archivePrefix = {arXiv},
       eprint = {astro-ph/0112448},
 primaryClass = {astro-ph},
       adsurl = {https://ui.adsabs.harvard.edu/abs/2002MNRAS.329..863C}
}

@ARTICLE{Mallik11997,
       author = {{Mallik}, S.~V.},
        title = "{The CA II triplet lines as diagnostics of luminosity, metallicity and chromospheric activity in cool stars}",
      journal = {\aaps},
         year = 1997,
        month = aug,
       volume = {124},
        pages = {359-384},
          doi = {10.1051/aas:1997199},
       adsurl = {https://ui.adsabs.harvard.edu/abs/1997A&AS..124..359M}
}

@ARTICLE{Bolton2012,
       author = {{Bolton}, Adam S. and {Schlegel}, David J. and {Aubourg}, {\'E}ric and {Bailey}, Stephen and {Bhardwaj}, Vaishali and {Brownstein}, Joel R. and {Burles}, Scott and {Chen}, Yan-Mei and {Dawson}, Kyle and {Eisenstein}, Daniel J. and {Gunn}, James E. and {Knapp}, G.~R. and {Loomis}, Craig P. and {Lupton}, Robert H. and {Maraston}, Claudia and {Muna}, Demitri and {Myers}, Adam D. and {Olmstead}, Matthew D. and {Padmanabhan}, Nikhil and {P{\^a}ris}, Isabelle and {Percival}, Will J. and {Petitjean}, Patrick and {Rockosi}, Constance M. and {Ross}, Nicholas P. and {Schneider}, Donald P. and {Shu}, Yiping and {Strauss}, Michael A. and {Thomas}, Daniel and {Tremonti}, Christy A. and {Wake}, David A. and {Weaver}, Benjamin A. and {Wood-Vasey}, W. Michael},
        title = "{Spectral Classification and Redshift Measurement for the SDSS-III Baryon Oscillation Spectroscopic Survey}",
      journal = {\aj},
         year = 2012,
        month = nov,
       volume = {144},
       number = {5},
          eid = {144},
        pages = {144},
          doi = {10.1088/0004-6256/144/5/144},
archivePrefix = {arXiv},
       eprint = {1207.7326},
 primaryClass = {astro-ph.CO},
       adsurl = {https://ui.adsabs.harvard.edu/abs/2012AJ....144..144B}
}

@ARTICLE{Prochaska2020,
       author = {{Prochaska}, J. and {Hennawi}, Joseph and {Westfall}, Kyle and {Cooke}, Ryan and {Wang}, Feige and {Hsyu}, Tiffany and {Davies}, Frederick and {Farina}, Emanuele and {Pelliccia}, Debora},
        title = "{PypeIt: The Python Spectroscopic Data Reduction Pipeline}",
      journal = {The Journal of Open Source Software},
         year = 2020,
        month = dec,
       volume = {5},
       number = {56},
          eid = {2308},
        pages = {2308},
          doi = {10.21105/joss.02308},
archivePrefix = {arXiv},
       eprint = {2005.06505},
 primaryClass = {astro-ph.IM},
       adsurl = {https://ui.adsabs.harvard.edu/abs/2020JOSS....5.2308P}
}

@software{Kansky2019,
       author = {{Kansky}, Jan and {Chilingarian}, Igor and {Fabricant}, Daniel and {Matthews}, Anne and {Moran}, Sean and {Paegert}, Martin and {Duane Gibson}, J. and {Porter}, Dallan and {Roll}, John},
        title = "{Binospec: Data reduction pipeline for the Binospec imaging spectrograph}",
 howpublished = {Astrophysics Source Code Library, record ascl:1905.004},
         year = 2019,
        month = may,
          eid = {ascl:1905.004},
       adsurl = {https://ui.adsabs.harvard.edu/abs/2019ascl.soft05004K}
}

@ARTICLE{Bianchi2020,
       author = {{Bianchi}, Luciana and {Shiao}, Bernard},
        title = "{Matched Photometric Catalogs of GALEX UV Sources with Gaia DR2 and SDSS DR14 Databases (GUVmatch)}",
      journal = {\apjs},
         year = 2020,
        month = oct,
       volume = {250},
       number = {2},
          eid = {36},
        pages = {36},
          doi = {10.3847/1538-4365/aba2d7},
archivePrefix = {arXiv},
       eprint = {2007.03808},
 primaryClass = {astro-ph.GA},
       adsurl = {https://ui.adsabs.harvard.edu/abs/2020ApJS..250...36B}
}

@ARTICLE{Flewelling2020,
       author = {{Flewelling}, H.~A. and {Magnier}, E.~A. and {Chambers}, K.~C. and {Heasley}, J.~N. and {Holmberg}, C. and {Huber}, M.~E. and {Sweeney}, W. and {Waters}, C.~Z. and {Calamida}, A. and {Casertano}, S. and {Chen}, X. and {Farrow}, D. and {Hasinger}, G. and {Henderson}, R. and {Long}, K.~S. and {Metcalfe}, N. and {Narayan}, G. and {Nieto-Santisteban}, M.~A. and {Norberg}, P. and {Rest}, A. and {Saglia}, R.~P. and {Szalay}, A. and {Thakar}, A.~R. and {Tonry}, J.~L. and {Valenti}, J. and {Werner}, S. and {White}, R. and {Denneau}, L. and {Draper}, P.~W. and {Hodapp}, K.~W. and {Jedicke}, R. and {Kaiser}, N. and {Kudritzki}, R.~P. and {Price}, P.~A. and {Wainscoat}, R.~J. and {Chastel}, S. and {McLean}, B. and {Postman}, M. and {Shiao}, B.},
        title = "{The Pan-STARRS1 Database and Data Products}",
      journal = {\apjs},
         year = 2020,
        month = nov,
       volume = {251},
       number = {1},
          eid = {7},
        pages = {7},
          doi = {10.3847/1538-4365/abb82d},
archivePrefix = {arXiv},
       eprint = {1612.05243},
 primaryClass = {astro-ph.IM},
       adsurl = {https://ui.adsabs.harvard.edu/abs/2020ApJS..251....7F}
}

@ARTICLE{Gaia2023,
       author = {{Gaia Collaboration} and {Vallenari}, A. and {Brown}, A.~G.~A. and {Prusti}, T. and {de Bruijne}, J.~H.~J. and {Arenou}, F. and {Babusiaux}, C. and {Biermann}, M. and {Creevey}, O.~L. and {Ducourant}, C. and {Evans}, D.~W. and {Eyer}, L. and {Guerra}, R. and {Hutton}, A. and {Jordi}, C. and {Klioner}, S.~A. and {Lammers}, U.~L. and {Lindegren}, L. and {Luri}, X. and {Mignard}, F. and {Panem}, C. and {Pourbaix}, D. and {Randich}, S. and {Sartoretti}, P. and {Soubiran}, C. and {Tanga}, P. and {Walton}, N.~A. and {Bailer-Jones}, C.~A.~L. and {Bastian}, U. and {Drimmel}, R. and {Jansen}, F. and {Katz}, D. and {Lattanzi}, M.~G. and {van Leeuwen}, F. and {Bakker}, J. and {Cacciari}, C. and {Casta{\~n}eda}, J. and {De Angeli}, F. and {Fabricius}, C. and {Fouesneau}, M. and {Fr{\'e}mat}, Y. and {Galluccio}, L. and {Guerrier}, A. and {Heiter}, U. and {Masana}, E. and {Messineo}, R. and {Mowlavi}, N. and {Nicolas}, C. and {Nienartowicz}, K. and {Pailler}, F. and {Panuzzo}, P. and {Riclet}, F. and {Roux}, W. and {Seabroke}, G.~M. and {Sordo}, R. and {Th{\'e}venin}, F. and {Gracia-Abril}, G. and {Portell}, J. and {Teyssier}, D. and {Altmann}, M. and {Andrae}, R. and {Audard}, M. and {Bellas-Velidis}, I. and {Benson}, K. and {Berthier}, J. and {Blomme}, R. and {Burgess}, P.~W. and {Busonero}, D. and {Busso}, G. and {C{\'a}novas}, H. and {Carry}, B. and {Cellino}, A. and {Cheek}, N. and {Clementini}, G. and {Damerdji}, Y. and {Davidson}, M. and {de Teodoro}, P. and {Nu{\~n}ez Campos}, M. and {Delchambre}, L. and {Dell'Oro}, A. and {Esquej}, P. and {Fern{\'a}ndez-Hern{\'a}ndez}, J. and {Fraile}, E. and {Garabato}, D. and {Garc{\'\i}a-Lario}, P. and {Gosset}, E. and {Haigron}, R. and {Halbwachs}, J. -L. and {Hambly}, N.~C. and {Harrison}, D.~L. and {Hern{\'a}ndez}, J. and {Hestroffer}, D. and {Hodgkin}, S.~T. and {Holl}, B. and {Jan{\ss}en}, K. and {Jevardat de Fombelle}, G. and {Jordan}, S. and {Krone-Martins}, A. and {Lanzafame}, A.~C. and {L{\"o}ffler}, W. and {Marchal}, O. and {Marrese}, P.~M. and {Moitinho}, A. and {Muinonen}, K. and {Osborne}, P. and {Pancino}, E. and {Pauwels}, T. and {Recio-Blanco}, A. and {Reyl{\'e}}, C. and {Riello}, M. and {Rimoldini}, L. and {Roegiers}, T. and {Rybizki}, J. and {Sarro}, L.~M. and {Siopis}, C. and {Smith}, M. and {Sozzetti}, A. and {Utrilla}, E. and {van Leeuwen}, M. and {Abbas}, U. and {{\'A}brah{\'a}m}, P. and {Abreu Aramburu}, A. and {Aerts}, C. and {Aguado}, J.~J. and {Ajaj}, M. and {Aldea-Montero}, F. and {Altavilla}, G. and {{\'A}lvarez}, M.~A. and {Alves}, J. and {Anders}, F. and {Anderson}, R.~I. and {Anglada Varela}, E. and {Antoja}, T. and {Baines}, D. and {Baker}, S.~G. and {Balaguer-N{\'u}{\~n}ez}, L. and {Balbinot}, E. and {Balog}, Z. and {Barache}, C. and {Barbato}, D. and {Barros}, M. and {Barstow}, M.~A. and {Bartolom{\'e}}, S. and {Bassilana}, J. -L. and {Bauchet}, N. and {Becciani}, U. and {Bellazzini}, M. and {Berihuete}, A. and {Bernet}, M. and {Bertone}, S. and {Bianchi}, L. and {Binnenfeld}, A. and {Blanco-Cuaresma}, S. and {Blazere}, A. and {Boch}, T. and {Bombrun}, A. and {Bossini}, D. and {Bouquillon}, S. and {Bragaglia}, A. and {Bramante}, L. and {Breedt}, E. and {Bressan}, A. and {Brouillet}, N. and {Brugaletta}, E. and {Bucciarelli}, B. and {Burlacu}, A. and {Butkevich}, A.~G. and {Buzzi}, R. and {Caffau}, E. and {Cancelliere}, R. and {Cantat-Gaudin}, T. and {Carballo}, R. and {Carlucci}, T. and {Carnerero}, M.~I. and {Carrasco}, J.~M. and {Casamiquela}, L. and {Castellani}, M. and {Castro-Ginard}, A. and {Chaoul}, L. and {Charlot}, P. and {Chemin}, L. and {Chiaramida}, V. and {Chiavassa}, A. and {Chornay}, N. and {Comoretto}, G. and {Contursi}, G. and {Cooper}, W.~J. and {Cornez}, T. and {Cowell}, S. and {Crifo}, F. and {Cropper}, M. and {Crosta}, M. and {Crowley}, C. and {Dafonte}, C. and {Dapergolas}, A. and {David}, M. and {David}, P. and {de Laverny}, P. and {De Luise}, F. and {De March}, R.},
        title = "{Gaia Data Release 3. Summary of the content and survey properties}",
      journal = {\aap},
         year = 2023,
        month = jun,
       volume = {674},
          eid = {A1},
        pages = {A1},
          doi = {10.1051/0004-6361/202243940},
archivePrefix = {arXiv},
       eprint = {2208.00211},
 primaryClass = {astro-ph.GA},
       adsurl = {https://ui.adsabs.harvard.edu/abs/2023A&A...674A...1G}
}

@ARTICLE{Lawrence2007,
       author = {{Lawrence}, A. and {Warren}, S.~J. and {Almaini}, O. and {Edge}, A.~C. and {Hambly}, N.~C. and {Jameson}, R.~F. and {Lucas}, P. and {Casali}, M. and {Adamson}, A. and {Dye}, S. and {Emerson}, J.~P. and {Foucaud}, S. and {Hewett}, P. and {Hirst}, P. and {Hodgkin}, S.~T. and {Irwin}, M.~J. and {Lodieu}, N. and {McMahon}, R.~G. and {Simpson}, C. and {Smail}, I. and {Mortlock}, D. and {Folger}, M.},
        title = "{The UKIRT Infrared Deep Sky Survey (UKIDSS)}",
      journal = {\mnras},
         year = 2007,
        month = aug,
       volume = {379},
       number = {4},
        pages = {1599-1617},
          doi = {10.1111/j.1365-2966.2007.12040.x},
archivePrefix = {arXiv},
       eprint = {astro-ph/0604426},
 primaryClass = {astro-ph},
       adsurl = {https://ui.adsabs.harvard.edu/abs/2007MNRAS.379.1599L}
}

@ARTICLE{Dey2019,
       author = {{Dey}, Arjun and {Schlegel}, David J. and {Lang}, Dustin and {Blum}, Robert and {Burleigh}, Kaylan and {Fan}, Xiaohui and {Findlay}, Joseph R. and {Finkbeiner}, Doug and {Herrera}, David and {Juneau}, St{\'e}phanie and {Landriau}, Martin and {Levi}, Michael and {McGreer}, Ian and {Meisner}, Aaron and {Myers}, Adam D. and {Moustakas}, John and {Nugent}, Peter and {Patej}, Anna and {Schlafly}, Edward F. and {Walker}, Alistair R. and {Valdes}, Francisco and {Weaver}, Benjamin A. and {Y{\`e}che}, Christophe and {Zou}, Hu and {Zhou}, Xu and {Abareshi}, Behzad and {Abbott}, T.~M.~C. and {Abolfathi}, Bela and {Aguilera}, C. and {Alam}, Shadab and {Allen}, Lori and {Alvarez}, A. and {Annis}, James and {Ansarinejad}, Behzad and {Aubert}, Marie and {Beechert}, Jacqueline and {Bell}, Eric F. and {BenZvi}, Segev Y. and {Beutler}, Florian and {Bielby}, Richard M. and {Bolton}, Adam S. and {Brice{\~n}o}, C{\'e}sar and {Buckley-Geer}, Elizabeth J. and {Butler}, Karen and {Calamida}, Annalisa and {Carlberg}, Raymond G. and {Carter}, Paul and {Casas}, Ricard and {Castander}, Francisco J. and {Choi}, Yumi and {Comparat}, Johan and {Cukanovaite}, Elena and {Delubac}, Timoth{\'e}e and {DeVries}, Kaitlin and {Dey}, Sharmila and {Dhungana}, Govinda and {Dickinson}, Mark and {Ding}, Zhejie and {Donaldson}, John B. and {Duan}, Yutong and {Duckworth}, Christopher J. and {Eftekharzadeh}, Sarah and {Eisenstein}, Daniel J. and {Etourneau}, Thomas and {Fagrelius}, Parker A. and {Farihi}, Jay and {Fitzpatrick}, Mike and {Font-Ribera}, Andreu and {Fulmer}, Leah and {G{\"a}nsicke}, Boris T. and {Gaztanaga}, Enrique and {George}, Koshy and {Gerdes}, David W. and {Gontcho}, Satya Gontcho A. and {Gorgoni}, Claudio and {Green}, Gregory and {Guy}, Julien and {Harmer}, Diane and {Hernandez}, M. and {Honscheid}, Klaus and {Huang}, Lijuan Wendy and {James}, David J. and {Jannuzi}, Buell T. and {Jiang}, Linhua and {Joyce}, Richard and {Karcher}, Armin and {Karkar}, Sonia and {Kehoe}, Robert and {Kneib}, Jean-Paul and {Kueter-Young}, Andrea and {Lan}, Ting-Wen and {Lauer}, Tod R. and {Le Guillou}, Laurent and {Le Van Suu}, Auguste and {Lee}, Jae Hyeon and {Lesser}, Michael and {Perreault Levasseur}, Laurence and {Li}, Ting S. and {Mann}, Justin L. and {Marshall}, Robert and {Mart{\'\i}nez-V{\'a}zquez}, C.~E. and {Martini}, Paul and {du Mas des Bourboux}, H{\'e}lion and {McManus}, Sean and {Meier}, Tobias Gabriel and {M{\'e}nard}, Brice and {Metcalfe}, Nigel and {Mu{\~n}oz-Guti{\'e}rrez}, Andrea and {Najita}, Joan and {Napier}, Kevin and {Narayan}, Gautham and {Newman}, Jeffrey A. and {Nie}, Jundan and {Nord}, Brian and {Norman}, Dara J. and {Olsen}, Knut A.~G. and {Paat}, Anthony and {Palanque-Delabrouille}, Nathalie and {Peng}, Xiyan and {Poppett}, Claire L. and {Poremba}, Megan R. and {Prakash}, Abhishek and {Rabinowitz}, David and {Raichoor}, Anand and {Rezaie}, Mehdi and {Robertson}, A.~N. and {Roe}, Natalie A. and {Ross}, Ashley J. and {Ross}, Nicholas P. and {Rudnick}, Gregory and {Safonova}, Sasha and {Saha}, Abhijit and {S{\'a}nchez}, F. Javier and {Savary}, Elodie and {Schweiker}, Heidi and {Scott}, Adam and {Seo}, Hee-Jong and {Shan}, Huanyuan and {Silva}, David R. and {Slepian}, Zachary and {Soto}, Christian and {Sprayberry}, David and {Staten}, Ryan and {Stillman}, Coley M. and {Stupak}, Robert J. and {Summers}, David L. and {Sien Tie}, Suk and {Tirado}, H. and {Vargas-Maga{\~n}a}, Mariana and {Vivas}, A. Katherina and {Wechsler}, Risa H. and {Williams}, Doug and {Yang}, Jinyi and {Yang}, Qian and {Yapici}, Tolga and {Zaritsky}, Dennis and {Zenteno}, A. and {Zhang}, Kai and {Zhang}, Tianmeng and {Zhou}, Rongpu and {Zhou}, Zhimin},
        title = "{Overview of the DESI Legacy Imaging Surveys}",
      journal = {\aj},
         year = 2019,
        month = may,
       volume = {157},
       number = {5},
          eid = {168},
        pages = {168},
          doi = {10.3847/1538-3881/ab089d},
archivePrefix = {arXiv},
       eprint = {1804.08657},
 primaryClass = {astro-ph.IM},
       adsurl = {https://ui.adsabs.harvard.edu/abs/2019AJ....157..168D}
}

@ARTICLE{Wright2010,
       author = {{Wright}, Edward L. and {Eisenhardt}, Peter R.~M. and {Mainzer}, Amy K. and {Ressler}, Michael E. and {Cutri}, Roc M. and {Jarrett}, Thomas and {Kirkpatrick}, J. Davy and {Padgett}, Deborah and {McMillan}, Robert S. and {Skrutskie}, Michael and {Stanford}, S.~A. and {Cohen}, Martin and {Walker}, Russell G. and {Mather}, John C. and {Leisawitz}, David and {Gautier}, III, Thomas N. and {McLean}, Ian and {Benford}, Dominic and {Lonsdale}, Carol J. and {Blain}, Andrew and {Mendez}, Bryan and {Irace}, William R. and {Duval}, Valerie and {Liu}, Fengchuan and {Royer}, Don and {Heinrichsen}, Ingolf and {Howard}, Joan and {Shannon}, Mark and {Kendall}, Martha and {Walsh}, Amy L. and {Larsen}, Mark and {Cardon}, Joel G. and {Schick}, Scott and {Schwalm}, Mark and {Abid}, Mohamed and {Fabinsky}, Beth and {Naes}, Larry and {Tsai}, Chao-Wei},
        title = "{The Wide-field Infrared Survey Explorer (WISE): Mission Description and Initial On-orbit Performance}",
      journal = {\aj},
         year = 2010,
        month = dec,
       volume = {140},
       number = {6},
        pages = {1868-1881},
          doi = {10.1088/0004-6256/140/6/1868},
archivePrefix = {arXiv},
       eprint = {1008.0031},
 primaryClass = {astro-ph.IM},
       adsurl = {https://ui.adsabs.harvard.edu/abs/2010AJ....140.1868W}
}

@ARTICLE{Ji2025,
       author = {{Ji}, Xihan and {Maiolino}, Roberto and {{\"U}bler}, Hannah and {Scholtz}, Jan and {D'Eugenio}, Francesco and {Sun}, Fengwu and {Perna}, Michele and {Turner}, Hannah and {Arribas}, Santiago and {Bennett}, Jake S. and {Bunker}, Andrew and {Carniani}, Stefano and {Charlot}, St{\'e}phane and {Cresci}, Giovanni and {Curti}, Mirko and {Egami}, Eiichi and {Fabian}, Andy and {Inayoshi}, Kohei and {Isobe}, Yuki and {Jones}, Gareth and {Juod{\v{z}}balis}, Ignas and {Kumari}, Nimisha and {Lyu}, Jianwei and {Mazzolari}, Giovanni and {Parlanti}, Eleonora and {Robertson}, Brant and {Rodr{\'\i}guez Del Pino}, Bruno and {Schneider}, Raffaella and {Sijacki}, Debora and {Tacchella}, Sandro and {Trinca}, Alessandro and {Valiante}, Rosa and {Venturi}, Giacomo and {Volonteri}, Marta and {Willott}, Chris and {Witten}, Callum and {Witstok}, Joris},
        title = "{BlackTHUNDER -- A non-stellar Balmer break in a black hole-dominated little red dot at $z=7.04$}",
      journal = {arXiv e-prints},
         year = 2025,
        month = jan,
          eid = {arXiv:2501.13082},
        pages = {arXiv:2501.13082},
          doi = {10.48550/arXiv.2501.13082},
archivePrefix = {arXiv},
       eprint = {2501.13082},
 primaryClass = {astro-ph.GA},
       adsurl = {https://ui.adsabs.harvard.edu/abs/2025arXiv250113082J}
}

@ARTICLE{GR2005,
       author = {{Garcia-Rissmann}, A. and {Vega}, L.~R. and {Asari}, N.~V. and {Cid Fernandes}, R. and {Schmitt}, H. and {Gonz{\'a}lez Delgado}, R.~M. and {Storchi-Bergmann}, T.},
        title = "{An atlas of calcium triplet spectra of active galaxies}",
      journal = {\mnras},
         year = 2005,
        month = may,
       volume = {359},
       number = {2},
        pages = {765-780},
          doi = {10.1111/j.1365-2966.2005.08957.x},
archivePrefix = {arXiv},
       eprint = {astro-ph/0502478},
 primaryClass = {astro-ph},
       adsurl = {https://ui.adsabs.harvard.edu/abs/2005MNRAS.359..765G}
}

@ARTICLE{Jin2012,
       author = {{Jin}, Chichuan and {Ward}, Martin and {Done}, Chris and {Gelbord}, Jonathan},
        title = "{A combined optical and X-ray study of unobscured type 1 active galactic nuclei - I. Optical spectra and spectral energy distribution modelling}",
      journal = {\mnras},
         year = 2012,
        month = mar,
       volume = {420},
       number = {3},
        pages = {1825-1847},
          doi = {10.1111/j.1365-2966.2011.19805.x},
archivePrefix = {arXiv},
       eprint = {1109.2069},
 primaryClass = {astro-ph.HE},
       adsurl = {https://ui.adsabs.harvard.edu/abs/2012MNRAS.420.1825J}
}

@ARTICLE{Sarkar2021,
       author = {{Sarkar}, A. and {Ferland}, G.~J. and {Chatzikos}, M. and {Guzm{\'a}n}, F. and {van Hoof}, P.~A.~M. and {Smyth}, R.~T. and {Ramsbottom}, C.~A. and {Keenan}, F.~P. and {Ballance}, C.~P.},
        title = "{Improved Fe II Emission-line Models for AGNs Using New Atomic Data Sets}",
      journal = {\apj},
         year = 2021,
        month = jan,
       volume = {907},
       number = {1},
          eid = {12},
        pages = {12},
          doi = {10.3847/1538-4357/abcaa6},
archivePrefix = {arXiv},
       eprint = {2011.09007},
 primaryClass = {astro-ph.GA},
       adsurl = {https://ui.adsabs.harvard.edu/abs/2021ApJ...907...12S}
}

@ARTICLE{Pandey2025,
       author = {{Pandey}, Ashwani and {Mart{\'\i}nez-Aldama}, Mary Loli and {Czerny}, Bo{\.z}ena and {Panda}, Swayamtrupta and {Zaja{\v{c}}ek}, Michal and {Wang}, Jian-Min and {Li}, Yan-Rong and {Du}, Pu},
        title = "{New Theoretical Fe II Templates for Bright Quasars}",
      journal = {\apjs},
         year = 2025,
        month = apr,
       volume = {277},
       number = {2},
          eid = {36},
        pages = {36},
          doi = {10.3847/1538-4365/adb427},
archivePrefix = {arXiv},
       eprint = {2401.18052},
 primaryClass = {astro-ph.GA},
       adsurl = {https://ui.adsabs.harvard.edu/abs/2025ApJS..277...36P}
}

@ARTICLE{deGraaff2025,
       author = {{de Graaff}, Anna and {Rix}, Hans-Walter and {Naidu}, Rohan P. and {Labbe}, Ivo and {Wang}, Bingjie and {Leja}, Joel and {Matthee}, Jorryt and {Katz}, Harley and {Greene}, Jenny E. and {Hviding}, Raphael E. and {Baggen}, Josephine and {Bezanson}, Rachel and {Boogaard}, Leindert A. and {Brammer}, Gabriel and {Dayal}, Pratika and {van Dokkum}, Pieter and {Goulding}, Andy D. and {Hirschmann}, Michaela and {Maseda}, Michael V. and {McConachie}, Ian and {Miller}, Tim B. and {Nelson}, Erica and {Oesch}, Pascal A. and {Setton}, David J. and {Shivaei}, Irene and {Weibel}, Andrea and {Whitaker}, Katherine E. and {Williams}, Christina C.},
        title = "{A remarkable Ruby: Absorption in dense gas, rather than evolved stars, drives the extreme Balmer break of a Little Red Dot at $z=3.5$}",
      journal = {arXiv e-prints},
         year = 2025,
        month = mar,
          eid = {arXiv:2503.16600},
        pages = {arXiv:2503.16600},
          doi = {10.48550/arXiv.2503.16600},
archivePrefix = {arXiv},
       eprint = {2503.16600},
 primaryClass = {astro-ph.GA},
       adsurl = {https://ui.adsabs.harvard.edu/abs/2025arXiv250316600D}
}

@ARTICLE{Setton2025,
       author = {{Setton}, David J. and {Greene}, Jenny E. and {Spilker}, Justin S. and {Williams}, Christina C. and {Labbe}, Ivo and {Ma}, Yilun and {Wang}, Bingjie and {Whitaker}, Katherine E. and {Leja}, Joel and {de Graaff}, Anna and {Alberts}, Stacey and {Bezanson}, Rachel and {Boogaard}, Leindert A. and {Brammer}, Gabriel and {Cutler}, Sam E. and {Cleri}, Nikko J. and {Cooper}, Olivia R. and {Dayal}, Pratika and {Fujimoto}, Seiji and {Furtak}, Lukas J. and {Goulding}, Andy D. and {Hirschmann}, Michaela and {Kokorev}, Vasily and {Maseda}, Michael V. and {McConachie}, Ian and {Matthee}, Jorryt and {Miller}, Tim B. and {Naidu}, Rohan P. and {Oesch}, Pascal A. and {Pan}, Richard and {Price}, Sedona H. and {Suess}, Katherine A. and {Weaver}, John R. and {Xiao}, Mengyuan and {Zhang}, Yunchong and {Zitrin}, Adi},
        title = "{A confirmed deficit of hot and cold dust emission in the most luminous Little Red Dots}",
      journal = {arXiv e-prints},
         year = 2025,
        month = mar,
          eid = {arXiv:2503.02059},
        pages = {arXiv:2503.02059},
          doi = {10.48550/arXiv.2503.02059},
archivePrefix = {arXiv},
       eprint = {2503.02059},
 primaryClass = {astro-ph.GA},
       adsurl = {https://ui.adsabs.harvard.edu/abs/2025arXiv250302059S}
}

@ARTICLE{Lyu2021,
       author = {{Lyu}, Jianwei and {Rieke}, George H.},
        title = "{The Dusty Heart of NGC 4151 Revealed by {\ensuremath{\lambda}} {\ensuremath{\sim}} 1-40 {\ensuremath{\mu}}m Reverberation Mapping and Variability: A Challenge to Current Clumpy Torus Models}",
      journal = {\apj},
         year = 2021,
        month = may,
       volume = {912},
       number = {2},
          eid = {126},
        pages = {126},
          doi = {10.3847/1538-4357/abee14},
archivePrefix = {arXiv},
       eprint = {2011.07638},
 primaryClass = {astro-ph.GA},
       adsurl = {https://ui.adsabs.harvard.edu/abs/2021ApJ...912..126L}
}

@ARTICLE{VMP2004,
       author = {{V{\'e}ron-Cetty}, M. -P. and {Joly}, M. and {V{\'e}ron}, P.},
        title = "{The unusual emission line spectrum of I Zw 1}",
      journal = {\aap},
         year = 2004,
        month = apr,
       volume = {417},
        pages = {515-525},
          doi = {10.1051/0004-6361:20035714},
archivePrefix = {arXiv},
       eprint = {astro-ph/0312654},
 primaryClass = {astro-ph},
       adsurl = {https://ui.adsabs.harvard.edu/abs/2004A&A...417..515V}
}

@ARTICLE{Naidu2025,
       author = {{Naidu}, Rohan P. and {Matthee}, Jorryt and {Katz}, Harley and {de Graaff}, Anna and {Oesch}, Pascal and {Smith}, Aaron and {Greene}, Jenny E. and {Brammer}, Gabriel and {Weibel}, Andrea and {Hviding}, Raphael and {Chisholm}, John and {Labb\textbackslash'e}, Ivo and {Simcoe}, Robert A. and {Witten}, Callum and {Atek}, Hakim and {Baggen}, Josephine F.~W. and {Belli}, Sirio and {Bezanson}, Rachel and {Boogaard}, Leindert A. and {Bose}, Sownak and {Covelo-Paz}, Alba and {Dayal}, Pratika and {Fudamoto}, Yoshinobu and {Furtak}, Lukas J. and {Giovinazzo}, Emma and {Goulding}, Andy and {Gronke}, Max and {Heintz}, Kasper E. and {Hirschmann}, Michaela and {Illingworth}, Garth and {Inoue}, Akio K. and {Johnson}, Benjamin D. and {Leja}, Joel and {Leonova}, Ecaterina and {McConachie}, Ian and {Maseda}, Michael V. and {Natarajan}, Priyamvada and {Nelson}, Erica and {Setton}, David J. and {Shivaei}, Irene and {Sobral}, David and {Stefanon}, Mauro and {Tacchella}, Sandro and {Toft}, Sune and {Torralba}, Alberto and {van Dokkum}, Pieter and {van der Wel}, Arjen and {Volonteri}, Marta and {Walter}, Fabian and {Wang}, Bingjie and {Watson}, Darach},
        title = "{A ``Black Hole Star'' Reveals the Remarkable Gas-Enshrouded Hearts of the Little Red Dots}",
      journal = {arXiv e-prints},
         year = 2025,
        month = mar,
          eid = {arXiv:2503.16596},
        pages = {arXiv:2503.16596},
          doi = {10.48550/arXiv.2503.16596},
archivePrefix = {arXiv},
       eprint = {2503.16596},
 primaryClass = {astro-ph.GA},
       adsurl = {https://ui.adsabs.harvard.edu/abs/2025arXiv250316596N}
}

@ARTICLE{Noll2009,
       author = {{Noll}, S. and {Burgarella}, D. and {Giovannoli}, E. and {Buat}, V. and {Marcillac}, D. and {Mu{\~n}oz-Mateos}, J.~C.},
        title = "{Analysis of galaxy spectral energy distributions from far-UV to far-IR with CIGALE: studying a SINGS test sample}",
      journal = {\aap},
         year = 2009,
        month = dec,
       volume = {507},
       number = {3},
        pages = {1793-1813},
          doi = {10.1051/0004-6361/200912497},
archivePrefix = {arXiv},
       eprint = {0909.5439},
 primaryClass = {astro-ph.CO},
       adsurl = {https://ui.adsabs.harvard.edu/abs/2009A&A...507.1793N}
}

@ARTICLE{Pei1992,
       author = {{Pei}, Yichuan C.},
        title = "{Interstellar Dust from the Milky Way to the Magellanic Clouds}",
      journal = {\apj},
         year = 1992,
        month = aug,
       volume = {395},
        pages = {130},
          doi = {10.1086/171637},
       adsurl = {https://ui.adsabs.harvard.edu/abs/1992ApJ...395..130P}
}

@ARTICLE{Burke2021,
       author = {{Burke}, Colin J. and {Liu}, Xin and {Chen}, Yu-Ching and {Shen}, Yue and {Guo}, Hengxiao},
        title = "{On the AGN nature of broad balmer emission in four low-redshift metal-poor galaxies}",
      journal = {\mnras},
         year = 2021,
        month = jun,
       volume = {504},
       number = {1},
        pages = {543-550},
          doi = {10.1093/mnras/stab912},
archivePrefix = {arXiv},
       eprint = {2011.10053},
 primaryClass = {astro-ph.GA},
       adsurl = {https://ui.adsabs.harvard.edu/abs/2021MNRAS.504..543B}
}

@ARTICLE{Drake2009,
       author = {{Drake}, A.~J. and {Djorgovski}, S.~G. and {Mahabal}, A. and {Beshore}, E. and {Larson}, S. and {Graham}, M.~J. and {Williams}, R. and {Christensen}, E. and {Catelan}, M. and {Boattini}, A. and {Gibbs}, A. and {Hill}, R. and {Kowalski}, R.},
        title = "{First Results from the Catalina Real-Time Transient Survey}",
      journal = {\apj},
         year = 2009,
        month = may,
       volume = {696},
       number = {1},
        pages = {870-884},
          doi = {10.1088/0004-637X/696/1/870},
archivePrefix = {arXiv},
       eprint = {0809.1394},
 primaryClass = {astro-ph},
       adsurl = {https://ui.adsabs.harvard.edu/abs/2009ApJ...696..870D}
}

@ARTICLE{Masci2019,
       author = {{Masci}, Frank J. and {Laher}, Russ R. and {Rusholme}, Ben and {Shupe}, David L. and {Groom}, Steven and {Surace}, Jason and {Jackson}, Edward and {Monkewitz}, Serge and {Beck}, Ron and {Flynn}, David and {Terek}, Scott and {Landry}, Walter and {Hacopians}, Eugean and {Desai}, Vandana and {Howell}, Justin and {Brooke}, Tim and {Imel}, David and {Wachter}, Stefanie and {Ye}, Quan-Zhi and {Lin}, Hsing-Wen and {Cenko}, S. Bradley and {Cunningham}, Virginia and {Rebbapragada}, Umaa and {Bue}, Brian and {Miller}, Adam A. and {Mahabal}, Ashish and {Bellm}, Eric C. and {Patterson}, Maria T. and {Juri{\'c}}, Mario and {Golkhou}, V. Zach and {Ofek}, Eran O. and {Walters}, Richard and {Graham}, Matthew and {Kasliwal}, Mansi M. and {Dekany}, Richard G. and {Kupfer}, Thomas and {Burdge}, Kevin and {Cannella}, Christopher B. and {Barlow}, Tom and {Van Sistine}, Angela and {Giomi}, Matteo and {Fremling}, Christoffer and {Blagorodnova}, Nadejda and {Levitan}, David and {Riddle}, Reed and {Smith}, Roger M. and {Helou}, George and {Prince}, Thomas A. and {Kulkarni}, Shrinivas R.},
        title = "{The Zwicky Transient Facility: Data Processing, Products, and Archive}",
      journal = {\pasp},
         year = 2019,
        month = jan,
       volume = {131},
       number = {995},
        pages = {018003},
          doi = {10.1088/1538-3873/aae8ac},
archivePrefix = {arXiv},
       eprint = {1902.01872},
 primaryClass = {astro-ph.IM},
       adsurl = {https://ui.adsabs.harvard.edu/abs/2019PASP..131a8003M}
}

@ARTICLE{Kelly2009,
       author = {{Kelly}, Brandon C. and {Bechtold}, Jill and {Siemiginowska}, Aneta},
        title = "{Are the Variations in Quasar Optical Flux Driven by Thermal Fluctuations?}",
      journal = {\apj},
         year = 2009,
        month = jun,
       volume = {698},
       number = {1},
        pages = {895-910},
          doi = {10.1088/0004-637X/698/1/895},
archivePrefix = {arXiv},
       eprint = {0903.5315},
 primaryClass = {astro-ph.CO},
       adsurl = {https://ui.adsabs.harvard.edu/abs/2009ApJ...698..895K}
}

@ARTICLE{MacLeod2010,
       author = {{MacLeod}, C.~L. and {Ivezi{\'c}}, {\v{Z}}. and {Kochanek}, C.~S. and {Koz{\l}owski}, S. and {Kelly}, B. and {Bullock}, E. and {Kimball}, A. and {Sesar}, B. and {Westman}, D. and {Brooks}, K. and {Gibson}, R. and {Becker}, A.~C. and {de Vries}, W.~H.},
        title = "{Modeling the Time Variability of SDSS Stripe 82 Quasars as a Damped Random Walk}",
      journal = {\apj},
         year = 2010,
        month = oct,
       volume = {721},
       number = {2},
        pages = {1014-1033},
          doi = {10.1088/0004-637X/721/2/1014},
archivePrefix = {arXiv},
       eprint = {1004.0276},
 primaryClass = {astro-ph.CO},
       adsurl = {https://ui.adsabs.harvard.edu/abs/2010ApJ...721.1014M}
}

@ARTICLE{Izotov2008,
       author = {{Izotov}, Yuri I. and {Thuan}, Trinh X.},
        title = "{Active Galactic Nuclei in Four Metal-poor Dwarf Emission-Line Galaxies}",
      journal = {\apj},
         year = 2008,
        month = nov,
       volume = {687},
       number = {1},
        pages = {133-140},
          doi = {10.1086/591660},
archivePrefix = {arXiv},
       eprint = {0807.2029},
 primaryClass = {astro-ph},
       adsurl = {https://ui.adsabs.harvard.edu/abs/2008ApJ...687..133I}
}

@ARTICLE{Simmonds2016,
       author = {{Simmonds}, C. and {Bauer}, F.~E. and {Thuan}, T.~X. and {Izotov}, Y.~I. and {Stern}, D. and {Harrison}, F.~A.},
        title = "{Do some AGN lack X-ray emission?}",
      journal = {\aap},
         year = 2016,
        month = dec,
       volume = {596},
          eid = {A64},
        pages = {A64},
          doi = {10.1051/0004-6361/201629310},
archivePrefix = {arXiv},
       eprint = {1609.07619},
 primaryClass = {astro-ph.GA},
       adsurl = {https://ui.adsabs.harvard.edu/abs/2016A&A...596A..64S}
}

@ARTICLE{Casey2025,
       author = {{Casey}, Caitlin M. and {Akins}, Hollis B. and {Finkelstein}, Steven L. and {Franco}, Maximilien and {Fujimoto}, Seiji and {Liu}, Daizhong and {Long}, Arianna S. and {Magdis}, Georgios and {Manning}, Sinclaire M. and {McKinney}, Jed and {Shuntov}, Marko and {Tanaka}, Takumi S.},
        title = "{An upper limit of 10$^6$ M$_\odot$ in dust from ALMA observations in 60 Little Red Dots}",
      journal = {arXiv e-prints},
         year = 2025,
        month = may,
          eid = {arXiv:2505.18873},
        pages = {arXiv:2505.18873},
          doi = {10.48550/arXiv.2505.18873},
archivePrefix = {arXiv},
       eprint = {2505.18873},
 primaryClass = {astro-ph.GA},
       adsurl = {https://ui.adsabs.harvard.edu/abs/2025arXiv250518873C}
}

@ARTICLE{Casey2024,
       author = {{Casey}, Caitlin M. and {Akins}, Hollis B. and {Kokorev}, Vasily and {McKinney}, Jed and {Cooper}, Olivia R. and {Long}, Arianna S. and {Franco}, Maximilien and {Manning}, Sinclaire M.},
        title = "{Dust in Little Red Dots}",
      journal = {\apjl},
         year = 2024,
        month = nov,
       volume = {975},
       number = {1},
          eid = {L4},
        pages = {L4},
          doi = {10.3847/2041-8213/ad7ba7},
archivePrefix = {arXiv},
       eprint = {2407.05094},
 primaryClass = {astro-ph.GA},
       adsurl = {https://ui.adsabs.harvard.edu/abs/2024ApJ...975L...4C}
}

@ARTICLE{Ma2025,
       author = {{Ma}, Yilun and {Greene}, Jenny E. and {Setton}, David J. and {Goulding}, Andy D. and {Annunziatella}, Marianna and {Fan}, Xiaohui and {Kokorev}, Vasily and {Labbe}, Ivo and {Li}, Jiaxuan and {Lin}, Xiaojing and {Marchesini}, Danilo and {Matthee}, Jorryt and {Robbins}, Luke and {Sajina}, Anna and {Sawicki}, Marcin and {Telford}, O. Grace},
        title = "{Counting Little Red Dots at $z<4$ with Ground-based Surveys and Spectroscopic Follow-up}",
      journal = {arXiv e-prints},
         year = 2025,
        month = apr,
          eid = {arXiv:2504.08032},
        pages = {arXiv:2504.08032},
          doi = {10.48550/arXiv.2504.08032},
archivePrefix = {arXiv},
       eprint = {2504.08032},
 primaryClass = {astro-ph.GA},
       adsurl = {https://ui.adsabs.harvard.edu/abs/2025arXiv250408032M}
}

@ARTICLE{LRD_EuclidCollaboration2025,
       author = {{Bisigello}, L. and {Rodighiero}, G. and {Fotopoulou}, S. and {Ricci}, F. and {Jahnke}, K. and {Feltre}, A. and {Allevato}, V. and {Shankar}, F. and {Cassata}, P. and {Dalla Bont{\`a}}, E. and {Gandolfi}, G. and {Girardi}, G. and {Giulietti}, M. and {Grazian}, A. and {Lovell}, C.~C. and {Maiolino}, R. and {Matamoro Zatarain}, T. and {Mezcua}, M. and {Prandoni}, I. and {Roberts}, D. and {Roster}, W. and {Salvato}, M. and {Siudek}, M. and {Tarsitano}, F. and {Toba}, Y. and {Vietri}, A. and {Wang}, L. and {Zamorani}, G. and {Baes}, M. and {Belladitta}, S. and {Nersesian}, A. and {Spinoglio}, L. and {Lopez Lopez}, X. and {Aghanim}, N. and {Altieri}, B. and {Amara}, A. and {Andreon}, S. and {Auricchio}, N. and {Aussel}, H. and {Baccigalupi}, C. and {Baldi}, M. and {Balestra}, A. and {Bardelli}, S. and {Basset}, A. and {Battaglia}, P. and {Bender}, R. and {Biviano}, A. and {Bonchi}, A. and {Branchini}, E. and {Brescia}, M. and {Brinchmann}, J. and {Camera}, S. and {Ca{\~n}as-Herrera}, G. and {Capobianco}, V. and {Carbone}, C. and {Carretero}, J. and {Casas}, S. and {Castellano}, M. and {Castignani}, G. and {Cavuoti}, S. and {Chambers}, K.~C. and {Cimatti}, A. and {Colodro-Conde}, C. and {Congedo}, G. and {Conselice}, C.~J. and {Conversi}, L. and {Copin}, Y. and {Courbin}, F. and {Courtois}, H.~M. and {Cropper}, M. and {Da Silva}, A. and {Degaudenzi}, H. and {De Lucia}, G. and {Di Giorgio}, A.~M. and {Dolding}, C. and {Dole}, H. and {Dubath}, F. and {Duncan}, C.~A.~J. and {Dupac}, X. and {Dusini}, S. and {Ealet}, A. and {Escoffier}, S. and {Farina}, M. and {Farinelli}, R. and {Faustini}, F. and {Ferriol}, S. and {Finelli}, F. and {Frailis}, M. and {Franceschi}, E. and {Galeotta}, S. and {George}, K. and {Gillard}, W. and {Gillis}, B. and {Giocoli}, C. and {G{\'o}mez-Alvarez}, P. and {Gracia-Carpio}, J. and {Granett}, B.~R. and {Grupp}, F. and {Gwyn}, S. and {Haugan}, S.~V.~H. and {Hoekstra}, H. and {Holmes}, W. and {Hook}, I.~M. and {Hormuth}, F. and {Hornstrup}, A. and {Hudelot}, P. and {Jhabvala}, M. and {Keih{\"a}nen}, E. and {Kermiche}, S. and {Kiessling}, A. and {Kubik}, B. and {K{\"u}mmel}, M. and {Kunz}, M. and {Kurki-Suonio}, H. and {Le Boulc'h}, Q. and {Le Brun}, A.~M.~C. and {Le Mignant}, D. and {Liebing}, P. and {Ligori}, S. and {Lilje}, P.~B. and {Lindholm}, V. and {Lloro}, I. and {Mainetti}, G. and {Maino}, D. and {Maiorano}, E. and {Mansutti}, O. and {Marcin}, S. and {Marggraf}, O. and {Martinelli}, M. and {Martinet}, N. and {Marulli}, F. and {Massey}, R. and {Maurogordato}, S. and {Medinaceli}, E. and {Mei}, S. and {Melchior}, M. and {Mellier}, Y. and {Meneghetti}, M. and {Merlin}, E. and {Meylan}, G. and {Mora}, A. and {Moresco}, M. and {Moscardini}, L. and {Nakajima}, R. and {Neissner}, C. and {Niemi}, S. -M. and {Nightingale}, J.~W. and {Padilla}, C. and {Paltani}, S. and {Pasian}, F. and {Pedersen}, K. and {Percival}, W.~J. and {Pettorino}, V. and {Pires}, S. and {Polenta}, G. and {Poncet}, M. and {Popa}, L.~A. and {Pozzetti}, L. and {Raison}, F. and {Rebolo}, R. and {Renzi}, A. and {Rhodes}, J. and {Riccio}, G. and {Romelli}, E. and {Roncarelli}, M. and {Rossetti}, E. and {Rottgering}, H.~J.~A. and {Rusholme}, B. and {Saglia}, R. and {Sakr}, Z. and {Sapone}, D. and {Sartoris}, B. and {Schewtschenko}, J.~A. and {Schirmer}, M. and {Schneider}, P. and {Schrabback}, T. and {Scodeggio}, M. and {Secroun}, A. and {Seidel}, G. and {Serrano}, S. and {Simon}, P. and {Sirignano}, C. and {Sirri}, G. and {Stanco}, L. and {Steinwagner}, J. and {Tallada-Cresp{\'\i}}, P. and {Taylor}, A.~N. and {Teplitz}, H.~I. and {Tereno}, I. and {Toft}, S. and {Toledo-Moreo}, R. and {Torradeflot}, F. and {Tutusaus}, I. and {Valenziano}, L. and {Valiviita}, J. and {Vassallo}, T. and {Verdoes Kleijn}, G. and {Veropalumbo}, A. and {Wang}, Y.},
        title = "{Euclid Quick Data Release (Q1). Extending the quest for little red dots to z<4}",
      journal = {arXiv e-prints},
         year = 2025,
        month = mar,
          eid = {arXiv:2503.15323},
        pages = {arXiv:2503.15323},
          doi = {10.48550/arXiv.2503.15323},
archivePrefix = {arXiv},
       eprint = {2503.15323},
 primaryClass = {astro-ph.GA},
       adsurl = {https://ui.adsabs.harvard.edu/abs/2025arXiv250315323E}
}

@ARTICLE{Juodzbalis2024,
       author = {{Juod{\v{z}}balis}, Ignas and {Ji}, Xihan and {Maiolino}, Roberto and {D'Eugenio}, Francesco and {Scholtz}, Jan and {Risaliti}, Guido and {Fabian}, Andrew C. and {Mazzolari}, Giovanni and {Gilli}, Roberto and {Prandoni}, Isabella and {Arribas}, Santiago and {Bunker}, Andrew J. and {Carniani}, Stefano and {Charlot}, St{\'e}phane and {Curtis-Lake}, Emma and {de Graaff}, Anna and {Hainline}, Kevin and {Parlanti}, Eleonora and {Perna}, Michele and {P{\'e}rez-Gonz{\'a}lez}, Pablo G. and {Robertson}, Brant and {Tacchella}, Sandro and {{\"U}bler}, Hannah and {Williams}, Christina C. and {Willott}, Chris and {Witstok}, Joris},
        title = "{JADES - the Rosetta stone of JWST-discovered AGN: deciphering the intriguing nature of early AGN}",
      journal = {\mnras},
         year = 2024,
        month = nov,
       volume = {535},
       number = {1},
        pages = {853-873},
          doi = {10.1093/mnras/stae2367},
archivePrefix = {arXiv},
       eprint = {2407.08643},
 primaryClass = {astro-ph.GA},
       adsurl = {https://ui.adsabs.harvard.edu/abs/2024MNRAS.535..853J}
}

@ARTICLE{Wang2025,
       author = {{Wang}, Bingjie and {de Graaff}, Anna and {Davies}, Rebecca L. and {Greene}, Jenny E. and {Leja}, Joel and {Brammer}, Gabriel B. and {Goulding}, Andy D. and {Miller}, Tim B. and {Suess}, Katherine A. and {Weibel}, Andrea and {Williams}, Christina C. and {Bezanson}, Rachel and {Boogaard}, Leindert A. and {Cleri}, Nikko J. and {Hirschmann}, Michaela and {Katz}, Harley and {Labb{\'e}}, Ivo and {Maseda}, Michael V. and {Matthee}, Jorryt and {McConachie}, Ian and {Naidu}, Rohan P. and {Oesch}, Pascal A. and {Rix}, Hans-Walter and {Setton}, David J. and {Whitaker}, Katherine E.},
        title = "{RUBIES: JWST/NIRSpec Confirmation of an Infrared-luminous, Broad-line Little Red Dot with an Ionized Outflow}",
      journal = {\apj},
         year = 2025,
        month = may,
       volume = {984},
       number = {2},
          eid = {121},
        pages = {121},
          doi = {10.3847/1538-4357/adc1ca},
archivePrefix = {arXiv},
       eprint = {2403.02304},
 primaryClass = {astro-ph.GA},
       adsurl = {https://ui.adsabs.harvard.edu/abs/2025ApJ...984..121W}
}

@ARTICLE{Leung2024,
       author = {{Leung}, Gene C.~K. and {Finkelstein}, Steven L. and {P{\'e}rez-Gonz{\'a}lez}, Pablo G. and {Morales}, Alexa M. and {Taylor}, Anthony J. and {Barro}, Guillermo and {Kocevski}, Dale D. and {Akins}, Hollis B. and {Carnall}, Adam C. and {Ch{\'a}vez Ortiz}, {\'O}scar A. and {Cleri}, Nikko J. and {Cullen}, Fergus and {Donnan}, Callum T. and {Dunlop}, James S. and {Ellis}, Richard S. and {Grogin}, Norman A. and {Hirschmann}, Michaela and {Koekemoer}, Anton M. and {Kokorev}, Vasily and {Lucas}, Ray A. and {McLeod}, Derek J. and {Papovich}, Casey and {Yung}, L.~Y. Aaron},
        title = "{Exploring the Nature of Little Red Dots: Constraints on AGN and Stellar Contributions from PRIMER MIRI Imaging}",
      journal = {arXiv e-prints},
         year = 2024,
        month = nov,
          eid = {arXiv:2411.12005},
        pages = {arXiv:2411.12005},
          doi = {10.48550/arXiv.2411.12005},
archivePrefix = {arXiv},
       eprint = {2411.12005},
 primaryClass = {astro-ph.GA},
       adsurl = {https://ui.adsabs.harvard.edu/abs/2024arXiv241112005L}
}

@ARTICLE{Taylor2025,
       author = {{Taylor}, Anthony J. and {Kokorev}, Vasily and {Kocevski}, Dale D. and {Akins}, Hollis B. and {Cullen}, Fergus and {Dickinson}, Mark and {Finkelstein}, Steven L. and {Arrabal Haro}, Pablo and {Bromm}, Volker and {Giavalisco}, Mauro and {Inayoshi}, Kohei and {Juneau}, Stephanie and {Leung}, Gene C.~K. and {Perez-Gonzalez}, Pablo G. and {Somerville}, Rachel S. and {Trump}, Jonathan R. and {Amorin}, Ricardo O. and {Barro}, Guillermo and {Burgarella}, Denis and {Brooks}, Madisyn and {Carnall}, Adam and {Casey}, Caitlin M. and {Cheng}, Yingjie and {Chisholm}, John and {Chworowsky}, Katherine and {Davis}, Kelcey and {Donnan}, Callum T. and {Dunlop}, James S. and {Ellis}, Richard S. and {Fernandez}, Vital and {Fujimoto}, Seiji and {Grogin}, Norman A. and {Gupta}, Ansh R. and {Hathi}, Nimish P. and {Jung}, Intae and {Hirschmann}, Michaela and {Kartaltepe}, Jeyhan S. and {Koekemoer}, Anton M. and {Larson}, Rebecca L. and {Leung}, Ho-Hin and {Llerena}, Mario and {Lucas}, Ray A. and {McLeod}, Derek J. and {McLure}, Ross and {Napolitano}, Lorenzo and {Papovich}, Casey and {Stanton}, Thomas M. and {Tripodi}, Roberta and {Wang}, Xin and {Wilkins}, Stephen M. and {Yung}, L.~Y. Aaron and {Zavala}, Jorge A.},
        title = "{CAPERS-LRD-z9: A Gas Enshrouded Little Red Dot Hosting a Broad-line AGN at z=9.288}",
      journal = {arXiv e-prints},
         year = 2025,
        month = may,
          eid = {arXiv:2505.04609},
        pages = {arXiv:2505.04609},
          doi = {10.48550/arXiv.2505.04609},
archivePrefix = {arXiv},
       eprint = {2505.04609},
 primaryClass = {astro-ph.GA},
       adsurl = {https://ui.adsabs.harvard.edu/abs/2025arXiv250504609T}
}

@ARTICLE{Lin2025,
       author = {{Lin}, Xiaojing and {Fan}, Xiaohui and {Wang}, Feige and {Sun}, Fengwu and {Champagne}, Jaclyn B. and {Egami}, Eiichi and {Kakiichi}, Koki and {Lyu}, Jianwei and {Tee}, Wei Leong and {Yang}, Jinyi and {Bian}, Fuyan and {Bosman}, Sarah E.~I. and {Cai}, Zheng and {Casey}, Caitlin M. and {Decarli}, Roberto and {Faisst}, Andreas L. and {Fujimoto}, Seiji and {Harish}, Santosh and {Ilbert}, Olivier and {Inoue}, Akio K. and {Jin}, Xiangyu and {Kartaltepe}, Jeyhan S. and {Kocevski}, Dale D. and {Li}, Mingyu and {Liu}, Weizhe and {Liu}, Yichen and {Schindler}, Jan-Torge and {Shuntov}, Marko and {Tanaka}, Takumi S. and {Vestergaard}, Marianne and {Wu}, Yunjing and {Zhang}, Haowen and {Zhang}, Zijian},
        title = "{Bridging Quasars and Little Red Dots: Insights into Broad-Line AGNs at $z=5-8$ from the First JWST COSMOS-3D Dataset}",
      journal = {arXiv e-prints},
         year = 2025,
        month = apr,
          eid = {arXiv:2504.08039},
        pages = {arXiv:2504.08039},
          doi = {10.48550/arXiv.2504.08039},
archivePrefix = {arXiv},
       eprint = {2504.08039},
 primaryClass = {astro-ph.GA},
       adsurl = {https://ui.adsabs.harvard.edu/abs/2025arXiv250408039L}
}

@ARTICLE{Greene2024,
       author = {{Greene}, Jenny E. and {Labbe}, Ivo and {Goulding}, Andy D. and {Furtak}, Lukas J. and {Chemerynska}, Iryna and {Kokorev}, Vasily and {Dayal}, Pratika and {Volonteri}, Marta and {Williams}, Christina C. and {Wang}, Bingjie and {Setton}, David J. and {Burgasser}, Adam J. and {Bezanson}, Rachel and {Atek}, Hakim and {Brammer}, Gabriel and {Cutler}, Sam E. and {Feldmann}, Robert and {Fujimoto}, Seiji and {Glazebrook}, Karl and {de Graaff}, Anna and {Khullar}, Gourav and {Leja}, Joel and {Marchesini}, Danilo and {Maseda}, Michael V. and {Matthee}, Jorryt and {Miller}, Tim B. and {Naidu}, Rohan P. and {Nanayakkara}, Themiya and {Oesch}, Pascal A. and {Pan}, Richard and {Papovich}, Casey and {Price}, Sedona H. and {van Dokkum}, Pieter and {Weaver}, John R. and {Whitaker}, Katherine E. and {Zitrin}, Adi},
        title = "{UNCOVER Spectroscopy Confirms the Surprising Ubiquity of Active Galactic Nuclei in Red Sources at z > 5}",
      journal = {\apj},
         year = 2024,
        month = mar,
       volume = {964},
       number = {1},
          eid = {39},
        pages = {39},
          doi = {10.3847/1538-4357/ad1e5f},
archivePrefix = {arXiv},
       eprint = {2309.05714},
 primaryClass = {astro-ph.GA},
       adsurl = {https://ui.adsabs.harvard.edu/abs/2024ApJ...964...39G}
}

@ARTICLE{Matthee2024,
       author = {{Matthee}, Jorryt and {Naidu}, Rohan P. and {Brammer}, Gabriel and {Chisholm}, John and {Eilers}, Anna-Christina and {Goulding}, Andy and {Greene}, Jenny and {Kashino}, Daichi and {Labbe}, Ivo and {Lilly}, Simon J. and {Mackenzie}, Ruari and {Oesch}, Pascal A. and {Weibel}, Andrea and {Wuyts}, Stijn and {Xiao}, Mengyuan and {Bordoloi}, Rongmon and {Bouwens}, Rychard and {van Dokkum}, Pieter and {Illingworth}, Garth and {Kramarenko}, Ivan and {Maseda}, Michael V. and {Mason}, Charlotte and {Meyer}, Romain A. and {Nelson}, Erica J. and {Reddy}, Naveen A. and {Shivaei}, Irene and {Simcoe}, Robert A. and {Yue}, Minghao},
        title = "{Little Red Dots: An Abundant Population of Faint Active Galactic Nuclei at z {\ensuremath{\sim}} 5 Revealed by the EIGER and FRESCO JWST Surveys}",
      journal = {\apj},
         year = 2024,
        month = mar,
       volume = {963},
       number = {2},
          eid = {129},
        pages = {129},
          doi = {10.3847/1538-4357/ad2345},
archivePrefix = {arXiv},
       eprint = {2306.05448},
 primaryClass = {astro-ph.GA},
       adsurl = {https://ui.adsabs.harvard.edu/abs/2024ApJ...963..129M}
}

@ARTICLE{Maiolino2024,
       author = {{Maiolino}, Roberto and {Scholtz}, Jan and {Curtis-Lake}, Emma and {Carniani}, Stefano and {Baker}, William and {de Graaff}, Anna and {Tacchella}, Sandro and {{\"U}bler}, Hannah and {D'Eugenio}, Francesco and {Witstok}, Joris and {Curti}, Mirko and {Arribas}, Santiago and {Bunker}, Andrew J. and {Charlot}, St{\'e}phane and {Chevallard}, Jacopo and {Eisenstein}, Daniel J. and {Egami}, Eiichi and {Ji}, Zhiyuan and {Jones}, Gareth C. and {Lyu}, Jianwei and {Rawle}, Tim and {Robertson}, Brant and {Rujopakarn}, Wiphu and {Perna}, Michele and {Sun}, Fengwu and {Venturi}, Giacomo and {Williams}, Christina C. and {Willott}, Chris},
        title = "{JADES: The diverse population of infant black holes at 4 < z < 11: Merging, tiny, poor, but mighty}",
      journal = {\aap},
         year = 2024,
        month = nov,
       volume = {691},
          eid = {A145},
        pages = {A145},
          doi = {10.1051/0004-6361/202347640},
archivePrefix = {arXiv},
       eprint = {2308.01230},
 primaryClass = {astro-ph.GA},
       adsurl = {https://ui.adsabs.harvard.edu/abs/2024A&A...691A.145M}
}

@ARTICLE{Lin2024,
       author = {{Lin}, Xiaojing and {Wang}, Feige and {Fan}, Xiaohui and {Cai}, Zheng and {Champagne}, Jaclyn B. and {Sun}, Fengwu and {Volonteri}, Marta and {Yang}, Jinyi and {Hennawi}, Joseph F. and {Ba{\~n}ados}, Eduardo and {Barth}, Aaron and {Eilers}, Anna-Christina and {Farina}, Emanuele Paolo and {Liu}, Weizhe and {Jin}, Xiangyu and {Jun}, Hyunsung D. and {Lupi}, Alessandro and {Kakiichi}, Koki and {Mazzucchelli}, Chiara and {Onoue}, Masafusa and {Pan}, Zhiwei and {Pizzati}, Elia and {Rojas-Ruiz}, Sof{\'\i}a and {Schindler}, Jan-Torge and {Trakhtenbrot}, Benny and {Shen}, Yue and {Trebitsch}, Maxime and {Zhuang}, Ming-Yang and {Endsley}, Ryan and {Meyer}, Romain A. and {Li}, Zihao and {Li}, Mingyu and {Pudoka}, Maria and {Tee}, Wei Leong and {Wu}, Yunjing and {Zhang}, Haowen},
        title = "{A SPectroscopic Survey of Biased Halos In the Reionization Era (ASPIRE): Broad-line AGN at z = 4‑5 Revealed by JWST/NIRCam WFSS}",
      journal = {\apj},
         year = 2024,
        month = oct,
       volume = {974},
       number = {1},
          eid = {147},
        pages = {147},
          doi = {10.3847/1538-4357/ad6565},
archivePrefix = {arXiv},
       eprint = {2407.17570},
 primaryClass = {astro-ph.GA},
       adsurl = {https://ui.adsabs.harvard.edu/abs/2024ApJ...974..147L}
}

@ARTICLE{Akins2024,
       author = {{Akins}, Hollis B. and {Casey}, Caitlin M. and {Lambrides}, Erini and {Allen}, Natalie and {Andika}, Irham T. and {Brinch}, Malte and {Champagne}, Jaclyn B. and {Cooper}, Olivia and {Ding}, Xuheng and {Drakos}, Nicole E. and {Faisst}, Andreas and {Finkelstein}, Steven L. and {Franco}, Maximilien and {Fujimoto}, Seiji and {Gentile}, Fabrizio and {Gillman}, Steven and {Gozaliasl}, Ghassem and {Harish}, Santosh and {Hayward}, Christopher C. and {Hirschmann}, Michaela and {Ilbert}, Olivier and {Kartaltepe}, Jeyhan S. and {Kocevski}, Dale D. and {Koekemoer}, Anton M. and {Kokorev}, Vasily and {Liu}, Daizhong and {Long}, Arianna S. and {McCracken}, Henry Joy and {McKinney}, Jed and {Onoue}, Masafusa and {Paquereau}, Louise and {Renzini}, Alvio and {Rhodes}, Jason and {Robertson}, Brant E. and {Shuntov}, Marko and {Silverman}, John D. and {Tanaka}, Takumi S. and {Toft}, Sune and {Trakhtenbrot}, Benny and {Valentino}, Francesco and {Zavala}, Jorge},
        title = "{COSMOS-Web: The over-abundance and physical nature of ``little red dots''--Implications for early galaxy and SMBH assembly}",
      journal = {arXiv e-prints},
         year = 2024,
        month = jun,
          eid = {arXiv:2406.10341},
        pages = {arXiv:2406.10341},
          doi = {10.48550/arXiv.2406.10341},
archivePrefix = {arXiv},
       eprint = {2406.10341},
 primaryClass = {astro-ph.GA},
       adsurl = {https://ui.adsabs.harvard.edu/abs/2024arXiv240610341A}
}

@ARTICLE{Rinaldi2024,
       author = {{Rinaldi}, P. and {Bonaventura}, N. and {Rieke}, G.~H. and {Alberts}, S. and {Caputi}, K.~I. and {Baker}, W.~M. and {Baum}, S. and {Bhatawdekar}, R. and {Bunker}, A.~J. and {Carniani}, S. and {Curtis-Lake}, E. and {D'Eugenio}, F. and {Egami}, E. and {Ji}, Z. and {Hainline}, K. and {Helton}, J.~M. and {Lin}, X. and {Lyu}, J. and {Johnson}, B.~D. and {Ma}, Z. and {Maiolino}, R. and {P{\'e}rez-Gonz{\'a}lez}, P.~G. and {Rieke}, M. and {Robertson}, B.~E. and {Shivaei}, I. and {Stone}, M. and {Sun}, Y. and {Tacchella}, S. and {{\"U}bler}, H. and {Williams}, C.~C. and {Willmer}, C.~N.~A. and {Willott}, C. and {Zhang}, J. and {Zhu}, Y.},
        title = "{Not Just a Dot: the complex UV morphology and underlying properties of Little Red Dots}",
      journal = {arXiv e-prints},
         year = 2024,
        month = nov,
          eid = {arXiv:2411.14383},
        pages = {arXiv:2411.14383},
          doi = {10.48550/arXiv.2411.14383},
archivePrefix = {arXiv},
       eprint = {2411.14383},
 primaryClass = {astro-ph.GA},
       adsurl = {https://ui.adsabs.harvard.edu/abs/2024arXiv241114383R}
}

@ARTICLE{Kocevski2024,
       author = {{Kocevski}, Dale D. and {Finkelstein}, Steven L. and {Barro}, Guillermo and {Taylor}, Anthony J. and {Calabr{\`o}}, Antonello and {Laloux}, Brivael and {Buchner}, Johannes and {Trump}, Jonathan R. and {Leung}, Gene C.~K. and {Yang}, Guang and {Dickinson}, Mark and {P{\'e}rez-Gonz{\'a}lez}, Pablo G. and {Pacucci}, Fabio and {Inayoshi}, Kohei and {Somerville}, Rachel S. and {McGrath}, Elizabeth J. and {Akins}, Hollis B. and {Bagley}, Micaela B. and {Bisigello}, Laura and {Bowler}, Rebecca A.~A. and {Carnall}, Adam and {Casey}, Caitlin M. and {Cheng}, Yingjie and {Cleri}, Nikko J. and {Costantin}, Luca and {Cullen}, Fergus and {Davis}, Kelcey and {Donnan}, Callum T. and {Dunlop}, James S. and {Ellis}, Richard S. and {Ferguson}, Henry C. and {Fujimoto}, Seiji and {Fontana}, Adriano and {Giavalisco}, Mauro and {Grazian}, Andrea and {Grogin}, Norman A. and {Hathi}, Nimish P. and {Hirschmann}, Michaela and {Huertas-Company}, Marc and {Holwerda}, Benne W. and {Illingworth}, Garth and {Juneau}, St{\'e}phanie and {Kartaltepe}, Jeyhan S. and {Koekemoer}, Anton M. and {Li}, Wenxiu and {Lucas}, Ray A. and {Magee}, Dan and {Mason}, Charlotte and {McLeod}, Derek J. and {McLure}, Ross J. and {Napolitano}, Lorenzo and {Papovich}, Casey and {Pirzkal}, Nor and {Rodighiero}, Giulia and {Santini}, Paola and {Wilkins}, Stephen M. and {Yung}, L.~Y. Aaron},
        title = "{The Rise of Faint, Red AGN at $z>4$: A Sample of Little Red Dots in the JWST Extragalactic Legacy Fields}",
      journal = {arXiv e-prints},
         year = 2024,
        month = apr,
          eid = {arXiv:2404.03576},
        pages = {arXiv:2404.03576},
          doi = {10.48550/arXiv.2404.03576},
archivePrefix = {arXiv},
       eprint = {2404.03576},
 primaryClass = {astro-ph.GA},
       adsurl = {https://ui.adsabs.harvard.edu/abs/2024arXiv240403576K}
}

@ARTICLE{Zhang2025,
       author = {{Zhang}, Junyu and {Egami}, Eiichi and {Sun}, Fengwu and {Lin}, Xiaojing and {Lyu}, Jianwei and {Zhu}, Yongda and {Rinaldi}, Pierluigi and {Sun}, Yang and {Bunker}, Andrew J. and {Bhatawdekar}, Rachana and {Helton}, Jakob M. and {Maiolino}, Roberto and {Ma}, Zheng and {Robertson}, Brant and {Tacchella}, Sandro and {Venturi}, Giacomo and {Williams}, Christina C. and {Willott}, Chris},
        title = "{Abundant Population of Broad H$α$ Emitters in the GOODS-N Field Revealed by CONGRESS, FRESCO, and JADES}",
      journal = {arXiv e-prints},
         year = 2025,
        month = may,
          eid = {arXiv:2505.02895},
        pages = {arXiv:2505.02895},
          doi = {10.48550/arXiv.2505.02895},
archivePrefix = {arXiv},
       eprint = {2505.02895},
 primaryClass = {astro-ph.GA},
       adsurl = {https://ui.adsabs.harvard.edu/abs/2025arXiv250502895Z}
}

@ARTICLE{Hainline2025,
       author = {{Hainline}, Kevin N. and {Maiolino}, Roberto and {Juod{\v{z}}balis}, Ignas and {Scholtz}, Jan and {{\"U}bler}, Hannah and {D'Eugenio}, Francesco and {Helton}, Jakob M. and {Sun}, Yang and {Sun}, Fengwu and {Robertson}, Brant and {Tacchella}, Sandro and {Bunker}, Andrew J. and {Carniani}, Stefano and {Charlot}, Stephane and {Curtis-Lake}, Emma and {Egami}, Eiichi and {Johnson}, Benjamin D. and {Lin}, Xiaojing and {Lyu}, Jianwei and {P{\'e}rez-Gonz{\'a}lez}, Pablo G. and {Rinaldi}, Pierluigi and {Silcock}, Maddie S. and {Venturi}, Giacomo and {Williams}, Christina C. and {Willmer}, Christopher N.~A. and {Willott}, Chris and {Zhang}, Junyu and {Zhu}, Yongda},
        title = "{An Investigation into the Selection and Colors of Little Red Dots and Active Galactic Nuclei}",
      journal = {\apj},
         year = 2025,
        month = feb,
       volume = {979},
       number = {2},
          eid = {138},
        pages = {138},
          doi = {10.3847/1538-4357/ad9920},
archivePrefix = {arXiv},
       eprint = {2410.00100},
 primaryClass = {astro-ph.GA},
       adsurl = {https://ui.adsabs.harvard.edu/abs/2025ApJ...979..138H}
}

@ARTICLE{DEugenio2025,
       author = {{D'Eugenio}, Francesco and {Juod{\v{z}}balis}, Ignas and {Ji}, Xihan and {Scholtz}, Jan and {Maiolino}, Roberto and {Carniani}, Stefano and {Perna}, Michele and {Mazzolari}, Giovanni and {{\"U}bler}, Hannah and {Arribas}, Santiago and {Bhatawdekar}, Rachana and {Bunker}, Andrew J. and {Cresci}, Giovanni and {Curtis-Lake}, Emma and {Hainline}, Kevin and {Inayoshi}, Kohei and {Isobe}, Yuki and {Johnson}, Benjamin D. and {Jones}, Gareth C. and {Looser}, Tobias J. and {Nelson}, Erica J. and {Parlanti}, Eleonora and {Pusk{\'a}s}, D{\'a}vid and {Rinaldi}, Pierluigi and {Robertson}, Brant and {Rodr{\'\i}guez Del Pino}, Bruno and {Shivaei}, Irene and {Sun}, Fengwu and {Tacchella}, Sandro and {Venturi}, Giacomo and {Volonteri}, Marta and {Williams}, Christina C. and {Willmer}, Christopher N.~A. and {Willott}, Chris and {Witstok}, Joris},
        title = "{JADES and BlackTHUNDER: rest-frame Balmer-line absorption and the local environment in a Little Red Dot at z = 5}",
      journal = {arXiv e-prints},
         year = 2025,
        month = jun,
          eid = {arXiv:2506.14870},
        pages = {arXiv:2506.14870},
archivePrefix = {arXiv},
       eprint = {2506.14870},
 primaryClass = {astro-ph.GA},
       adsurl = {https://ui.adsabs.harvard.edu/abs/2025arXiv250614870D}
}

@ARTICLE{Labbe2024,
       author = {{Labbe}, Ivo and {Greene}, Jenny E. and {Matthee}, Jorryt and {Treiber}, Helena and {Kokorev}, Vasily and {Miller}, Tim B. and {Kramarenko}, Ivan and {Setton}, David J. and {Ma}, Yilun and {Goulding}, Andy D. and {Bezanson}, Rachel and {Naidu}, Rohan P. and {Williams}, Christina C. and {Atek}, Hakim and {Brammer}, Gabriel and {Cutler}, Sam E. and {Chemerynska}, Iryna and {Cloonan}, Aidan P. and {Dayal}, Pratika and {de Graaff}, Anna and {Fudamoto}, Yoshinobu and {Fujimoto}, Seiji and {Furtak}, Lukas J. and {Glazebrook}, Karl and {Heintz}, Kasper E. and {Leja}, Joel and {Marchesini}, Danilo and {Nanayakkara}, Themiya and {Nelson}, Erica J. and {Oesch}, Pascal A. and {Pan}, Richard and {Price}, Sedona H. and {Shivaei}, Irene and {Sobral}, David and {Suess}, Katherine A. and {van Dokkum}, Pieter and {Wang}, Bingjie and {Weaver}, John R. and {Whitaker}, Katherine E. and {Zitrin}, Adi},
        title = "{An unambiguous AGN and a Balmer break in an Ultraluminous Little Red Dot at z=4.47 from Ultradeep UNCOVER and All the Little Things Spectroscopy}",
      journal = {arXiv e-prints},
         year = 2024,
        month = dec,
          eid = {arXiv:2412.04557},
        pages = {arXiv:2412.04557},
          doi = {10.48550/arXiv.2412.04557},
archivePrefix = {arXiv},
       eprint = {2412.04557},
 primaryClass = {astro-ph.GA},
       adsurl = {https://ui.adsabs.harvard.edu/abs/2024arXiv241204557L}
}

@ARTICLE{Wang2024,
       author = {{Wang}, Bingjie and {Leja}, Joel and {de Graaff}, Anna and {Brammer}, Gabriel B. and {Weibel}, Andrea and {van Dokkum}, Pieter and {Baggen}, Josephine F.~W. and {Suess}, Katherine A. and {Greene}, Jenny E. and {Bezanson}, Rachel and {Cleri}, Nikko J. and {Hirschmann}, Michaela and {Labb{\'e}}, Ivo and {Matthee}, Jorryt and {McConachie}, Ian and {Naidu}, Rohan P. and {Nelson}, Erica and {Oesch}, Pascal A. and {Setton}, David J. and {Williams}, Christina C.},
        title = "{RUBIES: Evolved Stellar Populations with Extended Formation Histories at z {\ensuremath{\sim}} 7{\textendash}8 in Candidate Massive Galaxies Identified with JWST/NIRSpec}",
      journal = {\apjl},
         year = 2024,
        month = jul,
       volume = {969},
       number = {1},
          eid = {L13},
        pages = {L13},
          doi = {10.3847/2041-8213/ad55f7},
archivePrefix = {arXiv},
       eprint = {2405.01473},
 primaryClass = {astro-ph.GA},
       adsurl = {https://ui.adsabs.harvard.edu/abs/2024ApJ...969L..13W}
}

@ARTICLE{Hviding2025,
       author = {{Hviding}, Raphael E. and {de Graaff}, Anna and {Miller}, Tim B. and {Setton}, David J. and {Greene}, Jenny E. and {Labb{\'e}}, Ivo and {Brammer}, Gabriel and {Bezanson}, Rachel and {Boogaard}, Leindert A. and {Cleri}, Nikko J. and {Leja}, Joel and {Maseda}, Michael V. and {McConachie}, Ian and {Matthee}, Jorryt and {Naidu}, Rohan P. and {Oesch}, Pascal A. and {Wang}, Bingjie and {Whitaker}, Katherine E. and {Williams}, Christina},
        title = "{RUBIES: A Spectroscopic Census of Little Red Dots; All V-Shaped Point Sources Have Broad Lines}",
      journal = {arXiv e-prints},
         year = 2025,
        month = jun,
          eid = {arXiv:2506.05459},
        pages = {arXiv:2506.05459},
          doi = {10.48550/arXiv.2506.05459},
archivePrefix = {arXiv},
       eprint = {2506.05459},
 primaryClass = {astro-ph.GA},
       adsurl = {https://ui.adsabs.harvard.edu/abs/2025arXiv250605459H}
}

@ARTICLE{Perez-Gonzalez2024,
       author = {{P{\'e}rez-Gonz{\'a}lez}, Pablo G. and {Barro}, Guillermo and {Rieke}, George H. and {Lyu}, Jianwei and {Rieke}, Marcia and {Alberts}, Stacey and {Williams}, Christina C. and {Hainline}, Kevin and {Sun}, Fengwu and {Pusk{\'a}s}, D{\'a}vid and {Annunziatella}, Marianna and {Baker}, William M. and {Bunker}, Andrew J. and {Egami}, Eiichi and {Ji}, Zhiyuan and {Johnson}, Benjamin D. and {Robertson}, Brant and {Rodr{\'\i}guez Del Pino}, Bruno and {Rujopakarn}, Wiphu and {Shivaei}, Irene and {Tacchella}, Sandro and {Willmer}, Christopher N.~A. and {Willott}, Chris},
        title = "{What Is the Nature of Little Red Dots and what Is Not, MIRI SMILES Edition}",
      journal = {\apj},
         year = 2024,
        month = jun,
       volume = {968},
       number = {1},
          eid = {4},
        pages = {4},
          doi = {10.3847/1538-4357/ad38bb},
archivePrefix = {arXiv},
       eprint = {2401.08782},
 primaryClass = {astro-ph.GA},
       adsurl = {https://ui.adsabs.harvard.edu/abs/2024ApJ...968....4P}
}

@ARTICLE{Williams2024,
       author = {{Williams}, Christina C. and {Alberts}, Stacey and {Ji}, Zhiyuan and {Hainline}, Kevin N. and {Lyu}, Jianwei and {Rieke}, George and {Endsley}, Ryan and {Suess}, Katherine A. and {Sun}, Fengwu and {Johnson}, Benjamin D. and {Florian}, Michael and {Shivaei}, Irene and {Rujopakarn}, Wiphu and {Baker}, William M. and {Bhatawdekar}, Rachana and {Boyett}, Kristan and {Bunker}, Andrew J. and {Cameron}, Alex J. and {Carniani}, Stefano and {Charlot}, Stephane and {Curtis-Lake}, Emma and {DeCoursey}, Christa and {de Graaff}, Anna and {Egami}, Eiichi and {Eisenstein}, Daniel J. and {Gibson}, Justus L. and {Hausen}, Ryan and {Helton}, Jakob M. and {Maiolino}, Roberto and {Maseda}, Michael V. and {Nelson}, Erica J. and {P{\'e}rez-Gonz{\'a}lez}, Pablo G. and {Rieke}, Marcia J. and {Robertson}, Brant E. and {Saxena}, Aayush and {Tacchella}, Sandro and {Willmer}, Christopher N.~A. and {Willott}, Chris J.},
        title = "{The Galaxies Missed by Hubble and ALMA: The Contribution of Extremely Red Galaxies to the Cosmic Census at 3 < z < 8}",
      journal = {\apj},
         year = 2024,
        month = jun,
       volume = {968},
       number = {1},
          eid = {34},
        pages = {34},
          doi = {10.3847/1538-4357/ad3f17},
archivePrefix = {arXiv},
       eprint = {2311.07483},
 primaryClass = {astro-ph.GA},
       adsurl = {https://ui.adsabs.harvard.edu/abs/2024ApJ...968...34W}
}

@ARTICLE{Li2025,
       author = {{Li}, Zhengrong and {Inayoshi}, Kohei and {Chen}, Kejian and {Ichikawa}, Kohei and {Ho}, Luis C.},
        title = "{Little Red Dots: Rapidly Growing Black Holes Reddened by Extended Dusty Flows}",
      journal = {\apj},
         year = 2025,
        month = feb,
       volume = {980},
       number = {1},
          eid = {36},
        pages = {36},
          doi = {10.3847/1538-4357/ada5fb},
archivePrefix = {arXiv},
       eprint = {2407.10760},
 primaryClass = {astro-ph.GA},
       adsurl = {https://ui.adsabs.harvard.edu/abs/2025ApJ...980...36L}
}

@ARTICLE{Inayoshi2024,
       author = {{Inayoshi}, Kohei and {Maiolino}, Roberto},
        title = "{Extremely Dense Gas around Little Red Dots and High-redshift Active Galactic Nuclei: A Non-stellar Origin of the Balmer Break and Absorption Features}",
      journal = {arXiv e-prints},
         year = 2024,
        month = sep,
          eid = {arXiv:2409.07805},
        pages = {arXiv:2409.07805},
          doi = {10.48550/arXiv.2409.07805},
archivePrefix = {arXiv},
       eprint = {2409.07805},
 primaryClass = {astro-ph.GA},
       adsurl = {https://ui.adsabs.harvard.edu/abs/2024arXiv240907805I}
}

@ARTICLE{Ma2024,
       author = {{Ma}, Yilun and {Greene}, Jenny E. and {Setton}, David J. and {Volonteri}, Marta and {Leja}, Joel and {Wang}, Bingjie and {Bezanson}, Rachel and {Brammer}, Gabriel and {Cutler}, Sam E. and {Dayal}, Pratika and {van Dokkum}, Pieter and {Furtak}, Lukas J. and {Glazebrook}, Karl and {Goulding}, Andy D. and {de Graaff}, Anna and {Kokorev}, Vasily and {Labbe}, Ivo and {Pan}, Richard and {Price}, Sedona H. and {Weaver}, John R. and {Williams}, Christina C. and {Whitaker}, Katherine E. and {Zitrin}, Adi},
        title = "{UNCOVER: 404 Error -- Models Not Found for the Triply Imaged Little Red Dot A2744-QSO1}",
      journal = {arXiv e-prints},
         year = 2024,
        month = oct,
          eid = {arXiv:2410.06257},
        pages = {arXiv:2410.06257},
          doi = {10.48550/arXiv.2410.06257},
archivePrefix = {arXiv},
       eprint = {2410.06257},
 primaryClass = {astro-ph.GA},
       adsurl = {https://ui.adsabs.harvard.edu/abs/2024arXiv241006257M}
}

@ARTICLE{Chen2025,
       author = {{Chen}, Kejian and {Li}, Zhengrong and {Inayoshi}, Kohei and {Ho}, Luis C.},
        title = "{Dust Budget Crisis in Little Red Dots}",
      journal = {arXiv e-prints},
         year = 2025,
        month = may,
          eid = {arXiv:2505.22600},
        pages = {arXiv:2505.22600},
          doi = {10.48550/arXiv.2505.22600},
archivePrefix = {arXiv},
       eprint = {2505.22600},
 primaryClass = {astro-ph.GA},
       adsurl = {https://ui.adsabs.harvard.edu/abs/2025arXiv250522600C}
}

@ARTICLE{Chen2010,
       author = {{Chen}, Yan-Mei and {Tremonti}, Christy A. and {Heckman}, Timothy M. and {Kauffmann}, Guinevere and {Weiner}, Benjamin J. and {Brinchmann}, Jarle and {Wang}, Jing},
        title = "{Absorption-line Probes of the Prevalence and Properties of Outflows in Present-day Star-forming Galaxies}",
      journal = {\aj},
         year = 2010,
        month = aug,
       volume = {140},
       number = {2},
        pages = {445-461},
          doi = {10.1088/0004-6256/140/2/445},
archivePrefix = {arXiv},
       eprint = {1003.5425},
 primaryClass = {astro-ph.GA},
       adsurl = {https://ui.adsabs.harvard.edu/abs/2010AJ....140..445C}
}

@ARTICLE{Rupke2021,
       author = {{Rupke}, David S.~N. and {Thomas}, Adam D. and {Dopita}, Michael A.},
        title = "{The spatially resolved gas and dust connection in neutral inflows and outflows in nearby AGN}",
      journal = {\mnras},
         year = 2021,
        month = jun,
       volume = {503},
       number = {4},
        pages = {4748-4766},
          doi = {10.1093/mnras/stab743},
archivePrefix = {arXiv},
       eprint = {2103.08502},
 primaryClass = {astro-ph.GA},
       adsurl = {https://ui.adsabs.harvard.edu/abs/2021MNRAS.503.4748R}
}

@ARTICLE{Cazzoli2016,
       author = {{Cazzoli}, S. and {Arribas}, S. and {Maiolino}, R. and {Colina}, L.},
        title = "{Neutral gas outflows in nearby [U]LIRGs via optical NaD feature}",
      journal = {\aap},
         year = 2016,
        month = may,
       volume = {590},
          eid = {A125},
        pages = {A125},
          doi = {10.1051/0004-6361/201526788},
archivePrefix = {arXiv},
       eprint = {1602.08505},
 primaryClass = {astro-ph.GA},
       adsurl = {https://ui.adsabs.harvard.edu/abs/2016A&A...590A.125C}
}

@ARTICLE{Welty2001,
       author = {{Welty}, Daniel E. and {Hobbs}, L.~M.},
        title = "{A High-Resolution Survey of Interstellar K I Absorption}",
      journal = {\apjs},
         year = 2001,
        month = apr,
       volume = {133},
       number = {2},
        pages = {345-393},
          doi = {10.1086/320354},
       adsurl = {https://ui.adsabs.harvard.edu/abs/2001ApJS..133..345W}
}

@ARTICLE{Sun2024,
       author = {{Sun}, Yang and {Lee}, Gwang-Ho and {Zabludoff}, Ann I. and {French}, K. Decker and {Helton}, Jakob M. and {Kerrison}, Nicole A. and {Tremonti}, Christy A. and {Yang}, Yujin},
        title = "{Evolution of gas flows along the starburst to post-starburst to quiescent galaxy sequence}",
      journal = {\mnras},
         year = 2024,
        month = mar,
       volume = {528},
       number = {4},
        pages = {5783-5803},
          doi = {10.1093/mnras/stae366},
archivePrefix = {arXiv},
       eprint = {2310.00424},
 primaryClass = {astro-ph.GA},
       adsurl = {https://ui.adsabs.harvard.edu/abs/2024MNRAS.528.5783S}
}

@ARTICLE{Rupke2005,
       author = {{Rupke}, David S. and {Veilleux}, Sylvain and {Sanders}, D.~B.},
        title = "{Outflows in Infrared-Luminous Starbursts at z < 0.5. I. Sample, Na I D Spectra, and Profile Fitting}",
      journal = {\apjs},
         year = 2005,
        month = sep,
       volume = {160},
       number = {1},
        pages = {87-114},
          doi = {10.1086/432886},
archivePrefix = {arXiv},
       eprint = {astro-ph/0506610},
 primaryClass = {astro-ph},
       adsurl = {https://ui.adsabs.harvard.edu/abs/2005ApJS..160...87R}
}

@ARTICLE{Pascucci2015,
       author = {{Pascucci}, I. and {Edwards}, S. and {Heyer}, M. and {Rigliaco}, E. and {Hillenbrand}, L. and {Gorti}, U. and {Hollenbach}, D. and {Simon}, M.~N.},
        title = "{Narrow Na and K Absorption Lines Toward T Tauri Stars: Tracing the Atomic Envelope of Molecular Clouds}",
      journal = {\apj},
         year = 2015,
        month = nov,
       volume = {814},
       number = {1},
          eid = {14},
        pages = {14},
          doi = {10.1088/0004-637X/814/1/14},
archivePrefix = {arXiv},
       eprint = {1510.02022},
 primaryClass = {astro-ph.EP},
       adsurl = {https://ui.adsabs.harvard.edu/abs/2015ApJ...814...14P}
}

@ARTICLE{Hobbs1975,
       author = {{Hobbs}, L.~M.},
        title = "{Interstellar K I and Na I absorption lines near Rho Ophiuchi.}",
      journal = {\apj},
         year = 1975,
        month = sep,
       volume = {200},
        pages = {621-624},
          doi = {10.1086/153829},
       adsurl = {https://ui.adsabs.harvard.edu/abs/1975ApJ...200..621H}
}

@ARTICLE{Ilic2012,
       author = {{Ili{\'c}}, D. and {Popovi{\'c}}, L. {\v{C}}. and {La Mura}, G. and {Ciroi}, S. and {Rafanelli}, P.},
        title = "{The analysis of the broad hydrogen Balmer line ratios: Possible implications for the physical properties of the broad line region of AGNs}",
      journal = {\aap},
         year = 2012,
        month = jul,
       volume = {543},
          eid = {A142},
        pages = {A142},
          doi = {10.1051/0004-6361/201219299},
archivePrefix = {arXiv},
       eprint = {1205.3950},
 primaryClass = {astro-ph.CO},
       adsurl = {https://ui.adsabs.harvard.edu/abs/2012A&A...543A.142I}
}

@ARTICLE{Kewley2006,
       author = {{Kewley}, Lisa J. and {Groves}, Brent and {Kauffmann}, Guinevere and {Heckman}, Tim},
        title = "{The host galaxies and classification of active galactic nuclei}",
      journal = {\mnras},
         year = 2006,
        month = nov,
       volume = {372},
       number = {3},
        pages = {961-976},
          doi = {10.1111/j.1365-2966.2006.10859.x},
archivePrefix = {arXiv},
       eprint = {astro-ph/0605681},
 primaryClass = {astro-ph},
       adsurl = {https://ui.adsabs.harvard.edu/abs/2006MNRAS.372..961K}
}

@ARTICLE{Kewley2013,
       author = {{Kewley}, Lisa J. and {Dopita}, Michael A. and {Leitherer}, Claus and {Dav{\'e}}, Romeel and {Yuan}, Tiantian and {Allen}, Mark and {Groves}, Brent and {Sutherland}, Ralph},
        title = "{Theoretical Evolution of Optical Strong Lines across Cosmic Time}",
      journal = {\apj},
         year = 2013,
        month = sep,
       volume = {774},
       number = {2},
          eid = {100},
        pages = {100},
          doi = {10.1088/0004-637X/774/2/100},
archivePrefix = {arXiv},
       eprint = {1307.0508},
 primaryClass = {astro-ph.CO},
       adsurl = {https://ui.adsabs.harvard.edu/abs/2013ApJ...774..100K}
}

@ARTICLE{Greene2005,
       author = {{Greene}, Jenny E. and {Ho}, Luis C.},
        title = "{Estimating Black Hole Masses in Active Galaxies Using the H{\ensuremath{\alpha}} Emission Line}",
      journal = {\apj},
         year = 2005,
        month = sep,
       volume = {630},
       number = {1},
        pages = {122-129},
          doi = {10.1086/431897},
archivePrefix = {arXiv},
       eprint = {astro-ph/0508335},
 primaryClass = {astro-ph},
       adsurl = {https://ui.adsabs.harvard.edu/abs/2005ApJ...630..122G}
}

@ARTICLE{Taylor2025a,
       author = {{Taylor}, Anthony J. and {Finkelstein}, Steven L. and {Kocevski}, Dale D. and {Jeon}, Junehyoung and {Bromm}, Volker and {Amor{\'\i}n}, Ricardo O. and {Arrabal Haro}, Pablo and {Backhaus}, Bren E. and {Bagley}, Micaela B. and {Banados}, Eduardo and {Bhatawdekar}, Rachana and {Brooks}, Madisyn and {Calabr{\`o}}, Antonello and {Ortiz}, {\'O}scar A. Ch{\'a}vez and {Cheng}, Yingjie and {Cleri}, Nikko J. and {Cole}, Justin W. and {Davis}, Kelcey and {Dickinson}, Mark and {Donnan}, Callum and {Dunlop}, James S. and {Ellis}, Richard S. and {Fern{\'a}ndez}, Vital and {Fontana}, Adriano and {Fujimoto}, Seiji and {Giavalisco}, Mauro and {Grazian}, Andrea and {Guo}, Jingsong and {Hathi}, Nimish P. and {Holwerda}, Benne W. and {Hirschmann}, Michaela and {Inayoshi}, Kohei and {Kartaltepe}, Jeyhan S. and {Khusanova}, Yana and {Koekemoer}, Anton M. and {Kokorev}, Vasily and {Larson}, Rebecca L. and {Leung}, Gene C.~K. and {Lucas}, Ray A. and {McLeod}, Derek J. and {Napolitano}, Lorenzo and {Onoue}, Masafusa and {Pacucci}, Fabio and {Papovich}, Casey and {P{\'e}rez-Gonz{\'a}lez}, Pablo G. and {Pirzkal}, Nor and {Somerville}, Rachel S. and {Trump}, Jonathan R. and {Wilkins}, Stephen M. and {Yung}, L.~Y. Aaron and {Zhang}, Haowen},
        title = "{Broad-line AGNs at 3.5 < z < 6: The Black Hole Mass Function and a Connection with Little Red Dots}",
      journal = {\apj},
         year = 2025,
        month = jun,
       volume = {986},
       number = {2},
          eid = {165},
        pages = {165},
          doi = {10.3847/1538-4357/add15b},
archivePrefix = {arXiv},
       eprint = {2409.06772},
 primaryClass = {astro-ph.GA},
       adsurl = {https://ui.adsabs.harvard.edu/abs/2025ApJ...986..165T}
}

@ARTICLE{Kocevski2025,
       author = {{Kocevski}, Dale D. and {Finkelstein}, Steven L. and {Barro}, Guillermo and {Taylor}, Anthony J. and {Calabr{\`o}}, Antonello and {Laloux}, Brivael and {Buchner}, Johannes and {Trump}, Jonathan R. and {Leung}, Gene C.~K. and {Yang}, Guang and {Dickinson}, Mark and {P{\'e}rez-Gonz{\'a}lez}, Pablo G. and {Pacucci}, Fabio and {Inayoshi}, Kohei and {Somerville}, Rachel S. and {McGrath}, Elizabeth J. and {Akins}, Hollis B. and {Bagley}, Micaela B. and {Bowler}, Rebecca A.~A. and {Bisigello}, Laura and {Carnall}, Adam and {Casey}, Caitlin M. and {Cheng}, Yingjie and {Cleri}, Nikko J. and {Costantin}, Luca and {Cullen}, Fergus and {Davis}, Kelcey and {Donnan}, Callum T. and {Dunlop}, James S. and {Ellis}, Richard S. and {Ferguson}, Henry C. and {Fujimoto}, Seiji and {Fontana}, Adriano and {Giavalisco}, Mauro and {Grazian}, Andrea and {Grogin}, Norman A. and {Hathi}, Nimish P. and {Hirschmann}, Michaela and {Huertas-Company}, Marc and {Holwerda}, Benne W. and {Illingworth}, Garth and {Juneau}, St{\'e}phanie and {Kartaltepe}, Jeyhan S. and {Koekemoer}, Anton M. and {Li}, Wenxiu and {Lucas}, Ray A. and {Magee}, Dan and {Mason}, Charlotte and {McLeod}, Derek J. and {McLure}, Ross J. and {Napolitano}, Lorenzo and {Papovich}, Casey and {Pirzkal}, Nor and {Rodighiero}, Giulia and {Santini}, Paola and {Wilkins}, Stephen M. and {Yung}, L.~Y. Aaron},
        title = "{The Rise of Faint, Red Active Galactic Nuclei at z > 4: A Sample of Little Red Dots in the JWST Extragalactic Legacy Fields}",
      journal = {\apj},
         year = 2025,
        month = jun,
       volume = {986},
       number = {2},
          eid = {126},
        pages = {126},
          doi = {10.3847/1538-4357/adbc7d},
archivePrefix = {arXiv},
       eprint = {2404.03576},
 primaryClass = {astro-ph.GA},
       adsurl = {https://ui.adsabs.harvard.edu/abs/2025ApJ...986..126K}
}

@ARTICLE{Izotov2006,
       author = {{Izotov}, Y.~I. and {Stasi{\'n}ska}, G. and {Meynet}, G. and {Guseva}, N.~G. and {Thuan}, T.~X.},
        title = "{The chemical composition of metal-poor emission-line galaxies in the Data Release 3 of the Sloan Digital Sky Survey}",
      journal = {\aap},
         year = 2006,
        month = mar,
       volume = {448},
       number = {3},
        pages = {955-970},
          doi = {10.1051/0004-6361:20053763},
archivePrefix = {arXiv},
       eprint = {astro-ph/0511644},
 primaryClass = {astro-ph},
       adsurl = {https://ui.adsabs.harvard.edu/abs/2006A&A...448..955I}
}

@ARTICLE{Tang2025,
       author = {{Tang}, Mengtao and {Stark}, Daniel P. and {Plat}, Ad{\`e}le and {Feltre}, Anna and {Katz}, Harley and {Senchyna}, Peter and {Mason}, Charlotte A. and {Whitler}, Lily and {Chen}, Zuyi and {Topping}, Michael W.},
        title = "{JWST/NIRSpec Observations of High Ionization Emission Lines in Galaxies at High Redshift}",
      journal = {arXiv e-prints},
         year = 2025,
        month = may,
          eid = {arXiv:2505.06359},
        pages = {arXiv:2505.06359},
          doi = {10.48550/arXiv.2505.06359},
archivePrefix = {arXiv},
       eprint = {2505.06359},
 primaryClass = {astro-ph.GA},
       adsurl = {https://ui.adsabs.harvard.edu/abs/2025arXiv250506359T}
}

@ARTICLE{Tan2024,
       author = {{Tan}, Shuyu and {Parker}, Quentin A. and {Zijlstra}, Albert A. and {Rees}, Bryan},
        title = "{A catalogue of planetary nebulae chemical abundances in the Galactic bulge}",
      journal = {\mnras},
         year = 2024,
        month = jan,
       volume = {527},
       number = {3},
        pages = {6363-6387},
          doi = {10.1093/mnras/stad3496},
archivePrefix = {arXiv},
       eprint = {2311.01836},
 primaryClass = {astro-ph.GA},
       adsurl = {https://ui.adsabs.harvard.edu/abs/2024MNRAS.527.6363T}
}

@ARTICLE{Liu2000,
       author = {{Liu}, X. -W. and {Storey}, P.~J. and {Barlow}, M.~J. and {Danziger}, I.~J. and {Cohen}, M. and {Bryce}, M.},
        title = "{NGC 6153: a super-metal-rich planetary nebula?}",
      journal = {\mnras},
         year = 2000,
        month = mar,
       volume = {312},
       number = {3},
        pages = {585-628},
          doi = {10.1046/j.1365-8711.2000.03167.x},
       adsurl = {https://ui.adsabs.harvard.edu/abs/2000MNRAS.312..585L}
}

@ARTICLE{Lyu2016,
       author = {{Lyu}, Jianwei and {Rieke}, G.~H. and {Alberts}, Stacey},
        title = "{The Contribution of Host Galaxies to the Infrared Energy Output of z {\ensuremath{\gtrsim}} 5.0 Quasars}",
      journal = {\apj},
         year = 2016,
        month = jan,
       volume = {816},
       number = {2},
          eid = {85},
        pages = {85},
          doi = {10.3847/0004-637X/816/2/85},
archivePrefix = {arXiv},
       eprint = {1511.05938},
 primaryClass = {astro-ph.GA},
       adsurl = {https://ui.adsabs.harvard.edu/abs/2016ApJ...816...85L}
}

@ARTICLE{Inayoshi2025b,
       author = {{Inayoshi}, Kohei and {Shangguan}, Jinyi and {Chen}, Xian and {Ho}, Luis C. and {Haiman}, Zoltan},
        title = "{The Emergence of Little Red Dots from Binary Massive Black Holes}",
      journal = {arXiv e-prints},
         year = 2025,
        month = may,
          eid = {arXiv:2505.05322},
        pages = {arXiv:2505.05322},
          doi = {10.48550/arXiv.2505.05322},
archivePrefix = {arXiv},
       eprint = {2505.05322},
 primaryClass = {astro-ph.HE},
       adsurl = {https://ui.adsabs.harvard.edu/abs/2025arXiv250505322I}
}

@ARTICLE{Inayoshi2025a,
       author = {{Inayoshi}, Kohei},
        title = "{Little Red Dots as the Very First Activity of Black Hole Growth}",
      journal = {arXiv e-prints},
         year = 2025,
        month = mar,
          eid = {arXiv:2503.05537},
        pages = {arXiv:2503.05537},
          doi = {10.48550/arXiv.2503.05537},
archivePrefix = {arXiv},
       eprint = {2503.05537},
 primaryClass = {astro-ph.GA},
       adsurl = {https://ui.adsabs.harvard.edu/abs/2025arXiv250305537I}
}

@ARTICLE{Kido2025,
       author = {{Kido}, Daisaburo and {Ioka}, Kunihito and {Hotokezaka}, Kenta and {Inayoshi}, Kohei and {Irwin}, Christopher M.},
        title = "{Black Hole Envelopes in Little Red Dots}",
      journal = {arXiv e-prints},
         year = 2025,
        month = may,
          eid = {arXiv:2505.06965},
        pages = {arXiv:2505.06965},
          doi = {10.48550/arXiv.2505.06965},
archivePrefix = {arXiv},
       eprint = {2505.06965},
 primaryClass = {astro-ph.HE},
       adsurl = {https://ui.adsabs.harvard.edu/abs/2025arXiv250506965K}
}

@ARTICLE{Inayoshi2024a,
       author = {{Inayoshi}, Kohei and {Kimura}, Shigeo S. and {Noda}, Hirofumi},
        title = "{Weakness of X-rays and Variability in High-redshift AGNs with Super-Eddington Accretion}",
      journal = {arXiv e-prints},
         year = 2024,
        month = dec,
          eid = {arXiv:2412.03653},
        pages = {arXiv:2412.03653},
          doi = {10.48550/arXiv.2412.03653},
archivePrefix = {arXiv},
       eprint = {2412.03653},
 primaryClass = {astro-ph.HE},
       adsurl = {https://ui.adsabs.harvard.edu/abs/2024arXiv241203653I}
}

@ARTICLE{Lambrides2024,
       author = {{Lambrides}, Erini and {Garofali}, Kristen and {Larson}, Rebecca and {Ptak}, Andrew and {Chiaberge}, Marco and {Long}, Arianna S. and {Hutchison}, Taylor A. and {Norman}, Colin and {McKinney}, Jed and {Akins}, Hollis B. and {Berg}, Danielle A. and {Chisholm}, John and {Civano}, Francesca and {Cloonan}, Aidan P. and {Endsley}, Ryan and {Faisst}, Andreas L. and {Gilli}, Roberto and {Gillman}, Steven and {Hirschmann}, Michaela and {Kartaltepe}, Jeyhan S. and {Kocevski}, Dale D. and {Kokorev}, Vasily and {Pacucci}, Fabio and {Richardson}, Chris T. and {Stiavelli}, Massimo and {Whalen}, Kelly E.},
        title = "{The Case for Super-Eddington Accretion: Connecting Weak X-ray and UV Line Emission in JWST Broad-Line AGN During the First Gyr of Cosmic Time}",
      journal = {arXiv e-prints},
         year = 2024,
        month = sep,
          eid = {arXiv:2409.13047},
        pages = {arXiv:2409.13047},
          doi = {10.48550/arXiv.2409.13047},
archivePrefix = {arXiv},
       eprint = {2409.13047},
 primaryClass = {astro-ph.HE},
       adsurl = {https://ui.adsabs.harvard.edu/abs/2024arXiv240913047L}
}

@ARTICLE{Dotan2011,
       author = {{Dotan}, Calanit and {Shaviv}, Nir J.},
        title = "{Super-Eddington slim accretion discs with winds}",
      journal = {\mnras},
         year = 2011,
        month = may,
       volume = {413},
       number = {3},
        pages = {1623-1632},
          doi = {10.1111/j.1365-2966.2011.18235.x},
archivePrefix = {arXiv},
       eprint = {1004.1797},
 primaryClass = {astro-ph.HE},
       adsurl = {https://ui.adsabs.harvard.edu/abs/2011MNRAS.413.1623D}
}

@ARTICLE{Pacucci2024,
       author = {{Pacucci}, Fabio and {Narayan}, Ramesh},
        title = "{Mildly Super-Eddington Accretion onto Slowly Spinning Black Holes Explains the X-Ray Weakness of the Little Red Dots}",
      journal = {\apj},
         year = 2024,
        month = nov,
       volume = {976},
       number = {1},
          eid = {96},
        pages = {96},
          doi = {10.3847/1538-4357/ad84f7},
archivePrefix = {arXiv},
       eprint = {2407.15915},
 primaryClass = {astro-ph.HE},
       adsurl = {https://ui.adsabs.harvard.edu/abs/2024ApJ...976...96P}
}

@ARTICLE{Filippenko2003,
       author = {{Filippenko}, Alexei V. and {Ho}, Luis C.},
        title = "{A Low-Mass Central Black Hole in the Bulgeless Seyfert 1 Galaxy NGC 4395}",
      journal = {\apjl},
         year = 2003,
        month = may,
       volume = {588},
       number = {1},
        pages = {L13-L16},
          doi = {10.1086/375361},
archivePrefix = {arXiv},
       eprint = {astro-ph/0303429},
 primaryClass = {astro-ph},
       adsurl = {https://ui.adsabs.harvard.edu/abs/2003ApJ...588L..13F}
}

@ARTICLE{Akins2025,
       author = {{Akins}, Hollis B. and {Casey}, Caitlin M. and {Berg}, Danielle A. and {Chisholm}, John and {Cloonan}, Aidan P. and {Franco}, Maximilien and {Finkelstein}, Steven L. and {Fujimoto}, Seiji and {Koekemoer}, Anton M. and {Kokorev}, Vasily and {Lambrides}, Erini and {Robertson}, Brant E. and {Taylor}, Anthony J. and {Coulter}, David A. and {Fox}, Ori and {Karmen}, Mitchell},
        title = "{Strong Rest-UV Emission Lines in a ``Little Red Dot'' Active Galactic Nucleus at z = 7: Early Supermassive Black Hole Growth alongside Compact Massive Star Formation?}",
      journal = {\apjl},
         year = 2025,
        month = feb,
       volume = {980},
       number = {2},
          eid = {L29},
        pages = {L29},
          doi = {10.3847/2041-8213/adab76},
archivePrefix = {arXiv},
       eprint = {2410.00949},
 primaryClass = {astro-ph.GA},
       adsurl = {https://ui.adsabs.harvard.edu/abs/2025ApJ...980L..29A}
}

@ARTICLE{Tee2025,
       author = {{Tee}, Wei Leong and {Fan}, Xiaohui and {Wang}, Feige and {Yang}, Jinyi},
        title = "{Lack of Rest-frame Ultraviolet Variability in Little Red Dots Based on HST and JWST Observations}",
      journal = {\apjl},
         year = 2025,
        month = apr,
       volume = {983},
       number = {1},
          eid = {L26},
        pages = {L26},
          doi = {10.3847/2041-8213/adc5e3},
archivePrefix = {arXiv},
       eprint = {2412.05242},
 primaryClass = {astro-ph.GA},
       adsurl = {https://ui.adsabs.harvard.edu/abs/2025ApJ...983L..26T}
}

@ARTICLE{Stalevski2016,
       author = {{Stalevski}, Marko and {Ricci}, Claudio and {Ueda}, Yoshihiro and {Lira}, Paulina and {Fritz}, Jacopo and {Baes}, Maarten},
        title = "{The dust covering factor in active galactic nuclei}",
      journal = {\mnras},
         year = 2016,
        month = may,
       volume = {458},
       number = {3},
        pages = {2288-2302},
          doi = {10.1093/mnras/stw444},
archivePrefix = {arXiv},
       eprint = {1602.06954},
 primaryClass = {astro-ph.GA},
       adsurl = {https://ui.adsabs.harvard.edu/abs/2016MNRAS.458.2288S}
}

@ARTICLE{BWang2025,
       author = {{Wang}, Ben and {Hennawi}, Joseph F. and {Cai}, Zheng and {Richards}, Gordon T. and {Schindler}, Jan-Torge and {Zakamska}, Nadia L. and {Ishikawa}, Yuzo and {Akins}, Hollis B. and {Sun}, Zechang},
        title = "{Luminous mid-IR-selected type 2 quasars at cosmic noon in SDSS Stripe 82 {\textendash} I. Selection, composite photometry, and spectral energy distributions}",
      journal = {\mnras},
         year = 2025,
        month = may,
       volume = {539},
       number = {2},
        pages = {1562-1594},
          doi = {10.1093/mnras/staf574},
archivePrefix = {arXiv},
       eprint = {2501.14026},
 primaryClass = {astro-ph.GA},
       adsurl = {https://ui.adsabs.harvard.edu/abs/2025MNRAS.539.1562W}
}

@ARTICLE{ZZhang2025,
       author = {{Zhang}, Zijian and {Jiang}, Linhua and {Liu}, Weiyang and {Ho}, Luis C.},
        title = "{Analysis of Multi-epoch JWST Images of {\ensuremath{\sim}}300 Little Red Dots: Tentative Detection of Variability in a Minority of Sources}",
      journal = {\apj},
         year = 2025,
        month = may,
       volume = {985},
       number = {1},
          eid = {119},
        pages = {119},
          doi = {10.3847/1538-4357/adcb3e},
archivePrefix = {arXiv},
       eprint = {2411.02729},
 primaryClass = {astro-ph.GA},
       adsurl = {https://ui.adsabs.harvard.edu/abs/2025ApJ...985..119Z}
}

@ARTICLE{Kokubo2024,
       author = {{Kokubo}, Mitsuru and {Harikane}, Yuichi},
        title = "{Challenging the AGN scenario for JWST/NIRSpec broad H$\alpha$ emitters/Little Red Dots in light of non-detection of NIRCam photometric variability and X-ray}",
      journal = {arXiv e-prints},
         year = 2024,
        month = jul,
          eid = {arXiv:2407.04777},
        pages = {arXiv:2407.04777},
          doi = {10.48550/arXiv.2407.04777},
archivePrefix = {arXiv},
       eprint = {2407.04777},
 primaryClass = {astro-ph.GA},
       adsurl = {https://ui.adsabs.harvard.edu/abs/2024arXiv240704777K}
}

@ARTICLE{RLin2025,
       author = {{Lin}, Ruqiu and {Zheng}, Zhen-Ya and {Jiang}, Chunyan and {Yuan}, Fang-Ting and {Ho}, Luis C. and {Wang}, Junxian and {Jiang}, Linhua and {Rhoads}, James E. and {Malhotra}, Sangeeta and {Barrientos}, L. Felipe and {Wold}, Isak and {Infante}, Leopoldo and {Zhu}, Shuairu and {Ji}, Xiang and {Fu}, Xiaodan},
        title = "{Discovery of Local Analogs to JWST's Little Red Dots}",
      journal = {\apjl},
         year = 2025,
        month = feb,
       volume = {980},
       number = {2},
          eid = {L34},
        pages = {L34},
          doi = {10.3847/2041-8213/adaaf1},
archivePrefix = {arXiv},
       eprint = {2412.08396},
 primaryClass = {astro-ph.GA},
       adsurl = {https://ui.adsabs.harvard.edu/abs/2025ApJ...980L..34L}
}

@ARTICLE{Richards2002,
       author = {{Richards}, Gordon T. and {Fan}, Xiaohui and {Newberg}, Heidi Jo and {Strauss}, Michael A. and {Vanden Berk}, Daniel E. and {Schneider}, Donald P. and {Yanny}, Brian and {Boucher}, Adam and {Burles}, Scott and {Frieman}, Joshua A. and {Gunn}, James E. and {Hall}, Patrick B. and {Ivezi{\'c}}, {\v{Z}}eljko and {Kent}, Stephen and {Loveday}, Jon and {Lupton}, Robert H. and {Rockosi}, Constance M. and {Schlegel}, David J. and {Stoughton}, Chris and {SubbaRao}, Mark and {York}, Donald G.},
        title = "{Spectroscopic Target Selection in the Sloan Digital Sky Survey: The Quasar Sample}",
      journal = {\aj},
         year = 2002,
        month = jun,
       volume = {123},
       number = {6},
        pages = {2945-2975},
          doi = {10.1086/340187},
archivePrefix = {arXiv},
       eprint = {astro-ph/0202251},
 primaryClass = {astro-ph},
       adsurl = {https://ui.adsabs.harvard.edu/abs/2002AJ....123.2945R}
}

@ARTICLE{Simcoe2013,
       author = {{Simcoe}, Robert A. and {Burgasser}, Adam J. and {Schechter}, Paul L. and {Fishner}, Jason and {Bernstein}, Rebecca A. and {Bigelow}, Bruce C. and {Pipher}, Judith L. and {Forrest}, William and {McMurtry}, Craig and {Smith}, Matthew J. and {Bochanski}, John J.},
        title = "{FIRE: A Facility Class Near-Infrared Echelle Spectrometer for the Magellan Telescopes}",
      journal = {\pasp},
         year = 2013,
        month = mar,
       volume = {125},
       number = {925},
        pages = {270},
          doi = {10.1086/670241},
       adsurl = {https://ui.adsabs.harvard.edu/abs/2013PASP..125..270S}
}

@ARTICLE{Fabricant2019,
       author = {{Fabricant}, Daniel and {Fata}, Robert and {Epps}, Harland and {Gauron}, Thomas and {Mueller}, Mark and {Zajac}, Joseph and {Amato}, Stephen and {Barberis}, Jack and {Bergner}, Henry and {Brennan}, Patricia and {Brown}, Warren and {Chilingarian}, Igor and {Geary}, John and {Kradinov}, Vladimir and {McLeod}, Brian and {Smith}, Matthew and {Woods}, Deborah},
        title = "{Binospec: A Wide-field Imaging Spectrograph for the MMT}",
      journal = {\pasp},
         year = 2019,
        month = jul,
       volume = {131},
       number = {1001},
        pages = {075004},
          doi = {10.1088/1538-3873/ab1d78},
archivePrefix = {arXiv},
       eprint = {1905.03320},
 primaryClass = {astro-ph.IM},
       adsurl = {https://ui.adsabs.harvard.edu/abs/2019PASP..131g5004F}
}

@ARTICLE{Fabricant2025,
       author = {{Fabricant}, Daniel and {Ben-Ami}, Sagi and {Chilingarian}, Igor V. and {Fata}, Robert and {Moran}, Sean and {Paegert}, Martin and {Smith}, Matthew and {Zajac}, Joseph},
        title = "{An Integral Field Unit for the Binospec Spectrograph}",
      journal = {\pasp},
         year = 2025,
        month = jan,
       volume = {137},
       number = {1},
          eid = {015002},
        pages = {015002},
          doi = {10.1088/1538-3873/ada701},
archivePrefix = {arXiv},
       eprint = {2501.01528},
 primaryClass = {astro-ph.IM},
       adsurl = {https://ui.adsabs.harvard.edu/abs/2025PASP..137a5002F}
}

@ARTICLE{Brooks2025,
       author = {{Brooks}, Madisyn and {Simons}, Raymond C. and {Trump}, Jonathan R. and {Taylor}, Anthony J. and {Bagley}, Micaela B. and {Backhaus}, Bren and {Davis}, Kelcey and {Buat}, V{\'e}ronique and {Cleri}, Nikko J. and {de la Vega}, Alexander and {Finkelstein}, Steven L. and {Hirschmann}, Michaela and {Holwerda}, Benne W. and {Kocevski}, Dale D. and {Koekemoer}, Anton M. and {Lucas}, Ray A. and {Pacucci}, Fabio and {Seill{\'e}}, Lise-Marie},
        title = "{Here There Be (Dusty) Monsters: High-redshift Active Galactic Nuclei Are Dustier than Their Hosts}",
      journal = {\apj},
         year = 2025,
        month = jun,
       volume = {986},
       number = {2},
          eid = {177},
        pages = {177},
          doi = {10.3847/1538-4357/addac4},
archivePrefix = {arXiv},
       eprint = {2410.07340},
 primaryClass = {astro-ph.GA},
       adsurl = {https://ui.adsabs.harvard.edu/abs/2025ApJ...986..177B}
}

@ARTICLE{Gunasekera2023,
       author = {{Gunasekera}, Chamani M. and {van Hoof}, Peter A.~M. and {Chatzikos}, Marios and {Ferland}, Gary J.},
        title = "{The 23.01 Release of Cloudy}",
      journal = {Research Notes of the American Astronomical Society},
         year = 2023,
        month = nov,
       volume = {7},
       number = {11},
          eid = {246},
        pages = {246},
          doi = {10.3847/2515-5172/ad0e75},
archivePrefix = {arXiv},
       eprint = {2311.10163},
 primaryClass = {astro-ph.GA},
       adsurl = {https://ui.adsabs.harvard.edu/abs/2023RNAAS...7..246G}
}

@ARTICLE{Smyth2019,
       author = {{Smyth}, R.~T. and {Ramsbottom}, C.~A. and {Keenan}, F.~P. and {Ferland}, G.~J. and {Ballance}, C.~P.},
        title = "{Towards converged electron-impact excitation calculations of low-lying transitions in Fe II}",
      journal = {\mnras},
         year = 2019,
        month = feb,
       volume = {483},
       number = {1},
        pages = {654-663},
          doi = {10.1093/mnras/sty3198},
       adsurl = {https://ui.adsabs.harvard.edu/abs/2019MNRAS.483..654S}
}

@ARTICLE{Baldwin1981,
       author = {{Baldwin}, J.~A. and {Phillips}, M.~M. and {Terlevich}, R.},
        title = "{Classification parameters for the emission-line spectra of extragalactic objects.}",
      journal = {\pasp},
         year = 1981,
        month = feb,
       volume = {93},
        pages = {5-19},
          doi = {10.1086/130766},
       adsurl = {https://ui.adsabs.harvard.edu/abs/1981PASP...93....5B}
}

@ARTICLE{Abdurrouf2022,
       author = {{Abdurro'uf} and {Accetta}, Katherine and {Aerts}, Conny and {Silva Aguirre}, V{\'\i}ctor and {Ahumada}, Romina and {Ajgaonkar}, Nikhil and {Filiz Ak}, N. and {Alam}, Shadab and {Allende Prieto}, Carlos and {Almeida}, Andr{\'e}s and {Anders}, Friedrich and {Anderson}, Scott F. and {Andrews}, Brett H. and {Anguiano}, Borja and {Aquino-Ort{\'\i}z}, Erik and {Arag{\'o}n-Salamanca}, Alfonso and {Argudo-Fern{\'a}ndez}, Maria and {Ata}, Metin and {Aubert}, Marie and {Avila-Reese}, Vladimir and {Badenes}, Carles and {Barb{\'a}}, Rodolfo H. and {Barger}, Kat and {Barrera-Ballesteros}, Jorge K. and {Beaton}, Rachael L. and {Beers}, Timothy C. and {Belfiore}, Francesco and {Bender}, Chad F. and {Bernardi}, Mariangela and {Bershady}, Matthew A. and {Beutler}, Florian and {Bidin}, Christian Moni and {Bird}, Jonathan C. and {Bizyaev}, Dmitry and {Blanc}, Guillermo A. and {Blanton}, Michael R. and {Boardman}, Nicholas Fraser and {Bolton}, Adam S. and {Boquien}, M{\'e}d{\'e}ric and {Borissova}, Jura and {Bovy}, Jo and {Brandt}, W.~N. and {Brown}, Jordan and {Brownstein}, Joel R. and {Brusa}, Marcella and {Buchner}, Johannes and {Bundy}, Kevin and {Burchett}, Joseph N. and {Bureau}, Martin and {Burgasser}, Adam and {Cabang}, Tuesday K. and {Campbell}, Stephanie and {Cappellari}, Michele and {Carlberg}, Joleen K. and {Wanderley}, F{\'a}bio Carneiro and {Carrera}, Ricardo and {Cash}, Jennifer and {Chen}, Yan-Ping and {Chen}, Wei-Huai and {Cherinka}, Brian and {Chiappini}, Cristina and {Choi}, Peter Doohyun and {Chojnowski}, S. Drew and {Chung}, Haeun and {Clerc}, Nicolas and {Cohen}, Roger E. and {Comerford}, Julia M. and {Comparat}, Johan and {da Costa}, Luiz and {Covey}, Kevin and {Crane}, Jeffrey D. and {Cruz-Gonzalez}, Irene and {Culhane}, Connor and {Cunha}, Katia and {Dai}, Y. Sophia and {Damke}, Guillermo and {Darling}, Jeremy and {Davidson}, Jr., James W. and {Davies}, Roger and {Dawson}, Kyle and {De Lee}, Nathan and {Diamond-Stanic}, Aleksandar M. and {Cano-D{\'\i}az}, Mariana and {S{\'a}nchez}, Helena Dom{\'\i}nguez and {Donor}, John and {Duckworth}, Chris and {Dwelly}, Tom and {Eisenstein}, Daniel J. and {Elsworth}, Yvonne P. and {Emsellem}, Eric and {Eracleous}, Mike and {Escoffier}, Stephanie and {Fan}, Xiaohui and {Farr}, Emily and {Feng}, Shuai and {Fern{\'a}ndez-Trincado}, Jos{\'e} G. and {Feuillet}, Diane and {Filipp}, Andreas and {Fillingham}, Sean P. and {Frinchaboy}, Peter M. and {Fromenteau}, Sebastien and {Galbany}, Llu{\'\i}s and {Garc{\'\i}a}, Rafael A. and {Garc{\'\i}a-Hern{\'a}ndez}, D.~A. and {Ge}, Junqiang and {Geisler}, Doug and {Gelfand}, Joseph and {G{\'e}ron}, Tobias and {Gibson}, Benjamin J. and {Goddy}, Julian and {Godoy-Rivera}, Diego and {Grabowski}, Kathleen and {Green}, Paul J. and {Greener}, Michael and {Grier}, Catherine J. and {Griffith}, Emily and {Guo}, Hong and {Guy}, Julien and {Hadjara}, Massinissa and {Harding}, Paul and {Hasselquist}, Sten and {Hayes}, Christian R. and {Hearty}, Fred and {Hern{\'a}ndez}, Jes{\'u}s and {Hill}, Lewis and {Hogg}, David W. and {Holtzman}, Jon A. and {Horta}, Danny and {Hsieh}, Bau-Ching and {Hsu}, Chin-Hao and {Hsu}, Yun-Hsin and {Huber}, Daniel and {Huertas-Company}, Marc and {Hutchinson}, Brian and {Hwang}, Ho Seong and {Ibarra-Medel}, H{\'e}ctor J. and {Chitham}, Jacob Ider and {Ilha}, Gabriele S. and {Imig}, Julie and {Jaekle}, Will and {Jayasinghe}, Tharindu and {Ji}, Xihan and {Johnson}, Jennifer A. and {Jones}, Amy and {J{\"o}nsson}, Henrik and {Katkov}, Ivan and {Khalatyan}, Dr., Arman and {Kinemuchi}, Karen and {Kisku}, Shobhit and {Knapen}, Johan H. and {Kneib}, Jean-Paul and {Kollmeier}, Juna A. and {Kong}, Miranda and {Kounkel}, Marina and {Kreckel}, Kathryn and {Krishnarao}, Dhanesh and {Lacerna}, Ivan and {Lane}, Richard R. and {Langgin}, Rachel and {Lavender}, Ramon and {Law}, David R. and {Lazarz}, Daniel and {Leung}, Henry W. and {Leung}, Ho-Hin and {Lewis}, Hannah M. and {Li}, Cheng and {Li}, Ran and {Lian}, Jianhui and {Liang}, Fu-Heng and {Lin}, Lihwai and {Lin}, Yen-Ting and {Lin}, Sicheng and {Lintott}, Chris and {Long}, Dan and {Longa-Pe{\~n}a}, Pen{\'e}lope and {L{\'o}pez-Cob{\'a}}, Carlos and {Lu}, Shengdong and {Lundgren}, Britt F. and {Luo}, Yuanze and {Mackereth}, J. Ted and {de la Macorra}, Axel and {Mahadevan}, Suvrath and {Majewski}, Steven R. and {Manchado}, Arturo and {Mandeville}, Travis and {Maraston}, Claudia and {Margalef-Bentabol}, Berta and {Masseron}, Thomas and {Masters}, Karen L. and {Mathur}, Savita and {McDermid}, Richard M. and {Mckay}, Myles and {Merloni}, Andrea and {Merrifield}, Michael and {Meszaros}, Szabolcs and {Miglio}, Andrea and {Di Mille}, Francesco and {Minniti}, Dante and {Minsley}, Rebecca and {Monachesi}, Antonela},
        title = "{The Seventeenth Data Release of the Sloan Digital Sky Surveys: Complete Release of MaNGA, MaStar, and APOGEE-2 Data}",
      journal = {\apjs},
         year = 2022,
        month = apr,
       volume = {259},
       number = {2},
          eid = {35},
        pages = {35},
          doi = {10.3847/1538-4365/ac4414},
archivePrefix = {arXiv},
       eprint = {2112.02026},
 primaryClass = {astro-ph.GA},
       adsurl = {https://ui.adsabs.harvard.edu/abs/2022ApJS..259...35A}
}

@ARTICLE{Petrosian1976,
       author = {{Petrosian}, V.},
        title = "{Surface Brightness and Evolution of Galaxies}",
      journal = {\apjl},
         year = 1976,
        month = dec,
       volume = {210},
        pages = {L53},
          doi = {10.1086/18230110.1086/182253},
       adsurl = {https://ui.adsabs.harvard.edu/abs/1976ApJ...209L...1P}
}

@ARTICLE{Lacy2020,
       author = {{Lacy}, M. and {Baum}, S.~A. and {Chandler}, C.~J. and {Chatterjee}, S. and {Clarke}, T.~E. and {Deustua}, S. and {English}, J. and {Farnes}, J. and {Gaensler}, B.~M. and {Gugliucci}, N. and {Hallinan}, G. and {Kent}, B.~R. and {Kimball}, A. and {Law}, C.~J. and {Lazio}, T.~J.~W. and {Marvil}, J. and {Mao}, S.~A. and {Medlin}, D. and {Mooley}, K. and {Murphy}, E.~J. and {Myers}, S. and {Osten}, R. and {Richards}, G.~T. and {Rosolowsky}, E. and {Rudnick}, L. and {Schinzel}, F. and {Sivakoff}, G.~R. and {Sjouwerman}, L.~O. and {Taylor}, R. and {White}, R.~L. and {Wrobel}, J. and {Andernach}, H. and {Beasley}, A.~J. and {Berger}, E. and {Bhatnager}, S. and {Birkinshaw}, M. and {Bower}, G.~C. and {Brandt}, W.~N. and {Brown}, S. and {Burke-Spolaor}, S. and {Butler}, B.~J. and {Comerford}, J. and {Demorest}, P.~B. and {Fu}, H. and {Giacintucci}, S. and {Golap}, K. and {G{\"u}th}, T. and {Hales}, C.~A. and {Hiriart}, R. and {Hodge}, J. and {Horesh}, A. and {Ivezi{\'c}}, {\v{Z}}. and {Jarvis}, M.~J. and {Kamble}, A. and {Kassim}, N. and {Liu}, X. and {Loinard}, L. and {Lyons}, D.~K. and {Masters}, J. and {Mezcua}, M. and {Moellenbrock}, G.~A. and {Mroczkowski}, T. and {Nyland}, K. and {O'Dea}, C.~P. and {O'Sullivan}, S.~P. and {Peters}, W.~M. and {Radford}, K. and {Rao}, U. and {Robnett}, J. and {Salcido}, J. and {Shen}, Y. and {Sobotka}, A. and {Witz}, S. and {Vaccari}, M. and {van Weeren}, R.~J. and {Vargas}, A. and {Williams}, P.~K.~G. and {Yoon}, I.},
        title = "{The Karl G. Jansky Very Large Array Sky Survey (VLASS). Science Case and Survey Design}",
      journal = {\pasp},
         year = 2020,
        month = mar,
       volume = {132},
       number = {1009},
          eid = {035001},
        pages = {035001},
          doi = {10.1088/1538-3873/ab63eb},
archivePrefix = {arXiv},
       eprint = {1907.01981},
 primaryClass = {astro-ph.IM},
       adsurl = {https://ui.adsabs.harvard.edu/abs/2020PASP..132c5001L}
}

@INPROCEEDINGS{Pogge2010,
       author = {{Pogge}, R.~W. and {Atwood}, B. and {Brewer}, D.~F. and {Byard}, P.~L. and {Derwent}, M.~A. and {Gonzalez}, R. and {Martini}, P. and {Mason}, J.~A. and {O'Brien}, T.~P. and {Osmer}, P.~S. and {Pappalardo}, D.~P. and {Steinbrecher}, D.~P. and {Teiga}, E.~J. and {Zhelem}, R.},
        title = "{The multi-object double spectrographs for the Large Binocular Telescope}",
    booktitle = {Ground-based and Airborne Instrumentation for Astronomy III},
         year = 2010,
       editor = {{McLean}, Ian S. and {Ramsay}, Suzanne K. and {Takami}, Hideki},
       series = {Society of Photo-Optical Instrumentation Engineers (SPIE) Conference Series},
       volume = {7735},
        month = jul,
          eid = {77350A},
        pages = {77350A},
          doi = {10.1117/12.857215},
       adsurl = {https://ui.adsabs.harvard.edu/abs/2010SPIE.7735E..0AP}
}

@ARTICLE{Intema2017,
       author = {{Intema}, H.~T. and {Jagannathan}, P. and {Mooley}, K.~P. and {Frail}, D.~A.},
        title = "{The GMRT 150 MHz all-sky radio survey. First alternative data release TGSS ADR1}",
      journal = {\aap},
         year = 2017,
        month = feb,
       volume = {598},
          eid = {A78},
        pages = {A78},
          doi = {10.1051/0004-6361/201628536},
archivePrefix = {arXiv},
       eprint = {1603.04368},
 primaryClass = {astro-ph.CO},
       adsurl = {https://ui.adsabs.harvard.edu/abs/2017A&A...598A..78I}
}

@ARTICLE{Becker1995,
       author = {{Becker}, Robert H. and {White}, Richard L. and {Helfand}, David J.},
        title = "{The FIRST Survey: Faint Images of the Radio Sky at Twenty Centimeters}",
      journal = {\apj},
         year = 1995,
        month = sep,
       volume = {450},
        pages = {559},
          doi = {10.1086/176166},
       adsurl = {https://ui.adsabs.harvard.edu/abs/1995ApJ...450..559B}
}

@ARTICLE{Yue2024,
       author = {{Yue}, Minghao and {Eilers}, Anna-Christina and {Ananna}, Tonima Tasnim and {Panagiotou}, Christos and {Kara}, Erin and {Miyaji}, Takamitsu},
        title = "{Stacking X-Ray Observations of ``Little Red Dots'': Implications for Their Active Galactic Nucleus Properties}",
      journal = {\apjl},
         year = 2024,
        month = oct,
       volume = {974},
       number = {2},
          eid = {L26},
        pages = {L26},
          doi = {10.3847/2041-8213/ad7eba},
archivePrefix = {arXiv},
       eprint = {2404.13290},
 primaryClass = {astro-ph.GA},
       adsurl = {https://ui.adsabs.harvard.edu/abs/2024ApJ...974L..26Y}
}

@ARTICLE{Ananna2024,
       author = {{Ananna}, Tonima Tasnim and {Bogd{\'a}n}, {\'A}kos and {Kov{\'a}cs}, Orsolya E. and {Natarajan}, Priyamvada and {Hickox}, Ryan C.},
        title = "{X-Ray View of Little Red Dots: Do They Host Supermassive Black Holes?}",
      journal = {\apjl},
         year = 2024,
        month = jul,
       volume = {969},
       number = {1},
          eid = {L18},
        pages = {L18},
          doi = {10.3847/2041-8213/ad5669},
archivePrefix = {arXiv},
       eprint = {2404.19010},
 primaryClass = {astro-ph.GA},
       adsurl = {https://ui.adsabs.harvard.edu/abs/2024ApJ...969L..18A}
}

@ARTICLE{Perger2025,
       author = {{Perger}, K. and {Fogasy}, J. and {Frey}, S. and {Gab{\'a}nyi}, K. {\'E}.},
        title = "{Deep silence: Radio properties of little red dots}",
      journal = {\aap},
         year = 2025,
        month = jan,
       volume = {693},
          eid = {L2},
        pages = {L2},
          doi = {10.1051/0004-6361/202452422},
archivePrefix = {arXiv},
       eprint = {2411.19518},
 primaryClass = {astro-ph.GA},
       adsurl = {https://ui.adsabs.harvard.edu/abs/2025A&A...693L...2P}
}

@ARTICLE{Mazzolari2024,
       author = {{Mazzolari}, G. and {Gilli}, R. and {Maiolino}, R. and {Prandoni}, I. and {Delvecchio}, I. and {Norman}, C. and {Jimenez-Andrade}, E.~F. and {Belladitta}, S. and {Vito}, F. and {Momjian}, E. and {Chiaberge}, M. and {Trefoloni}, B. and {Signorini}, M. and {Ji}, X. and {D'Amato}, Q. and {Risaliti}, G. and {Baldi}, R.~D. and {Fabian}, A. and {{\"U}bler}, H. and {D'Eugenio}, F. and {Scholtz}, J. and {Juod{\v{z}}balis}, I. and {Mignoli}, M. and {Brusa}, M. and {Murphy}, E. and {Muxlow}, T.~W.~B.},
        title = "{The radio properties of the JWST-discovered AGN}",
      journal = {arXiv e-prints},
         year = 2024,
        month = dec,
          eid = {arXiv:2412.04224},
        pages = {arXiv:2412.04224},
          doi = {10.48550/arXiv.2412.04224},
archivePrefix = {arXiv},
       eprint = {2412.04224},
 primaryClass = {astro-ph.GA},
       adsurl = {https://ui.adsabs.harvard.edu/abs/2024arXiv241204224M}
}

@ARTICLE{Baskin2018,
       author = {{Baskin}, Alexei and {Laor}, Ari},
        title = "{Dust inflated accretion disc as the origin of the broad line region in active galactic nuclei}",
      journal = {\mnras},
         year = 2018,
        month = feb,
       volume = {474},
       number = {2},
        pages = {1970-1994},
          doi = {10.1093/mnras/stx2850},
archivePrefix = {arXiv},
       eprint = {1711.00025},
 primaryClass = {astro-ph.GA},
       adsurl = {https://ui.adsabs.harvard.edu/abs/2018MNRAS.474.1970B}
}

@ARTICLE{Gaskell2018,
       author = {{Gaskell}, C. Martin and {Harrington}, P.~Z.},
        title = "{Partial dust obscuration in active galactic nuclei as a cause of broad-line profile and lag variability, and apparent accretion disc inhomogeneities}",
      journal = {\mnras},
         year = 2018,
        month = aug,
       volume = {478},
       number = {2},
        pages = {1660-1669},
          doi = {10.1093/mnras/sty848},
archivePrefix = {arXiv},
       eprint = {1704.06455},
 primaryClass = {astro-ph.HE},
       adsurl = {https://ui.adsabs.harvard.edu/abs/2018MNRAS.478.1660G}
}

@ARTICLE{Kishimoto2007,
       author = {{Kishimoto}, M. and {H{\"o}nig}, S.~F. and {Beckert}, T. and {Weigelt}, G.},
        title = "{The innermost region of AGN tori: implications from the HST/NICMOS type 1 point sources and near-IR reverberation}",
      journal = {\aap},
         year = 2007,
        month = dec,
       volume = {476},
       number = {2},
        pages = {713-721},
          doi = {10.1051/0004-6361:20077911},
archivePrefix = {arXiv},
       eprint = {0709.0431},
 primaryClass = {astro-ph},
       adsurl = {https://ui.adsabs.harvard.edu/abs/2007A&A...476..713K}
}

@ARTICLE{Martin1980,
       author = {{Martin}, P.~G. and {Ferland}, G.~J.},
        title = "{Far-ultraviolet dust opacity and photoionization in quasi-stellar objects}",
      journal = {\apjl},
         year = 1980,
        month = feb,
       volume = {235},
        pages = {L125-L128},
          doi = {10.1086/183174},
       adsurl = {https://ui.adsabs.harvard.edu/abs/1980ApJ...235L.125M}
}

@ARTICLE{Inayoshi2022,
       author = {{Inayoshi}, Kohei and {Onoue}, Masafusa and {Sugahara}, Yuma and {Inoue}, Akio K. and {Ho}, Luis C.},
        title = "{The Age of Discovery with the James Webb Space Telescope: Excavating the Spectral Signatures of the First Massive Black Holes}",
      journal = {\apjl},
         year = 2022,
        month = jun,
       volume = {931},
       number = {2},
          eid = {L25},
        pages = {L25},
          doi = {10.3847/2041-8213/ac6f01},
archivePrefix = {arXiv},
       eprint = {2204.09692},
 primaryClass = {astro-ph.GA},
       adsurl = {https://ui.adsabs.harvard.edu/abs/2022ApJ...931L..25I}
}

@BOOK{Hansen2004,
       author = {{Hansen}, Carl J. and {Kawaler}, Steven D. and {Trimble}, Virginia},
        title = "{Stellar interiors : physical principles, structure, and evolution}",
         year = 2004,
       adsurl = {https://ui.adsabs.harvard.edu/abs/2004sipp.book.....H}
}

@ARTICLE{Sikora2007,
       author = {{Sikora}, Marek and {Stawarz}, {\L}ukasz and {Lasota}, Jean-Pierre},
        title = "{Radio Loudness of Active Galactic Nuclei: Observational Facts and Theoretical Implications}",
      journal = {\apj},
         year = 2007,
        month = apr,
       volume = {658},
       number = {2},
        pages = {815-828},
          doi = {10.1086/511972},
archivePrefix = {arXiv},
       eprint = {astro-ph/0604095},
 primaryClass = {astro-ph},
       adsurl = {https://ui.adsabs.harvard.edu/abs/2007ApJ...658..815S}
}

@ARTICLE{Desroches2009,
       author = {{Desroches}, Louis-Benoit and {Greene}, Jenny E. and {Ho}, Luis C.},
        title = "{X-Ray Properties of Intermediate-Mass Black Holes in Active Galaxies. II. X-Ray-Bright Accretion and Possible Evidence for Slim Disks}",
      journal = {\apj},
         year = 2009,
        month = jun,
       volume = {698},
       number = {2},
        pages = {1515-1522},
          doi = {10.1088/0004-637X/698/2/1515},
archivePrefix = {arXiv},
       eprint = {0903.2257},
 primaryClass = {astro-ph.GA},
       adsurl = {https://ui.adsabs.harvard.edu/abs/2009ApJ...698.1515D}
}

@ARTICLE{Panessa2006,
       author = {{Panessa}, F. and {Bassani}, L. and {Cappi}, M. and {Dadina}, M. and {Barcons}, X. and {Carrera}, F.~J. and {Ho}, L.~C. and {Iwasawa}, K.},
        title = "{On the X-ray, optical emission line and black hole mass properties of local Seyfert galaxies}",
      journal = {\aap},
         year = 2006,
        month = aug,
       volume = {455},
       number = {1},
        pages = {173-185},
          doi = {10.1051/0004-6361:20064894},
archivePrefix = {arXiv},
       eprint = {astro-ph/0605236},
 primaryClass = {astro-ph},
       adsurl = {https://ui.adsabs.harvard.edu/abs/2006A&A...455..173P}
}

@ARTICLE{Greene2004,
       author = {{Greene}, Jenny E. and {Ho}, Luis C.},
        title = "{Active Galactic Nuclei with Candidate Intermediate-Mass Black Holes}",
      journal = {\apj},
         year = 2004,
        month = aug,
       volume = {610},
       number = {2},
        pages = {722-736},
          doi = {10.1086/421719},
archivePrefix = {arXiv},
       eprint = {astro-ph/0404110},
 primaryClass = {astro-ph},
       adsurl = {https://ui.adsabs.harvard.edu/abs/2004ApJ...610..722G}
}

@ARTICLE{Loiacono2025,
       author = {{Loiacono}, Federica and {Gilli}, Roberto and {Mignoli}, Marco and {Mazzolari}, Giovanni and {Decarli}, Roberto and {Brusa}, Marcella and {Calura}, Francesco and {Chiaberge}, Marco and {Comastri}, Andrea and {D'Amato}, Quirino and {Iwasawa}, Kazushi and {Juod{\v{z}}balis}, Ignas and {Lanzuisi}, Giorgio and {Maiolino}, Roberto and {Marchesi}, Stefano and {Norman}, Colin and {Peca}, Alessandro and {Prandoni}, Isabella and {Sapori}, Matteo and {Signorini}, Matilde and {Tozzi}, Paolo and {Vanzella}, Eros and {Vignali}, Cristian and {Vito}, Fabio and {Zamorani}, Gianni},
        title = "{A big red dot at cosmic noon}",
      journal = {arXiv e-prints},
         year = 2025,
        month = jun,
          eid = {arXiv:2506.12141},
        pages = {arXiv:2506.12141},
          doi = {10.48550/arXiv.2506.12141},
archivePrefix = {arXiv},
       eprint = {2506.12141},
 primaryClass = {astro-ph.GA},
       adsurl = {https://ui.adsabs.harvard.edu/abs/2025arXiv250612141L}
}

@ARTICLE{Dong2010,
       author = {{Dong}, Xiao-Bo and {Ho}, Luis C. and {Wang}, Jian-Guo and {Wang}, Ting-Gui and {Wang}, Huiyuan and {Fan}, Xiaohui and {Zhou}, Hongyan},
        title = "{The Prevalence of Narrow Optical Fe II Emission Lines in Type 1 Active Galactic Nuclei}",
      journal = {\apjl},
         year = 2010,
        month = oct,
       volume = {721},
       number = {2},
        pages = {L143-L147},
          doi = {10.1088/2041-8205/721/2/L143},
archivePrefix = {arXiv},
       eprint = {1009.2209},
 primaryClass = {astro-ph.GA},
       adsurl = {https://ui.adsabs.harvard.edu/abs/2010ApJ...721L.143D}
}

@ARTICLE{Schlafly2019,
       author = {{Schlafly}, Edward F. and {Meisner}, Aaron M. and {Green}, Gregory M.},
        title = "{The unWISE Catalog: Two Billion Infrared Sources from Five Years of WISE Imaging}",
      journal = {\apjs},
         year = 2019,
        month = feb,
       volume = {240},
       number = {2},
          eid = {30},
        pages = {30},
          doi = {10.3847/1538-4365/aafbea},
archivePrefix = {arXiv},
       eprint = {1901.03337},
 primaryClass = {astro-ph.IM},
       adsurl = {https://ui.adsabs.harvard.edu/abs/2019ApJS..240...30S}
}

@ARTICLE{Simm2015,
       author = {{Simm}, T. and {Saglia}, R. and {Salvato}, M. and {Bender}, R. and {Burgett}, W.~S. and {Chambers}, K.~C. and {Draper}, P.~W. and {Flewelling}, H. and {Kaiser}, N. and {Kudritzki}, R. -P. and {Magnier}, E.~A. and {Metcalfe}, N. and {Tonry}, J.~L. and {Wainscoat}, R.~J. and {Waters}, C.},
        title = "{Pan-STARRS1 variability of XMM-COSMOS AGN. I. Impact on photometric redshifts}",
      journal = {\aap},
         year = 2015,
        month = dec,
       volume = {584},
          eid = {A106},
        pages = {A106},
          doi = {10.1051/0004-6361/201526859},
archivePrefix = {arXiv},
       eprint = {1510.01739},
 primaryClass = {astro-ph.GA},
       adsurl = {https://ui.adsabs.harvard.edu/abs/2015A&A...584A.106S}
}

@ARTICLE{Abazajian2009,
       author = {{Abazajian}, Kevork N. and {Adelman-McCarthy}, Jennifer K. and {Ag{\"u}eros}, Marcel A. and {Allam}, Sahar S. and {Allende Prieto}, Carlos and {An}, Deokkeun and {Anderson}, Kurt S.~J. and {Anderson}, Scott F. and {Annis}, James and {Bahcall}, Neta A. and {Bailer-Jones}, C.~A.~L. and {Barentine}, J.~C. and {Bassett}, Bruce A. and {Becker}, Andrew C. and {Beers}, Timothy C. and {Bell}, Eric F. and {Belokurov}, Vasily and {Berlind}, Andreas A. and {Berman}, Eileen F. and {Bernardi}, Mariangela and {Bickerton}, Steven J. and {Bizyaev}, Dmitry and {Blakeslee}, John P. and {Blanton}, Michael R. and {Bochanski}, John J. and {Boroski}, William N. and {Brewington}, Howard J. and {Brinchmann}, Jarle and {Brinkmann}, J. and {Brunner}, Robert J. and {Budav{\'a}ri}, Tam{\'a}s and {Carey}, Larry N. and {Carliles}, Samuel and {Carr}, Michael A. and {Castander}, Francisco J. and {Cinabro}, David and {Connolly}, A.~J. and {Csabai}, Istv{\'a}n and {Cunha}, Carlos E. and {Czarapata}, Paul C. and {Davenport}, James R.~A. and {de Haas}, Ernst and {Dilday}, Ben and {Doi}, Mamoru and {Eisenstein}, Daniel J. and {Evans}, Michael L. and {Evans}, N.~W. and {Fan}, Xiaohui and {Friedman}, Scott D. and {Frieman}, Joshua A. and {Fukugita}, Masataka and {G{\"a}nsicke}, Boris T. and {Gates}, Evalyn and {Gillespie}, Bruce and {Gilmore}, G. and {Gonzalez}, Belinda and {Gonzalez}, Carlos F. and {Grebel}, Eva K. and {Gunn}, James E. and {Gy{\"o}ry}, Zsuzsanna and {Hall}, Patrick B. and {Harding}, Paul and {Harris}, Frederick H. and {Harvanek}, Michael and {Hawley}, Suzanne L. and {Hayes}, Jeffrey J.~E. and {Heckman}, Timothy M. and {Hendry}, John S. and {Hennessy}, Gregory S. and {Hindsley}, Robert B. and {Hoblitt}, J. and {Hogan}, Craig J. and {Hogg}, David W. and {Holtzman}, Jon A. and {Hyde}, Joseph B. and {Ichikawa}, Shin-ichi and {Ichikawa}, Takashi and {Im}, Myungshin and {Ivezi{\'c}}, {\v{Z}}eljko and {Jester}, Sebastian and {Jiang}, Linhua and {Johnson}, Jennifer A. and {Jorgensen}, Anders M. and {Juri{\'c}}, Mario and {Kent}, Stephen M. and {Kessler}, R. and {Kleinman}, S.~J. and {Knapp}, G.~R. and {Konishi}, Kohki and {Kron}, Richard G. and {Krzesinski}, Jurek and {Kuropatkin}, Nikolay and {Lampeitl}, Hubert and {Lebedeva}, Svetlana and {Lee}, Myung Gyoon and {Lee}, Young Sun and {French Leger}, R. and {L{\'e}pine}, S{\'e}bastien and {Li}, Nolan and {Lima}, Marcos and {Lin}, Huan and {Long}, Daniel C. and {Loomis}, Craig P. and {Loveday}, Jon and {Lupton}, Robert H. and {Magnier}, Eugene and {Malanushenko}, Olena and {Malanushenko}, Viktor and {Mandelbaum}, Rachel and {Margon}, Bruce and {Marriner}, John P. and {Mart{\'\i}nez-Delgado}, David and {Matsubara}, Takahiko and {McGehee}, Peregrine M. and {McKay}, Timothy A. and {Meiksin}, Avery and {Morrison}, Heather L. and {Mullally}, Fergal and {Munn}, Jeffrey A. and {Murphy}, Tara and {Nash}, Thomas and {Nebot}, Ada and {Neilsen}, Jr., Eric H. and {Newberg}, Heidi Jo and {Newman}, Peter R. and {Nichol}, Robert C. and {Nicinski}, Tom and {Nieto-Santisteban}, Maria and {Nitta}, Atsuko and {Okamura}, Sadanori and {Oravetz}, Daniel J. and {Ostriker}, Jeremiah P. and {Owen}, Russell and {Padmanabhan}, Nikhil and {Pan}, Kaike and {Park}, Changbom and {Pauls}, George and {Peoples}, Jr., John and {Percival}, Will J. and {Pier}, Jeffrey R. and {Pope}, Adrian C. and {Pourbaix}, Dimitri and {Price}, Paul A. and {Purger}, Norbert and {Quinn}, Thomas and {Raddick}, M. Jordan and {Re Fiorentin}, Paola and {Richards}, Gordon T. and {Richmond}, Michael W. and {Riess}, Adam G. and {Rix}, Hans-Walter and {Rockosi}, Constance M. and {Sako}, Masao and {Schlegel}, David J. and {Schneider}, Donald P. and {Scholz}, Ralf-Dieter and {Schreiber}, Matthias R. and {Schwope}, Axel D. and {Seljak}, Uro{\v{s}} and {Sesar}, Branimir and {Sheldon}, Erin and {Shimasaku}, Kazu and {Sibley}, Valena C. and {Simmons}, A.~E. and {Sivarani}, Thirupathi and {Allyn Smith}, J. and {Smith}, Martin C. and {Smol{\v{c}}i{\'c}}, Vernesa and {Snedden}, Stephanie A. and {Stebbins}, Albert and {Steinmetz}, Matthias and {Stoughton}, Chris and {Strauss}, Michael A. and {SubbaRao}, Mark and {Suto}, Yasushi and {Szalay}, Alexander S. and {Szapudi}, Istv{\'a}n and {Szkody}, Paula and {Tanaka}, Masayuki and {Tegmark}, Max and {Teodoro}, Luis F.~A. and {Thakar}, Aniruddha R. and {Tremonti}, Christy A. and {Tucker}, Douglas L. and {Uomoto}, Alan and {Vanden Berk}, Daniel E. and {Vandenberg}, Jan and {Vidrih}, S. and {Vogeley}, Michael S. and {Voges}, Wolfgang and {Vogt}, Nicole P. and {Wadadekar}, Yogesh and {Watters}, Shannon and {Weinberg}, David H. and {West}, Andrew A. and {White}, Simon D.~M. and {Wilhite}, Brian C. and {Wonders}, Alainna C. and {Yanny}, Brian and {Yocum}, D.~R.},
        title = "{The Seventh Data Release of the Sloan Digital Sky Survey}",
      journal = {\apjs},
         year = 2009,
        month = jun,
       volume = {182},
       number = {2},
        pages = {543-558},
          doi = {10.1088/0067-0049/182/2/543},
archivePrefix = {arXiv},
       eprint = {0812.0649},
 primaryClass = {astro-ph},
       adsurl = {https://ui.adsabs.harvard.edu/abs/2009ApJS..182..543A}
}

@article{NIKUTTA2020100411,
title = {Data Lab—A community science platform},
journal = {Astronomy and Computing},
volume = {33},
pages = {100411},
year = {2020},
issn = {2213-1337},
doi = {https://doi.org/10.1016/j.ascom.2020.100411},
url = {https://www.sciencedirect.com/science/article/pii/S2213133720300652},
author = {R. Nikutta and M. Fitzpatrick and A. Scott and B.A. Weaver}
}

@inproceedings{Fitzpatrick_SPIE,
author = {Michael J. Fitzpatrick and Knut Olsen and Frossie Economou and Elizabeth B. Stobie and T. C. Beers and Mark Dickinson and Patrick Norris and Abi Saha and Robert Seaman and David R. Silva and Robert A. Swaters and Brian Thomas and Francisco Valdes},
title = {{The NOAO Data Laboratory: a conceptual overview}},
volume = {9149},
booktitle = {Observatory Operations: Strategies, Processes, and Systems V},
editor = {Alison B. Peck and Chris R. Benn and Robert L. Seaman},
organization = {International Society for Optics and Photonics},
publisher = {SPIE},
pages = {91491T},
year = {2014},
doi = {10.1117/12.2057445},
URL = {https://doi.org/10.1117/12.2057445}
}

@ARTICLE{Peng2002,
       author = {{Peng}, Chien Y. and {Ho}, Luis C. and {Impey}, Chris D. and {Rix}, Hans-Walter},
        title = "{Detailed Structural Decomposition of Galaxy Images}",
      journal = {\aj},
         year = 2002,
        month = jul,
       volume = {124},
       number = {1},
        pages = {266-293},
          doi = {10.1086/340952},
archivePrefix = {arXiv},
       eprint = {astro-ph/0204182},
 primaryClass = {astro-ph},
       adsurl = {https://ui.adsabs.harvard.edu/abs/2002AJ....124..266P}
}

@ARTICLE{Liu2025,
       author = {{Liu}, Hanpu and {Jiang}, Yan-Fei and {Quataert}, Eliot and {Greene}, Jenny E. and {Ma}, Yilun},
        title = "{The Balmer Break and Optical Continuum of Little Red Dots From Super-Eddington Accretion}",
      journal = {arXiv e-prints},
         year = 2025,
        month = jul,
          eid = {arXiv:2507.07190},
        pages = {arXiv:2507.07190},
archivePrefix = {arXiv},
       eprint = {2507.07190},
 primaryClass = {astro-ph.GA},
       adsurl = {https://ui.adsabs.harvard.edu/abs/2025arXiv250707190L}
}

@ARTICLE{Jorgensen1992,
       author = {{Jorgensen}, U.~G. and {Carlsson}, M. and {Johnson}, H.~R.},
        title = "{The calcium infrared triplet lines in stellar spectra.}",
      journal = {\aap},
         year = 1992,
        month = feb,
       volume = {254},
        pages = {258-265},
       adsurl = {https://ui.adsabs.harvard.edu/abs/1992A&A...254..258J}
}

@ARTICLE{Evans2024,
       author = {{Evans}, Ian N. and {Evans}, Janet D. and {Mart{\'\i}nez-Galarza}, J. Rafael and {Miller}, Joseph B. and {Primini}, Francis A. and {Azadi}, Mojegan and {Burke}, Douglas J. and {Civano}, Francesca M. and {D'Abrusco}, Raffaele and {Fabbiano}, Giuseppina and {Graessle}, Dale E. and {Grier}, John D. and {Houck}, John C. and {Lauer}, Jennifer and {McCollough}, Michael L. and {Nowak}, Michael A. and {Plummer}, David A. and {Rots}, Arnold H. and {Siemiginowska}, Aneta and {Tibbetts}, Michael S.},
        title = "{The Chandra Source Catalog Release 2 Series}",
      journal = {\apjs},
         year = 2024,
        month = oct,
       volume = {274},
       number = {2},
          eid = {22},
        pages = {22},
          doi = {10.3847/1538-4365/ad6319},
archivePrefix = {arXiv},
       eprint = {2407.10799},
 primaryClass = {astro-ph.HE},
       adsurl = {https://ui.adsabs.harvard.edu/abs/2024ApJS..274...22E}
}

@ARTICLE{Labbe2025,
       author = {{Labbe}, Ivo and {Greene}, Jenny E. and {Bezanson}, Rachel and {Fujimoto}, Seiji and {Furtak}, Lukas J. and {Goulding}, Andy D. and {Matthee}, Jorryt and {Naidu}, Rohan P. and {Oesch}, Pascal A. and {Atek}, Hakim and {Brammer}, Gabriel and {Chemerynska}, Iryna and {Coe}, Dan and {Cutler}, Sam E. and {Dayal}, Pratika and {Feldmann}, Robert and {Franx}, Marijn and {Glazebrook}, Karl and {Leja}, Joel and {Maseda}, Michael and {Marchesini}, Danilo and {Nanayakkara}, Themiya and {Nelson}, Erica J. and {Pan}, Richard and {Papovich}, Casey and {Price}, Sedona H. and {Suess}, Katherine A. and {Wang}, Bingjie and {Weaver}, John R. and {Whitaker}, Katherine E. and {Williams}, Christina C. and {Zitrin}, Adi},
        title = "{UNCOVER: Candidate Red Active Galactic Nuclei at 3 < z < 7 with JWST and ALMA}",
      journal = {\apj},
         year = 2025,
        month = jan,
       volume = {978},
       number = {1},
          eid = {92},
        pages = {92},
          doi = {10.3847/1538-4357/ad3551},
archivePrefix = {arXiv},
       eprint = {2306.07320},
 primaryClass = {astro-ph.GA},
       adsurl = {https://ui.adsabs.harvard.edu/abs/2025ApJ...978...92L}
}

@ARTICLE{Xiao2025,
       author = {{Xiao}, Mengyuan and {Oesch}, Pascal A. and {Bing}, Longji and {Elbaz}, David and {Matthee}, Jorryt and {Fudamoto}, Yoshinobu and {Fujimoto}, Seiji and {Marques-Chaves}, Rui and {Williams}, Christina C. and {Dessauges-Zavadsky}, Miroslava and {Valentino}, Francesco and {Brammer}, Gabriel and {Covelo-Paz}, Alba and {Daddi}, Emanuele and {Fynbo}, Johan P.~U. and {Gillman}, Steven and {Ginolfi}, Michele and {Giovinazzo}, Emma and {Greene}, Jenny E. and {Gu}, Qiusheng and {Illingworth}, Garth and {Inayoshi}, Kohei and {Kokorev}, Vasily and {Meyer}, Romain A. and {Naidu}, Rohan P. and {Reddy}, Naveen A. and {Schaerer}, Daniel and {Shapley}, Alice and {Stefanon}, Mauro and {Steinhardt}, Charles L. and {Setton}, David J. and {Vestergaard}, Marianne and {Wang}, Tao},
        title = "{No [CII] or dust detection in two Little Red Dots at z$_{\rm spec}$ > 7}",
      journal = {arXiv e-prints},
         year = 2025,
        month = mar,
          eid = {arXiv:2503.01945},
        pages = {arXiv:2503.01945},
          doi = {10.48550/arXiv.2503.01945},
archivePrefix = {arXiv},
       eprint = {2503.01945},
 primaryClass = {astro-ph.GA},
       adsurl = {https://ui.adsabs.harvard.edu/abs/2025arXiv250301945X}
}

@ARTICLE{Zhuang2025,
       author = {{Zhuang}, Ming-Yang and {Li}, Junyao and {Shen}, Yue and {Lin}, Xiaojing and {Shapley}, Alice E. and {Wang}, Feige and {Wu}, Qiaoya and {Yang}, Qian},
        title = "{NEXUS: A Spectroscopic Census of Broad-line AGNs and Little Red Dots at $3\lesssim z\lesssim 6$}",
      journal = {arXiv e-prints},
         year = 2025,
        month = may,
          eid = {arXiv:2505.20393},
        pages = {arXiv:2505.20393},
          doi = {10.48550/arXiv.2505.20393},
archivePrefix = {arXiv},
       eprint = {2505.20393},
 primaryClass = {astro-ph.GA},
       adsurl = {https://ui.adsabs.harvard.edu/abs/2025arXiv250520393Z}
}

@ARTICLE{Piskunov1995,
       author = {{Piskunov}, N.~E. and {Kupka}, F. and {Ryabchikova}, T.~A. and {Weiss}, W.~W. and {Jeffery}, C.~S.},
        title = "{VALD: The Vienna Atomic Line Data Base.}",
      journal = {\aaps},
         year = 1995,
        month = sep,
       volume = {112},
        pages = {525},
       adsurl = {https://ui.adsabs.harvard.edu/abs/1995A&AS..112..525P}
}

@ARTICLE{Ryabchikova2015,
       author = {{Ryabchikova}, T. and {Piskunov}, N. and {Kurucz}, R.~L. and {Stempels}, H.~C. and {Heiter}, U. and {Pakhomov}, Yu and {Barklem}, P.~S.},
        title = "{A major upgrade of the VALD database}",
      journal = {\physscr},
         year = 2015,
        month = may,
       volume = {90},
       number = {5},
          eid = {054005},
        pages = {054005},
          doi = {10.1088/0031-8949/90/5/054005},
       adsurl = {https://ui.adsabs.harvard.edu/abs/2015PhyS...90e4005R}
}

@ARTICLE{Struve1934,
       author = {{Struve}, Otto and {Elvey}, C.~T.},
        title = "{The Intensities of Stellar Absorption Lines}",
      journal = {\apj},
         year = 1934,
        month = may,
       volume = {79},
        pages = {409},
          doi = {10.1086/143551},
       adsurl = {https://ui.adsabs.harvard.edu/abs/1934ApJ....79..409S}
}

@BOOK{Moore1966,
       author = {{Moore}, Charlotte E. and {Minnaert}, M.~G.~J. and {Houtgast}, J.},
        title = "{The solar spectrum 2935 A to 8770 A}",
         year = 1966,
       adsurl = {https://ui.adsabs.harvard.edu/abs/1966sst..book.....M}
}

@ARTICLE{Heiter2021,
       author = {{Heiter}, U. and {Lind}, K. and {Bergemann}, M. and {Asplund}, M. and {Mikolaitis}, {\v{S}}. and {Barklem}, P.~S. and {Masseron}, T. and {de Laverny}, P. and {Magrini}, L. and {Edvardsson}, B. and {J{\"o}nsson}, H. and {Pickering}, J.~C. and {Ryde}, N. and {Bayo Ar{\'a}n}, A. and {Bensby}, T. and {Casey}, A.~R. and {Feltzing}, S. and {Jofr{\'e}}, P. and {Korn}, A.~J. and {Pancino}, E. and {Damiani}, F. and {Lanzafame}, A. and {Lardo}, C. and {Monaco}, L. and {Morbidelli}, L. and {Smiljanic}, R. and {Worley}, C. and {Zaggia}, S. and {Randich}, S. and {Gilmore}, G.~F.},
        title = "{Atomic data for the Gaia-ESO Survey}",
      journal = {\aap},
         year = 2021,
        month = jan,
       volume = {645},
          eid = {A106},
        pages = {A106},
          doi = {10.1051/0004-6361/201936291},
archivePrefix = {arXiv},
       eprint = {2011.02049},
 primaryClass = {astro-ph.IM},
       adsurl = {https://ui.adsabs.harvard.edu/abs/2021A&A...645A.106H}
}

@ARTICLE{Ji2025_lord,
       author = {{Ji}, Xihan and {D'Eugenio}, Francesco and {Juod{\v{z}}balis}, Ignas and {Walton}, Dominic J. and {Fabian}, Andrew C. and {Maiolino}, Roberto and {Ramos Almeida}, Cristina and {Acosta Pulido}, Jose A. and {Belokurov}, Vasily A. and {Isobe}, Yuki and {Jones}, Gareth and {Maraston}, Claudia and {Scholtz}, Jan and {Simmonds}, Charlotte and {Tacchella}, Sandro and {Terlevich}, Elena and {Terlevich}, Roberto},
        title = "{Lord of LRDs: Insights into a ``Little Red Dot'' with a low-ionization spectrum at z = 0.1}",
      journal = {arXiv e-prints},
         year = 2025,
        month = jul,
          eid = {arXiv:2507.23774},
        pages = {arXiv:2507.23774},
          doi = {10.48550/arXiv.2507.23774},
archivePrefix = {arXiv},
       eprint = {2507.23774},
 primaryClass = {astro-ph.GA},
       adsurl = {https://ui.adsabs.harvard.edu/abs/2025arXiv250723774J}
}

@ARTICLE{Bae2014,
       author = {{Bae}, Hyun-Jin and {Woo}, Jong-Hak},
        title = "{A Census of Gas Outflows in Type 2 Active Galactic Nuclei}",
      journal = {\apj},
         year = 2014,
        month = nov,
       volume = {795},
       number = {1},
          eid = {30},
        pages = {30},
          doi = {10.1088/0004-637X/795/1/30},
archivePrefix = {arXiv},
       eprint = {1409.1580},
 primaryClass = {astro-ph.GA},
       adsurl = {https://ui.adsabs.harvard.edu/abs/2014ApJ...795...30B}
}

@ARTICLE{Shen2016,
       author = {{Shen}, Yue and {Brandt}, W.~N. and {Richards}, Gordon T. and {Denney}, Kelly D. and {Greene}, Jenny E. and {Grier}, C.~J. and {Ho}, Luis C. and {Peterson}, Bradley M. and {Petitjean}, Patrick and {Schneider}, Donald P. and {Tao}, Charling and {Trump}, Jonathan R.},
        title = "{The Sloan Digital Sky Survey Reverberation Mapping Project: Velocity Shifts of Quasar Emission Lines}",
      journal = {\apj},
         year = 2016,
        month = nov,
       volume = {831},
       number = {1},
          eid = {7},
        pages = {7},
          doi = {10.3847/0004-637X/831/1/7},
archivePrefix = {arXiv},
       eprint = {1602.03894},
 primaryClass = {astro-ph.GA},
       adsurl = {https://ui.adsabs.harvard.edu/abs/2016ApJ...831....7S}
}

@ARTICLE{Moullet2023,
       author = {{Moullet}, A. and {Kataria}, T. and {Lis}, D. and {Unwin}, S. and {Hasegawa}, Y. and {Mills}, E. and {Battersby}, C. and {Roc}, A. and {Meixner}, M.},
        title = "{PRIMA General Observer Science Book}",
      journal = {arXiv e-prints},
         year = 2023,
        month = oct,
          eid = {arXiv:2310.20572},
        pages = {arXiv:2310.20572},
          doi = {10.48550/arXiv.2310.20572},
archivePrefix = {arXiv},
       eprint = {2310.20572},
 primaryClass = {astro-ph.IM},
       adsurl = {https://ui.adsabs.harvard.edu/abs/2023arXiv231020572M}
}

@ARTICLE{Maiolino2025,
       author = {{Maiolino}, Roberto and {Risaliti}, Guido and {Signorini}, Matilde and {Trefoloni}, Bartolomeo and {Juod{\v{z}}balis}, Ignas and {Scholtz}, Jan and {{\"U}bler}, Hannah and {D'Eugenio}, Francesco and {Carniani}, Stefano and {Fabian}, Andy and {Ji}, Xihan and {Mazzolari}, Giovanni and {Bertola}, Elena and {Brusa}, Marcella and {Bunker}, Andrew J. and {Charlot}, Stephane and {Comastri}, Andrea and {Cresci}, Giovanni and {DeCoursey}, Christa Noel and {Egami}, Eiichi and {Fiore}, Fabrizio and {Gilli}, Roberto and {Perna}, Michele and {Tacchella}, Sandro and {Venturi}, Giacomo},
        title = "{JWST meets Chandra: a large population of Compton thick, feedback-free, and intrinsically X-ray weak AGN, with a sprinkle of SNe}",
      journal = {\mnras},
         year = 2025,
        month = apr,
       volume = {538},
       number = {3},
        pages = {1921-1943},
          doi = {10.1093/mnras/staf359},
archivePrefix = {arXiv},
       eprint = {2405.00504},
 primaryClass = {astro-ph.GA},
       adsurl = {https://ui.adsabs.harvard.edu/abs/2025MNRAS.538.1921M}
}

@ARTICLE{Sacchi2025,
       author = {{Sacchi}, Andrea and {Bogd{\'a}n}, {\'A}kos},
        title = "{Chandra Rules Out Super-Eddington Accretion Models for Little Red Dots}",
      journal = {\apjl},
         year = 2025,
        month = aug,
       volume = {989},
       number = {2},
          eid = {L30},
        pages = {L30},
          doi = {10.3847/2041-8213/adf5c8},
archivePrefix = {arXiv},
       eprint = {2505.09669},
 primaryClass = {astro-ph.GA},
       adsurl = {https://ui.adsabs.harvard.edu/abs/2025ApJ...989L..30S}
}

@ARTICLE{Sanders2023,
       author = {{Sanders}, Ryan L. and {Shapley}, Alice E. and {Topping}, Michael W. and {Reddy}, Naveen A. and {Brammer}, Gabriel B.},
        title = "{Excitation and Ionization Properties of Star-forming Galaxies at z = 2.0-9.3 with JWST/NIRSpec}",
      journal = {\apj},
         year = 2023,
        month = sep,
       volume = {955},
       number = {1},
          eid = {54},
        pages = {54},
          doi = {10.3847/1538-4357/acedad},
archivePrefix = {arXiv},
       eprint = {2301.06696},
 primaryClass = {astro-ph.GA},
       adsurl = {https://ui.adsabs.harvard.edu/abs/2023ApJ...955...54S}
}

@ARTICLE{Shapley2025,
       author = {{Shapley}, Alice E. and {Sanders}, Ryan L. and {Topping}, Michael W. and {Reddy}, Naveen A. and {Berg}, Danielle A. and {Bouwens}, Rychard J. and {Brammer}, Gabriel and {Carnall}, Adam C. and {Cullen}, Fergus and {Dav{\'e}}, Romeel and {Dunlop}, James S. and {Ellis}, Richard S. and {F{\"o}rster Schreiber}, N.~M. and {Furlanetto}, Steven R. and {Glazebrook}, Karl and {Illingworth}, Garth D. and {Jones}, Tucker and {Kriek}, Mariska and {McLeod}, Derek J. and {McLure}, Ross J. and {Narayanan}, Desika and {Oesch}, Pascal and {Pahl}, Anthony J. and {Pettini}, Max and {Schaerer}, Daniel and {Stark}, Daniel P. and {Steidel}, Charles C. and {Tang}, Mengtao and {Clarke}, Leonardo and {Donnan}, Callum T. and {Kehoe}, Emily},
        title = "{The AURORA Survey: A New Era of Emission-line Diagrams with JWST/NIRSpec}",
      journal = {\apj},
         year = 2025,
        month = feb,
       volume = {980},
       number = {2},
          eid = {242},
        pages = {242},
          doi = {10.3847/1538-4357/adad68},
archivePrefix = {arXiv},
       eprint = {2407.00157},
 primaryClass = {astro-ph.GA},
       adsurl = {https://ui.adsabs.harvard.edu/abs/2025ApJ...980..242S}
}

@ARTICLE{Shapley2023,
       author = {{Shapley}, Alice E. and {Reddy}, Naveen A. and {Sanders}, Ryan L. and {Topping}, Michael W. and {Brammer}, Gabriel B.},
        title = "{JWST/NIRSpec Measurements of the Relationships between Nebular Emission-line Ratios and Stellar Mass at z   3-6}",
      journal = {\apjl},
         year = 2023,
        month = jun,
       volume = {950},
       number = {1},
          eid = {L1},
        pages = {L1},
          doi = {10.3847/2041-8213/acd939},
archivePrefix = {arXiv},
       eprint = {2303.00410},
 primaryClass = {astro-ph.GA},
       adsurl = {https://ui.adsabs.harvard.edu/abs/2023ApJ...950L...1S}
}

@ARTICLE{Juodzbalis2025,
       author = {{Juod{\v{z}}balis}, Ignas and {Maiolino}, Roberto and {Baker}, William M. and {Lake}, Emma Curtis and {Scholtz}, Jan and {D'Eugenio}, Francesco and {Trefoloni}, Bartolomeo and {Isobe}, Yuki and {Tacchella}, Sandro and {Bunker}, Andrew J. and {Carniani}, Stefano and {Charlot}, St{\'e}phane and {Jones}, Gareth C. and {Parlanti}, Eleonora and {Perna}, Michele and {Rinaldi}, Pierluigi and {Robertson}, Brant and {{\"U}bler}, Hannah and {Venturi}, Giacomo and {Willott}, Chris},
        title = "{JADES: comprehensive census of broad-line AGN from Reionization to Cosmic Noon revealed by JWST}",
      journal = {arXiv e-prints},
         year = 2025,
        month = apr,
          eid = {arXiv:2504.03551},
        pages = {arXiv:2504.03551},
          doi = {10.48550/arXiv.2504.03551},
archivePrefix = {arXiv},
       eprint = {2504.03551},
 primaryClass = {astro-ph.GA},
       adsurl = {https://ui.adsabs.harvard.edu/abs/2025arXiv250403551J}
}

@ARTICLE{Brocklehurst1971,
       author = {{Brocklehurst}, M.},
        title = "{Calculations of level populations for the low levels of hydrogenic ions in gaseous nebulae.}",
      journal = {\mnras},
         year = 1971,
        month = jan,
       volume = {153},
        pages = {471},
          doi = {10.1093/mnras/153.4.471},
       adsurl = {https://ui.adsabs.harvard.edu/abs/1971MNRAS.153..471B}
}

@BOOK{Osterbrock2006,
       author = {{Osterbrock}, Donald E. and {Ferland}, Gary J.},
        title = "{Astrophysics of gaseous nebulae and active galactic nuclei}",
         year = 2006,
       adsurl = {https://ui.adsabs.harvard.edu/abs/2006agna.book.....O}
}

@ARTICLE{Hummer1987,
       author = {{Hummer}, D.~G. and {Storey}, P.~J.},
        title = "{Recombination-line intensities for hydrogenic ions - I. Case B calculations for H I and He II.}",
      journal = {\mnras},
         year = 1987,
        month = feb,
       volume = {224},
        pages = {801-820},
          doi = {10.1093/mnras/224.3.801},
       adsurl = {https://ui.adsabs.harvard.edu/abs/1987MNRAS.224..801H}
}

@ARTICLE{Rinaldi2025,
       author = {{Rinaldi}, Pierluigi and {Rieke}, George H. and {Wu}, Zihao and {Gilbert}, Carys J.~E. and {Pacucci}, Fabio and {Barchiesi}, Luigi and {Alberts}, Stacey and {Carniani}, Stefano and {Bunker}, Andrew J. and {Bhatawdekar}, Rachana and {D'Eugenio}, Francesco and {Ji}, Zhiyuan and {Johnson}, Benjamin D. and {Hainline}, Kevin and {Kokorev}, Vasily and {Kumari}, Nimisha and {Iani}, Edoardo and {Lyu}, Jianwei and {Maiolino}, Roberto and {Parlanti}, Eleonora and {Robertson}, Brant E. and {Sun}, Yang and {Vignali}, Cristian and {Williams}, Christina C. and {Willmer}, Christopher N.~A. and {Zhu}, Yongda},
        title = "{Beyond the Dot: an LRD-like nucleus at the Heart of an IR-Bright Galaxy and its implications for high-redshift LRDs}",
      journal = {arXiv e-prints},
         year = 2025,
        month = jul,
          eid = {arXiv:2507.17738},
        pages = {arXiv:2507.17738},
          doi = {10.48550/arXiv.2507.17738},
archivePrefix = {arXiv},
       eprint = {2507.17738},
 primaryClass = {astro-ph.GA},
       adsurl = {https://ui.adsabs.harvard.edu/abs/2025arXiv250717738R}
}

@ARTICLE{Gutkin2016,
       author = {{Gutkin}, Julia and {Charlot}, St{\'e}phane and {Bruzual}, Gustavo},
        title = "{Modelling the nebular emission from primeval to present-day star-forming galaxies}",
      journal = {\mnras},
         year = 2016,
        month = oct,
       volume = {462},
       number = {2},
        pages = {1757-1774},
          doi = {10.1093/mnras/stw1716},
archivePrefix = {arXiv},
       eprint = {1607.06086},
 primaryClass = {astro-ph.GA},
       adsurl = {https://ui.adsabs.harvard.edu/abs/2016MNRAS.462.1757G}
}

@ARTICLE{Feltre2016,
       author = {{Feltre}, A. and {Charlot}, S. and {Gutkin}, J.},
        title = "{Nuclear activity versus star formation: emission-line diagnostics at ultraviolet and optical wavelengths}",
      journal = {\mnras},
         year = 2016,
        month = mar,
       volume = {456},
       number = {3},
        pages = {3354-3374},
          doi = {10.1093/mnras/stv2794},
archivePrefix = {arXiv},
       eprint = {1511.08217},
 primaryClass = {astro-ph.GA},
       adsurl = {https://ui.adsabs.harvard.edu/abs/2016MNRAS.456.3354F}
}

@ARTICLE{Mingozzi2022,
       author = {{Mingozzi}, Matilde and {James}, Bethan L. and {Arellano-C{\'o}rdova}, Karla Z. and {Berg}, Danielle A. and {Senchyna}, Peter and {Chisholm}, John and {Brinchmann}, Jarle and {Aloisi}, Alessandra and {Amor{\'\i}n}, Ricardo O. and {Charlot}, St{\'e}phane and {Feltre}, Anna and {Hayes}, Matthew and {Heckman}, Timothy and {Henry}, Alaina and {Hernandez}, Svea and {Kumari}, Nimisha and {Leitherer}, Claus and {Llerena}, Mario and {Martin}, Crystal L. and {Nanayakkara}, Themiya and {Ravindranath}, Swara and {Skillman}, Evan D. and {Sugahara}, Yuma and {Wofford}, Aida and {Xu}, Xinfeng},
        title = "{CLASSY IV. Exploring UV Diagnostics of the Interstellar Medium in Local High-z Analogs at the Dawn of the JWST Era}",
      journal = {\apj},
         year = 2022,
        month = nov,
       volume = {939},
       number = {2},
          eid = {110},
        pages = {110},
          doi = {10.3847/1538-4357/ac952c},
archivePrefix = {arXiv},
       eprint = {2209.09047},
 primaryClass = {astro-ph.GA},
       adsurl = {https://ui.adsabs.harvard.edu/abs/2022ApJ...939..110M}
}

@ARTICLE{Madau2025,
       author = {{Madau}, Piero},
        title = "{Chasing the Light: Shadowing, Collimation, and the Super-Eddington Growth of Infant Black Holes in JWST-Discovered AGNs}",
      journal = {arXiv e-prints},
         year = 2025,
        month = jan,
          eid = {arXiv:2501.09854},
        pages = {arXiv:2501.09854},
          doi = {10.48550/arXiv.2501.09854},
archivePrefix = {arXiv},
       eprint = {2501.09854},
 primaryClass = {astro-ph.HE},
       adsurl = {https://ui.adsabs.harvard.edu/abs/2025arXiv250109854M}
}

@misc{GUVcat_DOI,
  doi = {10.17909/T9-PYXY-KG53},
  url = {http://archive.stsci.edu/doi/resolve/resolve.html?doi=10.17909/t9-pyxy-kg53},
  author = {Bianchi,  Luciana},
  title = {GALEX UV Unique Source Catalogs ("GUVcat") and Cross-Matches With Gaia and SDSS ("GUVmatch")},
  publisher = {STScI/MAST},
  year = {2020}
}

@ARTICLE{Reines2015,
       author = {{Reines}, Amy E. and {Volonteri}, Marta},
        title = "{Relations between Central Black Hole Mass and Total Galaxy Stellar Mass in the Local Universe}",
      journal = {\apj},
         year = 2015,
        month = nov,
       volume = {813},
       number = {2},
          eid = {82},
        pages = {82},
          doi = {10.1088/0004-637X/813/2/82},
archivePrefix = {arXiv},
       eprint = {1508.06274},
 primaryClass = {astro-ph.GA},
       adsurl = {https://ui.adsabs.harvard.edu/abs/2015ApJ...813...82R}
}

@ARTICLE{Brazzini2025,
       author = {{Brazzini}, Matilde and {D'Eugenio}, Francesco and {Maiolino}, Roberto and {Juod{\v{z}}balis}, Ignas and {Ji}, Xihan and {Scholtz}, Jan},
        title = "{Ruling out dominant electron scattering in Little Red Dots' Rosetta Stone using multiple hydrogen lines}",
      journal = {arXiv e-prints},
         year = 2025,
        month = jul,
          eid = {arXiv:2507.08929},
        pages = {arXiv:2507.08929},
          doi = {10.48550/arXiv.2507.08929},
archivePrefix = {arXiv},
       eprint = {2507.08929},
 primaryClass = {astro-ph.GA},
       adsurl = {https://ui.adsabs.harvard.edu/abs/2025arXiv250708929B}
}

@ARTICLE{Rusakov2025,
       author = {{Rusakov}, V. and {Watson}, D. and {Nikopoulos}, G.~P. and {Brammer}, G. and {Gottumukkala}, R. and {Harvey}, T. and {Heintz}, K.~E. and {Nielsen}, R.~D. and {Sim}, S.~A. and {Sneppen}, A. and {Vijayan}, A.~P. and {Adams}, N. and {Austin}, D. and {Conselice}, C.~J. and {Goolsby}, C.~M. and {Toft}, S. and {Witstok}, J.},
        title = "{JWST's little red dots: an emerging population of young, low-mass AGN cocooned in dense ionized gas}",
      journal = {arXiv e-prints},
         year = 2025,
        month = mar,
          eid = {arXiv:2503.16595},
        pages = {arXiv:2503.16595},
          doi = {10.48550/arXiv.2503.16595},
archivePrefix = {arXiv},
       eprint = {2503.16595},
 primaryClass = {astro-ph.GA},
       adsurl = {https://ui.adsabs.harvard.edu/abs/2025arXiv250316595R}
}

@ARTICLE{Greene2025,
       author = {{Greene}, Jenny E. and {Setton}, David J. and {Furtak}, Lukas J. and {Naidu}, Rohan P. and {Volonteri}, Marta and {Dayal}, Pratika and {Labbe}, Ivo and {van Dokkum}, Pieter and {Bezanson}, Rachel and {Brammer}, Gabriel and {Cutler}, Sam E. and {Glazebrook}, Karl and {de Graaff}, Anna and {Hirschmann}, Michaela and {Hviding}, Raphael E. and {Kokorev}, Vasily and {Leja}, Joel and {Liu}, Hanpu and {Ma}, Yilun and {Matthee}, Jorryt and {Nanayakkara}, Themiya and {Oesch}, Pascal A. and {Pan}, Richard and {Price}, Sedona H. and {Spilker}, Justin S. and {Wang}, Bingjie and {Weaver}, John R. and {Whitaker}, Katherine E. and {Williams}, Christina C. and {Zitrin}, Adi},
        title = "{What you see is what you get: empirically measured bolometric luminosities of Little Red Dots}",
      journal = {arXiv e-prints},
         year = 2025,
        month = sep,
          eid = {arXiv:2509.05434},
        pages = {arXiv:2509.05434},
          doi = {10.48550/arXiv.2509.05434},
archivePrefix = {arXiv},
       eprint = {2509.05434},
 primaryClass = {astro-ph.GA},
       adsurl = {https://ui.adsabs.harvard.edu/abs/2025arXiv250905434G}
}

@ARTICLE{Stern2012,
       author = {{Stern}, Jonathan and {Laor}, Ari},
        title = "{Type 1 AGN at low z - II. The relative strength of narrow lines and the nature of intermediate type AGN}",
      journal = {\mnras},
         year = 2012,
        month = nov,
       volume = {426},
       number = {4},
        pages = {2703-2718},
          doi = {10.1111/j.1365-2966.2012.21772.x},
archivePrefix = {arXiv},
       eprint = {1207.5543},
 primaryClass = {astro-ph.CO},
       adsurl = {https://ui.adsabs.harvard.edu/abs/2012MNRAS.426.2703S}
}

@ARTICLE{Zhou2019,
       author = {{Zhou}, Yu and {Feng}, Hua and {Ho}, Luis C. and {Yao}, Yuhan},
        title = "{Evidence for Optically Thick, Eddington-limited Winds Driven by Supercritical Accretion}",
      journal = {\apj},
         year = 2019,
        month = jan,
       volume = {871},
       number = {1},
          eid = {115},
        pages = {115},
          doi = {10.3847/1538-4357/aaf724},
archivePrefix = {arXiv},
       eprint = {1812.02923},
 primaryClass = {astro-ph.HE},
       adsurl = {https://ui.adsabs.harvard.edu/abs/2019ApJ...871..115Z}
}

@ARTICLE{Lyu2019,
       author = {{Lyu}, Jianwei and {Rieke}, George H. and {Smith}, Paul S.},
        title = "{Mid-IR Variability and Dust Reverberation Mapping of Low-z Quasars. I. Data, Methods, and Basic Results}",
      journal = {\apj},
         year = 2019,
        month = nov,
       volume = {886},
       number = {1},
          eid = {33},
        pages = {33},
          doi = {10.3847/1538-4357/ab481d},
archivePrefix = {arXiv},
       eprint = {1909.11101},
 primaryClass = {astro-ph.GA},
       adsurl = {https://ui.adsabs.harvard.edu/abs/2019ApJ...886...33L}
}

@ARTICLE{Mainzer2011,
       author = {{Mainzer}, A. and {Bauer}, J. and {Grav}, T. and {Masiero}, J. and {Cutri}, R.~M. and {Dailey}, J. and {Eisenhardt}, P. and {McMillan}, R.~S. and {Wright}, E. and {Walker}, R. and {Jedicke}, R. and {Spahr}, T. and {Tholen}, D. and {Alles}, R. and {Beck}, R. and {Brandenburg}, H. and {Conrow}, T. and {Evans}, T. and {Fowler}, J. and {Jarrett}, T. and {Marsh}, K. and {Masci}, F. and {McCallon}, H. and {Wheelock}, S. and {Wittman}, M. and {Wyatt}, P. and {DeBaun}, E. and {Elliott}, G. and {Elsbury}, D. and {Gautier}, IV, T. and {Gomillion}, S. and {Leisawitz}, D. and {Maleszewski}, C. and {Micheli}, M. and {Wilkins}, A.},
        title = "{Preliminary Results from NEOWISE: An Enhancement to the Wide-field Infrared Survey Explorer for Solar System Science}",
      journal = {\apj},
         year = 2011,
        month = apr,
       volume = {731},
       number = {1},
          eid = {53},
        pages = {53},
          doi = {10.1088/0004-637X/731/1/53},
archivePrefix = {arXiv},
       eprint = {1102.1996},
 primaryClass = {astro-ph.EP},
       adsurl = {https://ui.adsabs.harvard.edu/abs/2011ApJ...731...53M}
}
